\documentclass[pdflatex,sn-nature,twocolumn]{sn-jnl}

\usepackage{graphicx}%
\usepackage{multirow}%
\usepackage{amsmath,amssymb,amsfonts}%
\usepackage{amsthm}%
\usepackage{mathrsfs}%
\usepackage[title]{appendix}%
\usepackage{xcolor}%
\usepackage{textcomp}%
\usepackage{manyfoot}%
\usepackage{booktabs}%
\usepackage{algorithm}%
\usepackage{algorithmicx}%
\usepackage{algpseudocode}%
\usepackage{listings}%
\usepackage{caption}
\usepackage{physics}
\usepackage{float}

\renewcommand{\sectionfont}{%
  \normalfont\fontsize{12bp}{14bp}\bfseries
  \selectfont\raggedright\boldmath
}

\renewcommand{\subsectionfont}{%
  \normalfont\fontsize{10.5bp}{12.5bp}\bfseries
  \selectfont\raggedright\boldmath
}

\theoremstyle{thmstyleone}%
\theoremstyle{thmstyletwo}%

\theoremstyle{thmstylethree}%

\unnumbered% uncomment this for unnumbered level heads

\makeatletter

\renewcommand{\section}{%
  \@startsection{section}{1}{\z@}%
  {-12pt \@plus -4pt \@minus -2pt}%
  {4pt}%
  {\sectionfont}%
}

\renewcommand{\subsection}{%
  \@startsection{subsection}{2}{\z@}%
  {-12pt \@plus -4pt \@minus -2pt}%
  {3pt}%
  {\subsectionfont}%
}

\makeatother

\setcitestyle{super,open={},close={}}

\definecolor{nphysblue}{RGB}{0,115,170}

\hypersetup{
    colorlinks=true,
    citecolor=nphysblue,
    linkcolor=nphysblue,
    urlcolor=nphysblue
}

\begin{document}

\title[Engineering Bosonic Steady States]{Engineering Bosonic Steady States through Fock-Space Currents}

\author*{Yan Liu}\email{\textcolor{nphysblue}{neonorth@foxmail.com}}
\author*{Jiahua Li}\email{\textcolor{nphysblue}{huajia\_li@163.com}}
% Remove the trailing semicolon in the displayed email list
\makeatletter
\gdef\corrauthemail{%
  \textcolor{nphysblue}{neonorth@foxmail.com};\ 
  \textcolor{nphysblue}{huajia\_li@163.com}%
}
\affil[]{%
  \orgdiv{School of Physics}, 
  \orgname{Huazhong University of Science and Technology, Wuhan 430074, People's Republic of China}}

\abstract{
Fock states are valuable resources for quantum technologies, but intrinsic loss makes their stabilization challenging.
Whereas existing stabilization schemes are typically formulated for specific microscopic processes, here we develop a unified Fock-space current-engineering framework for the autonomous stabilization of selectable Fock states.
We derive a stationary current-balance law that predicts a population background scaling inversely with Fock number and quantifies how a local current selectively enhances the target Fock-state population.
This law reveals how distinct microscopic implementations obey the same steady-state design rule, allowing different target states to be selected by moving a local transition without changing the overall stabilization architecture.
Full quantum simulations further support these predictions and demonstrate the autonomous buildup of target-state populations from the vacuum.
Our results provide a common physical picture of Fock-state stabilization and establish a transferable design principle for engineering selectable bosonic steady states. 
}

\maketitle

In an open quantum system, coupling to the environment can be engineered to select a state rather than merely destroy its coherence~\cite{Diehl2008a,Verstraete2009_DissipativeQE,Poyatos1996,Kraus2008,Kienzler2015}. 
Engineered dissipation has consequently become a tool for autonomous quantum-state stabilization~\cite{Shankar2013,LeghtasScience2015,Gertler2021AQEC}, while programmable implementations extend its scope from fixed targets to controllable sets of states~\cite{Harrington2022,Li2024}.
This development invites a broader design perspective: can selected transitions between quantum levels serve as building blocks for generating a family of steady states within a common dissipative architecture?

Bosonic Fock states provide a natural setting for this perspective.
They are nonclassical resources for quantum information processing and quantum-enhanced sensing~\cite{Hofheinz2008,Deng2024}, while their discrete ladder of excitation-number states provides a natural basis for spectrally selective control.
Fock states have been generated through coherent control~\cite{Hofheinz2008}, maintained by real-time quantum feedback~\cite{Sayrin2011}, and autonomously stabilized by engineered reservoirs~\cite{PhysRevLett.115.180501}.
More recently, cascaded selective photon addition has enabled stabilization following the loss of multiple photons~\cite{PhysRevLett.132.203602}.
Related theoretical studies have explored number-selective population-raising processes based on incoherently pumped Jaynes--Cummings (JC) interactions in Kerr photon and dispersively coupled magnon systems~\cite{LiLiu2026PhotonClimbing,Liu2026WignerNegative}, as well as relaxation-assisted anti-Jaynes--Cummings (AJC) cascades~\cite{LiuLi2026AJCCascades}.
Together, these studies raise the question of whether the steady-state stabilization of Fock states in open bosonic systems can be understood through a common underlying physical mechanism, despite the different microscopic processes involved.

Here we develop a Fock-space current-engineering framework in which selected transitions establish distinct pathways for population replenishment and local capture, while the balance between engineered and loss-induced population currents determines the steady-state population distribution.
This architecture combines a spectrally selected multi-quantum global connector between the vacuum and an upper Fock state, a downward cascade induced by intrinsic single-quantum loss, and a movable local connector acting on a selected transition between neighboring Fock states.
Tuning the local resonance enables the autonomous stabilization of different Fock states.

We derive an effective Fock-space population-rate equation and obtain closed-form stationary populations in terms of the global and local currents. 
The solution shows that the global current and intrinsic single-quantum loss establish a population background that decreases inversely with Fock number within the loss cascade, while the local current produces an additional contribution at the selected target state.
To test these predictions, we consider three global implementations: pumped multi-quantum JC interactions, relaxation-assisted multi-quantum AJC interactions, and coherent multi-quantum driving. 
Each is combined with a local connector implemented through pumped single-quantum JC or relaxation-assisted single-quantum AJC interactions. 
Across the Kerr- and dispersively resolved implementations considered, full microscopic Lindblad simulations reproduce the predicted inverse-number background and show that the target-state population enhancement follows the selected local transition.
The effective population-rate model also captures the main features of the autonomous buildup of the target-state population from the vacuum.
Together, these results demonstrate that the same Fock-space current-engineering architecture can be realized through distinct microscopic interactions, providing multiple physical routes to the stabilization of selectable Fock states.

\begin{figure*}[t]
\centering
\includegraphics[width=0.96\textwidth]{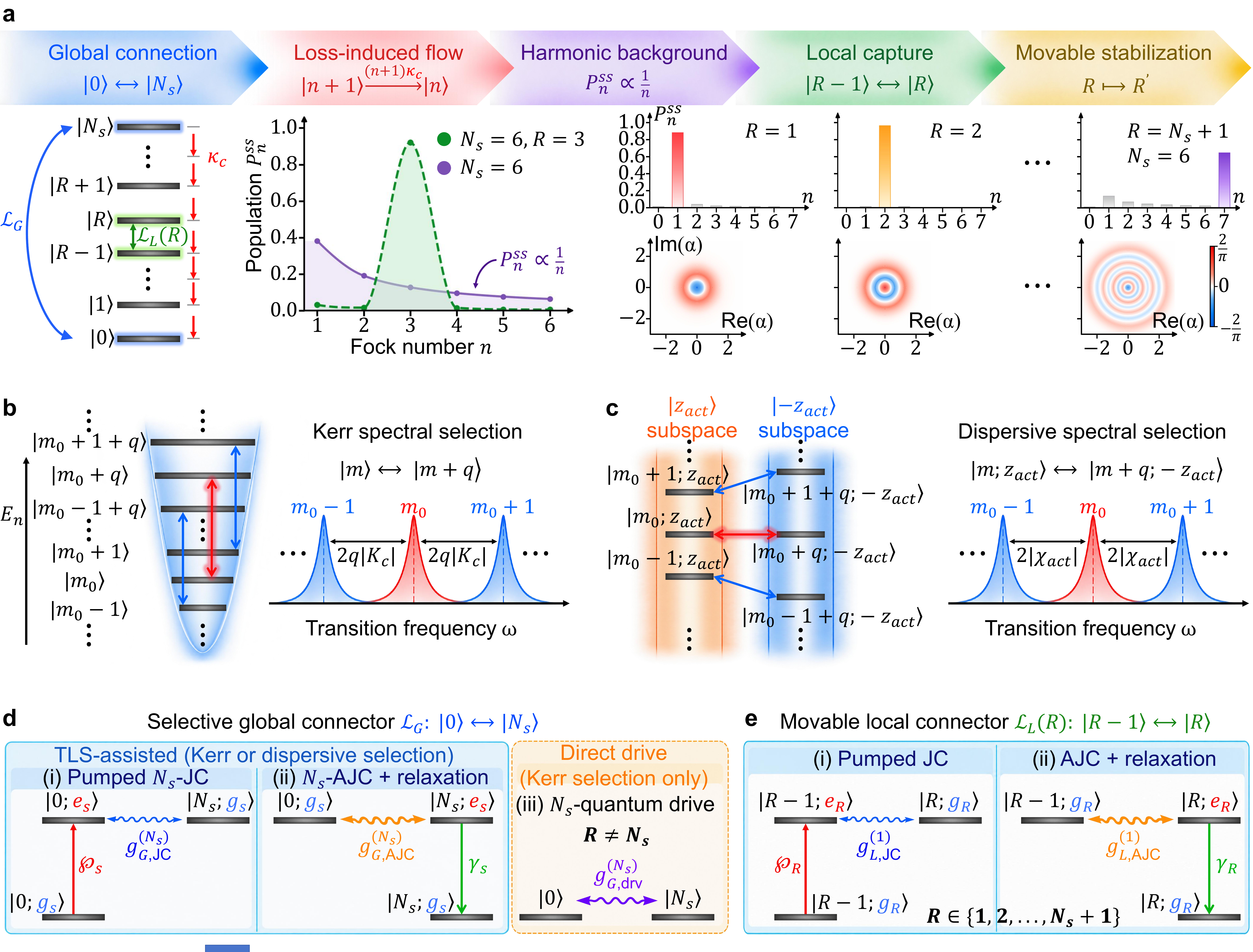}
\caption{\textbf{Architecture and microscopic implementations of autonomous Fock-state stabilization.}
\textbf{a,} Stabilization scheme combining a global connection, intrinsic loss and movable local capture.
For $N_s=6$, the population curves compare a coherent $N_s$-quantum global drive alone (purple) with the same drive supplemented by local AJC capture at $R=3$ (green).
The distributions and Wigner functions on the right illustrate the target positions $R=1$, $R=2$ and $R=N_s+1$.
\textbf{b,} Kerr-resolved spectral selection, with $K_c$ denoting the bosonic Kerr nonlinearity.
\textbf{c,} Dispersively resolved spectral selection, with $\chi_{\mathrm{act}}$ denoting the dispersive coupling of the active TLS.
Here $m_0$ denotes the initial Fock number and $q$ the number of exchanged bosonic quanta for the selected transition in \textbf{b,c}.
In \textbf{c}, $z_{\mathrm{act}}$ denotes an eigenvalue of $\hat{\sigma}_{\mathrm a}^{z}$ for the active TLS ($\mathrm a\in\{s,R\}$).
The labels $z_{\mathrm{act}}$ and $-z_{\mathrm{act}}$ correspond to the excited and ground states $\ket{e_{\mathrm a}}$ and $\ket{g_{\mathrm a}}$, in either order.
Accordingly, $\ket{m;z_{\mathrm{act}}}$ denotes the tensor-product state of $\ket{m}$ and the corresponding TLS eigenstate.
Red and blue arrows and spectral peaks in \textbf{b,c} indicate the selected and neighboring transitions, respectively.
\textbf{d,} Global connector implementations: pumped $N_s$-quantum JC, relaxation-assisted $N_s$-quantum AJC, and a coherent $N_s$-quantum drive (Kerr selection only).
\textbf{e,} Local connector implementations: pumped single-quantum JC and relaxation-assisted single-quantum AJC.
The numerical populations in \textbf{a} use $K_c/\kappa_c=10^4$ and $g_{G,\mathrm{drv}}^{(N_s)}/\kappa_c=5$; the local-capture curve additionally uses $g_{L,\mathrm{AJC}}^{(1)}/\kappa_c=100$ and $(\wp_R, \gamma_R)/\kappa_c=(0, 300)$.
The rotating-frame detunings in \textbf{a} are $\Delta_c/\kappa_c=-6\times10^4$ and, for local AJC capture, $\Delta_R/\kappa_c=(7-2R)\times10^4$ ($\Delta_R/\kappa_c=10^4$ at $R=3$), which make the selected transitions resonant.
}
\label{fig1}
\end{figure*}

\section{Fock-space stabilization architecture}

We propose an autonomous scheme for the steady-state preparation of a selectable Fock state $\ket{R}$ in a bosonic system.
The central strategy, schematically illustrated in Fig.~\ref{fig1}\textcolor{nphysblue}{a}, is to engineer population transport through Fock space by combining a selective global connection between $\ket{0}$ and $\ket{N_s}$, intrinsic loss-driven flow along the Fock ladder, and a movable local connection addressing $\ket{R-1}\leftrightarrow\ket{R}$.
Here, $N_s$ denotes the upper Fock number addressed by the global connector, while $R$ specifies the target Fock-state index selected by the local connector.
The global connection replenishes population at the upper end of the loss cascade, while the local connection selectively supplies population to the target state $\ket{R}$, compensating for its intrinsic loss and enabling steady-state stabilization.
In the $N_s=6$ example, adding local capture at $R=3$ reshapes the global-only steady-state distribution into a pronounced target-state population peak.

The microscopic dynamics of the stabilization scheme are governed by the master equation ($\hbar=1$)
\begin{equation}
    \frac{d\hat{\rho}}{dt}
    =
    \left[
        \mathcal{L}_{c}
        +
        \mathcal{L}_{G}
        +
        \mathcal{L}_{L}(R)
    \right]\hat{\rho},
    \label{eq:master}
\end{equation}
where $\hat{\rho}$ denotes the joint density operator of the bosonic mode and the auxiliary two-level systems (TLSs), when present.
Here, $\mathcal{L}_{c}$ accounts for the intrinsic bosonic loss and the Hamiltonian terms responsible for spectral selectivity, while $\mathcal{L}_{G}$ and $\mathcal{L}_{L}(R)$ describe the engineered global and movable local connectors, respectively.
Here $\kappa_c$ denotes the intrinsic single-quantum loss rate of the bosonic mode.
Kerr anharmonicity and dispersive couplings provide spectral selectivity (Fig.~\ref{fig1}\textcolor{nphysblue}{b,c}), while three global and two local microscopic implementations are illustrated in Fig.~\ref{fig1}\textcolor{nphysblue}{d,e}. 
Their detailed models are provided in Methods.

\begin{figure*}[tbp]
\centering
\includegraphics[width=0.96\textwidth]{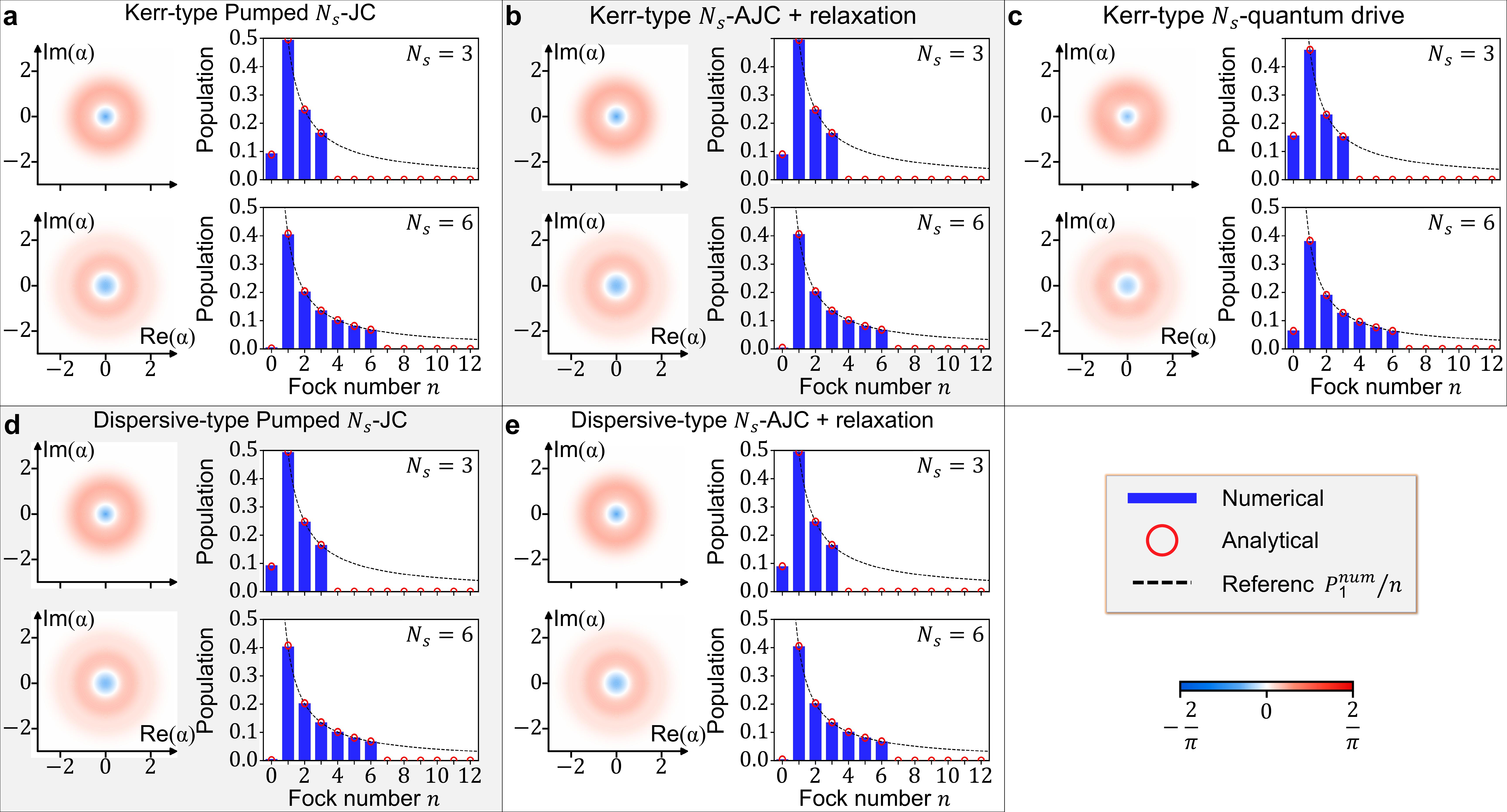}
\caption{\textbf{Common harmonic Fock-state populations generated by selective global connections.}
\textbf{a--c,} Kerr-resolved global implementations: (a) pumped $N_s$-quantum JC, (b) relaxation-assisted $N_s$-quantum AJC, and (c) a coherent $N_s$-quantum drive.
\textbf{d, e,} Dispersively resolved counterparts of (a, b), respectively.
The upper (lower) row in each panel corresponds to $N_s=3$ ($N_s=6$).
The Wigner functions (left) are computed numerically from the reduced bosonic steady states obtained by solving the full microscopic Lindblad master equations; the corresponding Fock-state populations are shown on the right.
Blue bars show full microscopic Lindblad results, red open circles the analytical effective-current predictions, and black dashed curves the reference $P_1^{\mathrm{num}}/n$, where $P_1^{\mathrm{num}}$ is the numerically obtained population at $n=1$. 
The reference curves are extended beyond $N_s$ only as a guide.
The ideal analytical populations vanish for $n>N_s$.
All Wigner functions use the same linear color scale, $[-2/\pi,2/\pi]$.
No local connector is applied.
The global coupling strengths satisfy $g_{G,\mathrm{JC}}^{(N_s)}/\kappa_c=g_{G,\mathrm{AJC}}^{(N_s)}/\kappa_c=g_{G,\mathrm{drv}}^{(N_s)}/\kappa_c=5$.
The pumped-JC cases (a, d) use $\wp_s/\kappa_c=100$ and $\gamma_s/\kappa_c=1$; the AJC cases (b, e) use $(\wp_s,\gamma_s)/\kappa_c=(0,100)$.
For (a--c), $K_c/\kappa_c=10^4$ and $\Delta_c/\kappa_c=-10^4N_s$, with $\Delta_s=0$ in (a, b).
For (d, e), $K_c=\Delta_c=0$, $\chi_s/\kappa_c=3\times10^4$ and $\Delta_s/\kappa_c=-3\times10^4N_s$.
These choices place the selected global transitions on resonance.
}
\label{fig2}
\end{figure*}

\section{A universal stationary population law}
\label{sec:fock_space_currents}

To understand how the engineered connections enable steady-state Fock-state stabilization, we examine the resulting population dynamics in Fock space.
In the spectrally selective regime, where the engineered processes admit an effective population-rate description, the global and local connectors generate net population currents $J_{\mathrm G}(t)$ and $J_{\mathrm L}(R,t)$ across their respective selected transitions.
Together with intrinsic loss, these currents govern the Fock-state populations $P_n(t)=\bra{n}\hat{\rho}_c(t)\ket{n}$, where $\hat{\rho}_c(t)=\operatorname{Tr}_{\mathrm{TLS}}\hat{\rho}(t)$ is the reduced bosonic density operator obtained by tracing out the auxiliary TLSs.
The corresponding population equation is
\begin{align}
\frac{dP_n}{dt}
=&
(n+1)\kappa_c P_{n+1}
-n\kappa_c P_n
\nonumber\\
&+
(\delta_{n,N_s}-\delta_{n,0})J_{\mathrm G}
\nonumber\\
&+
(\delta_{n,R}-\delta_{n,R-1})J_{\mathrm L}(R),
\label{eq:population_dynamics}
\end{align}
where $\delta_{i,j}$ is the Kronecker delta.
Here, $J_{\mathrm G}$ denotes the net population current from $\ket{0}$ to $\ket{N_s}$, while $J_{\mathrm L}(R)$ denotes the net current from $\ket{R-1}$ to $\ket{R}$.
The effective population currents are defined as
\begin{align}
    J_{\mathrm G}(t)
    &=
    \mathcal R_{\mathrm G}^{\uparrow}P_0(t)
    -
    \mathcal R_{\mathrm G}^{\downarrow}P_{N_s}(t),
    \label{eq:global_current}
    \\
    J_{\mathrm L}(R,t)
    &=
    \mathcal R_{\mathrm L}^{\uparrow}(R)P_{R-1}(t)
    -
    \mathcal R_{\mathrm L}^{\downarrow}(R)P_R(t),
    \label{eq:local_current}
\end{align}
where $\mathcal R_{\mathrm G}^{\uparrow}$ and $\mathcal R_{\mathrm G}^{\downarrow}$ denote the effective upward and downward population-transfer rates of the global connector, while $\mathcal R_{\mathrm L}^{\uparrow}(R)$ and $\mathcal R_{\mathrm L}^{\downarrow}(R)$ are the corresponding rates of the local connector.
Positive $J_{\mathrm G}$ and $J_{\mathrm L}(R)$ represent net population transfer toward $\ket{N_s}$ and $\ket{R}$, respectively.
The effective rates depend on the chosen microscopic implementations; their explicit expressions and mappings onto the global and local connectors are given in Supplementary Sects.~S2 and S3 and summarized in Supplementary Table~2.

At steady state, denoted by the superscript $\mathrm{ss}$, the balance between the engineered population currents and intrinsic loss determines the Fock-state population distribution.
Setting $\dot{P}_m=0$ and summing Eq.~\eqref{eq:population_dynamics} over the Fock-number index $m$ from $m=n$ to infinity yields
\begin{equation}
    n\kappa_c P_n^{\mathrm{ss}}
    =
    \Theta_{\mathrm d}(N_s-n)J_{\mathrm G}^{\mathrm{ss}}
    +
    \delta_{n,R}J_{\mathrm L}^{\mathrm{ss}}(R),
    \qquad n\geq 1,
    \label{eq:steady_state_current_balance}
\end{equation}
where $\Theta_{\mathrm d}(j)$ is the discrete step function, equal to unity for $j\geq 0$ and zero otherwise.
Here, $J_{\mathrm G}^{\mathrm{ss}}$ and $J_{\mathrm L}^{\mathrm{ss}}(R)$ denote the stationary global and local population currents, respectively.

Equation~\eqref{eq:steady_state_current_balance} reveals the population distribution sustained by the global connection.
For non-target Fock states within the intrinsic-loss cascade, the steady-state populations satisfy
\begin{equation}
    P_n^{\mathrm{ss}}
    =
    \frac{J_{\mathrm G}^{\mathrm{ss}}}{n\kappa_c},
    \qquad
    1\leq n\leq N_s,\quad n\neq R.
    \label{eq:harmonic_population}
\end{equation}
The number-dependent intrinsic loss therefore produces a harmonic population profile $P_n^{\mathrm{ss}}\propto 1/n$, sustained by the stationary global current.
This profile provides the population background upon which the movable local connector acts.

At the selected target state $\ket{R}$, the local current modifies the stationary population balance, giving
\begin{equation}
    P_R^{\mathrm{ss}}
    =
    \begin{cases}
        \displaystyle
        \frac{
            J_{\mathrm G}^{\mathrm{ss}}
            +
            J_{\mathrm L}^{\mathrm{ss}}(R)
        }{
            R\kappa_c
        },
        & 1\leq R\leq N_s,\\[10pt]
        \displaystyle
        \frac{
            J_{\mathrm L}^{\mathrm{ss}}(R)
        }{
            R\kappa_c
        },
        & R=N_s+1.
    \end{cases}
    \label{eq:target_population}
\end{equation}
For $R\leq N_s$, a positive local current enhances the target-state population relative to the harmonic profile supported by the same stationary global current.
For $R=N_s+1$, the target lies above the global connection and is populated directly through the local current.
For $R\leq N_s$, reversing the local current, $J_{\mathrm L}^{\mathrm{ss}}(R)<0$, instead depletes the selected Fock-state population relative to the harmonic background.
Such Fock-selective depletion is demonstrated using relaxation-dominated JC and pumping-dominated AJC local connectors in Supplementary Sect.~S9 and Supplementary Fig.~4.
The same current-balance relation thus accommodates both enhancement and depletion, while the main-text realizations below focus on positive-current stabilization.
The matrix formulation of the effective population dynamics and the full closed-form stationary solutions, including special cases, are provided in Supplementary Sects.~S4 and S5.
The achievable target-state population, however, depends on the microscopic interactions, auxiliary-TLS pumping and relaxation, and spectral selectivity.
We therefore investigate the preparation performance using the full microscopic Lindblad dynamics.

\begin{figure*}[t]
\centering
\includegraphics[width=0.96\textwidth]{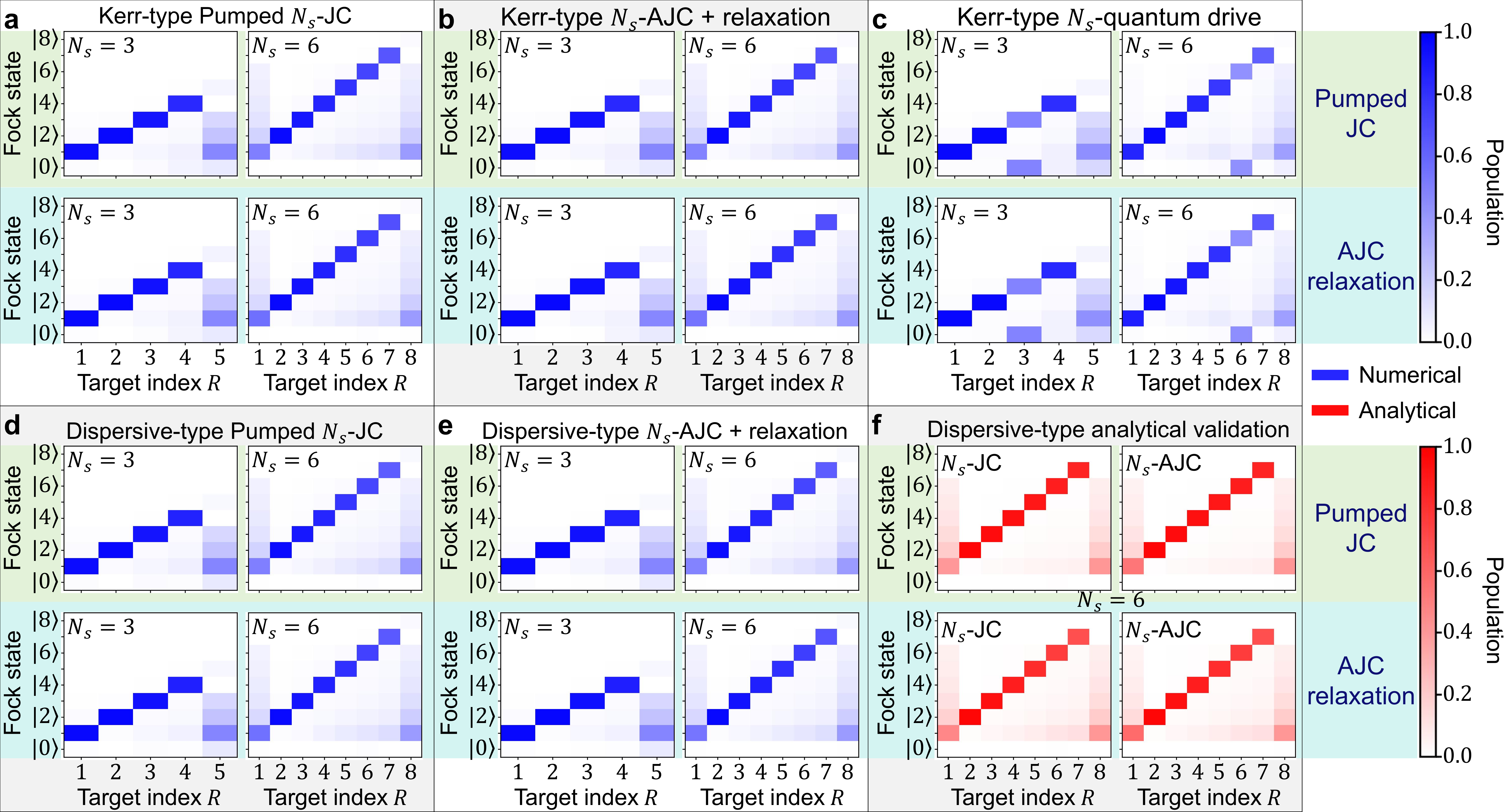}
\caption{\textbf{Movable steady-state Fock-state stabilization across Kerr- and dispersively resolved implementations.}
\textbf{a--c,} Full microscopic Lindblad steady-state populations for Kerr-resolved global connectors: (a) pumped $N_s$-quantum JC, (b) relaxation-assisted $N_s$-quantum AJC, and (c) a coherent $N_s$-quantum drive.
\textbf{d,e,} Corresponding numerical results for dispersively resolved pumped JC and relaxation-assisted AJC global connectors.
In \textbf{a--e}, the upper (lower) row uses a pumped single-quantum JC (relaxation-assisted AJC) local connector, and the left (right) column shows $N_s=3$ ($N_s=6$).
\textbf{f,} Analytical population maps from the stationary effective-current solution for the four dispersive global--local combinations at $N_s=6$, arranged with global JC (AJC) in the left (right) columns and local JC (AJC) in the upper (lower) rows.
All maps plot $P_n^{\mathrm{ss}}$ versus the target index $R$ and Fock number $n$ on a common linear scale from $0$ to $1$ (blue, microscopic numerics; red, analytical predictions).
We use $g_{G,\mathrm{JC}}^{(N_s)}=g_{G,\mathrm{AJC}}^{(N_s)}=g_{G,\mathrm{drv}}^{(N_s)}=5\kappa_c$ and $g_{L,\mathrm{JC}}^{(1)}=g_{L,\mathrm{AJC}}^{(1)}=100\kappa_c$.
For pumped global JC, $(\wp_s, \gamma_s)/\kappa_c=(100, 1)$; for global AJC, $(\wp_s, \gamma_s)/\kappa_c=(0, 100)$.
For pumped local JC, $(\wp_R, \gamma_R)/\kappa_c=(300, 1)$; for local AJC, $(\wp_R, \gamma_R)/\kappa_c=(0, 300)$.
For \textbf{a--c}, $K_c/\kappa_c=10^4$, $\Delta_c/\kappa_c=-10^4N_s$, and $\Delta_s=0$ when a global TLS is present; the local detuning is $\Delta_R/\kappa_c=\pm10^4(2R-1-N_s)$ for JC (+) and AJC (-) local capture.
For \textbf{d--f}, $K_c=\Delta_c=0$, $\chi_s/\kappa_c=3\times10^4$, and $\chi_R/\kappa_c=1.2\times10^4$.
The dispersive detunings $(\Delta_s/\kappa_c, \Delta_R/\kappa_c)$ are $(-1.8\times10^4N_s,\;3\times10^4-1.2\times10^4(2R-1))$ for JC--JC and AJC--AJC global--local combinations, and $(-4.2\times10^4N_s,\;-3\times10^4-1.2\times10^4(2R-1))$ for the mixed combinations.
These detunings select the reference spectator-TLS subspaces associated with the pumped-JC (excited) or relaxation-assisted AJC (ground) realizations.
}
\label{fig3}
\end{figure*}

\section{Microscopic realizations of current control}

We test the predicted harmonic population background in the global-only case by numerically solving the full microscopic master equation, Eq.~\eqref{eq:master}, with $\mathcal{L}_{L}(R)=0$.
Figure~\ref{fig2} shows that five Kerr- and dispersively resolved microscopic implementations reproduce the harmonic $1/n$ population profile predicted by Eq.~\eqref{eq:harmonic_population} for $1\leq n\leq N_s$.
The overall population scale nevertheless differs between implementations through the stationary global current $J_{\mathrm G}^{\mathrm{ss}}$.

Building on this global-only baseline, we include the movable local connector and vary its target index $R$.
Figure~\ref{fig3} compares the steady-state Fock-state populations for $N_s=3$ and $N_s=6$ across five global implementations, each combined with either pumped JC or relaxation-assisted AJC local capture.
For both Kerr- and dispersively resolved implementations, the population enhancement follows the selected target position over $1\leq R\leq N_s+1$, consistent with the stationary target-population relation in Eq.~\eqref{eq:target_population}.
The extra setting $R=N_s+2$ serves as an out-of-range control: the ideal effective-current solution then reduces to the global-only population distribution rather than producing a target-selective peak.

The target-state population nevertheless depends on the microscopic implementation.
Under the parameters considered, the coherent global drive yields a lower target population near $R=N_s$ than the TLS-assisted global connectors, consistent with its reversible population transfer.
The symmetric effective population-transfer structure of the coherent global drive can also be realized with balanced pumping and relaxation of the global auxiliary TLS (Supplementary Sect.~S8 and Supplementary Fig.~3).
For $N_s=6$, the four analytical population maps in Fig.~\ref{fig3}\textcolor{nphysblue}{f} reproduce the main target-dependent features of the corresponding
dispersively resolved microscopic results in Fig.~\ref{fig3}\textcolor{nphysblue}{d,e}.
This comparison supports the effective-current description while retaining the implementation-dependent differences in the attainable target populations.

\section{Autonomous population buildup}

\begin{figure*}[t]
\centering
\includegraphics[width=0.96\textwidth]{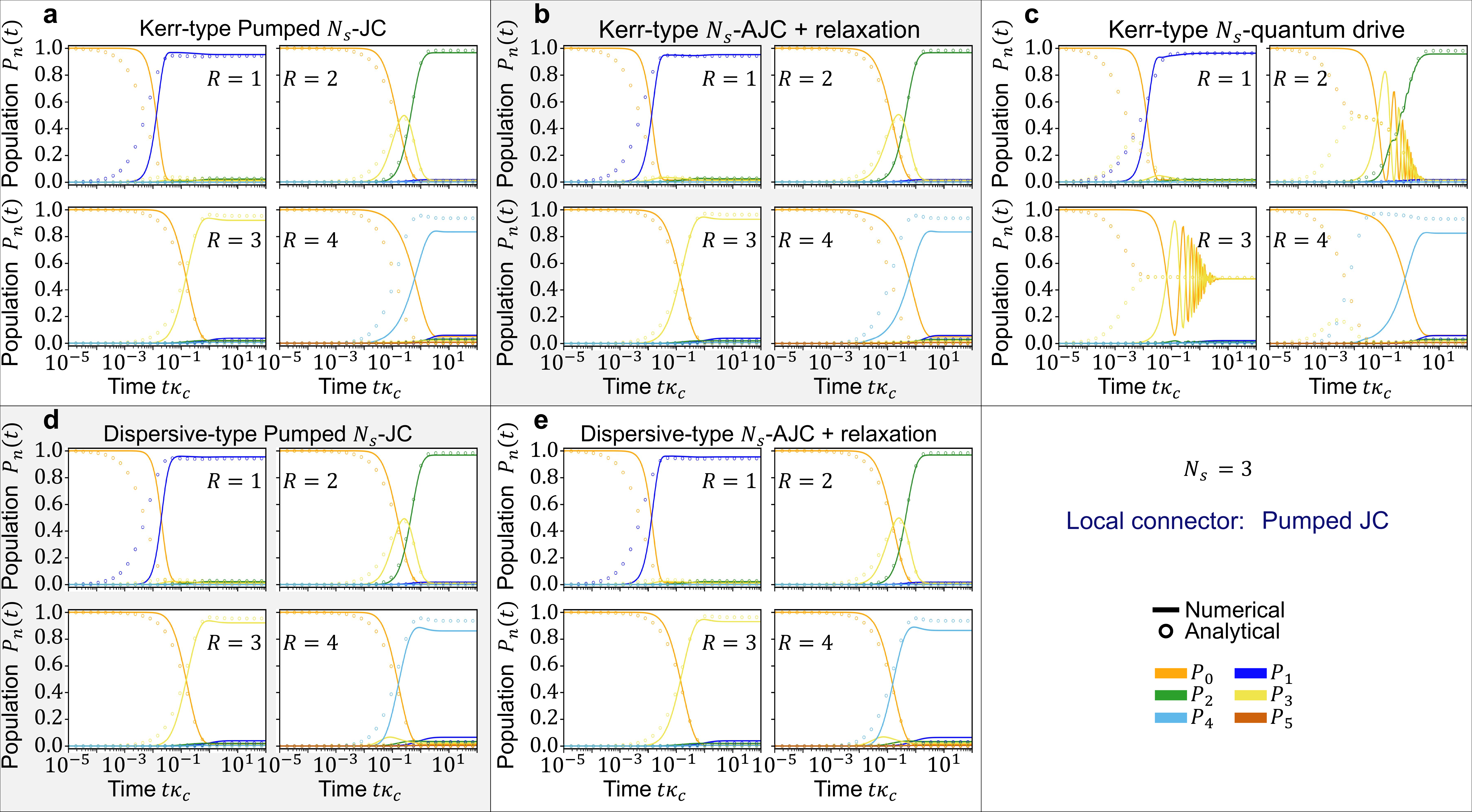}
\caption{\textbf{Autonomous Fock-state population buildup with pumped-JC local capture.}
\textbf{a--c,} Kerr-resolved global connectors: (a) pumped $N_s$-quantum JC, (b) relaxation-assisted $N_s$-quantum AJC, and (c) a coherent $N_s$-quantum drive.
\textbf{d,e,} Dispersively resolved pumped-JC (d) and relaxation-assisted AJC (e) global connectors.
All panels use $N_s=3$, a pumped single-quantum JC local connector, and target indices $R=1,2,3,4$.
The bosonic populations $P_n(t)$ ($n=0,\ldots,5$) start from the vacuum, with all auxiliary TLSs initially in their ground states.
Colored solid curves are numerical results from the full microscopic Lindblad master equations; corresponding open circles show predictions obtained from the analytically derived effective population-rate equation, Eq.~\eqref{eq:population_dynamics}.
The global couplings satisfy $g_{G,\mathrm{JC}}^{(N_s)}=g_{G,\mathrm{AJC}}^{(N_s)}=g_{G,\mathrm{drv}}^{(N_s)}=5\kappa_c$, and $g_{L,\mathrm{JC}}^{(1)}=100\kappa_c$.
For pumped global JC, $(\wp_s, \gamma_s)/\kappa_c=(100, 1)$; for global AJC, $(\wp_s, \gamma_s)/\kappa_c=(0, 100)$.
The pumped local JC connector uses $(\wp_R, \gamma_R)/\kappa_c=(300,1)$.
For \textbf{a--c}, $K_c/\kappa_c=10^4$, $\Delta_c/\kappa_c=-3\times10^4$, and $\Delta_R/\kappa_c=10^4(2R-4)$, with $\Delta_s=0$ where a global TLS is present.
For \textbf{d,e}, $K_c=\Delta_c=0$, $\chi_s/\kappa_c=3\times10^4$, and $\chi_R/\kappa_c=1.2\times10^4$.
The dispersive detunings $(\Delta_s/\kappa_c,\Delta_R/\kappa_c)$ are
$(-5.4\times10^4,\;3\times10^4-1.2\times10^4(2R-1))$ in \textbf{d} and
$(-1.26\times10^5,\;-3\times10^4-1.2\times10^4(2R-1))$ in \textbf{e}.
These detunings select the transitions in the corresponding reference spectator-TLS subspaces.
}
\label{fig4}
\end{figure*}

\begin{figure*}[t]
\centering
\includegraphics[width=0.96\textwidth]{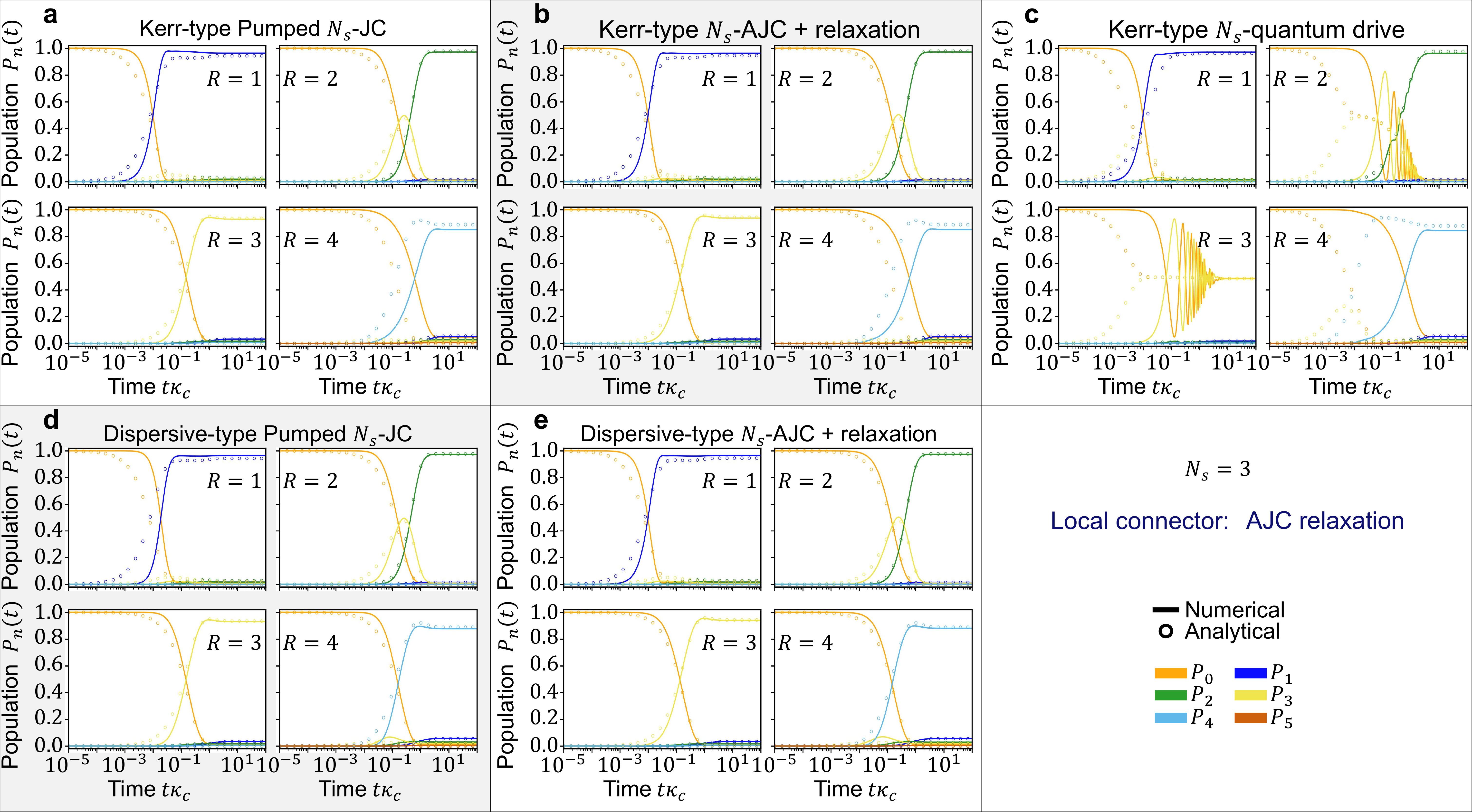}
\caption{\textbf{Autonomous Fock-state population buildup with relaxation-assisted AJC local capture.}
\textbf{a--c,} Kerr-resolved global connectors: (a) pumped $N_s$-quantum JC, (b) relaxation-assisted $N_s$-quantum AJC, and (c) a coherent $N_s$-quantum drive.
\textbf{d,e,} Dispersively resolved pumped-JC (d) and relaxation-assisted AJC (e) global connectors.
All panels use $N_s=3$, a relaxation-assisted single-quantum AJC local connector, and target indices $R=1,2,3,4$.
The bosonic populations $P_n(t)$ ($n=0,\ldots,5$) start from the vacuum, with all auxiliary TLSs initially in their ground states.
Colored solid curves show numerical solutions of the full microscopic Lindblad master equations; open circles show predictions obtained from the analytically derived effective population-rate equation, Eq.~\eqref{eq:population_dynamics}.
The global connector remains present at $R=1$.
The global couplings satisfy $g_{G,\mathrm{JC}}^{(N_s)}=g_{G,\mathrm{AJC}}^{(N_s)}=g_{G,\mathrm{drv}}^{(N_s)}=5\kappa_c$, while $g_{L,\mathrm{AJC}}^{(1)}=100\kappa_c$.
For pumped global JC, $(\wp_s,\gamma_s)/\kappa_c=(100,1)$; for relaxation-assisted global AJC, $(\wp_s,\gamma_s)/\kappa_c=(0,100)$.
The local AJC connector uses $\gamma_R/\kappa_c=300$, without auxiliary-TLS pumping.
For \textbf{a--c}, $K_c/\kappa_c=10^4$, $\Delta_c/\kappa_c=-3\times10^4$, and $\Delta_R/\kappa_c=-10^4(2R-4)$, with $\Delta_s=0$ in \textbf{a,b}.
For \textbf{d,e}, $K_c=\Delta_c=0$, $\chi_s/\kappa_c=3\times10^4$, and $\chi_R/\kappa_c=1.2\times10^4$.
The dispersive detunings $(\Delta_s/\kappa_c,\Delta_R/\kappa_c)$ are $(-1.26\times10^5,\;-3\times10^4-1.2\times10^4(2R-1))$ in \textbf{d} and $(-5.4\times10^4,\;3\times10^4-1.2\times10^4(2R-1))$ in \textbf{e}.
These values select the transitions in the corresponding reference spectator-TLS subspaces.
}
\label{fig5}
\end{figure*}

Beyond stationary populations, we examine the transient dynamics of autonomous Fock-state preparation.
Figures~\ref{fig4} and \ref{fig5} compare five microscopic global implementations for $N_s=3$ and $R=1,2,3,4$, combined with pumped-JC and relaxation-assisted AJC local connectors, respectively.
The bosonic mode is initially in the vacuum state and all auxiliary TLSs, when present, are initialized in their ground states.

For pumped-JC local capture, Fig.~\ref{fig4} shows population redistribution from the vacuum and the buildup of the selected target-state population.
Figure~\ref{fig5} demonstrates the corresponding preparation dynamics when local capture is instead implemented through an AJC interaction followed by auxiliary-TLS relaxation.

The time-resolved results also distinguish effective population transport from microscopic transients.
The effective population-rate equation captures the principal population redistribution and long-time populations, whereas the full Lindblad trajectories may show differences in their transient onset and damped coherent oscillations, particularly for the Kerr-resolved coherent global drive at $R=3$ in Fig.~\ref{fig5}\textcolor{nphysblue}{c}.
These differences reflect coherent and auxiliary-TLS dynamics that are not explicitly resolved in the reduced population-rate description.
Together, these results complement the movable steady-state stabilization shown in Fig.~\ref{fig3} by demonstrating the autonomous buildup of selected Fock-state populations from the vacuum.

Additional parameter scans of the stationary target-state population for representative Kerr- and dispersively resolved configurations ($N_s=2$, $R=3$) probe the effects of spectral selectivity, coupling strengths, auxiliary-TLS dissipation and detuning (Supplementary Sects.~S6 and S7 and Supplementary Figs.~1 and 2).
They reveal finite operating regions with high target-state populations and tolerance to nonzero detunings, while identifying conditions under which the stabilization performance deteriorates.

\section{Experimental prospects}

Superconducting circuit quantum electrodynamics (QED) offers a promising platform for experimentally exploring the proposed Fock-space current-engineering scheme~\cite{RevModPhys.93.025005,Kjaergaard2020}.
Ref.~\cite{Kirchmair2013} demonstrated a Kerr nonlinearity of $325\,\mathrm{kHz}$ [$|K_c|/(2\pi)\simeq0.163\,\mathrm{MHz}$], a photon-number-dependent qubit dispersive shift of $9.4\,\mathrm{MHz}$ [$|\chi_{\mathrm{act}}|/(2\pi)\simeq4.7\,\mathrm{MHz}$], and a single-photon cavity decay rate of $\kappa_c/(2\pi)\simeq10\,\mathrm{kHz}$ in the same device.
Ref.~\cite{Lu2023SNAIL} reported a Kerr-induced resonator-frequency shift of $(-236\pm65)\,\mathrm{kHz}$ per photon [$K_c/(2\pi)\simeq(-118\pm33)\,\mathrm{kHz}$] and a photon-number-resolved qubit spectral splitting of $(3.143\pm0.021)\,\mathrm{MHz}$ [$|\chi_{\mathrm{act}}|/(2\pi)\simeq(1.572\pm0.011)\,\mathrm{MHz}$].
Flux-tunable Kerr nonlinearities ranging approximately from $-5$ to $6\,\mathrm{MHz}$ [$|K_c|/(2\pi)\leq3\,\mathrm{MHz}$] have also been demonstrated~\cite{He2023}.
Separately, superconducting cavities have reached single-photon lifetimes of $25.6\,\mathrm{ms}$ [$\kappa_c\simeq(25.6\,\mathrm{ms})^{-1}\simeq39.1\,\mathrm{s}^{-1}$, or $\kappa_c/(2\pi)\simeq6.2\,\mathrm{Hz}$]~\cite{PRXQuantum.4.030336}.
These results demonstrate the relevant spectral and coherence ingredients, but the largest nonlinearities and longest cavity lifetimes were achieved in different devices. 

Turning to coherent interactions, single-mode three-photon parametric down-conversion has been demonstrated in a flux-pumped superconducting cavity~\cite{PhysRevX.10.011011}, providing an experimental precedent for the three-quantum coherent-drive interaction ($N_s=3$).
Two-photon coherent drives with amplitudes of $2\pi\times15.5\,\mathrm{MHz}$ [$g_{G,\mathrm{drv}}^{(2)}/(2\pi)\simeq15.5\,\mathrm{MHz}$] have been demonstrated in superconducting Kerr resonators~\cite{GrimmNature2020}.
A separate experiment employed a two-photon drive with a Hamiltonian coefficient of $2\pi\times\frac{1}{2}\times3.13\,\mathrm{MHz}$ [$g_{G,\mathrm{drv}}^{(2)}/(2\pi)\simeq1.565\,\mathrm{MHz}$]~\cite{Iyama2024KPO}.
Single-quantum JC coupling between a superconducting qubit and a resonator has been experimentally demonstrated with a vacuum Rabi splitting of $11.6\,\mathrm{MHz}$ [$g_{L,\mathrm{JC}}^{(1)}/(2\pi)\simeq5.8\,\mathrm{MHz}$]~\cite{Wallraff2004}, and stronger coupling of $2\pi\times154\,\mathrm{MHz}$ [$g_{L,\mathrm{JC}}^{(1)}/(2\pi)\simeq154\,\mathrm{MHz}$] has been reported~\cite{Fink2008Nature}.
A driven single-quantum AJC-type transition between a transmon and a resonator has been demonstrated at a Rabi frequency of $0.49\,\mathrm{MHz}$ [$g_{L,\mathrm{AJC}}^{(1)}/(2\pi)\simeq0.245\,\mathrm{MHz}$]~\cite{Li2024Sidebands}, while a theoretical scheme for experimentally realizing effective $n$-photon JC and AJC interactions through parametric modulation of a cyclic qutrit in a circuit-QED setup has been proposed~\cite{Dodonov2019MultiPhotonJC}.
In a related experiment, a detuned flux qubit mediated coherent conversion between a two-photon state in one resonator mode and a single-photon state in another, with an effective coupling strength of $2\pi\times59\,\mathrm{MHz}$ [$g_{G,\mathrm{JC}}^{(2)}/(2\pi)\simeq41.7\,\mathrm{MHz}$ in a two-level projection of the latter mode]~\cite{Wang2025TwoPhotonConversion}.
A two-photon qubit--resonator interaction mediated by a superconducting quantum interference device has also been theoretically proposed, with a predicted coupling strength of up to $2\pi\times18\,\mathrm{MHz}$ [$g_{G,\mathrm{JC}}^{(2)}/(2\pi)\simeq18\,\mathrm{MHz}$ under a two-level approximation]~\cite{Stolyarov2025TwoPhotonCoupling}.
A two-tone-driven circuit-QED scheme has also been theoretically proposed to generate effective two-photon JC and AJC terms, with an estimated underlying coupling of $2\pi\times(25\text{--}50)\,\mathrm{MHz}$ [$g_{G,\mathrm{JC}}^{(2)}/(2\pi)\simeq g_{G,\mathrm{AJC}}^{(2)}/(2\pi)\simeq(12.5\text{--}25)\,\mathrm{MHz}$ for both components of the driven effective Hamiltonian]~\cite{Ayyash2024DrivenMultiphoton}.

Parametrically engineered reservoirs have enabled controlled incoherent pumping and relaxation of superconducting qubits~\cite{Cao2025TransmonBath,Guo2026LocalReservoirs}. 
A bath engineered with a lossy superconducting nonlinear asymmetric inductive element yielded fitted qubit excitation and relaxation rates of $2\pi\times54\,\mathrm{kHz}$ each [$\wp_{\mathrm a}/(2\pi)\simeq\gamma_{\mathrm a}/(2\pi)\simeq54\,\mathrm{kHz}$]~\cite{Cao2025TransmonBath}.
More recently, parametrically driven readout resonators were used to realize energy-selective qubit pumping and loss, with an estimated on-resonance pumping rate of $2\pi\times1.5\,\mathrm{MHz}$ [$\wp_{\mathrm a}/(2\pi)\simeq1.5\,\mathrm{MHz}$]~\cite{Guo2026LocalReservoirs}.
For waveguide-coupled transmons, radiative relaxation rates of $2\pi\times4.4\,\mathrm{MHz}$ [$\gamma_{\mathrm a}/(2\pi)\simeq4.4\,\mathrm{MHz}$] have been extracted from measured transmission spectra~\cite{Redchenko2023Waveguide}.
Low background qubit relaxation has also been demonstrated, with transmon energy-relaxation times reaching $1.68\,\mathrm{ms}$ [$\gamma_{\mathrm a}/(2\pi)\simeq95\,\mathrm{Hz}$]~\cite{Bland2025Transmon}.

Taking an illustrative intrinsic cavity loss rate of $\kappa_c/(2\pi)=0.1\,\mathrm{kHz}$, corresponding to a single-photon lifetime of $1/\kappa_c\simeq1.6\,\mathrm{ms}$, the numerical parameters translate to a Kerr nonlinearity of $|K_c|/(2\pi)=1\,\mathrm{MHz}$ or dispersive couplings of $(|\chi_s|,|\chi_R|)/(2\pi)=(3,1.2)\,\mathrm{MHz}$.
The global $N_s$-quantum JC, AJC and coherent-drive strengths are each $2\pi\times0.5\,\mathrm{kHz}$, and the local single-quantum JC and AJC coupling strengths are $2\pi\times10\,\mathrm{kHz}$.
The corresponding global and local auxiliary-qubit incoherent pumping or engineered relaxation rates are $2\pi\times10\,\mathrm{kHz}$ and $2\pi\times30\,\mathrm{kHz}$, respectively, while the weak auxiliary-qubit relaxation rates used in the pumped-JC implementations are $2\pi\times0.1\,\mathrm{kHz}$.
These comparisons suggest that several required parameter scales have precedents in existing experiments; however, simultaneously achieving strong spectral selectivity, low cavity loss and higher-order $N_s$-quantum interactions in a single device remains an open challenge.

\section{Discussion and outlook}

We have developed a Fock-space current-engineering framework for the autonomous stabilization of selectable Fock states. 
The stationary current-balance solution shows that local population transfer can selectively stabilize a target Fock state by harnessing the population flow sustained by the global connector and intrinsic loss.
Despite their different microscopic mechanisms, the implementations considered are described by a common effective population-rate equation and exhibit the same analytically predicted population structure, demonstrating that the same stabilization architecture can be realized through distinct physical processes.
Reversing the local current also enables selective Fock-state population depletion, illustrating a broader form of steady-state control within the same framework.
More broadly, this framework provides a starting point for extending autonomous stabilization to large Fock states through additional selective global connections, potentially implemented using further auxiliary TLSs or multiple coherent multi-quantum drives, and for investigating the effects of multiphoton loss on such stabilization.

\bibliography{WF-sub}

\clearpage
\section*{Methods}

\subsection{Microscopic Lindblad model and spectral selection}

Here, $\hat{a}_c$ ($\hat{a}_c^\dagger$) is the bosonic annihilation (creation) operator, and $\hat{n}_c=\hat{a}_c^\dagger\hat{a}_c$ is the corresponding number operator.
The Liouvillian $\mathcal{L}_{c}$ is given by
\begin{equation}
    \mathcal{L}_{c}\hat{\rho}
    =
    -i[\hat{H}_{c},\hat{\rho}]
    +
    \kappa_{c}
    \mathcal{D}[\hat{a}_{c}]\hat{\rho},
    \label{eq:background_liouvillian}
\end{equation}
where $\kappa_c$ is the intrinsic single-quantum decay rate of the bosonic mode, and $\mathcal{D}[\hat{O}]\hat{\rho}=\hat{O}\hat{\rho}\hat{O}^{\dagger}-\frac{1}{2}\{\hat{O}^{\dagger}\hat{O},\hat{\rho}\}$ is the Lindblad dissipator. 
We use $\kappa_c$ as the common reference frequency scale when reporting numerical parameters for the different implementations.
For the two spectral-selectivity mechanisms illustrated in Fig.~\ref{fig1}\textcolor{nphysblue}{b,c}, the Hamiltonian $\hat{H}_{c}$ takes the respective forms
\begin{align}
    \hat{H}_{c}^{(K)}
    &=
    \Delta_{c}\hat{n}_{c}
    +
    K_{c}\hat{n}_{c}^{2},
    \label{eq:kerr_hamiltonian}\\
    \hat{H}_{c}^{(D)}
    &=
    \Delta_{c}\hat{n}_{c}
    +
    \chi_{s}\hat{n}_{c}\hat{\sigma}_{s}^{z}
    +
    \chi_{R}\hat{n}_{c}\hat{\sigma}_{R}^{z}.
    \label{eq:dispersive_hamiltonian}
\end{align}
The detuning $\Delta_c=\omega_c-\omega_r$ is defined relative to the rotating-frame reference frequency $\omega_r$, where $\omega_c$ is the bare frequency of the bosonic mode, and $K_c$ denotes the Kerr nonlinearity.
The parameters $\chi_s$ and $\chi_R$ characterize the dispersive couplings between the bosonic mode and the auxiliary TLSs associated with the global and local connectors, respectively.
The corresponding Pauli operators are defined as $\hat{\sigma}_{\mathrm a}^z=\ket{e_{\mathrm a}}\bra{e_{\mathrm a}}-\ket{g_{\mathrm a}}\bra{g_{\mathrm a}}$, where $\mathrm a\in\{s,R\}$ and $\ket{g_{\mathrm a}}$ ($\ket{e_{\mathrm a}}$) denotes the ground (excited) state of the respective TLS.
Dispersive terms associated with absent auxiliary TLSs are omitted.
The Kerr nonlinearity produces anharmonic Fock-state level spacings, whereas the dispersive couplings induce Fock-number-dependent shifts of the joint boson--TLS transition frequencies, thereby enabling selective addressing of the desired transitions.
For a $q$-quantum transition, neighboring Kerr-resolved resonances are separated by $2q|K_c|$, whereas the dispersively resolved separation is $2|\chi_{\mathrm{act}}|$, independent of $q$, within a fixed spectator-TLS subspace when present.
Here $q$ denotes the number of exchanged bosonic quanta and $\chi_{\mathrm{act}}\in\{\chi_s,\chi_R\}$ is the dispersive coupling of the active TLS.

We next specify the microscopic realizations of the engineered global and local connectors, $\mathcal{L}_G$ and $\mathcal{L}_L(R)$, respectively, as illustrated in Fig.~\ref{fig1}\textcolor{nphysblue}{d,e}.
The global connector can be implemented through three alternative $N_s$-quantum interactions, whereas the movable local connector is realized through two single-quantum interactions.

\subsection{Global connector implementations}

For the TLS-assisted global implementations, the raising and lowering operators of the auxiliary TLS are defined as $\hat{\sigma}_s^+=\ket{e_s}\bra{g_s}$ and $\hat{\sigma}_s^-=\ket{g_s}\bra{e_s}$, respectively.
The TLS Hamiltonian in the rotating frame is $\hat{H}_s=\Delta_s\hat{\sigma}_s^+\hat{\sigma}_s^-$, where $\Delta_s=\omega_s-N_s\omega_r$ is the TLS detuning and $\omega_s$ denotes its bare transition frequency.

The three global implementations are described by the interaction Hamiltonians
\begin{equation}
\begin{aligned}
\hat{H}_{G}^{\mathrm{JC}}
&=
g_{G,\mathrm{JC}}^{(N_s)}
\left[
(\hat{a}_c^\dagger)^{N_s}\hat{\sigma}_s^-
+\mathrm{H.c.}
\right],\\
\hat{H}_{G}^{\mathrm{AJC}}
&=
g_{G,\mathrm{AJC}}^{(N_s)}
\left[
(\hat{a}_c^\dagger)^{N_s}\hat{\sigma}_s^+
+\mathrm{H.c.}
\right],\\
\hat{H}_{G}^{\mathrm{drv}}
&=
g_{G,\mathrm{drv}}^{(N_s)}
\left[
(\hat{a}_c^\dagger)^{N_s}
+\hat{a}_c^{N_s}
\right],
\end{aligned}
\label{eq:global_interactions}
\end{equation}
where $g_{G,\mathrm{JC}}^{(N_s)}$, $g_{G,\mathrm{AJC}}^{(N_s)}$, and $g_{G,\mathrm{drv}}^{(N_s)}$ denote the corresponding $N_s$-quantum coupling strengths, and H.c. denotes the Hermitian conjugate.
For the TLS-assisted global implementations, we denote the joint boson--TLS basis states by $\ket{n;q_s}\equiv\ket{n}\otimes\ket{q_s}$, where $n$ is the bosonic excitation number and $q_s\in\{g_s,e_s\}$ labels the state of the global auxiliary TLS.
Under the selected resonance conditions, the JC and AJC interactions resonantly couple the joint states $\ket{0;e_s}\leftrightarrow\ket{N_s;g_s}$ and $\ket{0;g_s}\leftrightarrow\ket{N_s;e_s}$, respectively.
The coherent $N_s$-quantum drive instead directly addresses $\ket{0}\leftrightarrow\ket{N_s}$ without requiring an auxiliary TLS.
The $R\neq N_s$ label in Fig.~\ref{fig1}\textcolor{nphysblue}{d(iii)} indicates only the illustrated configuration; when $R=N_s$, the reversible coherent drive can transfer population from the target state back to the vacuum, competing with local capture.
In the implementations considered here, the TLS-assisted global interactions admit either Kerr or dispersive spectral selection, whereas the direct coherent drive is restricted to Kerr-based selectivity.
Related nonlinear and multiphoton spin--boson models have been investigated~\cite{Cheng2018,Puebla2019}, while other reservoir-engineering approaches have targeted Fock-state preparation~\cite{PhysRevA.93.060301}.

The corresponding global Liouvillian is written as
\begin{equation}
\mathcal{L}_{G}\hat{\rho}
=
-i\left[
\hat{H}_s+\hat{H}_{G},
\hat{\rho}
\right]
+
\wp_s\mathcal{D}[\hat{\sigma}_s^+]\hat{\rho}
+
\gamma_s\mathcal{D}[\hat{\sigma}_s^-]\hat{\rho}.
\label{eq:global_liouvillian}
\end{equation}
Here, $\hat{H}_{G}$ is chosen from Eq.~\eqref{eq:global_interactions}, while $\wp_s$ and $\gamma_s$ denote the incoherent pumping and relaxation rates of the global auxiliary TLS, respectively.
For the direct coherent-drive implementation, the auxiliary TLS is absent, so $\hat{H}_s$ and the associated TLS dissipative terms are omitted.
The resonance conditions for the selected transitions are detailed in Supplementary Sects.~S2 and S3.

\subsection{Movable local connector implementations}

The movable local connector is implemented through single-quantum JC or AJC interactions between the bosonic mode and a local auxiliary TLS.
The raising and lowering operators of the local TLS are defined as $\hat{\sigma}_R^+=\ket{e_R}\bra{g_R}$ and $\hat{\sigma}_R^-=\ket{g_R}\bra{e_R}$, respectively.
Its Hamiltonian is $\hat{H}_R=\Delta_R\hat{\sigma}_R^+\hat{\sigma}_R^-$, where $\Delta_R=\omega_R-\omega_r$ is the rotating-frame detuning and $\omega_R$ denotes the bare transition frequency of the local TLS.
The subscript $R$ on the local-TLS operators and parameters serves as a label for the local auxiliary TLS, whereas $R$ in $\ket{R}$ and $\mathcal L_L(R)$ denotes the selected target Fock-state index.

The two local implementations are described by the interaction Hamiltonians
\begin{equation}
\begin{aligned}
\hat{H}_{L}^{\mathrm{JC}}(R)
&=
g_{L,\mathrm{JC}}^{(1)}
\left[
\hat{a}_c^\dagger\hat{\sigma}_R^-
+\mathrm{H.c.}
\right],\\
\hat{H}_{L}^{\mathrm{AJC}}(R)
&=
g_{L,\mathrm{AJC}}^{(1)}
\left[
\hat{a}_c^\dagger\hat{\sigma}_R^+
+\mathrm{H.c.}
\right],
\end{aligned}
\label{eq:local_interactions}
\end{equation}
where $g_{L,\mathrm{JC}}^{(1)}$ and $g_{L,\mathrm{AJC}}^{(1)}$ denote the single-quantum JC and AJC coupling strengths, respectively.
Using the joint-state notation $\ket{n;q_R}\equiv\ket{n}\otimes\ket{q_R}$, with $q_R\in\{g_R,e_R\}$, the JC and AJC interactions resonantly couple the selected states $\ket{R-1;e_R}\leftrightarrow\ket{R;g_R}$ and $\ket{R-1;g_R}\leftrightarrow\ket{R;e_R}$, respectively.

The corresponding local Liouvillian is given by
\begin{align}
\mathcal{L}_{L}(R)\hat{\rho}
=&
-i\left[
\hat{H}_R+\hat{H}_{L}(R),
\hat{\rho}
\right]
+
\wp_R\mathcal{D}[\hat{\sigma}_R^+]\hat{\rho}\nonumber\\&
+
\gamma_R\mathcal{D}[\hat{\sigma}_R^-]\hat{\rho},
\label{eq:local_liouvillian}
\end{align}
where $\hat{H}_{L}(R)$ is chosen from Eq.~\eqref{eq:local_interactions}, and $\wp_R$ and $\gamma_R$ denote the incoherent pumping and relaxation rates of the local auxiliary TLS, respectively.
The target Fock state $\ket{R}$ can be changed by tuning the selected local resonance, without modifying the form of the interaction Hamiltonian.

Numerical simulations of the full Lindblad master equation were performed using QuTiP~\cite{JOHANSSON20121760}.
The bosonic Hilbert space was truncated to $20$ Fock levels for the reported simulations, and convergence was independently verified by repeating the calculations with a cutoff of $40$ Fock levels.

\section*{Data availability}

The data supporting the figures and findings of this study are available from the corresponding author upon reasonable request.

\section*{Code availability}

The code is available from the authors upon reasonable request.

\section*{Acknowledgments}
We acknowledge Ying Wu and Xin-You L\"{u} for useful discussions. 
The present research is supported in part by the National Natural Science Foundation of China through Grant No.\,12275092 and by the National Key Research and Development Program of China under Contract No.\,2021YFA1400700.

\section*{Author contributions}

All the authors made critical contributions to this work.

\section*{Competing interests}

The authors declare no competing interests.

\section*{Additional information}

\noindent\textbf{Supplementary information} Supplementary information is included with this manuscript.

\noindent\textbf{Correspondence and requests for materials} should be addressed to Yan Liu or Jiahua Li.

\end{document}

\title{Supplementary Information for ``Engineering Bosonic Steady States through Fock-Space Currents''}
\author{Yan Liu}\email{neonorth@foxmail.com}
\author{Jiahua Li}\email{huajia\_li@163.com}
\affiliation{School of Physics, Huazhong University of Science and
Technology, Wuhan 430074, People's Republic of China}

\date{\today}

\maketitle

\setcounter{tocdepth}{2}
\tableofcontents
\clearpage

This Supplementary Information provides detailed theoretical derivations and additional numerical results supporting the Fock-space current-engineering framework presented in the main text. 
Supplementary Sec.~\ref{sec:supp_notation_scope} summarizes the notation and microscopic model, including the global and local connectors and their auxiliary two-level systems (TLSs).
Supplementary Sec.~\ref{sec:supp_effective_connectors} derives the effective population-transfer rates for Kerr-resolved $q$-quantum Jaynes--Cummings (JC), anti-Jaynes--Cummings (AJC), and coherent-drive realizations and maps them onto the global and movable local connectors.
Here, $q\geq1$ denotes the number of exchanged bosonic quanta.
Supplementary Sec.~\ref{sec:supp_dispersive_connectors} extends the effective-rate derivations to dispersively resolved transitions, including the dependence on spectator-TLS states.
Supplementary Sec.~\ref{sec:supp_steady_state} formulates the unified effective Fock-space population equation and its matrix representation, while Supplementary Sec.~\ref{sec:supp_closed_form_stationary} derives the closed-form stationary populations and examines the nonzero- and zero-global-current cases, including local-only stabilization and coincident global and local connectors.

Supplementary Sec.~\ref{sec:supp_figS1_parameter_dependence} examines the parameter dependence of Kerr-resolved Fock-state stabilization, whereas Supplementary Sec.~\ref{sec:supp_figS2_parameter_dependence} presents the corresponding analysis for dispersively resolved stabilization. Supplementary Sec.~\ref{sec:supp_figS3_balanced_transport} investigates symmetric global population transport under balanced auxiliary-TLS pumping and relaxation. 
Finally, Supplementary Sec.~\ref{sec:supp_figS4_depletion} examines Fock-selective population depletion through reversed local transport and establishes the unitary correspondence between the local JC and pumped AJC realizations.

\onecolumngrid

\section*{List of Symbols}

\begingroup
\small
\setlength{\tabcolsep}{6pt}
\renewcommand{\arraystretch}{1.18}

\begin{center}
\begin{tabular}{
@{}p{0.30\textwidth}p{0.62\textwidth}@{}
}
\hline
\textbf{Symbol} & \textbf{Description} \\
\hline

$N_s$
& Upper Fock-state index of the global connector. \\

$R$
& Target Fock-state index of the movable local connector. \\

$\mathcal L_c,\mathcal L_G,\mathcal L_L(R)$
& Background, global, and local Liouvillians. \\

$\kappa_c$
& Intrinsic single-quantum loss rate. \\

$K_c$
& Kerr nonlinearity. \\

$\chi_s,\chi_R$
& Dispersive coupling strengths of the global and local auxiliary TLSs. \\

$\Delta_c,\Delta_s,\Delta_R$
& Rotating-frame detunings of the bosonic mode, global TLS, and local TLS. \\

$\wp_s,\wp_R$
& Incoherent pumping rates of the global and local TLSs. \\

$\gamma_s,\gamma_R$
& Relaxation rates of the global and local TLSs. \\

$g_{G,\mathrm{JC}}^{(N_s)}$
& Global $N_s$-quantum JC coupling strength. \\

$g_{G,\mathrm{AJC}}^{(N_s)}$
& Global $N_s$-quantum AJC coupling strength. \\

$g_{G,\mathrm{drv}}^{(N_s)}$
& Coherent global $N_s$-quantum drive strength. \\

$g_{L,\mathrm{JC}}^{(1)}$
& Local JC coupling strength. \\

$g_{L,\mathrm{AJC}}^{(1)}$
& Local AJC coupling strength. \\

$P_n(t),P_n^{\mathrm{ss}}$
& Time-dependent and steady-state Fock-state populations. \\

$J_{\mathrm G},J_{\mathrm L}(R)$
& Net global and local Fock-space population currents. \\

$\mathsf M(R)$
& Effective Fock-space population rate matrix. \\

$\hat U_R$
& Local-TLS basis-swap operator in the JC and AJC correspondence. \\

\hline
\end{tabular}
\end{center}

\endgroup

\twocolumngrid

\section{Notation and scope}
\label{sec:supp_notation_scope}

We follow the notation and rotating-frame conventions of the main text and set $\hbar=1$ throughout.
The bosonic annihilation (creation) and number operators are $\hat a_c$ ($\hat a_c^\dagger$) and $\hat n_c=\hat a_c^\dagger\hat a_c$, respectively.
The global connector addresses $\ket{0}\leftrightarrow\ket{N_s}$, whereas the movable local connector addresses $\ket{R-1}\leftrightarrow\ket{R}$.
Their auxiliary two-level systems, when present, are labeled by $s$ and $R$, respectively.

Consistent with the main text, the microscopic dynamics obey the three-part Liouvillian master equation
\begin{equation}
\frac{d\hat\rho}{dt}
=
\left[
\mathcal L_c+\mathcal L_G+\mathcal L_L(R)
\right]\hat\rho,
\label{eq:supp_general_master}
\end{equation}
where $\hat\rho$ is the joint density operator of the bosonic mode and the auxiliary TLSs present in the chosen realization.
The background Liouvillian is
\begin{equation}
\mathcal L_c\hat\rho
=
-i[\hat H_c,\hat\rho]
+\kappa_c\mathcal D[\hat a_c]\hat\rho,
\end{equation}
where $\kappa_c$ denotes the intrinsic single-quantum loss rate and $\mathcal D[\hat O]\hat\rho=\hat O\hat\rho\hat O^\dagger-\frac12\{\hat O^\dagger\hat O,\hat\rho\}$.
The background Hamiltonian $\hat H_c$ describes Kerr-resolved or dispersively resolved spectral selection, while $\mathcal L_G$ and $\mathcal L_L(R)$ describe the chosen global and local connector implementations.
A connector or auxiliary TLS absent from a particular realization contributes no corresponding terms.

The rotating-frame detunings are $\Delta_c=\omega_c-\omega_r$, $\Delta_s=\omega_s-N_s\omega_r$, and $\Delta_R=\omega_R-\omega_r$.
For $\mathrm a\in\{s,R\}$, the auxiliary TLS operators are $\hat\sigma_{\mathrm a}^+=\ket{e_{\mathrm a}}\bra{g_{\mathrm a}}$, $\hat\sigma_{\mathrm a}^-=\ket{g_{\mathrm a}}\bra{e_{\mathrm a}}$, and $\hat\sigma_{\mathrm a}^z=\ket{e_{\mathrm a}}\bra{e_{\mathrm a}}-\ket{g_{\mathrm a}}\bra{g_{\mathrm a}}$; $\wp_{\mathrm a}$ and $\gamma_{\mathrm a}$ denote the respective incoherent pumping and relaxation rates.

Supplementary Table~\ref{tab:supp_model_overview} summarizes the Hamiltonian terms and dissipative configurations corresponding to the three Liouvillian contributions.
The Kerr and dispersive Hamiltonians define two alternative mechanisms for spectral selectivity.
The global connector is realized by a pumped $N_s$-quantum JC interaction, an $N_s$-quantum AJC interaction with auxiliary-TLS relaxation, or a coherent $N_s$-quantum drive, whereas the local connector employs either a pumped single-quantum JC interaction or a single-quantum AJC interaction with auxiliary-TLS relaxation.

\begin{table*}[t]
\caption{Microscopic model components and connector realizations.
The first two rows specify Kerr- and dispersively resolved spectral selection.
The remaining rows summarize the $N_s$-quantum global and single-quantum local connector realizations, specifying their interaction Hamiltonians and the associated auxiliary-TLS pumping and relaxation rates, where applicable.
Here, $\mathrm{H.c.}$ denotes the Hermitian conjugate.
}
\label{tab:supp_model_overview}
\centering
\small
\setlength{\tabcolsep}{4pt}
\renewcommand{\arraystretch}{1.35}

\begin{ruledtabular}
\begin{tabular}{
@{}
p{0.105\textwidth}
p{0.215\textwidth}
p{0.425\textwidth}
p{0.19\textwidth}
}
\textbf{Block}
& \textbf{Realization}
& \textbf{Hamiltonian contribution}
& \textbf{TLS dissipative rates}
\\
\hline

$\mathcal L_c$
& Kerr-resolved
& $\hat H_c^{(K)}
=\Delta_c\hat n_c+K_c\hat n_c^2$
& ---
\\

$\mathcal L_c$
& Dispersively resolved
& $\hat H_c^{(D)}
=\Delta_c\hat n_c
+\chi_s\hat n_c\hat\sigma_s^z
+\chi_R\hat n_c\hat\sigma_R^z$
& ---
\\
\hline

$\mathcal L_G$
& Pumped $N_s$-JC
& $\hat H_G^{\mathrm{JC}}
=g_{G,\mathrm{JC}}^{(N_s)}
[(\hat a_c^\dagger)^{N_s}
\hat\sigma_s^-+\mathrm{H.c.}]$
& $\wp_s,\ \gamma_s$
\\

$\mathcal L_G$
& $N_s$-AJC+relaxation
& $\hat H_G^{\mathrm{AJC}}
=g_{G,\mathrm{AJC}}^{(N_s)}
[(\hat a_c^\dagger)^{N_s}
\hat\sigma_s^++\mathrm{H.c.}]$
& $\wp_s=0,\ \gamma_s>0$
\\

$\mathcal L_G$
& Coherent $N_s$ drive
& $\hat H_G^{\mathrm{drv}}
=g_{G,\mathrm{drv}}^{(N_s)}
[(\hat a_c^\dagger)^{N_s}
+\hat a_c^{N_s}]$
& No global TLS
\\
\hline

$\mathcal L_L(R)$
& Pumped JC
& $\hat H_L^{\mathrm{JC}}(R)
=g_{L,\mathrm{JC}}^{(1)}
[\hat a_c^\dagger\hat\sigma_R^-
+\mathrm{H.c.}]$
& $\wp_R,\ \gamma_R$
\\

$\mathcal L_L(R)$
& AJC+relaxation
& $\hat H_L^{\mathrm{AJC}}(R)
=g_{L,\mathrm{AJC}}^{(1)}
[\hat a_c^\dagger\hat\sigma_R^+
+\mathrm{H.c.}]$
& $\wp_R=0,\ \gamma_R>0$
\\

\end{tabular}
\end{ruledtabular}
\end{table*}

The tabulated global and local connector Hamiltonians are interaction terms: each TLS-assisted connector also includes the auxiliary Hamiltonian $\hat H_{\mathrm a}=\Delta_{\mathrm a}\hat\sigma_{\mathrm a}^+\hat\sigma_{\mathrm a}^-$ and the dissipators $\wp_{\mathrm a}\mathcal D[\hat\sigma_{\mathrm a}^+]$ and $\gamma_{\mathrm a}\mathcal D[\hat\sigma_{\mathrm a}^-]$, where $\mathrm a=s$ or $R$ as appropriate.
The global and local interactions change the bosonic excitation number by $N_s$ and one, respectively.
In the implementations considered here, the coherent global drive is used only with Kerr-based spectral selection.

For the standard AJC+relaxation effective-rate derivations, the active AJC TLS has no incoherent pumping, $\wp_{\mathrm a}=0$ with $\gamma_{\mathrm a}>0$.
By contrast, the pumped-JC derivations retain both $\wp_{\mathrm a}$ and $\gamma_{\mathrm a}$; the AJC zero-pumping assumption does not constrain an independently present spectator TLS.
The pumping-dominated local AJC realization in Supplementary Sec.~S9 uses the same AJC interaction Hamiltonian with $\wp_R>0$ and is evaluated using the full microscopic master equation, rather than the zero-pumping AJC effective-rate formula.

The reduced bosonic populations are $P_n(t)=\bra n\operatorname{Tr}_{\mathrm{TLS}}[\hat\rho(t)]\ket n$, and $P_n^{\mathrm{ss}}$ denotes their stationary values.
In the effective population model, $\mathcal R_{\mathrm G}^{\uparrow}$ and $\mathcal R_{\mathrm G}^{\downarrow}$ denote the effective upward and downward population-transfer rates of the global connector, while $\mathcal R_{\mathrm L}^{\uparrow}(R)$ and $\mathcal R_{\mathrm L}^{\downarrow}(R)$ denote the corresponding rates of the local connector.
Upward transfer is defined as $\ket{0}\to\ket{N_s}$ and $\ket{R-1}\to\ket{R}$ for the global and local connectors, respectively.
These rates describe an effective reduction of the microscopic dynamics rather than an exact identity for arbitrary parameters.

In the generic selected-transition derivations, $m$ denotes the lower Fock number and $q\geq1$ the number of exchanged quanta, with $(m,q)=(0,N_s)$ for the global connector and $(m,q)=(R-1,1)$ for the local connector.
The generic auxiliary index $\mathrm a$ and the dispersive active or spectator labels $\mathrm{act}$ and $\mathrm{sp}$ are mapped to $s$ or $R$ as appropriate.
For the numerical calculations, $N_{\mathrm{cut}}$ denotes the bosonic Hilbert-space dimension, so the retained basis runs from $\ket0$ to $\ket{N_{\mathrm{cut}}-1}$.
The numerical results are obtained with $N_{\mathrm{cut}}=20$, and convergence is verified by repeating the calculations with larger cutoff values.

\section{Effective connector dynamics with Kerr-resolved spectral selection}
\label{sec:supp_effective_connectors}

For the Kerr-resolved implementation considered in the main text, we introduced three microscopic realizations of the selective global connector and two TLS-assisted realizations of the movable local connector.
In this section, we derive the reduced effective dynamics of these microscopic realizations, assuming that the desired Fock-state transitions are spectrally selected by the intrinsic Kerr anharmonicity of the bosonic mode.
Rather than treating the global and local connectors separately from the outset, we first formulate the spectral selection of a generic transition $\ket{m}\leftrightarrow\ket{m+q}$, where $m$ denotes the Fock number of the lower state and $q\geq1$ is the number of bosonic excitations exchanged in the selected transition.
We then treat separately three microscopic mechanisms acting on this selected transition: an incoherently pumped $q$-quantum JC coupling, a coherent $q$-quantum drive, and a $q$-quantum AJC coupling with fast auxiliary-TLS relaxation.
The global and local connectors of the stabilization architecture follow only afterward as specific choices of $m$ and $q$.

The resulting effective connector rates enter the unified Fock-state population model formulated in Supplementary Sec.~\ref{sec:supp_steady_state}, where $P_n$ denotes the population of the Fock state $\ket{n}$, $N_s$ specifies the upper endpoint of the global connector, and $R$ specifies the position of the local connector.
Within this unified description, the steady-state behavior is governed by the position $R$ of the local connector relative to the upper endpoint $N_s$ of the global connector, while the microscopic realization enters only through the corresponding effective connector rates.

\subsection{Spectral selection of a generic Fock-state transition}
\label{subsec:supp_generic_transition}

Using the Kerr spectrum defined in the main text, the rotating-frame energy of the Fock state $\ket{m}$ is
\begin{equation}
    \varepsilon_m
    =
    m\Delta_c
    +
    m^2 K_c.
    \label{eq:supp_fock_energy}
\end{equation}
The energy difference associated with a generic $q$-quantum transition $\ket{m}\leftrightarrow\ket{m+q}$ is therefore
\begin{equation}
    \varepsilon_{m+q}-\varepsilon_m
    =
    q\Delta_c
    +
    (2mq+q^2)K_c.
    \label{eq:supp_q_transition}
\end{equation}

We select $\ket{m_0}\leftrightarrow\ket{m_0+q}$ as the resonant transition. 
For another $q$-quantum transition $\ket{m}\leftrightarrow\ket{m+q}$, the detuning relative to the selected transition is
\begin{equation}
\begin{aligned}
    \delta_q(m;m_0)
    &\equiv
    \left(
        \varepsilon_{m+q}-\varepsilon_m
    \right)
    -
    \left(
        \varepsilon_{m_0+q}-\varepsilon_{m_0}
    \right)
    \\&=
    2qK_c(m-m_0).
    \label{eq:supp_selective_detuning}
\end{aligned}
\end{equation}
Thus, the Kerr nonlinearity makes transitions of the same order spectrally distinguishable according to their initial Fock states. 
When $|\delta_q(m;m_0)|$ is large compared with the relevant coupling and dissipation scales for all unwanted transitions, the off-resonant processes can be neglected and the dynamics reduce to the selected Fock-state transition.

\subsection{Effective dynamics of a pumped JC realization}
\label{subsec:supp_generic_JC}

We first consider a generic $q$-quantum JC interaction between the bosonic mode and an auxiliary TLS, labeled by the subscript $\mathrm a$, with ground state $\ket{g_{\mathrm a}}$ and excited state $\ket{e_{\mathrm a}}$.
Its raising and lowering operators are defined as
\begin{equation}
    \hat{\sigma}_{\mathrm a}^{+}
    =
    \ket{e_{\mathrm a}}\!\bra{g_{\mathrm a}},
    \qquad
    \hat{\sigma}_{\mathrm a}^{-}
    =
    \ket{g_{\mathrm a}}\!\bra{e_{\mathrm a}}.
    \label{eq:supp_generic_auxiliary_operators}
\end{equation}
Let $\omega_{\mathrm a}$ denote the bare transition frequency of the auxiliary TLS. 
Consistent with the rotating-frame convention used in the main text, we describe the auxiliary TLS in a frame rotating at $q\omega_r$, in which its free Hamiltonian is
\begin{equation}
    \hat{H}_{\mathrm a}
    =
    \Delta_{\mathrm a}
    \hat{\sigma}_{\mathrm a}^{+}\hat{\sigma}_{\mathrm a}^{-},
    \qquad
    \Delta_{\mathrm a}
    =
    \omega_{\mathrm a}
    -
    q\omega_r.
    \label{eq:supp_generic_auxiliary_H}
\end{equation}
The $q$-quantum JC interaction is described by
\begin{equation}
    \hat{H}_{\mathrm{JC}}^{(q)}
    =
    g_{\mathrm{JC}}^{(q)}
    \left[
        (\hat{a}_c^{\dagger})^q
        \hat{\sigma}_{\mathrm a}^{-}
        +
        \hat{a}_c^{q}
        \hat{\sigma}_{\mathrm a}^{+}
    \right],
    \label{eq:supp_generic_JC_H}
\end{equation}
where $g_{\mathrm{JC}}^{(q)}$ denotes the corresponding $q$-quantum JC coupling strength.

For the selected Fock-state transition $\ket{m}\leftrightarrow\ket{m+q}$, the $q$-quantum JC interaction couples the two joint states
\begin{equation}
    \ket{A}
    =
    \ket{m;e_{\mathrm a}},
    \qquad
    \ket{B}
    =
    \ket{m+q;g_{\mathrm a}},
    \label{eq:supp_JC_states}
\end{equation}
where $\ket{n;q_{\mathrm a}}\equiv\ket{n}\otimes\ket{q_{\mathrm a}}$ denotes a joint state with the bosonic mode in the Fock state $\ket{n}$ and the auxiliary TLS in the state $\ket{q_{\mathrm a}}$, with $q_{\mathrm a}\in\{g_{\mathrm a},e_{\mathrm a}\}$.

Using the bosonic ladder relation
\begin{equation}
    (\hat{a}_c^{\dagger})^q\ket{m}
    =
    \sqrt{\frac{(m+q)!}{m!}}
    \ket{m+q},
    \label{eq:supp_generic_JC_ladder}
\end{equation}
the coherent coupling between the two joint states $\ket{A}$ and $\ket{B}$ is
\begin{equation}
    \Omega_{\mathrm{JC}}^{(m,q)}
    \equiv
    \bra{B}
    \hat{H}_{\mathrm{JC}}^{(q)}
    \ket{A}
    =
    g_{\mathrm{JC}}^{(q)}
    \sqrt{\frac{(m+q)!}{m!}}.
    \label{eq:supp_generic_JC_matrix}
\end{equation}
The detuning between these two states is defined as
\begin{equation}
    \delta_{\mathrm{JC}}^{(m,q)}
    \equiv
    \varepsilon_{m+q}
    -
    \left(
        \varepsilon_m+\Delta_{\mathrm a}
    \right)
    =
    q\Delta_c
    +
    (2mq+q^2)K_c
    -
    \Delta_{\mathrm a}.
    \label{eq:supp_generic_JC_detuning}
\end{equation}
The target JC transition is resonant when
\begin{equation}
    \delta_{\mathrm{JC}}^{(m,q)}
    =
    0,
    \label{eq:supp_generic_JC_resonance}
\end{equation}
or equivalently when $\Delta_{\mathrm a}=\varepsilon_{m+q}-\varepsilon_m$.

We next incorporate the incoherent pumping and relaxation of the auxiliary TLS into the open-system description of the selected JC process.
We denote by $\wp_{\mathrm a}$ the incoherent pumping rate for the transition $\ket{g_{\mathrm a}}\rightarrow\ket{e_{\mathrm a}}$ and by $\gamma_{\mathrm a}$ the relaxation rate for the reverse transition $\ket{e_{\mathrm a}}\rightarrow\ket{g_{\mathrm a}}$. 
Let $\hat{\rho}_{\mathrm{ca}}$ denote the joint density operator of the bosonic mode and the auxiliary TLS. 
Including single-quantum loss of the bosonic mode at a total rate $\kappa_c$, the joint dynamics in the rotating frame defined above are governed by
\begin{equation}
    \begin{aligned}
        \frac{d\hat{\rho}_{\mathrm{ca}}}{dt}
        &=
        -i
        \left[
            \hat{H}_c
            +
            \hat{H}_{\mathrm a}
            +
            \hat{H}_{\mathrm{JC}}^{(q)},
            \hat{\rho}_{\mathrm{ca}}
        \right]
        +
        \kappa_c
        \mathcal D
        [
            \hat{a}_c
        ]
        \hat{\rho}_{\mathrm{ca}}
        \\
        &\quad
        +
        \wp_{\mathrm a}
        \mathcal D
        [
            \hat{\sigma}_{\mathrm a}^{+}
        ]
        \hat{\rho}_{\mathrm{ca}}
        +
        \gamma_{\mathrm a}
        \mathcal D
        [
            \hat{\sigma}_{\mathrm a}^{-}
        ]
        \hat{\rho}_{\mathrm{ca}}.
    \end{aligned}
    \label{eq:supp_JC_joint_master_equation}
\end{equation}

Under the spectral-selection condition established above, off-resonant $q$-quantum JC transitions can be neglected, allowing the coherent JC dynamics associated with the selected Fock-state transition to be reduced to the two joint states $\ket{A}$ and $\ket{B}$.
To derive the corresponding population transfer, we therefore retain the populations of these two states together with the coherence generated by their JC coupling.
We introduce the corresponding joint-state populations and coherence as
\begin{equation}
    P_A
    \equiv
    \bra{A}
    \hat{\rho}_{\mathrm{ca}}
    \ket{A},
    P_B
    \equiv
    \bra{B}
    \hat{\rho}_{\mathrm{ca}}
    \ket{B},
    C_{\mathrm{JC}}^{(m,q)}
    \equiv
    \bra{A}
    \hat{\rho}_{\mathrm{ca}}
    \ket{B}.
    \label{eq:supp_JC_variables}
\end{equation}

To identify the contribution of each term in Eq.~\eqref{eq:supp_JC_joint_master_equation} to the evolution of the selected coherence, we project the joint master equation onto the matrix element $\bra{A}\hat{\rho}_{\mathrm{ca}}\ket{B}$. 
Throughout this work, an overdot denotes differentiation with respect to time, i.e., $\dot X\equiv dX/dt$ for any time-dependent quantity $X$.
The resulting time derivative of $C_{\mathrm{JC}}^{(m,q)}$ is decomposed as
\begin{equation}
    \begin{aligned}
        \dot C_{\mathrm{JC}}^{(m,q)}
        &=
        \left.
            \dot C_{\mathrm{JC}}^{(m,q)}
        \right|_{\mathrm{free}}
        +
        \left.
            \dot C_{\mathrm{JC}}^{(m,q)}
        \right|_{\mathrm{JC}}
        +
        \left.
            \dot C_{\mathrm{JC}}^{(m,q)}
        \right|_{\mathrm{pump}}
        \\
        &\quad
        +
        \left.
            \dot C_{\mathrm{JC}}^{(m,q)}
        \right|_{\mathrm{relax}}
        +
        \left.
            \dot C_{\mathrm{JC}}^{(m,q)}
        \right|_{\mathrm{loss}}.
    \end{aligned}
    \label{eq:supp_JC_coherence_decomposition}
\end{equation}
Here, $\left.\dot X\right|_{Y}$ denotes the contribution to the time derivative of a dynamical quantity $X$ arising from the term or physical process labeled by $Y$. 
Specifically, the labels $\mathrm{free}$, $\mathrm{JC}$, $\mathrm{pump}$, $\mathrm{relax}$, and $\mathrm{loss}$ refer to the free Hamiltonian $\hat{H}_c+\hat{H}_{\mathrm a}$, the $q$-quantum JC interaction $\hat{H}_{\mathrm{JC}}^{(q)}$, the auxiliary pumping channel $\wp_{\mathrm a}\mathcal D[\hat{\sigma}_{\mathrm a}^{+}]$, the auxiliary relaxation channel $\gamma_{\mathrm a}\mathcal D[\hat{\sigma}_{\mathrm a}^{-}]$, and the single-quantum-loss channel $\kappa_c\mathcal D[\hat{a}_c]$, respectively.
The contribution from the free Hamiltonians is
\begin{equation}
    \begin{aligned}
        \left.
            \dot C_{\mathrm{JC}}^{(m,q)}
        \right|_{\mathrm{free}}
        &=
        -i
        \bra{A}
        \left[
            \hat{H}_c
            +
            \hat{H}_{\mathrm a},
            \hat{\rho}_{\mathrm{ca}}
        \right]
        \ket{B}
        \\
        &=
        i\delta_{\mathrm{JC}}^{(m,q)}
        C_{\mathrm{JC}}^{(m,q)}.
    \end{aligned}
    \label{eq:supp_JC_coherence_free}
\end{equation}
Here, $\delta_{\mathrm{JC}}^{(m,q)}$ is the detuning between the joint states $\ket{A}$ and $\ket{B}$ defined in Eq.~\eqref{eq:supp_generic_JC_detuning}.
The contribution of the $q$-quantum JC interaction to $\dot C_{\mathrm{JC}}^{(m,q)}$ is
\begin{equation}
    \begin{aligned}
        \left.
            \dot C_{\mathrm{JC}}^{(m,q)}
        \right|_{\mathrm{JC}}
        &=
        -i
        \bra{A}
        \left[
            \hat{H}_{\mathrm{JC}}^{(q)},
            \hat{\rho}_{\mathrm{ca}}
        \right]
        \ket{B}
        \\
        &=
        -
        i\Omega_{\mathrm{JC}}^{(m,q)}
        \left(
            P_B-P_A
        \right).
    \end{aligned}
    \label{eq:supp_JC_coherence_interaction}
\end{equation}
Here, $\Omega_{\mathrm{JC}}^{(m,q)}$ is the coherent coupling matrix element between $\ket{A}$ and $\ket{B}$ defined in Eq.~\eqref{eq:supp_generic_JC_matrix}.
The contribution from the incoherent pumping channel of the auxiliary TLS is
\begin{equation}
    \begin{aligned}
        \left.
            \dot C_{\mathrm{JC}}^{(m,q)}
        \right|_{\mathrm{pump}}
        &=
        \wp_{\mathrm a}
        \bra{A}
        \mathcal D
        [
            \hat{\sigma}_{\mathrm a}^{+}
        ]
        \hat{\rho}_{\mathrm{ca}}
        \ket{B}
        \\
        &=
        -
        \frac{\wp_{\mathrm a}}{2}
        C_{\mathrm{JC}}^{(m,q)}.
    \end{aligned}
    \label{eq:supp_JC_coherence_pumping}
\end{equation}
The relaxation channel of the auxiliary TLS contributes to the coherence dynamics as
\begin{equation}
    \begin{aligned}
        \left.
            \dot C_{\mathrm{JC}}^{(m,q)}
        \right|_{\mathrm{relax}}
        &=
        \gamma_{\mathrm a}
        \bra{A}
        \mathcal D
        [
            \hat{\sigma}_{\mathrm a}^{-}
        ]
        \hat{\rho}_{\mathrm{ca}}
        \ket{B}
        \\
        &=
        -
        \frac{\gamma_{\mathrm a}}{2}
        C_{\mathrm{JC}}^{(m,q)}.
    \end{aligned}
    \label{eq:supp_JC_coherence_relaxation}
\end{equation}

The contribution from the single-quantum-loss channel of the bosonic mode is
\begin{equation}
    \begin{aligned}
        \left.
            \dot C_{\mathrm{JC}}^{(m,q)}
        \right|_{\mathrm{loss}}
        &=
        \kappa_c
        \bra{A}
        \mathcal D
        [
            \hat{a}_c
        ]
        \hat{\rho}_{\mathrm{ca}}
        \ket{B}
        \\
        &=
        \kappa_c
        \bra{A}
        \hat{a}_c
        \hat{\rho}_{\mathrm{ca}}
        \hat{a}_c^{\dagger}
        \ket{B}
        -
        \frac{\kappa_c}{2}
        \bra{A}
        \left\{
            \hat{a}_c^{\dagger}\hat{a}_c,
            \hat{\rho}_{\mathrm{ca}}
        \right\}
        \ket{B}
        \\
        &=
        \kappa_c
        \sqrt{
            (m+1)(m+q+1)
        }
        C_{\mathrm{JC}}^{(m+1,q)}
        \\&\quad-
        \frac{
            (2m+q)\kappa_c
        }{2}
        C_{\mathrm{JC}}^{(m,q)},
    \end{aligned}
    \label{eq:supp_JC_coherence_cavity_loss}
\end{equation}
where
\begin{equation}
    C_{\mathrm{JC}}^{(m+1,q)}
    \equiv
    \bra{m+1;e_{\mathrm a}}
    \hat{\rho}_{\mathrm{ca}}
    \ket{m+q+1;g_{\mathrm a}}
    \label{eq:supp_JC_neighboring_coherence}
\end{equation}
denotes the coherence between the two joint states obtained by increasing the bosonic Fock-state index in both $\ket{A}$ and $\ket{B}$ by one.

Combining Eqs.~\eqref{eq:supp_JC_coherence_free}--\eqref{eq:supp_JC_coherence_cavity_loss}, the equation of motion for the coherence $C_{\mathrm{JC}}^{(m,q)}$ is
\begin{equation}
    \begin{aligned}
        \dot C_{\mathrm{JC}}^{(m,q)}
        &=
        -
        \left[
            \Gamma_{\mathrm{JC}}^{(m,q)}
            -
            i\delta_{\mathrm{JC}}^{(m,q)}
        \right]
        C_{\mathrm{JC}}^{(m,q)}
        -
        i\Omega_{\mathrm{JC}}^{(m,q)}
        \left(
            P_B-P_A
        \right)
        \\
        &\quad
        +
        \kappa_c
        \sqrt{
            (m+1)(m+q+1)
        }
        C_{\mathrm{JC}}^{(m+1,q)},
    \end{aligned}
    \label{eq:supp_JC_coherence_full}
\end{equation}
where the total coherence-decay rate of the selected JC coherence is defined as
\begin{equation}
    \Gamma_{\mathrm{JC}}^{(m,q)}
    \equiv
    \frac{
        \wp_{\mathrm a}
        +
        \gamma_{\mathrm a}
        +
        (2m+q)\kappa_c
    }{2}.
    \label{eq:supp_JC_total_coherence_decay}
\end{equation}

If pure dephasing of the auxiliary TLS and the bosonic mode is taken into account, we introduce the corresponding Lindblad channels with rates $\gamma_{\mathrm a,\phi}$ and $\kappa_{c,\phi}$, respectively. 
The dephasing operator of the auxiliary TLS is defined as
\begin{equation}
    \hat{\sigma}_{\mathrm a}^{z}
    \equiv
    \ket{e_{\mathrm a}}\!\bra{e_{\mathrm a}}
    -
    \ket{g_{\mathrm a}}\!\bra{g_{\mathrm a}},
    \label{eq:supp_auxiliary_sigma_z}
\end{equation}
and the additional dissipative terms in the joint master equation are
\begin{equation}
    \gamma_{\mathrm a,\phi}
    \mathcal D
    [
        \hat{\sigma}_{\mathrm a}^{z}
    ]
    \hat{\rho}_{\mathrm{ca}}
    +
    \kappa_{c,\phi}
    \mathcal D
    [
        \hat{a}_c^{\dagger}\hat{a}_c
    ]
    \hat{\rho}_{\mathrm{ca}}.
    \label{eq:supp_JC_pure_dephasing_channels}
\end{equation}
The pure-dephasing channel of the auxiliary TLS contributes
\begin{equation}
    \begin{aligned}
        \left.
            \dot C_{\mathrm{JC}}^{(m,q)}
        \right|_{\mathrm{deph,a}}
        &=
        \gamma_{\mathrm a,\phi}
        \bra{A}
        \mathcal D
        [
            \hat{\sigma}_{\mathrm a}^{z}
        ]
        \hat{\rho}_{\mathrm{ca}}
        \ket{B}
        \\
        &=
        -
        2\gamma_{\mathrm a,\phi}
        C_{\mathrm{JC}}^{(m,q)}.
    \end{aligned}
    \label{eq:supp_JC_coherence_auxiliary_dephasing}
\end{equation}
The pure-dephasing channel of the bosonic mode contributes
\begin{equation}
    \begin{aligned}
        \left.
            \dot C_{\mathrm{JC}}^{(m,q)}
        \right|_{\mathrm{deph,c}}
        &=
        \kappa_{c,\phi}
        \bra{A}
        \mathcal D
        [
            \hat{a}_c^{\dagger}\hat{a}_c
        ]
        \hat{\rho}_{\mathrm{ca}}
        \ket{B}
        \\
        &=
        -
        \frac{q^{2}\kappa_{c,\phi}}{2}
        C_{\mathrm{JC}}^{(m,q)}.
    \end{aligned}
    \label{eq:supp_JC_coherence_bosonic_dephasing}
\end{equation}
When both pure-dephasing channels are included, their effect on Eq.~\eqref{eq:supp_JC_coherence_full} is incorporated through a modification of the total coherence-decay rate. 
Specifically, $\Gamma_{\mathrm{JC}}^{(m,q)}$ is replaced by
\begin{equation}
    \Gamma_{\mathrm{JC}}^{(m,q)}
    =
    \frac{
        \wp_{\mathrm a}
        +
        \gamma_{\mathrm a}
        +
        (2m+q)\kappa_c
    }{2}
    +
    2\gamma_{\mathrm a,\phi}
    +
    \frac{
        q^{2}\kappa_{c,\phi}
    }{2}.
    \label{eq:supp_JC_total_coherence_decay_with_dephasing}
\end{equation}
Equation~\eqref{eq:supp_JC_coherence_full} therefore retains the same form, with $\Gamma_{\mathrm{JC}}^{(m,q)}$ understood as the total coherence-decay rate given in Eq.~\eqref{eq:supp_JC_total_coherence_decay_with_dephasing}.

The remaining term proportional to $C_{\mathrm{JC}}^{(m+1,q)}$ couples the selected JC coherence to the neighboring $q$-quantum coherence associated with the transition $\ket{m+1}\leftrightarrow\ket{m+q+1}$. 
According to Eq.~\eqref{eq:supp_selective_detuning}, this neighboring transition is detuned from the selected transition by
\begin{equation}
    \delta_q(m+1;m)
    =
    2qK_c.
    \label{eq:supp_JC_neighboring_detuning}
\end{equation}
A sufficiently large Kerr-induced spectral separation therefore suppresses the neighboring coherence $C_{\mathrm{JC}}^{(m+1,q)}$. 
In addition, when the single-quantum-loss rate is sufficiently small, the corresponding feeding term $\kappa_c\sqrt{(m+1)(m+q+1)}C_{\mathrm{JC}}^{(m+1,q)}$ provides only a weak correction to the dynamics of $C_{\mathrm{JC}}^{(m,q)}$.
A convenient sufficient criterion for neglecting this coupling is
\begin{equation}
    \begin{aligned}
        &2q|K_c|
        \gg
        \max
        \left\{
            \left|
                \Omega_{\mathrm{JC}}^{(m+1,q)}
            \right|,
            \Gamma_{\mathrm{JC}}^{(m+1,q)}
        \right\},
        \\
        &\kappa_c
        \sqrt{
            (m+1)(m+q+1)
        }
        \ll
        \Gamma_{\mathrm{JC}}^{(m,q)}.
    \end{aligned}
    \label{eq:supp_JC_closed_coherence_condition}
\end{equation}
These inequalities provide a sufficient condition rather than a necessary one; the neighboring-coherence contribution can also remain negligible outside this regime, depending on the population of the neighboring Fock states and the resulting magnitude of $C_{\mathrm{JC}}^{(m+1,q)}$.
Under the condition that this contribution is negligible, Eq.~\eqref{eq:supp_JC_coherence_full} reduces to the closed equation
\begin{equation}
    \dot C_{\mathrm{JC}}^{(m,q)}
    =
    -
    \left[
        \Gamma_{\mathrm{JC}}^{(m,q)}
        -
        i\delta_{\mathrm{JC}}^{(m,q)}
    \right]
    C_{\mathrm{JC}}^{(m,q)}
    -
    i\Omega_{\mathrm{JC}}^{(m,q)}
    \left(
        P_B-P_A
    \right).
    \label{eq:supp_JC_coherence_closed}
\end{equation}

For the time-dependent dynamics, $C_{\mathrm{JC}}^{(m,q)}$ can be adiabatically eliminated when its evolution is fast compared with the population dynamics. 
A convenient sufficient condition for this time-scale separation is
\begin{equation}
    \sqrt{
        \left[
            \Gamma_{\mathrm{JC}}^{(m,q)}
        \right]^2
        +
        \left[
            \delta_{\mathrm{JC}}^{(m,q)}
        \right]^2
    }
    \gg
    \left|
        \Omega_{\mathrm{JC}}^{(m,q)}
    \right|.
    \label{eq:supp_JC_adiabatic_condition}
\end{equation}
Near resonance, this condition reduces to
\begin{equation}
    \Gamma_{\mathrm{JC}}^{(m,q)}
    \gg
    \left|
        \Omega_{\mathrm{JC}}^{(m,q)}
    \right|.
    \label{eq:supp_JC_adiabatic_condition_resonant}
\end{equation}
Under this condition, setting $\dot C_{\mathrm{JC}}^{(m,q)}\simeq0$ gives
\begin{equation}
    C_{\mathrm{JC}}^{(m,q)}
    \simeq
    -
    \frac{
        i\Omega_{\mathrm{JC}}^{(m,q)}
    }{
        \Gamma_{\mathrm{JC}}^{(m,q)}
        -
        i\delta_{\mathrm{JC}}^{(m,q)}
    }
    \left(
        P_B-P_A
    \right).
    \label{eq:supp_JC_coherence_adiabatic}
\end{equation}

For the stationary solution of the reduced model described by Eq.~\eqref{eq:supp_JC_coherence_closed}, no adiabatic condition is required.
Setting $\dot C_{\mathrm{JC}}^{(m,q)}=0$ gives the exact stationary relation
\begin{equation}
    C_{\mathrm{JC}}^{(m,q)}
    =
    -
    \frac{
        i\Omega_{\mathrm{JC}}^{(m,q)}
    }{
        \Gamma_{\mathrm{JC}}^{(m,q)}
        -
        i\delta_{\mathrm{JC}}^{(m,q)}
    }
    \left(
        P_B-P_A
    \right).
    \label{eq:supp_JC_coherence_stationary}
\end{equation}
Thus, the time-dependent and stationary treatments yield the same algebraic form for the eliminated coherence, but the former relies on the adiabatic approximation, whereas the latter is exact within the reduced model.
If the adiabatic-elimination condition is not well satisfied, the resulting analytical description of the time-dependent dynamics may deviate appreciably from direct numerical solutions of the reduced model in which the dynamics of $C_{\mathrm{JC}}^{(m,q)}$ are retained explicitly.

We next determine the coherent population transfer between the two selected joint states $\ket{A}$ and $\ket{B}$ by projecting the Hamiltonian part of Eq.~\eqref{eq:supp_JC_joint_master_equation} onto the corresponding diagonal matrix elements. 
Since $\ket{A}$ and $\ket{B}$ are eigenstates of the free Hamiltonian $\hat{H}_c+\hat{H}_{\mathrm a}$, the free evolution does not change their populations. 
The Hamiltonian contributions to $P_A$ and $P_B$ therefore reduce to
\begin{equation}
    \begin{aligned}
        \left.
            \dot P_A
        \right|_{\mathrm{free+JC}}
        &=
        -i
        \bra{A}
        \left[
            \hat{H}_c
            +
            \hat{H}_{\mathrm a}
            +
            \hat{H}_{\mathrm{JC}}^{(q)},
            \hat{\rho}_{\mathrm{ca}}
        \right]
        \ket{A}
        \\
        &=
        -i
        \bra{A}
        \left[
            \hat{H}_{\mathrm{JC}}^{(q)},
            \hat{\rho}_{\mathrm{ca}}
        \right]
        \ket{A}
        =
        -2
        \Omega_{\mathrm{JC}}^{(m,q)}
        \operatorname{Im}
        C_{\mathrm{JC}}^{(m,q)},
        \\
        \left.
            \dot P_B
        \right|_{\mathrm{free+JC}}
        &=
        -i
        \bra{B}
        \left[
            \hat{H}_c
            +
            \hat{H}_{\mathrm a}
            +
            \hat{H}_{\mathrm{JC}}^{(q)},
            \hat{\rho}_{\mathrm{ca}}
        \right]
        \ket{B}
        \\
        &=
        -i
        \bra{B}
        \left[
            \hat{H}_{\mathrm{JC}}^{(q)},
            \hat{\rho}_{\mathrm{ca}}
        \right]
        \ket{B}
        =
        2
        \Omega_{\mathrm{JC}}^{(m,q)}
        \operatorname{Im}
        C_{\mathrm{JC}}^{(m,q)}.
    \end{aligned}
    \label{eq:supp_JC_population_exchange}
\end{equation}
Here, the label $\mathrm{free+JC}$ in $\left.\dot X\right|_{\mathrm{free+JC}}$ denotes the contribution generated by the full Hamiltonian part of the joint master equation, while $\operatorname{Im}(X)$ denotes the imaginary part of a complex quantity $X$.
In obtaining the final equalities, we have used the JC matrix element $\Omega_{\mathrm{JC}}^{(m,q)}$ defined in Eq.~\eqref{eq:supp_generic_JC_matrix} together with $\bra{B}\hat{\rho}_{\mathrm{ca}}\ket{A}=\left[C_{\mathrm{JC}}^{(m,q)}\right]^{*}$.

The Hamiltonian-induced population transfer between $\ket{A}$ and $\ket{B}$ can be expressed in the rate form
\begin{equation}
    \begin{aligned}
        \left.
            \dot P_A
        \right|_{\mathrm{free+JC}}
        &=
        W_{\mathrm{JC}}^{(m,q)}
        \left(
            P_B-P_A
        \right),
        \\
        \left.
            \dot P_B
        \right|_{\mathrm{free+JC}}
        &=
        W_{\mathrm{JC}}^{(m,q)}
        \left(
            P_A-P_B
        \right),
    \end{aligned}
    \label{eq:supp_JC_effective_population_exchange}
\end{equation}
where $W_{\mathrm{JC}}^{(m,q)}$ denotes the population-exchange rate associated with the selected $q$-quantum JC process. 
The population-exchange rate $W_{\mathrm{JC}}^{(m,q)}$ follows from Eq.~\eqref{eq:supp_JC_coherence_adiabatic} for the time-dependent dynamics and from Eq.~\eqref{eq:supp_JC_coherence_stationary} for the stationary solution. 
In both cases, the resulting rate takes the same form,
\begin{equation}
    W_{\mathrm{JC}}^{(m,q)}
    =
    \frac{
        2
        \left|
            \Omega_{\mathrm{JC}}^{(m,q)}
        \right|^{2}
        \Gamma_{\mathrm{JC}}^{(m,q)}
    }{
        \left[
            \Gamma_{\mathrm{JC}}^{(m,q)}
        \right]^{2}
        +
        \left[
            \delta_{\mathrm{JC}}^{(m,q)}
        \right]^{2}
    }.
    \label{eq:supp_JC_effective_exchange_rate}
\end{equation}
The two population-exchange terms in Eq.~\eqref{eq:supp_JC_effective_population_exchange} are equal in magnitude and opposite in sign, so the Hamiltonian evolution transfers population between $\ket{A}$ and $\ket{B}$ without changing their total population.

To reduce the joint-state population transfer to an effective population dynamics of the bosonic mode, we first determine the stationary state of the auxiliary TLS in isolation, with all terms involving the bosonic mode excluded from its dynamics.
Its density matrix $\hat{\rho}_{\mathrm a}$ therefore obeys
\begin{equation}
    \frac{d\hat{\rho}_{\mathrm a}}{dt}
    =
    -i
    \left[
        \hat{H}_{\mathrm a},
        \hat{\rho}_{\mathrm a}
    \right]
    +
    \wp_{\mathrm a}
    \mathcal D
    \left[
        \hat{\sigma}_{\mathrm a}^{+}
    \right]
    \hat{\rho}_{\mathrm a}
    +
    \gamma_{\mathrm a}
    \mathcal D
    \left[
        \hat{\sigma}_{\mathrm a}^{-}
    \right]
    \hat{\rho}_{\mathrm a}.
    \label{eq:supp_auxiliary_master_equation}
\end{equation}
Defining the excited- and ground-state populations as
\begin{equation}
    p_{e,\mathrm a}
    \equiv
    \bra{e_{\mathrm a}}
    \hat{\rho}_{\mathrm a}
    \ket{e_{\mathrm a}},
    \qquad
    p_{g,\mathrm a}
    \equiv
    \bra{g_{\mathrm a}}
    \hat{\rho}_{\mathrm a}
    \ket{g_{\mathrm a}},
    \label{eq:supp_auxiliary_populations}
\end{equation}
the master equation gives
\begin{equation}
    \dot p_{e,\mathrm a}
    =
    \wp_{\mathrm a}p_{g,\mathrm a}
    -
    \gamma_{\mathrm a}p_{e,\mathrm a},
    \qquad
    p_{e,\mathrm a}
    +
    p_{g,\mathrm a}
    =
    1.
    \label{eq:supp_auxiliary_population_equation}
\end{equation}
In the stationary state, the excited-state population satisfies $\dot p_{e,\mathrm a}=0$. 
We denote the stationary excited- and ground-state populations by $p_{e,\mathrm a}^{\mathrm{ss}}$ and $p_{g,\mathrm a}^{\mathrm{ss}}$, respectively. 
Substituting the stationary condition into Eq.~\eqref{eq:supp_auxiliary_population_equation} yields
\begin{equation}
    p_{e,\mathrm a}^{\mathrm{ss}}
    =
    \frac{\wp_{\mathrm a}}
    {\wp_{\mathrm a}+\gamma_{\mathrm a}},
    \qquad
    p_{g,\mathrm a}^{\mathrm{ss}}
    =
    \frac{\gamma_{\mathrm a}}
    {\wp_{\mathrm a}+\gamma_{\mathrm a}}.
    \label{eq:supp_auxiliary_stationary_populations}
\end{equation}
Since the incoherent pumping and relaxation do not generate coherence between $\ket{g_{\mathrm a}}$ and $\ket{e_{\mathrm a}}$, the stationary state of the isolated auxiliary TLS is
\begin{equation}
    \hat{\rho}_{\mathrm a}^{\mathrm{ss}}
    =
    p_{g,\mathrm a}^{\mathrm{ss}}
    \ket{g_{\mathrm a}}\!\bra{g_{\mathrm a}}
    +
    p_{e,\mathrm a}^{\mathrm{ss}}
    \ket{e_{\mathrm a}}\!\bra{e_{\mathrm a}}.
    \label{eq:supp_auxiliary_stationary_state}
\end{equation}

To reduce the joint-state population transfer to an effective population dynamics of the bosonic mode, we assume that the auxiliary TLS relaxes rapidly toward the stationary population distribution determined by its incoherent pumping and relaxation and remains close to this isolated-TLS distribution in the presence of its coupling to the bosonic mode.
In particular, its population-relaxation time scale $(\wp_{\mathrm a}+\gamma_{\mathrm a})^{-1}$ is assumed to be short compared with the characteristic time scale of the bosonic population dynamics, while correlations between the bosonic Fock-state populations and the auxiliary-TLS populations induced by the coupling remain weak.
Under this separation of time scales, the conditional auxiliary-TLS populations are approximately independent of the bosonic Fock state and are given by their stationary values $p_{e,\mathrm a}^{\mathrm{ss}}$ and $p_{g,\mathrm a}^{\mathrm{ss}}$.
Defining the bosonic Fock-state populations as
\begin{equation}
    P_n
    \equiv
    \bra{n}
    \hat{\rho}_c
    \ket{n},
    \label{eq:supp_bosonic_fock_populations}
\end{equation}
The relevant joint-state populations can therefore be approximated as
\begin{equation}
    P_A
    \simeq
    P_m
    p_{e,\mathrm a}^{\mathrm{ss}},
    \qquad
    P_B
    \simeq
    P_{m+q}
    p_{g,\mathrm a}^{\mathrm{ss}}.
    \label{eq:supp_JC_factorized_populations}
\end{equation}
This population-level approximation does not set the JC-induced coherence $C_{\mathrm{JC}}^{(m,q)}$ to zero; its effect has already been incorporated into the exchange rate $W_{\mathrm{JC}}^{(m,q)}$ through the preceding coherence elimination.

Each population-transfer event from $\ket{A}=\ket{m;e_{\mathrm a}}$ to $\ket{B}=\ket{m+q;g_{\mathrm a}}$ changes the bosonic state from $\ket{m}$ to $\ket{m+q}$, while the reverse event produces the opposite transition. 
Using the approximate joint-state populations in Eq.~\eqref{eq:supp_JC_factorized_populations}, the Hamiltonian-induced contributions to the bosonic populations can be written as
\begin{equation}
    \begin{aligned}
        &\left.
            \dot P_m
        \right|_{\mathrm{free+JC}}
        =
        -
        \mathcal R_{\mathrm{JC}}^{\uparrow}(m,q)
        P_m
        +
        \mathcal R_{\mathrm{JC}}^{\downarrow}(m,q)
        P_{m+q},
        \\
        &\left.
            \dot P_{m+q}
        \right|_{\mathrm{free+JC}}
        =
        \mathcal R_{\mathrm{JC}}^{\uparrow}(m,q)
        P_m
        -
        \mathcal R_{\mathrm{JC}}^{\downarrow}(m,q)
        P_{m+q},
    \end{aligned}
    \label{eq:supp_JC_population_rates}
\end{equation}
where $\mathcal R_{\mathrm{JC}}^{\uparrow}(m,q)$ and $\mathcal R_{\mathrm{JC}}^{\downarrow}(m,q)$ denote the effective bosonic population-transfer rates for the transitions $\ket{m}\rightarrow\ket{m+q}$ and $\ket{m+q}\rightarrow\ket{m}$, respectively. They are given by
\begin{equation}
    \begin{aligned}
        \mathcal R_{\mathrm{JC}}^{\uparrow}(m,q)
        &=
        W_{\mathrm{JC}}^{(m,q)}
        p_{e,\mathrm a}^{\mathrm{ss}},
        \\
        \mathcal R_{\mathrm{JC}}^{\downarrow}(m,q)
        &=
        W_{\mathrm{JC}}^{(m,q)}
        p_{g,\mathrm a}^{\mathrm{ss}}.
    \end{aligned}
    \label{eq:supp_JC_general_rates}
\end{equation}
At resonance, where $\delta_{\mathrm{JC}}^{(m,q)}=0$, Eq.~\eqref{eq:supp_JC_effective_exchange_rate} gives
\begin{equation}
    W_{\mathrm{JC}}^{(m,q)}
    =
    \frac{
        2
        \left|
            \Omega_{\mathrm{JC}}^{(m,q)}
        \right|^2
    }{
        \Gamma_{\mathrm{JC}}^{(m,q)}
    }.
    \label{eq:supp_JC_exchange_rate_resonance}
\end{equation}
Using the stationary populations in Eq.~\eqref{eq:supp_auxiliary_stationary_populations}, the corresponding upward and downward effective bosonic population-transfer rates are
\begin{equation}
    \begin{aligned}
        \mathcal R_{\mathrm{JC}}^{\uparrow}(m,q)
        &=
        \frac{
            2
            \left|
                \Omega_{\mathrm{JC}}^{(m,q)}
            \right|^2
        }{
            \Gamma_{\mathrm{JC}}^{(m,q)}
        }
        \frac{
            \wp_{\mathrm a}
        }{
            \wp_{\mathrm a}
            +
            \gamma_{\mathrm a}
        },
        \\
        \mathcal R_{\mathrm{JC}}^{\downarrow}(m,q)
        &=
        \frac{
            2
            \left|
                \Omega_{\mathrm{JC}}^{(m,q)}
            \right|^2
        }{
            \Gamma_{\mathrm{JC}}^{(m,q)}
        }
        \frac{
            \gamma_{\mathrm a}
        }{
            \wp_{\mathrm a}
            +
            \gamma_{\mathrm a}
        }.
    \end{aligned}
    \label{eq:supp_JC_resonant_directional_rates}
\end{equation}
Their ratio is therefore
\begin{equation}
    \frac{
        \mathcal R_{\mathrm{JC}}^{\uparrow}(m,q)
    }{
        \mathcal R_{\mathrm{JC}}^{\downarrow}(m,q)
    }
    =
    \frac{
        p_{e,\mathrm a}^{\mathrm{ss}}
    }{
        p_{g,\mathrm a}^{\mathrm{ss}}
    }
    =
    \frac{
        \wp_{\mathrm a}
    }{
        \gamma_{\mathrm a}
    }.
    \label{eq:supp_JC_directionality}
\end{equation}
Thus, the relative strength of the upward and downward population transfer is determined by the stationary population ratio of the auxiliary TLS, while the magnitudes of both effective bosonic population-transfer rates are governed by the common population-exchange rate $W_{\mathrm{JC}}^{(m,q)}$, which depends on the coherent coupling $\Omega_{\mathrm{JC}}^{(m,q)}$, the detuning $\delta_{\mathrm{JC}}^{(m,q)}$, and the total coherence-decay rate $\Gamma_{\mathrm{JC}}^{(m,q)}$.

Importantly, the increase in directionality with the auxiliary pumping rate $\wp_{\mathrm a}$ does not imply an indefinite increase in the upward effective bosonic population-transfer rate.
At resonance, substituting the coherence-decay rate in Eq.~\eqref{eq:supp_JC_total_coherence_decay} into Eq.~\eqref{eq:supp_JC_resonant_directional_rates} gives
\begin{equation}
    \mathcal R_{\mathrm{JC}}^{\uparrow}(m,q)
    =
    \frac{
        4
        \left|
            \Omega_{\mathrm{JC}}^{(m,q)}
        \right|^2
        \wp_{\mathrm a}
    }{
        \left(
            \wp_{\mathrm a}
            +
            \gamma_{\mathrm a}
        \right)
        \left[
            \wp_{\mathrm a}
            +
            \gamma_{\mathrm a}
            +
            (2m+q)\kappa_c
        \right]
    }.
    \label{eq:supp_JC_upward_rate_pump_dependence}
\end{equation}
Although increasing $\wp_{\mathrm a}$ increases the excited-state population of the auxiliary TLS, it simultaneously broadens the selected JC coherence through $\Gamma_{\mathrm{JC}}^{(m,q)}$.
Consequently, $\mathcal R_{\mathrm{JC}}^{\uparrow}(m,q)$ is maximized at the finite pumping rate
\begin{equation}
    \wp_{\mathrm a}^{\mathrm{opt}}
    =
    \sqrt{
        \gamma_{\mathrm a}
        \left[
            \gamma_{\mathrm a}
            +
            (2m+q)\kappa_c
        \right]
    }.
    \label{eq:supp_JC_optimal_pump}
\end{equation}
The corresponding maximal upward transfer rate is
\begin{equation}
    \mathcal R_{\mathrm{JC}}^{\uparrow,\mathrm{max}}(m,q)
    =
    \frac{
        4
        \left|
            \Omega_{\mathrm{JC}}^{(m,q)}
        \right|^2
    }{
        \left[
            \sqrt{\gamma_{\mathrm a}}
            +
            \sqrt{
                \gamma_{\mathrm a}
                +
                (2m+q)\kappa_c
            }
        \right]^2
    }.
    \label{eq:supp_JC_maximum_upward_rate}
\end{equation}
In the strong-pumping regime,
$\wp_{\mathrm a}\gg\{\gamma_{\mathrm a},(2m+q)\kappa_c\}$,
the upward transfer rate instead decreases as
\begin{equation}
    \mathcal R_{\mathrm{JC}}^{\uparrow}(m,q)
    \simeq
    \frac{
        4
        \left|
            \Omega_{\mathrm{JC}}^{(m,q)}
        \right|^2
    }{
        \wp_{\mathrm a}
    }
    \propto
    \frac{1}{\wp_{\mathrm a}}.
    \label{eq:supp_JC_upward_rate_strong_pump}
\end{equation}
Thus, stronger incoherent pumping enhances the directionality of the JC-assisted population transfer, but an excessively large pumping rate reduces the transfer speed because of pump-induced broadening of the selected JC coherence.

\subsection{Effective dynamics of a coherent realization}
\label{subsec:supp_generic_coherent}

We next consider a generic coherent $q$-quantum drive acting directly on the bosonic mode. In the rotating frame defined above, the drive Hamiltonian is
\begin{equation}
    \hat{H}_{\mathrm{drv}}^{(q)}
    =
    g_{\mathrm{drv}}^{(q)}
    \left[
        \left(
            \hat{a}_c^{\dagger}
        \right)^q
        +
        \hat{a}_c^{q}
    \right],
    \label{eq:supp_generic_coherent_H}
\end{equation}
where $g_{\mathrm{drv}}^{(q)}$ denotes the amplitude of the coherent $q$-quantum drive.

For the selected Fock-state transition $\ket{m}\leftrightarrow\ket{m+q}$, the corresponding coherent coupling is
\begin{equation}
    \Omega_{\mathrm{drv}}^{(m,q)}
    \equiv
    \bra{m+q}
    \hat{H}_{\mathrm{drv}}^{(q)}
    \ket{m}
    =
    g_{\mathrm{drv}}^{(q)}
    \sqrt{
        \frac{
            (m+q)!
        }{
            m!
        }
    }.
    \label{eq:supp_generic_coherent_matrix}
\end{equation}
The detuning of the selected transition is defined as
\begin{equation}
    \delta_{\mathrm{drv}}^{(m,q)}
    \equiv
    \varepsilon_{m+q}
    -
    \varepsilon_m
    =
    q\Delta_c
    +
    \left(
        2mq+q^2
    \right)
    K_c.
    \label{eq:supp_generic_coherent_detuning}
\end{equation}
The selected transition is resonant when
\begin{equation}
    \delta_{\mathrm{drv}}^{(m,q)}
    =
    0.
    \label{eq:supp_generic_coherent_resonance}
\end{equation}

Let $\hat{\rho}_c$ denote the density operator of the bosonic mode. Including single-quantum loss at a total rate $\kappa_c$, its dynamics are governed by
\begin{equation}
    \frac{d\hat{\rho}_c}{dt}
    =
    -i
    \left[
        \hat{H}_c
        +
        \hat{H}_{\mathrm{drv}}^{(q)},
        \hat{\rho}_c
    \right]
    +
    \kappa_c
    \mathcal D
    [
        \hat{a}_c
    ]
    \hat{\rho}_c.
    \label{eq:supp_coherent_master_equation}
\end{equation}

Under the spectral-selection condition established above, off-resonant $q$-quantum transitions can be neglected, allowing the coherent dynamics associated with the selected Fock-state transition to be reduced to the two-state subspace spanned by $\ket{m}$ and $\ket{m+q}$.
Using the bosonic Fock-state populations $P_n$ defined in Eq.~\eqref{eq:supp_bosonic_fock_populations}, we introduce the coherence between the selected states as
\begin{equation}
    C_{\mathrm{drv}}^{(m,q)}
    \equiv
    \bra{m}
    \hat{\rho}_c
    \ket{m+q}.
    \label{eq:supp_coherent_selected_coherence}
\end{equation}

To identify the contribution of each term in Eq.~\eqref{eq:supp_coherent_master_equation} to the evolution of the selected coherence, we project the master equation onto the matrix element $\bra{m}\hat{\rho}_c\ket{m+q}$. 
The resulting equation of motion can be decomposed as
\begin{equation}
    \dot C_{\mathrm{drv}}^{(m,q)}
    =
    \left.
        \dot C_{\mathrm{drv}}^{(m,q)}
    \right|_{\mathrm{free}}
    +
    \left.
        \dot C_{\mathrm{drv}}^{(m,q)}
    \right|_{\mathrm{drv}}
    +
    \left.
        \dot C_{\mathrm{drv}}^{(m,q)}
    \right|_{\mathrm{loss}}.
    \label{eq:supp_coherent_coherence_decomposition}
\end{equation}
The contribution from the free Hamiltonian is
\begin{equation}
    \begin{aligned}
        \left.
            \dot C_{\mathrm{drv}}^{(m,q)}
        \right|_{\mathrm{free}}
        &=
        -i
        \bra{m}
        \left[
            \hat{H}_c,
            \hat{\rho}_c
        \right]
        \ket{m+q}
        \\
        &=
        i
        \delta_{\mathrm{drv}}^{(m,q)}
        C_{\mathrm{drv}}^{(m,q)},
    \end{aligned}
    \label{eq:supp_coherent_coherence_free}
\end{equation}
where $\delta_{\mathrm{drv}}^{(m,q)}$ is the detuning defined in Eq.~\eqref{eq:supp_generic_coherent_detuning}.
The contribution from the coherent $q$-quantum drive is
\begin{equation}
    \begin{aligned}
        \left.
            \dot C_{\mathrm{drv}}^{(m,q)}
        \right|_{\mathrm{drv}}
        &=
        -i
        \bra{m}
        \left[
            \hat{H}_{\mathrm{drv}}^{(q)},
            \hat{\rho}_c
        \right]
        \ket{m+q}
        \\
        &=
        -
        i
        \Omega_{\mathrm{drv}}^{(m,q)}
        \left(
            P_{m+q}
            -
            P_m
        \right),
    \end{aligned}
    \label{eq:supp_coherent_coherence_drive}
\end{equation}
where $\Omega_{\mathrm{drv}}^{(m,q)}$ is the coherent coupling defined in Eq.~\eqref{eq:supp_generic_coherent_matrix}.
The contribution from the single-quantum-loss channel is
\begin{equation}
    \begin{aligned}
        \left.
            \dot C_{\mathrm{drv}}^{(m,q)}
        \right|_{\mathrm{loss}}
        &=
        \kappa_c
        \bra{m}
        \mathcal D
        [
            \hat{a}_c
        ]
        \hat{\rho}_c
        \ket{m+q}
        \\
       & =
        \kappa_c
        \sqrt{
            (m+1)(m+q+1)
        }
        C_{\mathrm{drv}}^{(m+1,q)}
        \\&\quad-
        \frac{
            (2m+q)\kappa_c
        }{2}
        C_{\mathrm{drv}}^{(m,q)},
    \end{aligned}
    \label{eq:supp_coherent_coherence_loss}
\end{equation}
where
\begin{equation}
    C_{\mathrm{drv}}^{(m+1,q)}
    \equiv
    \bra{m+1}
    \hat{\rho}_c
    \ket{m+q+1}
    \label{eq:supp_coherent_neighboring_coherence}
\end{equation}
denotes the coherence between the corresponding joint Fock states with both Fock indices increased by one.

Combining Eqs.~\eqref{eq:supp_coherent_coherence_free}--\eqref{eq:supp_coherent_coherence_loss}, the equation of motion for the selected coherence becomes
\begin{equation}
    \begin{aligned}
        \dot C_{\mathrm{drv}}^{(m,q)}
        &=
        -
        \left[
            \Gamma_{\mathrm{drv}}^{(m,q)}
            -
            i\delta_{\mathrm{drv}}^{(m,q)}
        \right]
        C_{\mathrm{drv}}^{(m,q)}
        -
        i
        \Omega_{\mathrm{drv}}^{(m,q)}
        \left(
            P_{m+q}
            -
            P_m
        \right)
        \\
        &\quad
        +
        \kappa_c
        \sqrt{
            (m+1)(m+q+1)
        }
        C_{\mathrm{drv}}^{(m+1,q)},
    \end{aligned}
    \label{eq:supp_coherent_coherence_full}
\end{equation}
where the coherence-decay rate associated with single-quantum loss is defined as
\begin{equation}
    \Gamma_{\mathrm{drv}}^{(m,q)}
    \equiv
    \frac{
        (2m+q)\kappa_c
    }{2}.
    \label{eq:supp_coherent_total_coherence_decay}
\end{equation}

If pure dephasing of the bosonic mode is taken into account, we introduce the corresponding Lindblad channel with rate $\kappa_{c,\phi}$,
\begin{equation}
    \kappa_{c,\phi}
    \mathcal D
    [
        \hat{a}_c^{\dagger}
        \hat{a}_c
    ]
    \hat{\rho}_c.
    \label{eq:supp_coherent_bosonic_dephasing_channel}
\end{equation}
Its contribution to the selected coherence is
\begin{equation}
    \begin{aligned}
        \left.
            \dot C_{\mathrm{drv}}^{(m,q)}
        \right|_{\mathrm{deph,c}}
        &=
        \kappa_{c,\phi}
        \bra{m}
        \mathcal D
        [
            \hat{a}_c^{\dagger}
            \hat{a}_c
        ]
        \hat{\rho}_c
        \ket{m+q}
        \\
        &=
        -
        \frac{
            q^2\kappa_{c,\phi}
        }{2}
        C_{\mathrm{drv}}^{(m,q)}.
    \end{aligned}
    \label{eq:supp_coherent_bosonic_dephasing}
\end{equation}
When this pure-dephasing channel is included, its effect on Eq.~\eqref{eq:supp_coherent_coherence_full} is incorporated through the coherence-decay rate, which becomes
\begin{equation}
    \Gamma_{\mathrm{drv}}^{(m,q)}
    =
    \frac{
        (2m+q)\kappa_c
        +
        q^2\kappa_{c,\phi}
    }{2}.
    \label{eq:supp_coherent_total_coherence_decay_with_dephasing}
\end{equation}

The remaining term proportional to $C_{\mathrm{drv}}^{(m+1,q)}$ couples the selected coherence to the neighboring $q$-quantum coherence associated with the transition $\ket{m+1}\leftrightarrow\ket{m+q+1}$. 
According to Eq.~\eqref{eq:supp_selective_detuning}, this neighboring transition is detuned from the selected transition by
\begin{equation}
    \delta_q(m+1;m)
    =
    2qK_c.
    \label{eq:supp_coherent_neighboring_detuning}
\end{equation}
A sufficiently large Kerr-induced spectral separation therefore suppresses the neighboring coherence. 
A convenient sufficient criterion for neglecting its contribution is
\begin{equation}
    \begin{aligned}
        2q
        \left|
            K_c
        \right|
        \gg
        \max
        \Bigl\{
            &\left|
                \Omega_{\mathrm{drv}}^{(m+1,q)}
            \right|,
            \Gamma_{\mathrm{drv}}^{(m+1,q)},
            \\
            &\kappa_c
            \sqrt{
                (m+1)(m+q+1)
            }
        \Bigr\}.
    \end{aligned}
    \label{eq:supp_coherent_closed_coherence_condition}
\end{equation}
This criterion is sufficient rather than necessary; the neighboring-coherence contribution can also remain negligible outside this regime, depending on the populations of the neighboring Fock states and the resulting magnitude of $C_{\mathrm{drv}}^{(m+1,q)}$.

When the neighboring-coherence contribution is negligible, Eq.~\eqref{eq:supp_coherent_coherence_full} reduces to the closed equation
\begin{equation}
    \dot C_{\mathrm{drv}}^{(m,q)}
    =
    -
    [
        \Gamma_{\mathrm{drv}}^{(m,q)}
        -
        i\delta_{\mathrm{drv}}^{(m,q)}
    ]
    C_{\mathrm{drv}}^{(m,q)}
    -
    i
    \Omega_{\mathrm{drv}}^{(m,q)}
    (
        P_{m+q}
        -
        P_m
    ).
    \label{eq:supp_coherent_coherence_closed}
\end{equation}

For the time-dependent dynamics, $C_{\mathrm{drv}}^{(m,q)}$ can be adiabatically eliminated when its evolution is fast compared with the population dynamics. 
A convenient sufficient condition for this time-scale separation is
\begin{equation}
    \sqrt{
        \left[
            \Gamma_{\mathrm{drv}}^{(m,q)}
        \right]^2
        +
        \left[
            \delta_{\mathrm{drv}}^{(m,q)}
        \right]^2
    }
    \gg
    \left|
        \Omega_{\mathrm{drv}}^{(m,q)}
    \right|.
    \label{eq:supp_coherent_adiabatic_condition}
\end{equation}
Near resonance, this condition reduces to
\begin{equation}
    \Gamma_{\mathrm{drv}}^{(m,q)}
    \gg
    \left|
        \Omega_{\mathrm{drv}}^{(m,q)}
    \right|.
    \label{eq:supp_coherent_adiabatic_condition_resonant}
\end{equation}
Under this condition, setting $\dot C_{\mathrm{drv}}^{(m,q)}\simeq0$ gives
\begin{equation}
    C_{\mathrm{drv}}^{(m,q)}
    \simeq
    -
    \frac{
        i\Omega_{\mathrm{drv}}^{(m,q)}
    }{
        \Gamma_{\mathrm{drv}}^{(m,q)}
        -
        i\delta_{\mathrm{drv}}^{(m,q)}
    }
    \left(
        P_{m+q}
        -
        P_m
    \right).
    \label{eq:supp_coherent_coherence_adiabatic}
\end{equation}

For the stationary solution of the reduced model described by Eq.~\eqref{eq:supp_coherent_coherence_closed}, no adiabatic condition is required. 
Setting $\dot C_{\mathrm{drv}}^{(m,q)}=0$ gives the exact stationary relation
\begin{equation}
    C_{\mathrm{drv}}^{(m,q)}
    =
    -
    \frac{
        i\Omega_{\mathrm{drv}}^{(m,q)}
    }{
        \Gamma_{\mathrm{drv}}^{(m,q)}
        -
        i\delta_{\mathrm{drv}}^{(m,q)}
    }
    \left(
        P_{m+q}
        -
        P_m
    \right).
    \label{eq:supp_coherent_coherence_stationary}
\end{equation}
Thus, the time-dependent and stationary treatments yield the same algebraic form for the eliminated coherence, but the former relies on the adiabatic approximation, whereas the latter is exact within the reduced model.

We next determine the coherent population transfer between the selected Fock states $\ket{m}$ and $\ket{m+q}$ by projecting the Hamiltonian part of Eq.~\eqref{eq:supp_coherent_master_equation} onto the corresponding diagonal matrix elements. 
Since the selected Fock states are eigenstates of $\hat{H}_c$, the free Hamiltonian does not change their populations. 
The Hamiltonian-induced contributions are therefore
\begin{equation}
    \begin{aligned}
        \left.
            \dot P_m
        \right|_{\mathrm{free+drv}}
        &=
        -i
        \bra{m}
        \left[
            \hat{H}_c
            +
            \hat{H}_{\mathrm{drv}}^{(q)},
            \hat{\rho}_c
        \right]
        \ket{m}
        \\
        &=
        -2
        \Omega_{\mathrm{drv}}^{(m,q)}
        \operatorname{Im}
        \left(
            C_{\mathrm{drv}}^{(m,q)}
        \right),
        \\
        \left.
            \dot P_{m+q}
        \right|_{\mathrm{free+drv}}
        &=
        -i
        \bra{m+q}
        \left[
            \hat{H}_c
            +
            \hat{H}_{\mathrm{drv}}^{(q)},
            \hat{\rho}_c
        \right]
        \ket{m+q}
        \\
        &=
        2
        \Omega_{\mathrm{drv}}^{(m,q)}
        \operatorname{Im}
        \left(
            C_{\mathrm{drv}}^{(m,q)}
        \right).
    \end{aligned}
    \label{eq:supp_coherent_population_exchange}
\end{equation}
Here, the label $\mathrm{free+drv}$ in $\left.\dot X\right|_{\mathrm{free+drv}}$ denotes the contribution generated by the full Hamiltonian part of the master equation.

The Hamiltonian-induced population transfer between $\ket{m}$ and $\ket{m+q}$ can be expressed in the rate form
\begin{equation}
    \begin{aligned}
        \left.
            \dot P_m
        \right|_{\mathrm{free+drv}}
        &=
        W_{\mathrm{drv}}^{(m,q)}
        \left(
            P_{m+q}
            -
            P_m
        \right),
        \\
        \left.
            \dot P_{m+q}
        \right|_{\mathrm{free+drv}}
        &=
        W_{\mathrm{drv}}^{(m,q)}
        \left(
            P_m
            -
            P_{m+q}
        \right).
    \end{aligned}
    \label{eq:supp_coherent_effective_population_exchange}
\end{equation}
The population-exchange rate $W_{\mathrm{drv}}^{(m,q)}$ follows from Eq.~\eqref{eq:supp_coherent_coherence_adiabatic} for the time-dependent dynamics and from Eq.~\eqref{eq:supp_coherent_coherence_stationary} for the stationary solution. 
In both cases, the resulting rate takes the same form,
\begin{equation}
    W_{\mathrm{drv}}^{(m,q)}
    =
    \frac{
        2
        \left|
            \Omega_{\mathrm{drv}}^{(m,q)}
        \right|^2
        \Gamma_{\mathrm{drv}}^{(m,q)}
    }{
        \left[
            \Gamma_{\mathrm{drv}}^{(m,q)}
        \right]^2
        +
        \left[
            \delta_{\mathrm{drv}}^{(m,q)}
        \right]^2
    }.
    \label{eq:supp_coherent_effective_exchange_rate}
\end{equation}
The two population-exchange terms in Eq.~\eqref{eq:supp_coherent_effective_population_exchange} are equal in magnitude and opposite in sign, so the Hamiltonian evolution transfers population between the two selected Fock states without changing their total population.

Accordingly, the Hamiltonian-induced population transfer generated by the selected coherent $q$-quantum drive can be represented by the symmetric effective bosonic population-transfer rates
\begin{equation}
    \mathcal R_{\mathrm{drv}}^{\uparrow}(m,q)
    =
    \mathcal R_{\mathrm{drv}}^{\downarrow}(m,q)
    =
    W_{\mathrm{drv}}^{(m,q)},
    \label{eq:supp_coherent_symmetric_rates}
\end{equation}
where $\mathcal R_{\mathrm{drv}}^{\uparrow}(m,q)$ and $\mathcal R_{\mathrm{drv}}^{\downarrow}(m,q)$ correspond to the transitions $\ket{m}\rightarrow\ket{m+q}$ and $\ket{m+q}\rightarrow\ket{m}$, respectively. 
These effective bosonic population-transfer rates describe the population transfer obtained after eliminating the selected coherence and do not identify the underlying coherent drive with an irreversible microscopic transition.

\subsection{Effective dynamics of an AJC+relaxation realization}
\label{subsec:supp_generic_AJC}

We next consider a generic $q$-quantum AJC interaction between the bosonic mode and the auxiliary TLS introduced above.
Throughout this standard AJC+relaxation derivation, we assume $\wp_{\mathrm a}=0$ and $\gamma_{\mathrm a}>0$; the auxiliary TLS has relaxation but no incoherent pumping.
The interaction Hamiltonian is
\begin{equation}
    \hat{H}_{\mathrm{AJC}}^{(q)}
    =
    g_{\mathrm{AJC}}^{(q)}
    \left[
        (\hat{a}_c^{\dagger})^q
        \hat{\sigma}_{\mathrm a}^{+}
        +
        \hat{a}_c^{q}
        \hat{\sigma}_{\mathrm a}^{-}
    \right],
    \label{eq:supp_generic_AJC_H}
\end{equation}
where $g_{\mathrm{AJC}}^{(q)}$ denotes the corresponding $q$-quantum AJC coupling strength.

For the selected Fock-state transition $\ket{m}\leftrightarrow\ket{m+q}$, the $q$-quantum AJC interaction couples the joint states
\begin{equation}
    \ket{D}
    =
    \ket{m;g_{\mathrm a}},
    \qquad
    \ket{E}
    =
    \ket{m+q;e_{\mathrm a}}.
    \label{eq:supp_AJC_coupled_states}
\end{equation}
Relaxation of the auxiliary TLS subsequently connects $\ket{E}$ to
\begin{equation}
    \ket{B}
    =
    \ket{m+q;g_{\mathrm a}},
    \label{eq:supp_AJC_relaxed_state}
\end{equation}
which is the same joint state $\ket{B}$ introduced for the JC realization above.
The resulting sequence
$\ket{D}\leftrightarrow\ket{E}\rightarrow\ket{B}$
therefore changes the bosonic Fock state from $\ket{m}$ to $\ket{m+q}$ when relaxation from $\ket{E}$ dominates the decay of the intermediate state.

Using the bosonic ladder relation in Eq.~\eqref{eq:supp_generic_JC_ladder}, the coherent coupling between $\ket{D}$ and $\ket{E}$ is
\begin{equation}
    \Omega_{\mathrm{AJC}}^{(m,q)}
    \equiv
    \bra{E}
    \hat{H}_{\mathrm{AJC}}^{(q)}
    \ket{D}
    =
    g_{\mathrm{AJC}}^{(q)}
    \sqrt{
        \frac{(m+q)!}{m!}
    }.
    \label{eq:supp_AJC_matrix}
\end{equation}
The detuning between the joint states $\ket{D}$ and $\ket{E}$ is defined as
\begin{equation}
    \begin{aligned}
        \delta_{\mathrm{AJC}}^{(m,q)}
        &\equiv
        \left(
            \varepsilon_{m+q}
            +
            \Delta_{\mathrm a}
        \right)
        -
        \varepsilon_m
        \\
        &=
        q\Delta_c
        +
        (2mq+q^2)K_c
        +
        \Delta_{\mathrm a}.
    \end{aligned}
    \label{eq:supp_generic_AJC_detuning}
\end{equation}
The selected AJC transition is resonant when
\begin{equation}
    \delta_{\mathrm{AJC}}^{(m,q)}
    =
    0,
    \label{eq:supp_generic_AJC_resonance}
\end{equation}
or equivalently when
$\Delta_{\mathrm a}=-(\varepsilon_{m+q}-\varepsilon_m)$.

We now include relaxation of the auxiliary TLS at rate $\gamma_{\mathrm a}$ and single-quantum loss of the bosonic mode at the total rate $\kappa_c$. 
Let $\hat{\rho}_{\mathrm{ca}}$ denote the joint density operator of the bosonic mode and the auxiliary TLS. 
In the rotating frame defined above, the joint dynamics are governed by
\begin{equation}
    \begin{aligned}
        \frac{d\hat{\rho}_{\mathrm{ca}}}{dt}
        &=
        -i
        \left[
            \hat{H}_c
            +
            \hat{H}_{\mathrm a}
            +
            \hat{H}_{\mathrm{AJC}}^{(q)},
            \hat{\rho}_{\mathrm{ca}}
        \right]
        +
        \kappa_c
        \mathcal D
        [
            \hat{a}_c
        ]
        \hat{\rho}_{\mathrm{ca}}
        \\
        &\quad
        +
        \gamma_{\mathrm a}
        \mathcal D
        [
            \hat{\sigma}_{\mathrm a}^{-}
        ]
        \hat{\rho}_{\mathrm{ca}}.
    \end{aligned}
    \label{eq:supp_AJC_joint_master_equation}
\end{equation}

Under the spectral-selection condition established above, off-resonant $q$-quantum AJC transitions can be neglected, allowing the coherent AJC dynamics associated with the selected Fock-state transition to be reduced to the joint states $\ket{D}$ and $\ket{E}$. 
We introduce their populations as
\begin{equation}
    P_D
    \equiv
    \bra{D}
    \hat{\rho}_{\mathrm{ca}}
    \ket{D},
    \qquad
    P_E
    \equiv
    \bra{E}
    \hat{\rho}_{\mathrm{ca}}
    \ket{E},
    \label{eq:supp_AJC_populations}
\end{equation}
and define the corresponding selected AJC coherence as
\begin{equation}
    C_{\mathrm{AJC}}^{(m,q)}
    \equiv
    \bra{D}
    \hat{\rho}_{\mathrm{ca}}
    \ket{E}
    =
    \bra{m;g_{\mathrm a}}
    \hat{\rho}_{\mathrm{ca}}
    \ket{m+q;e_{\mathrm a}}.
    \label{eq:supp_AJC_selected_coherence}
\end{equation}

To identify the contribution of each term in Eq.~\eqref{eq:supp_AJC_joint_master_equation} to the evolution of the selected AJC coherence, we project the master equation onto the matrix element $\bra{D}\hat{\rho}_{\mathrm{ca}}\ket{E}$.
The resulting equation of motion can be decomposed as
\begin{equation}
\begin{aligned}
    \dot C_{\mathrm{AJC}}^{(m,q)}
    &=
    \left.
        \dot C_{\mathrm{AJC}}^{(m,q)}
    \right|_{\mathrm{free}}
    +
    \left.
        \dot C_{\mathrm{AJC}}^{(m,q)}
    \right|_{\mathrm{AJC}}
    +
    \left.
        \dot C_{\mathrm{AJC}}^{(m,q)}
    \right|_{\mathrm{loss}}
    \\&\quad+
    \left.
        \dot C_{\mathrm{AJC}}^{(m,q)}
    \right|_{\mathrm{relax}}.
\end{aligned}
    \label{eq:supp_AJC_coherence_decomposition}
\end{equation}
The contribution from the free Hamiltonians is
\begin{equation}
    \begin{aligned}
        \left.
            \dot C_{\mathrm{AJC}}^{(m,q)}
        \right|_{\mathrm{free}}
        &=
        -i
        \bra{D}
        \left[
            \hat{H}_c
            +
            \hat{H}_{\mathrm a},
            \hat{\rho}_{\mathrm{ca}}
        \right]
        \ket{E}
        \\
        &=
        i
        \delta_{\mathrm{AJC}}^{(m,q)}
        C_{\mathrm{AJC}}^{(m,q)}.
    \end{aligned}
    \label{eq:supp_AJC_coherence_free}
\end{equation}
Here, $\delta_{\mathrm{AJC}}^{(m,q)}$ is the detuning between the joint states $\ket{D}$ and $\ket{E}$ defined in Eq.~\eqref{eq:supp_generic_AJC_detuning}.
The contribution from the AJC interaction is
\begin{equation}
    \begin{aligned}
        \left.
            \dot C_{\mathrm{AJC}}^{(m,q)}
        \right|_{\mathrm{AJC}}
        &=
        -i
        \bra{D}
        \left[
            \hat{H}_{\mathrm{AJC}}^{(q)},
            \hat{\rho}_{\mathrm{ca}}
        \right]
        \ket{E}
        \\
        &=
        -i
        \Omega_{\mathrm{AJC}}^{(m,q)}
        \left(
            P_E-P_D
        \right).
    \end{aligned}
    \label{eq:supp_AJC_coherence_interaction}
\end{equation}
Here, $\Omega_{\mathrm{AJC}}^{(m,q)}$ is the coherent coupling matrix element between $\ket{D}$ and $\ket{E}$ defined in Eq.~\eqref{eq:supp_AJC_matrix}.
The contribution from relaxation of the auxiliary TLS is
\begin{equation}
    \begin{aligned}
        \left.
            \dot C_{\mathrm{AJC}}^{(m,q)}
        \right|_{\mathrm{relax}}
        &=
        \gamma_{\mathrm a}
        \bra{D}
        \mathcal D
        [
            \hat{\sigma}_{\mathrm a}^{-}
        ]
        \hat{\rho}_{\mathrm{ca}}
        \ket{E}
        \\
        &=
        -
        \frac{\gamma_{\mathrm a}}{2}
        C_{\mathrm{AJC}}^{(m,q)}.
    \end{aligned}
    \label{eq:supp_AJC_coherence_relaxation}
\end{equation}
The single-quantum-loss contribution is
\begin{equation}
    \begin{aligned}
        \left.
            \dot C_{\mathrm{AJC}}^{(m,q)}
        \right|_{\mathrm{loss}}
        &=
        \kappa_c
        \bra{D}
        \mathcal D
        [
            \hat{a}_c
        ]
        \hat{\rho}_{\mathrm{ca}}
        \ket{E}
        \\
        &=
        \kappa_c
        \sqrt{
            (m+1)(m+q+1)
        }
        C_{\mathrm{AJC}}^{(m+1,q)}
        \\&\quad-
        \frac{
            (2m+q)\kappa_c
        }{2}
        C_{\mathrm{AJC}}^{(m,q)},
    \end{aligned}
    \label{eq:supp_AJC_coherence_cavity_loss}
\end{equation}
where
\begin{equation}
    C_{\mathrm{AJC}}^{(m+1,q)}
    \equiv
    \bra{m+1;g_{\mathrm a}}
    \hat{\rho}_{\mathrm{ca}}
    \ket{m+q+1;e_{\mathrm a}}
    \label{eq:supp_AJC_neighboring_coherence}
\end{equation}
denotes the neighboring AJC coherence associated with the transition $\ket{m+1;g_{\mathrm a}}\leftrightarrow\ket{m+q+1;e_{\mathrm a}}$.

Combining these contributions, the equation of motion for the selected AJC coherence is
\begin{equation}
    \begin{aligned}
        \dot C_{\mathrm{AJC}}^{(m,q)}
        &=
        -
        \left[
            \Gamma_{\mathrm{AJC}}^{(m,q)}
            -
            i\delta_{\mathrm{AJC}}^{(m,q)}
        \right]
        C_{\mathrm{AJC}}^{(m,q)}
        -
        i
        \Omega_{\mathrm{AJC}}^{(m,q)}
        \left(
            P_E-P_D
        \right)
        \\
        &\quad
        +
        \kappa_c
        \sqrt{
            (m+1)(m+q+1)
        }
        C_{\mathrm{AJC}}^{(m+1,q)},
    \end{aligned}
    \label{eq:supp_AJC_coherence_full}
\end{equation}
where the total coherence-decay rate of the selected AJC coherence is
\begin{equation}
    \Gamma_{\mathrm{AJC}}^{(m,q)}
    \equiv
    \frac{
        \gamma_{\mathrm a}
        +
        (2m+q)\kappa_c
    }{2}.
    \label{eq:supp_AJC_total_coherence_decay}
\end{equation}

If pure dephasing of the auxiliary TLS and the bosonic mode is included, we introduce the corresponding Lindblad channels with rates $\gamma_{\mathrm a,\phi}$ and $\kappa_{c,\phi}$,
\begin{equation}
\begin{aligned}
    \gamma_{\mathrm a,\phi}
    \mathcal D
    [
        \hat{\sigma}_{\mathrm a}^{z}
    ]
    \hat{\rho}_{\mathrm{ca}},\quad\kappa_{c,\phi}
    \mathcal D
    [
        \hat{a}_c^{\dagger}
        \hat{a}_c
    ]
    \hat{\rho}_{\mathrm{ca}}.
\end{aligned}
    \label{eq:supp_AJC_pure_dephasing_channels}
\end{equation}
Their corresponding contributions to the selected AJC coherence are
\begin{equation}
\begin{aligned}
    \left.
        \dot C_{\mathrm{AJC}}^{(m,q)}
    \right|_{\mathrm{deph,a}}
    &=
        \gamma_{\mathrm a,\phi}
        \bra{D}
        \mathcal D
        [
            \hat{\sigma}_{\mathrm a}^{z}
        ]
        \hat{\rho}_{\mathrm{ca}}
        \ket{E}
        \\
        &=
    -2\gamma_{\mathrm a,\phi}
    C_{\mathrm{AJC}}^{(m,q)},
\end{aligned}
    \label{eq:supp_AJC_coherence_auxiliary_dephasing}
\end{equation}
and
\begin{equation}
\begin{aligned}
    \left.
        \dot C_{\mathrm{AJC}}^{(m,q)}
    \right|_{\mathrm{deph,c}}
    &=
        \kappa_{c,\phi}
        \bra{D}
        \mathcal D
        [
            \hat{a}_c^{\dagger}
            \hat{a}_c
        ]
        \hat{\rho}_{\mathrm{ca}}
        \ket{E}
        \\
        &=
    -
    \frac{q^{2}\kappa_{c,\phi}}{2}
    C_{\mathrm{AJC}}^{(m,q)}.
\end{aligned}
    \label{eq:supp_AJC_coherence_bosonic_dephasing}
\end{equation}
Accordingly, the total coherence-decay rate becomes
\begin{equation}
    \Gamma_{\mathrm{AJC}}^{(m,q)}
    =
    \frac{
        \gamma_{\mathrm a}
        +
        (2m+q)\kappa_c
    }{2}
    +
    2\gamma_{\mathrm a,\phi}
    +
    \frac{
        q^{2}\kappa_{c,\phi}
    }{2}.
    \label{eq:supp_AJC_total_coherence_decay_with_dephasing}
\end{equation}
The equation of motion for $C_{\mathrm{AJC}}^{(m,q)}$ therefore retains the same form, with $\Gamma_{\mathrm{AJC}}^{(m,q)}$ understood as the total coherence-decay rate given in Eq.~\eqref{eq:supp_AJC_total_coherence_decay_with_dephasing}.

The remaining term proportional to $C_{\mathrm{AJC}}^{(m+1,q)}$ couples the selected AJC coherence to the neighboring $q$-quantum coherence associated with the transition $\ket{m+1;g_{\mathrm a}}\leftrightarrow\ket{m+q+1;e_{\mathrm a}}$.
According to Eq.~\eqref{eq:supp_selective_detuning}, this neighboring transition is detuned from the selected transition by
\begin{equation}
    \delta_q(m+1;m)
    =
    2qK_c.
    \label{eq:supp_AJC_neighboring_detuning}
\end{equation}
A sufficiently large Kerr-induced spectral separation therefore suppresses the neighboring coherence $C_{\mathrm{AJC}}^{(m+1,q)}$.
In addition, when the single-quantum-loss rate is sufficiently small, the corresponding feeding term $\kappa_c\sqrt{(m+1)(m+q+1)}C_{\mathrm{AJC}}^{(m+1,q)}$ provides only a weak correction to the dynamics of $C_{\mathrm{AJC}}^{(m,q)}$.
A convenient sufficient criterion for neglecting this contribution is
\begin{equation}
    \begin{aligned}
        &2q|K_c|
        \gg
        \max
        \left\{
            \left|
                \Omega_{\mathrm{AJC}}^{(m+1,q)}
            \right|,
            \Gamma_{\mathrm{AJC}}^{(m+1,q)}
        \right\},
        \\
        &\kappa_c
        \sqrt{
            (m+1)(m+q+1)
        }
        \ll
        \Gamma_{\mathrm{AJC}}^{(m,q)}.
    \end{aligned}
    \label{eq:supp_AJC_closed_coherence_condition}
\end{equation}
These inequalities provide a sufficient condition rather than a necessary one.
The neighboring-coherence contribution can also remain negligible outside this regime, depending on the population of the neighboring Fock states and hence on the magnitude of $C_{\mathrm{AJC}}^{(m+1,q)}$.

Under this condition, Eq.~\eqref{eq:supp_AJC_coherence_full} reduces to the closed equation
\begin{equation}
    \dot C_{\mathrm{AJC}}^{(m,q)}
    =
    -
    \left[
        \Gamma_{\mathrm{AJC}}^{(m,q)}
        -
        i\delta_{\mathrm{AJC}}^{(m,q)}
    \right]
    C_{\mathrm{AJC}}^{(m,q)}
    -
    i
    \Omega_{\mathrm{AJC}}^{(m,q)}
    \left(
        P_E-P_D
    \right).
    \label{eq:supp_AJC_coherence_closed}
\end{equation}

For the time-dependent dynamics, $C_{\mathrm{AJC}}^{(m,q)}$ can be adiabatically eliminated when its evolution is fast compared with the population dynamics.
A convenient sufficient condition for this time-scale separation is
\begin{equation}
    \sqrt{
        \left[
            \Gamma_{\mathrm{AJC}}^{(m,q)}
        \right]^2
        +
        \left[
            \delta_{\mathrm{AJC}}^{(m,q)}
        \right]^2
    }
    \gg
    \left|
        \Omega_{\mathrm{AJC}}^{(m,q)}
    \right|.
    \label{eq:supp_AJC_coherence_adiabatic_condition}
\end{equation}
Near resonance, this condition reduces to
\begin{equation}
    \Gamma_{\mathrm{AJC}}^{(m,q)}
    \gg
    \left|
        \Omega_{\mathrm{AJC}}^{(m,q)}
    \right|.
    \label{eq:supp_AJC_coherence_adiabatic_condition_resonant}
\end{equation}
Under this condition, setting
$\dot C_{\mathrm{AJC}}^{(m,q)}\simeq 0$
gives
\begin{equation}
    C_{\mathrm{AJC}}^{(m,q)}
    \simeq
    -
    \frac{
        i\Omega_{\mathrm{AJC}}^{(m,q)}
    }{
        \Gamma_{\mathrm{AJC}}^{(m,q)}
        -
        i\delta_{\mathrm{AJC}}^{(m,q)}
    }
    \left(
        P_E-P_D
    \right).
    \label{eq:supp_AJC_coherence_adiabatic}
\end{equation}

For the stationary solution of the reduced model described by Eq.~\eqref{eq:supp_AJC_coherence_closed}, no adiabatic condition is required.
Setting $\dot C_{\mathrm{AJC}}^{(m,q)}=0$ gives the exact stationary relation
\begin{equation}
    C_{\mathrm{AJC}}^{(m,q)}
    =
    -
    \frac{
        i\Omega_{\mathrm{AJC}}^{(m,q)}
    }{
        \Gamma_{\mathrm{AJC}}^{(m,q)}
        -
        i\delta_{\mathrm{AJC}}^{(m,q)}
    }
    \left(
        P_E-P_D
    \right).
    \label{eq:supp_AJC_coherence_stationary}
\end{equation}
Thus, the time-dependent and stationary treatments lead to the same algebraic form for the coherence, but the former relies on the adiabatic approximation, whereas the latter is exact within the reduced model.

We next determine the coherent population transfer between the two selected joint states $\ket{D}$ and $\ket{E}$ by projecting the Hamiltonian part of Eq.~\eqref{eq:supp_AJC_joint_master_equation} onto the corresponding diagonal matrix elements.
Since $\ket{D}$ and $\ket{E}$ are eigenstates of the free Hamiltonian $\hat{H}_c+\hat{H}_{\mathrm a}$, the free evolution does not change their populations.
The Hamiltonian contributions to $P_D$ and $P_E$ therefore reduce to
\begin{equation}
    \begin{aligned}
        \left.
            \dot P_D
        \right|_{\mathrm{free+AJC}}
        &=
        -i
        \bra{D}
        \left[
            \hat{H}_c
            +
            \hat{H}_{\mathrm a}
            +
            \hat{H}_{\mathrm{AJC}}^{(q)},
            \hat{\rho}_{\mathrm{ca}}
        \right]
        \ket{D}
        \\
        &=
        -i
        \bra{D}
        \left[
            \hat{H}_{\mathrm{AJC}}^{(q)},
            \hat{\rho}_{\mathrm{ca}}
        \right]
        \ket{D}
        =
        -2
        \Omega_{\mathrm{AJC}}^{(m,q)}
        \operatorname{Im}
        C_{\mathrm{AJC}}^{(m,q)},
        \\
        \left.
            \dot P_E
        \right|_{\mathrm{free+AJC}}
        &=
        -i
        \bra{E}
        \left[
            \hat{H}_c
            +
            \hat{H}_{\mathrm a}
            +
            \hat{H}_{\mathrm{AJC}}^{(q)},
            \hat{\rho}_{\mathrm{ca}}
        \right]
        \ket{E}
        \\
        &=
        -i
        \bra{E}
        \left[
            \hat{H}_{\mathrm{AJC}}^{(q)},
            \hat{\rho}_{\mathrm{ca}}
        \right]
        \ket{E}
        =
        2
        \Omega_{\mathrm{AJC}}^{(m,q)}
        \operatorname{Im}
        C_{\mathrm{AJC}}^{(m,q)}.
    \end{aligned}
    \label{eq:supp_AJC_population_exchange}
\end{equation}
Here, the label $\mathrm{free+AJC}$ in $\left.\dot X\right|_{\mathrm{free+AJC}}$ denotes the contribution generated by the full Hamiltonian part of the joint master equation.
In obtaining the final equalities, we have used the AJC matrix element $\Omega_{\mathrm{AJC}}^{(m,q)}$ defined in Eq.~\eqref{eq:supp_AJC_matrix} together with $\bra{E}\hat{\rho}_{\mathrm{ca}}\ket{D}=[C_{\mathrm{AJC}}^{(m,q)}]^{*}$.

The Hamiltonian-induced population transfer between $\ket{D}$ and $\ket{E}$ can then be expressed in the rate form
\begin{equation}
    \begin{aligned}
        \left.
            \dot P_D
        \right|_{\mathrm{free+AJC}}
        &=
        W_{\mathrm{AJC}}^{(m,q)}
        \left(
            P_E-P_D
        \right),
        \\
        \left.
            \dot P_E
        \right|_{\mathrm{free+AJC}}
        &=
        W_{\mathrm{AJC}}^{(m,q)}
        \left(
            P_D-P_E
        \right),
    \end{aligned}
    \label{eq:supp_AJC_effective_population_exchange}
\end{equation}
where $W_{\mathrm{AJC}}^{(m,q)}$ denotes the population-exchange rate between the two joint states coupled by the selected $q$-quantum AJC process.
The population-exchange rate $W_{\mathrm{AJC}}^{(m,q)}$ follows from Eq.~\eqref{eq:supp_AJC_coherence_adiabatic} for the time-dependent dynamics and from Eq.~\eqref{eq:supp_AJC_coherence_stationary} for the stationary solution.
In both cases, the resulting rate takes the same form,
\begin{equation}
    W_{\mathrm{AJC}}^{(m,q)}
    =
    \frac{
        2
        \left|
            \Omega_{\mathrm{AJC}}^{(m,q)}
        \right|^{2}
        \Gamma_{\mathrm{AJC}}^{(m,q)}
    }{
        \left[
            \Gamma_{\mathrm{AJC}}^{(m,q)}
        \right]^{2}
        +
        \left[
            \delta_{\mathrm{AJC}}^{(m,q)}
        \right]^{2}
    }.
    \label{eq:supp_AJC_effective_exchange_rate}
\end{equation}
The two population-exchange terms in Eq.~\eqref{eq:supp_AJC_effective_population_exchange} are equal in magnitude and opposite in sign, so the AJC Hamiltonian alone produces coherent population exchange between $\ket{D}$ and $\ket{E}$ without changing their total population.

Auxiliary relaxation transfers $\ket{E}=\ket{m+q;e_{\mathrm a}}$ to $\ket{B}=\ket{m+q;g_{\mathrm a}}$ while preserving the bosonic Fock number.
We recall that the corresponding population is
\begin{equation}
    P_B
    =
    \bra{B}
    \hat{\rho}_{\mathrm{ca}}
    \ket{B}.
    \label{eq:supp_AJC_relaxed_population}
\end{equation}
Within the selected-transition description, the bosonic populations associated with the two relevant Fock states can therefore be written as
\begin{equation}
    P_m
    \simeq
    P_D,
    \qquad
    P_{m+q}
    =
    P_E
    +
    P_B.
    \label{eq:supp_AJC_bosonic_populations}
\end{equation}
Here, the first relation uses the fact that no incoherent auxiliary-pumping channel acts on the auxiliary TLS, so that the population associated with the bosonic state $\ket{m}$ is predominantly contained in $\ket{D}=\ket{m;g_{\mathrm a}}$.
The second relation is exact upon tracing over the auxiliary TLS, since $\ket{E}$ and $\ket{B}$ are the excited- and ground-state components, respectively, associated with the same bosonic Fock state $\ket{m+q}$.

We first consider the combined contributions of the selected AJC Hamiltonian and auxiliary relaxation to the relevant joint-state populations. 
For each population, we write
\begin{equation}
    \left.
        \dot P_\mu
    \right|_{\mathrm{free+AJC+relax}}
    =
    \left.
        \dot P_\mu
    \right|_{\mathrm{free+AJC}}
    +
    \left.
        \dot P_\mu
    \right|_{\mathrm{relax}},
    \qquad
    \mu=D,E,B.
    \label{eq:supp_AJC_H_relax_decomposition}
\end{equation}
The Hamiltonian contributions for $P_D$ and $P_E$ have already been obtained in
Eq.~\eqref{eq:supp_AJC_effective_population_exchange}.
By contrast, $\ket{B}=\ket{m+q;g_{\mathrm a}}$ does not participate in the selected resonant AJC transition and therefore has no Hamiltonian-induced population change within the selected-transition description,
\begin{equation}
    \left.
        \dot P_B
    \right|_{\mathrm{free+AJC}}
    \simeq
    0.
    \label{eq:supp_AJC_B_H_population}
\end{equation}

The auxiliary-relaxation contributions follow from projecting
$\gamma_{\mathrm a}\mathcal D[\hat{\sigma}_{\mathrm a}^{-}]\hat{\rho}_{\mathrm{ca}}$
onto the corresponding joint states. 
For $\ket{D}=\ket{m;g_{\mathrm a}}$, the auxiliary-relaxation contribution is
\begin{equation}
    \begin{aligned}
        \left.
            \dot P_D
        \right|_{\mathrm{relax}}
        &=
        \gamma_{\mathrm a}
        \bra{D}
        \mathcal D
        [
            \hat{\sigma}_{\mathrm a}^{-}
        ]
        \hat{\rho}_{\mathrm{ca}}
        \ket{D}
        \\
        &=
        \gamma_{\mathrm a}
        P_{m,e_{\mathrm a}}
        \simeq
        0.
    \end{aligned}
    \label{eq:supp_AJC_D_relax_population}
\end{equation}
Here, $P_{m,e_{\mathrm a}}\equiv\bra{m;e_{\mathrm a}}\hat{\rho}_{\mathrm{ca}}\ket{m;e_{\mathrm a}}$ denotes the population of the joint state $\ket{m;e_{\mathrm a}}$.
The last approximation follows from the selected-transition description, in which the off-resonant state $\ket{m;e_{\mathrm a}}$ is not appreciably populated.
For the intermediate state $\ket{E}=\ket{m+q;e_{\mathrm a}}$, the auxiliary-relaxation contribution is
\begin{equation}
    \begin{aligned}
        \left.
            \dot P_E
        \right|_{\mathrm{relax}}
        &=
        \gamma_{\mathrm a}
        \bra{E}
        \mathcal D
        [
            \hat{\sigma}_{\mathrm a}^{-}
        ]
        \hat{\rho}_{\mathrm{ca}}
        \ket{E}
        \\
        &=
        -
        \gamma_{\mathrm a}P_E,
    \end{aligned}
    \label{eq:supp_AJC_E_relax_population}
\end{equation}
whereas for $\ket{B}=\ket{m+q;g_{\mathrm a}}$, it is
\begin{equation}
    \begin{aligned}
        \left.
            \dot P_B
        \right|_{\mathrm{relax}}
        &=
        \gamma_{\mathrm a}
        \bra{B}
        \mathcal D
        [
            \hat{\sigma}_{\mathrm a}^{-}
        ]
        \hat{\rho}_{\mathrm{ca}}
        \ket{B}
        \\
        &=
        \gamma_{\mathrm a}P_E.
    \end{aligned}
    \label{eq:supp_AJC_B_relax_population}
\end{equation}
Thus, auxiliary relaxation removes population from $\ket{E}$ and transfers it to $\ket{B}$ without changing the bosonic Fock number.

Combining the Hamiltonian and relaxation contributions gives
\begin{equation}
    \begin{aligned}
        \left.
            \dot P_D
        \right|_{\mathrm{free+AJC+relax}}
        &\simeq
        W_{\mathrm{AJC}}^{(m,q)}
        \left(
            P_E-P_D
        \right),
        \\
        \left.
            \dot P_E
        \right|_{\mathrm{free+AJC+relax}}
        &=
        W_{\mathrm{AJC}}^{(m,q)}
        \left(
            P_D-P_E
        \right)
        -
        \gamma_{\mathrm a}P_E,
        \\
        \left.
            \dot P_B
        \right|_{\mathrm{free+AJC+relax}}
        &\simeq
        \gamma_{\mathrm a}P_E.
    \end{aligned}
    \label{eq:supp_AJC_joint_population_dynamics}
\end{equation}

Using the relations between the bosonic and joint-state populations given above, together with the combined Hamiltonian and auxiliary-relaxation contributions obtained above, the population changes of the two relevant bosonic Fock states are
\begin{equation}
    \begin{aligned}
        \left.
            \dot P_m
        \right|_{\mathrm{free+AJC+relax}}
        &\simeq
        \left.
            \dot P_D
        \right|_{\mathrm{free+AJC+relax}}
        \\
        &\simeq
        -
        W_{\mathrm{AJC}}^{(m,q)}
        \left(
            P_D-P_E
        \right),
        \\
        \left.
            \dot P_{m+q}
        \right|_{\mathrm{free+AJC+relax}}
        &=
        \left.
            \dot P_E
        \right|_{\mathrm{free+AJC+relax}}
        +
        \left.
            \dot P_B
        \right|_{\mathrm{free+AJC+relax}}
        \\
        &\simeq
        W_{\mathrm{AJC}}^{(m,q)}
        \left(
            P_D-P_E
        \right)
        -
        \gamma_{\mathrm a}P_E
        +
        \gamma_{\mathrm a}P_E
        \\
        &=
        W_{\mathrm{AJC}}^{(m,q)}
        \left(
            P_D-P_E
        \right).
    \end{aligned}
    \label{eq:supp_AJC_bosonic_population_exchange}
\end{equation}
Thus, although auxiliary relaxation transfers population from $\ket{E}$ to $\ket{B}$, its loss and gain contributions cancel in $P_{m+q}=P_E+P_B$, leaving the population exchange between the two bosonic Fock states in the same form as the Hamiltonian-induced exchange derived above.

The population-exchange equations above still depend on the intermediate-state population $P_E$ and therefore do not yet provide a closed description of the bosonic population dynamics.
To obtain an effective bosonic population-transfer rate expressed solely in terms of the retained bosonic populations, we next eliminate $P_E$ by relating it to $P_D\simeq P_m$.

For this purpose, the intrinsic single-quantum loss of the bosonic mode must also be included in the dynamics of $\ket{E}=\ket{m+q;e_{\mathrm a}}$.
Projecting the full joint master equation, Eq.~\eqref{eq:supp_AJC_joint_master_equation}, onto $\bra{E}\hat{\rho}_{\mathrm{ca}}\ket{E}$ gives
\begin{equation}
    \dot P_E
    =
    \left.
        \dot P_E
    \right|_{\mathrm{free+AJC+relax}}
    +
    \left.
        \dot P_E
    \right|_{\mathrm{loss}}.
    \label{eq:supp_AJC_intermediate_population_decomposition}
\end{equation}
The intrinsic single-quantum-loss channel contributes
\begin{equation}
    \begin{aligned}
        \left.
            \dot P_E
        \right|_{\mathrm{loss}}
        &=
        \kappa_c
        \bra{E}
        \mathcal D
        [
            \hat{a}_c
        ]
        \hat{\rho}_{\mathrm{ca}}
        \ket{E}
        \\
        &=
        (m+q+1)
        \kappa_c
        P_{m+q+1,e_{\mathrm a}}
        -
        (m+q)
        \kappa_c
        P_E,
    \end{aligned}
    \label{eq:supp_AJC_intermediate_population_loss}
\end{equation}
where
\begin{equation}
    P_{n,e_{\mathrm a}}
    \equiv
    \bra{n;e_{\mathrm a}}
    \hat{\rho}_{\mathrm{ca}}
    \ket{n;e_{\mathrm a}}
    \label{eq:supp_AJC_excited_auxiliary_population}
\end{equation}
denotes the population of the joint state $\ket{n;e_{\mathrm a}}$.
The first term in Eq.~\eqref{eq:supp_AJC_intermediate_population_loss} describes population feeding into $\ket{E}$ from $\ket{m+q+1;e_{\mathrm a}}$, whereas the second term describes loss from $\ket{E}$ to $\ket{m+q-1;e_{\mathrm a}}$.
If the pure-dephasing channels introduced above are included, they do not directly modify the population equation for $P_E$; their effect is already contained in $W_{\mathrm{AJC}}^{(m,q)}$ through the total coherence-decay rate $\Gamma_{\mathrm{AJC}}^{(m,q)}$.

Combining the Hamiltonian and auxiliary-relaxation contribution in Eq.~\eqref{eq:supp_AJC_joint_population_dynamics} with the single-quantum-loss contribution in Eq.~\eqref{eq:supp_AJC_intermediate_population_loss}, we obtain the full population equation for the intermediate state,
\begin{equation}
    \begin{aligned}
        \dot P_E
        &=
        W_{\mathrm{AJC}}^{(m,q)}
        P_D
        +
        (m+q+1)
        \kappa_c
        P_{m+q+1,e_{\mathrm a}}
        \\
        &\quad
        -
        \left[
            W_{\mathrm{AJC}}^{(m,q)}
            +
            \gamma_{\mathrm a}
            +
            (m+q)\kappa_c
        \right]
        P_E.
    \end{aligned}
    \label{eq:supp_AJC_intermediate_population_dynamics}
\end{equation}

The feeding term proportional to $P_{m+q+1,e_{\mathrm a}}$ originates from single-quantum loss from the neighboring state $\ket{m+q+1;e_{\mathrm a}}$.
Because this state lies outside the selected resonant AJC transition, its population arises only from weak off-resonant processes within the selected-transition description and is therefore expected to remain small.
Accordingly, we assume that this feeding term is negligible compared with the AJC-induced population transfer into $\ket{E}$,
\begin{equation}
    (m+q+1)
    \kappa_c
    P_{m+q+1,e_{\mathrm a}}
    \ll
    W_{\mathrm{AJC}}^{(m,q)}
    P_D.
    \label{eq:supp_AJC_upper_feeding_condition}
\end{equation}
Equation~\eqref{eq:supp_AJC_intermediate_population_dynamics} then reduces to
\begin{equation}
    \dot P_E
    =
    W_{\mathrm{AJC}}^{(m,q)}
    P_D
    -
    \Lambda_E^{(m,q)}
    P_E,
    \label{eq:supp_AJC_intermediate_population_reduced}
\end{equation}
where we have introduced the total removal rate of the intermediate-state population,
\begin{equation}
    \Lambda_E^{(m,q)}
    \equiv
    \gamma_{\mathrm a}
    +
    (m+q)\kappa_c
    +
    W_{\mathrm{AJC}}^{(m,q)}.
    \label{eq:supp_AJC_intermediate_removal_rate}
\end{equation}

For the time-dependent dynamics, $P_E$ can be adiabatically eliminated when its relaxation is fast compared with the population dynamics of the retained bosonic states.
A convenient sufficient condition is
\begin{equation}
    \gamma_{\mathrm a}
    +
    (m+q)\kappa_c
    \gg
    W_{\mathrm{AJC}}^{(m,q)}.
    \label{eq:supp_AJC_population_adiabatic_condition}
\end{equation}
Under this separation of time scales, setting $\dot P_E\simeq0$ in Eq.~\eqref{eq:supp_AJC_intermediate_population_reduced} gives
\begin{equation}
    P_E
    \simeq
    \frac{
        W_{\mathrm{AJC}}^{(m,q)}
    }{
        \gamma_{\mathrm a}
        +
        (m+q)\kappa_c
        +
        W_{\mathrm{AJC}}^{(m,q)}
    }
    P_D.
    \label{eq:supp_AJC_intermediate_population_adiabatic}
\end{equation}
For the stationary solution of the reduced population model, the same relation in Eq.~\eqref{eq:supp_AJC_intermediate_population_adiabatic} follows directly from $\dot P_E=0$, without invoking the time-scale-separation condition.
Under the condition in Eq.~\eqref{eq:supp_AJC_population_adiabatic_condition}, this relation further reduces to
\begin{equation}
    P_E
    \simeq
    \frac{
        W_{\mathrm{AJC}}^{(m,q)}
    }{
        \gamma_{\mathrm a}
        +
        (m+q)\kappa_c
    }
    P_D
    \simeq
    \frac{
        W_{\mathrm{AJC}}^{(m,q)}
    }{
        \gamma_{\mathrm a}
        +
        (m+q)\kappa_c
    }
    P_m.
    \label{eq:supp_AJC_intermediate_population_fast}
\end{equation}

Substituting the eliminated intermediate-state population in Eq.~\eqref{eq:supp_AJC_intermediate_population_adiabatic} into the bosonic population-exchange equations in Eq.~\eqref{eq:supp_AJC_bosonic_population_exchange}, and using $P_D\simeq P_m$, yields the reduced bosonic population dynamics
\begin{equation}
    \begin{aligned}
        \left.
            \dot P_m
        \right|_{\mathrm{free+AJC+relax}}
        &\simeq
        -
        \mathcal R_{\mathrm{AJC}}(m,q)
        P_m,
        \\
        \left.
            \dot P_{m+q}
        \right|_{\mathrm{free+AJC+relax}}
        &\simeq
        \mathcal R_{\mathrm{AJC}}(m,q)
        P_m.
    \end{aligned}
    \label{eq:supp_AJC_reduced_population_dynamics}
\end{equation}
Here, the effective bosonic population-transfer rate between the bosonic Fock states $\ket{m}$ and $\ket{m+q}$ is defined as
\begin{equation}
    \mathcal R_{\mathrm{AJC}}(m,q)
    =
    \frac{
        W_{\mathrm{AJC}}^{(m,q)}
        \left[
            \gamma_{\mathrm a}
            +
            (m+q)\kappa_c
        \right]
    }{
        W_{\mathrm{AJC}}^{(m,q)}
        +
        \gamma_{\mathrm a}
        +
        (m+q)\kappa_c
    }.
    \label{eq:supp_AJC_effective_rate}
\end{equation}
Thus, $\mathcal R_{\mathrm{AJC}}(m,q)$ is the effective bosonic population-transfer rate from $\ket{m}$ to $\ket{m+q}$ in the reduced description where the auxiliary degree of freedom has been eliminated.

It is useful to distinguish this effective bosonic population-transfer rate from the rate at which population completes the specific sequential pathway
\begin{equation}
    \ket{D}
    \rightarrow
    \ket{E}
    \rightarrow
    \ket{B}.
\end{equation}
The latter is determined by the auxiliary-relaxation flux out of the intermediate state, given in Eq.~\eqref{eq:supp_AJC_B_relax_population} as $\left.\dot P_B\right|_{\mathrm{relax}}=\gamma_{\mathrm a}P_E$.
Using the adiabatically eliminated population $P_E$ and $P_D\simeq P_m$, this flux becomes
\begin{equation}
    \left.
        \dot P_B
    \right|_{\mathrm{relax}}
    \simeq
    \mathcal R_{\mathrm{AJC}}^{\mathrm{path}}(m,q)
    P_m,
\end{equation}
where
\begin{equation}
    \mathcal R_{\mathrm{AJC}}^{\mathrm{path}}(m,q)
    =
    \frac{
        \gamma_{\mathrm a}
        W_{\mathrm{AJC}}^{(m,q)}
    }{
        W_{\mathrm{AJC}}^{(m,q)}
        +
        \gamma_{\mathrm a}
        +
        (m+q)\kappa_c
    }.
    \label{eq:supp_AJC_path_rate}
\end{equation}
The two rates are therefore related by
\begin{equation}
    \mathcal R_{\mathrm{AJC}}^{\mathrm{path}}(m,q)
    =
    \frac{
        \gamma_{\mathrm a}
    }{
        \gamma_{\mathrm a}
        +
        (m+q)\kappa_c
    }
    \mathcal R_{\mathrm{AJC}}(m,q).
    \label{eq:supp_AJC_path_rate_relation}
\end{equation}
The prefactor
\begin{equation}
    \frac{
        \gamma_{\mathrm a}
    }{
        \gamma_{\mathrm a}
        +
        (m+q)\kappa_c
    }
\end{equation}
is the branching fraction of the irreversible decay from the intermediate state $\ket{E}$ that proceeds through auxiliary relaxation, rather than through the competing single-quantum-loss channel. 
The former transfers the population to $\ket{B}$ and thereby completes the desired sequential pathway $\ket{D}\rightarrow\ket{E}\rightarrow\ket{B}$.

Under the time-scale-separation condition in Eq.~\eqref{eq:supp_AJC_population_adiabatic_condition}, the effective bosonic population-transfer rate for the AJC realization reduces to
\begin{equation}
    \mathcal R_{\mathrm{AJC}}(m,q)
    \simeq
    W_{\mathrm{AJC}}^{(m,q)},
    \label{eq:supp_AJC_rate_fast}
\end{equation}
whereas the rate for completing the desired sequential pathway becomes
\begin{equation}
    \mathcal R_{\mathrm{AJC}}^{\mathrm{path}}(m,q)
    \simeq
    \frac{
        \gamma_{\mathrm a}
    }{
        \gamma_{\mathrm a}
        +
        (m+q)\kappa_c
    }
    W_{\mathrm{AJC}}^{(m,q)}.
    \label{eq:supp_AJC_path_rate_fast}
\end{equation}
For the AJC-assisted process to predominantly follow the desired pathway, auxiliary relaxation must further dominate the competing single-quantum loss of the intermediate state,
\begin{equation}
    \gamma_{\mathrm a}
    \gg
    (m+q)\kappa_c.
    \label{eq:supp_AJC_relaxation_dominance}
\end{equation}
In this limit, the branching fraction approaches unity and the two rates coincide to leading order,
\begin{equation}
    \mathcal R_{\mathrm{AJC}}^{\mathrm{path}}(m,q)
    \simeq
    \mathcal R_{\mathrm{AJC}}(m,q)
    \simeq
    W_{\mathrm{AJC}}^{(m,q)}.
    \label{eq:supp_AJC_rates_rapid_relaxation}
\end{equation}

We now specialize the effective bosonic population-transfer rate of the AJC realization to the resonant, relaxation-dominated regime, in which it can be expressed directly in terms of the AJC coupling strength and the auxiliary relaxation rate.
At resonance, the selected AJC transition satisfies $\delta_{\mathrm{AJC}}^{(m,q)}=0$.
If auxiliary relaxation further dominates the bosonic-mode loss contribution to the decay of the selected AJC coherence,
\begin{equation}
    \gamma_{\mathrm a}
    \gg
    (2m+q)\kappa_c,
    \label{eq:supp_AJC_relaxation_dominated_coherence}
\end{equation}
the coherence-decay rate given in Eq.~\eqref{eq:supp_AJC_total_coherence_decay} reduces to
\begin{equation}
        \Gamma_{\mathrm{AJC}}^{(m,q)}
        =
        \frac{
            \gamma_{\mathrm a}
            +
            (2m+q)\kappa_c
        }{2}
        \simeq
        \frac{\gamma_{\mathrm a}}{2}.
    \label{eq:supp_AJC_coherence_decay_rapid_relaxation}
\end{equation}
If pure dephasing is included, its contribution to $\Gamma_{\mathrm{AJC}}^{(m,q)}$ is also assumed to be small compared with $\gamma_{\mathrm a}/2$ in this limit.
Combining Eq.~\eqref{eq:supp_AJC_effective_exchange_rate} with the resonant condition and Eq.~\eqref{eq:supp_AJC_coherence_decay_rapid_relaxation}, the population-exchange rate becomes
\begin{equation}
    W_{\mathrm{AJC}}^{(m,q)}
    \simeq
    \frac{
        4
        \left|
            \Omega_{\mathrm{AJC}}^{(m,q)}
        \right|^2
    }{
        \gamma_{\mathrm a}
    }.
    \label{eq:supp_AJC_exchange_rate_resonance}
\end{equation}
Substituting Eq.~\eqref{eq:supp_AJC_exchange_rate_resonance} into the effective bosonic population-transfer rate defined in Eq.~\eqref{eq:supp_AJC_effective_rate} gives
\begin{equation}
    \begin{aligned}
        \mathcal R_{\mathrm{AJC}}(m,q)
        &\simeq
        \frac{
            4
            \left|
                \Omega_{\mathrm{AJC}}^{(m,q)}
            \right|^2
            \left[
                \gamma_{\mathrm a}
                +
                (m+q)\kappa_c
            \right]
        }{
            4
            \left|
                \Omega_{\mathrm{AJC}}^{(m,q)}
            \right|^2
            +
            \gamma_{\mathrm a}
            \left[
                \gamma_{\mathrm a}
                +
                (m+q)\kappa_c
            \right]
        }
        \\
        &\simeq
        \frac{
            4\gamma_{\mathrm a}
            \left|
                \Omega_{\mathrm{AJC}}^{(m,q)}
            \right|^2
        }{
            \gamma_{\mathrm a}^2
            +
            4
            \left|
                \Omega_{\mathrm{AJC}}^{(m,q)}
            \right|^2
        },
    \end{aligned}
    \label{eq:supp_AJC_eff_rate_resonance}
\end{equation}
where the second approximation follows from the relaxation-dominated condition in Eq.~\eqref{eq:supp_AJC_relaxation_dominated_coherence}.

The effective bosonic population-transfer rate in Eq.~\eqref{eq:supp_AJC_eff_rate_resonance} also provides a direct characterization of the population retained in the intermediate state.
For the stationary solution, Eq.~\eqref{eq:supp_AJC_intermediate_population_adiabatic} provides the relation between the intermediate-state population and the effective bosonic population-transfer rate, yielding
\begin{equation}
    \begin{aligned}
        \frac{P_E}{P_m}
        &\simeq
        \frac{
            W_{\mathrm{AJC}}^{(m,q)}
        }{
            W_{\mathrm{AJC}}^{(m,q)}
            +
            \gamma_{\mathrm a}
            +
            (m+q)\kappa_c
        }
        \\
        &=
        \frac{
            \mathcal R_{\mathrm{AJC}}(m,q)
        }{
            \gamma_{\mathrm a}
            +
            (m+q)\kappa_c
        }
        \\
        &\simeq
        \frac{
            4
            \left|
                \Omega_{\mathrm{AJC}}^{(m,q)}
            \right|^2
        }{
            \gamma_{\mathrm a}^2
            +
            4
            \left|
                \Omega_{\mathrm{AJC}}^{(m,q)}
            \right|^2
        }.
    \end{aligned}
    \label{eq:supp_AJC_intermediate_population_resonance}
\end{equation}
Here, the equality follows from the definition of $\mathcal R_{\mathrm{AJC}}(m,q)$ in Eq.~\eqref{eq:supp_AJC_effective_rate}, while the last approximation follows from Eqs.~\eqref{eq:supp_AJC_exchange_rate_resonance} and \eqref{eq:supp_AJC_relaxation_dominated_coherence}.

Equation~\eqref{eq:supp_AJC_intermediate_population_resonance} shows that the population retained in the intermediate state $\ket{E}$ decreases as the auxiliary relaxation rate $\gamma_{\mathrm a}$ increases.
Thus, increasing $\gamma_{\mathrm a}$ reduces the population available for coherent back-transfer from $\ket{E}$ to $\ket{D}$.
At the same time, auxiliary relaxation irreversibly transfers population from $\ket{E}$ to $\ket{B}=\ket{m+q;g_{\mathrm a}}$, which does not participate in the selected resonant AJC transition.
Consequently, population transferred to $\ket{B}$ is removed from the coherent AJC exchange, producing a net bosonic population current from $\ket{m}$ to $\ket{m+q}$ characterized by the effective bosonic population-transfer rate $\mathcal R_{\mathrm{AJC}}(m,q)$ in Eq.~\eqref{eq:supp_AJC_eff_rate_resonance}.
The selected AJC realization therefore generates an effectively directional population transfer
$\ket{m}\rightarrow\ket{m+q}$, with the directionality enhanced as the intermediate-state population is suppressed.

This enhanced directionality, however, does not imply a monotonic increase in the transfer speed.
Increasing the auxiliary relaxation rate $\gamma_{\mathrm a}$ also broadens the selected AJC coherence, and this effect is captured by the effective bosonic population-transfer rate in Eq.~\eqref{eq:supp_AJC_eff_rate_resonance}.
In particular, for $\gamma_{\mathrm a}\gg 2\left|\Omega_{\mathrm{AJC}}^{(m,q)}\right|$, the effective bosonic population-transfer rate decreases as
\begin{equation}
    \mathcal R_{\mathrm{AJC}}(m,q)
    \simeq
    \frac{
        4
        \left|
            \Omega_{\mathrm{AJC}}^{(m,q)}
        \right|^2
    }{
        \gamma_{\mathrm a}
    }
    \propto
    \frac{1}{\gamma_{\mathrm a}}.
    \label{eq:supp_AJC_eff_inverse_relaxation_scaling}
\end{equation}
Thus, auxiliary relaxation promotes directional AJC-assisted population transfer, but increasing $\gamma_{\mathrm a}$ beyond the required regime ultimately reduces the transfer rate through coherence broadening.

\subsection{Mapping onto the global and local connectors}
\label{subsec:supp_mapping}

We now apply the generic effective bosonic population-transfer rates derived above to the selective global and movable local connectors introduced in the main text.
The mapping is obtained by specifying the lower Fock-state index $m$, the transition order $q$, and the corresponding auxiliary-system parameters for each connector.

For the global connector, the selected transition is $\ket{0}\leftrightarrow\ket{N_s}$, corresponding to
\begin{equation}
    m=0,
    \qquad
    q=N_s.
    \label{eq:supp_global_mapping}
\end{equation}
The corresponding coherent matrix elements are therefore
\begin{align}
    \Omega_{\mathrm{JC}}^{(0,N_s)}
    &=
    g_{G,\mathrm{JC}}^{(N_s)}
    \sqrt{N_s!},
    \\
    \Omega_{\mathrm{AJC}}^{(0,N_s)}
    &=
    g_{G,\mathrm{AJC}}^{(N_s)}
    \sqrt{N_s!},
    \\
    \Omega_{\mathrm{drv}}^{(0,N_s)}
    &=
    g_{G,\mathrm{drv}}^{(N_s)}
    \sqrt{N_s!}.
    \label{eq:supp_global_matrix_elements}
\end{align}

For the pumped $N_s$-JC realization, the auxiliary pumping and relaxation rates in the generic JC description are identified as $\wp_{\mathrm a}\rightarrow\wp_s$ and $\gamma_{\mathrm a}\rightarrow\gamma_s$.
We denote the corresponding effective bosonic population-transfer rates of the global connector by $\mathcal R_{\mathrm{G,JC}}^{\uparrow}$ and $\mathcal R_{\mathrm{G,JC}}^{\downarrow}$, for the transitions $\ket{0}\rightarrow\ket{N_s}$ and $\ket{N_s}\rightarrow\ket{0}$, respectively.
Under the mapping $m=0$ and $q=N_s$ in Eq.~\eqref{eq:supp_global_mapping}, these rates are identified with $\mathcal R_{\mathrm{JC}}^{\uparrow}(0,N_s)$ and $\mathcal R_{\mathrm{JC}}^{\downarrow}(0,N_s)$.
Substituting this mapping into the general rates in Eq.~\eqref{eq:supp_JC_general_rates}, together with the stationary auxiliary populations in Eq.~\eqref{eq:supp_auxiliary_stationary_populations}, gives
\begin{equation}
    \begin{aligned}
        \mathcal R_{\mathrm{G,JC}}^{\uparrow}
        &=
        \mathcal R_{\mathrm{JC}}^{\uparrow}(0,N_s)
        =
        W_{\mathrm{JC}}^{(0,N_s)}
        \frac{
            \wp_s
        }{
            \wp_s
            +
            \gamma_s
        },
        \\
        \mathcal R_{\mathrm{G,JC}}^{\downarrow}
        &=
        \mathcal R_{\mathrm{JC}}^{\downarrow}(0,N_s)
        =
        W_{\mathrm{JC}}^{(0,N_s)}
        \frac{
            \gamma_s
        }{
            \wp_s
            +
            \gamma_s
        }.
    \end{aligned}
    \label{eq:supp_global_JC_general_rates}
\end{equation}
Their ratio is therefore
\begin{equation}
    \frac{
        \mathcal R_{\mathrm{G,JC}}^{\uparrow}
    }{
        \mathcal R_{\mathrm{G,JC}}^{\downarrow}
    }
    =
    \frac{
        \wp_s
    }{
        \gamma_s
    }.
    \label{eq:supp_global_JC_directionality}
\end{equation}
The corresponding population-exchange rate follows from Eq.~\eqref{eq:supp_JC_effective_exchange_rate}.
Under the global mapping, it becomes
\begin{equation}
    W_{\mathrm{JC}}^{(0,N_s)}
    =
    \frac{
        2N_s!
        \left|
            g_{G,\mathrm{JC}}^{(N_s)}
        \right|^2
        \Gamma_{\mathrm{JC}}^{(0,N_s)}
    }{
        \left[
            \Gamma_{\mathrm{JC}}^{(0,N_s)}
        \right]^2
        +
        \left[
            \delta_{\mathrm{JC}}^{(0,N_s)}
        \right]^2
    }.
    \label{eq:supp_global_JC_exchange_rate}
\end{equation}
Here, $\delta_{\mathrm{JC}}^{(0,N_s)}$ and $\Gamma_{\mathrm{JC}}^{(0,N_s)}$ are, respectively, the detuning and total coherence-decay rate of the selected global JC transition $\ket{0;e_s}\leftrightarrow\ket{N_s;g_s}$.
At resonance, the selected global JC transition satisfies $\delta_{\mathrm{JC}}^{(0,N_s)}=0$.
For the pumped $N_s$-JC realization, the corresponding total coherence-decay rate follows from Eq.~\eqref{eq:supp_JC_total_coherence_decay} as
\begin{equation}
    \Gamma_{\mathrm{JC}}^{(0,N_s)}
    =
    \frac{
        \wp_s
        +
        \gamma_s
        +
        N_s\kappa_c
    }{2}.
    \label{eq:supp_global_JC_coherence_decay}
\end{equation}
Substituting this resonant condition and Eq.~\eqref{eq:supp_global_JC_coherence_decay} into Eq.~\eqref{eq:supp_global_JC_general_rates} gives
\begin{align}
    \mathcal R_{\mathrm{G,JC}}^{\uparrow}
    &=
    \frac{
        4N_s!
        \left|
            g_{G,\mathrm{JC}}^{(N_s)}
        \right|^2
        \wp_s
    }{
        \left(
            \wp_s
            +
            \gamma_s
        \right)
        \left[
            \wp_s
            +
            \gamma_s
            +
            N_s\kappa_c
        \right]
    },
    \label{eq:supp_global_JC_up}
    \\
    \mathcal R_{\mathrm{G,JC}}^{\downarrow}
    &=
    \frac{
        4N_s!
        \left|
            g_{G,\mathrm{JC}}^{(N_s)}
        \right|^2
        \gamma_s
    }{
        \left(
            \wp_s
            +
            \gamma_s
        \right)
        \left[
            \wp_s
            +
            \gamma_s
            +
            N_s\kappa_c
        \right]
    }.
    \label{eq:supp_global_JC_down}
\end{align}

For the $N_s$-quantum-drive realization, we denote the corresponding effective bosonic population-transfer rates of the global connector by $\mathcal R_{\mathrm{G,drv}}^{\uparrow}$ and $\mathcal R_{\mathrm{G,drv}}^{\downarrow}$, for the transitions $\ket{0}\rightarrow\ket{N_s}$ and $\ket{N_s}\rightarrow\ket{0}$, respectively.
Under the mapping $m=0$ and $q=N_s$ in Eq.~\eqref{eq:supp_global_mapping}, these rates are identified with $\mathcal R_{\mathrm{drv}}^{\uparrow}(0,N_s)$ and $\mathcal R_{\mathrm{drv}}^{\downarrow}(0,N_s)$.
Equation~\eqref{eq:supp_coherent_symmetric_rates} then gives
\begin{equation}
    \mathcal R_{\mathrm{G,drv}}^{\uparrow}
    =
    \mathcal R_{\mathrm{drv}}^{\uparrow}(0,N_s)
    =
    \mathcal R_{\mathrm{G,drv}}^{\downarrow}
    =
    \mathcal R_{\mathrm{drv}}^{\downarrow}(0,N_s)
    =
    W_{\mathrm{drv}}^{(0,N_s)}.
    \label{eq:supp_global_drive_rates}
\end{equation}
Thus, unlike the pumped $N_s$-JC realization, the coherent-drive realization produces equal effective bosonic population-transfer rates in the two directions.

Substituting the global mapping into the population-exchange rate in Eq.~\eqref{eq:supp_coherent_effective_exchange_rate} gives
\begin{equation}
    W_{\mathrm{drv}}^{(0,N_s)}
    =
    \frac{
        2N_s!
        \left|
            g_{G,\mathrm{drv}}^{(N_s)}
        \right|^2
        \Gamma_{\mathrm{drv}}^{(0,N_s)}
    }{
        \left[
            \Gamma_{\mathrm{drv}}^{(0,N_s)}
        \right]^2
        +
        \left[
            \delta_{\mathrm{drv}}^{(0,N_s)}
        \right]^2
    }.
    \label{eq:supp_global_drive_exchange_rate}
\end{equation}
Here, $\delta_{\mathrm{drv}}^{(0,N_s)}$ and $\Gamma_{\mathrm{drv}}^{(0,N_s)}$ are, respectively, the detuning and coherence-decay rate of the selected global transition $\ket{0}\leftrightarrow\ket{N_s}$.
At resonance, the selected global transition satisfies $\delta_{\mathrm{drv}}^{(0,N_s)}=0$.
The corresponding coherence-decay rate follows from Eq.~\eqref{eq:supp_coherent_total_coherence_decay} as
\begin{equation}
    \Gamma_{\mathrm{drv}}^{(0,N_s)}
    =
    \frac{
        N_s\kappa_c
    }{2}.
    \label{eq:supp_global_drive_coherence_decay}
\end{equation}

For the $N_s$-AJC+relaxation realization, the auxiliary relaxation rate in the generic AJC description is identified as $\gamma_{\mathrm a}\rightarrow\gamma_s$.
We denote the corresponding effective bosonic population-transfer rates of the global connector by $\mathcal R_{\mathrm{G,AJC}}^{\uparrow}$ and $\mathcal R_{\mathrm{G,AJC}}^{\downarrow}$, for the transitions $\ket{0}\rightarrow\ket{N_s}$ and $\ket{N_s}\rightarrow\ket{0}$, respectively.
Under the mapping $m=0$ and $q=N_s$ in Eq.~\eqref{eq:supp_global_mapping}, the forward rate is identified with $\mathcal R_{\mathrm{AJC}}(0,N_s)$, while the reduced bosonic dynamics in Eq.~\eqref{eq:supp_AJC_reduced_population_dynamics} give $\mathcal R_{\mathrm{G,AJC}}^{\downarrow}\simeq0$.
Substituting the global mapping into the effective bosonic population-transfer rate in Eq.~\eqref{eq:supp_AJC_effective_rate} gives
\begin{equation}
    \begin{aligned}
        \mathcal R_{\mathrm{G,AJC}}^{\uparrow}
        &=
        \mathcal R_{\mathrm{AJC}}(0,N_s)
        \\
        &=
        \frac{
            W_{\mathrm{AJC}}^{(0,N_s)}
            \left[
                \gamma_s
                +
                N_s\kappa_c
            \right]
        }{
            W_{\mathrm{AJC}}^{(0,N_s)}
            +
            \gamma_s
            +
            N_s\kappa_c
        },
        \\
        \mathcal R_{\mathrm{G,AJC}}^{\downarrow}
        &\simeq
        0.
    \end{aligned}
    \label{eq:supp_global_AJC_rates}
\end{equation}
The corresponding population-exchange rate follows from Eq.~\eqref{eq:supp_AJC_effective_exchange_rate}.
Under the global mapping, it becomes
\begin{equation}
    W_{\mathrm{AJC}}^{(0,N_s)}
    =
    \frac{
        2N_s!
        \left|
            g_{G,\mathrm{AJC}}^{(N_s)}
        \right|^2
        \Gamma_{\mathrm{AJC}}^{(0,N_s)}
    }{
        \left[
            \Gamma_{\mathrm{AJC}}^{(0,N_s)}
        \right]^2
        +
        \left[
            \delta_{\mathrm{AJC}}^{(0,N_s)}
        \right]^2
    }.
    \label{eq:supp_global_AJC_exchange_rate}
\end{equation}
Here, $\delta_{\mathrm{AJC}}^{(0,N_s)}$ and $\Gamma_{\mathrm{AJC}}^{(0,N_s)}$ are, respectively, the detuning and total coherence-decay rate of the selected global AJC transition $\ket{0;g_s}\leftrightarrow\ket{N_s;e_s}$.
At resonance, the selected global AJC transition satisfies $\delta_{\mathrm{AJC}}^{(0,N_s)}=0$.
For the $N_s$-AJC+relaxation realization, the corresponding total coherence-decay rate follows from Eq.~\eqref{eq:supp_AJC_total_coherence_decay} as
\begin{equation}
    \Gamma_{\mathrm{AJC}}^{(0,N_s)}
    =
    \frac{
        \gamma_s
        +
        N_s\kappa_c
    }{2}.
    \label{eq:supp_global_AJC_coherence_decay}
\end{equation}
The resonant population-exchange rate is therefore
\begin{equation}
    W_{\mathrm{AJC}}^{(0,N_s)}
    =
    \frac{
        4N_s!
        \left|
            g_{G,\mathrm{AJC}}^{(N_s)}
        \right|^2
    }{
        \gamma_s
        +
        N_s\kappa_c
    }.
    \label{eq:supp_global_AJC_exchange_rate_resonance}
\end{equation}
Substituting Eq.~\eqref{eq:supp_global_AJC_exchange_rate_resonance} into Eq.~\eqref{eq:supp_global_AJC_rates} gives the explicit upward effective bosonic population-transfer rate
\begin{equation}
    \mathcal R_{\mathrm{G,AJC}}^{\uparrow}
    =
    \frac{
        4N_s!
        \left|
            g_{G,\mathrm{AJC}}^{(N_s)}
        \right|^2
        \left[
            \gamma_s
            +
            N_s\kappa_c
        \right]
    }{
        4N_s!
        \left|
            g_{G,\mathrm{AJC}}^{(N_s)}
        \right|^2
        +
        \left[
            \gamma_s
            +
            N_s\kappa_c
        \right]^2
    }.
    \label{eq:supp_global_AJC_effective_rate}
\end{equation}

For the movable local connector, the selected transition is $\ket{R-1}\leftrightarrow\ket{R}$, so that
\begin{equation}
    m=R-1,
    \qquad
    q=1.
    \label{eq:supp_local_mapping}
\end{equation}
The corresponding matrix elements are
\begin{align}
    \Omega_{\mathrm{JC}}^{(R-1,1)}
    &=
    g_{L,\mathrm{JC}}^{(1)}
    \sqrt{R},
    \\
    \Omega_{\mathrm{AJC}}^{(R-1,1)}
    &=
    g_{L,\mathrm{AJC}}^{(1)}
    \sqrt{R}.
    \label{eq:supp_local_matrix_elements}
\end{align}

For the pumped-JC realization of the movable local connector, the auxiliary pumping and relaxation rates in the generic JC description are identified as $\wp_{\mathrm a}\rightarrow\wp_R$ and $\gamma_{\mathrm a}\rightarrow\gamma_R$.
We denote the corresponding effective bosonic population-transfer rates of the local connector by $\mathcal R_{\mathrm{L,JC}}^{\uparrow}(R)$ and $\mathcal R_{\mathrm{L,JC}}^{\downarrow}(R)$, for the transitions $\ket{R-1}\rightarrow\ket{R}$ and $\ket{R}\rightarrow\ket{R-1}$, respectively.
Under the mapping $m=R-1$ and $q=1$ in Eq.~\eqref{eq:supp_local_mapping}, these rates are identified with $\mathcal R_{\mathrm{JC}}^{\uparrow}(R-1,1)$ and $\mathcal R_{\mathrm{JC}}^{\downarrow}(R-1,1)$.
Substituting this mapping into the general rates in Eq.~\eqref{eq:supp_JC_general_rates}, together with the stationary auxiliary populations in Eq.~\eqref{eq:supp_auxiliary_stationary_populations}, gives
\begin{equation}
    \begin{aligned}
        \mathcal R_{\mathrm{L,JC}}^{\uparrow}(R)
        &=
        \mathcal R_{\mathrm{JC}}^{\uparrow}(R-1,1)
        =
        W_{\mathrm{JC}}^{(R-1,1)}
        \frac{
            \wp_R
        }{
            \wp_R
            +
            \gamma_R
        },
        \\
        \mathcal R_{\mathrm{L,JC}}^{\downarrow}(R)
        &=
        \mathcal R_{\mathrm{JC}}^{\downarrow}(R-1,1)
        =
        W_{\mathrm{JC}}^{(R-1,1)}
        \frac{
            \gamma_R
        }{
            \wp_R
            +
            \gamma_R
        }.
    \end{aligned}
    \label{eq:supp_local_JC_general_rates}
\end{equation}
Their ratio is therefore
\begin{equation}
    \frac{
        \mathcal R_{\mathrm{L,JC}}^{\uparrow}(R)
    }{
        \mathcal R_{\mathrm{L,JC}}^{\downarrow}(R)
    }
    =
    \frac{
        \wp_R
    }{
        \gamma_R
    }.
    \label{eq:supp_local_JC_directionality}
\end{equation}
The corresponding population-exchange rate follows from Eq.~\eqref{eq:supp_JC_effective_exchange_rate}.
Under the local mapping, it becomes
\begin{equation}
    W_{\mathrm{JC}}^{(R-1,1)}
    =
    \frac{
        2R
        \left|
            g_{L,\mathrm{JC}}^{(1)}
        \right|^2
        \Gamma_{\mathrm{JC}}^{(R-1,1)}
    }{
        \left[
            \Gamma_{\mathrm{JC}}^{(R-1,1)}
        \right]^2
        +
        \left[
            \delta_{\mathrm{JC}}^{(R-1,1)}
        \right]^2
    }.
    \label{eq:supp_local_JC_exchange_rate}
\end{equation}
Here, $\delta_{\mathrm{JC}}^{(R-1,1)}$ and $\Gamma_{\mathrm{JC}}^{(R-1,1)}$ are, respectively, the detuning and total coherence-decay rate of the selected local JC transition $\ket{R-1;e_R}\leftrightarrow\ket{R;g_R}$.
At resonance, the selected local JC transition satisfies $\delta_{\mathrm{JC}}^{(R-1,1)}=0$.
For the pumped-JC realization of the movable local connector, the corresponding total coherence-decay rate follows from Eq.~\eqref{eq:supp_JC_total_coherence_decay} as
\begin{equation}
    \Gamma_{\mathrm{JC}}^{(R-1,1)}
    =
    \frac{
        \wp_R
        +
        \gamma_R
        +
        (2R-1)\kappa_c
    }{2}.
    \label{eq:supp_local_JC_coherence_decay}
\end{equation}
Substituting this resonant condition and Eq.~\eqref{eq:supp_local_JC_coherence_decay} into Eq.~\eqref{eq:supp_local_JC_general_rates} gives the upward effective bosonic population-transfer rate
\begin{equation}
    \begin{aligned}
        \mathcal R_{\mathrm{L,JC}}^{\uparrow}(R)
        &=
        \frac{
            4R
            \left|
                g_{L,\mathrm{JC}}^{(1)}
            \right|^2
            \wp_R
        }{
            \left(
                \wp_R
                +
                \gamma_R
            \right)
            \left[
                \wp_R
                +
                \gamma_R
                +
                (2R-1)\kappa_c
            \right]
        }.
    \end{aligned}
    \label{eq:supp_local_JC_up}
\end{equation}
The corresponding downward effective bosonic population-transfer rate is
\begin{equation}
    \begin{aligned}
        \mathcal R_{\mathrm{L,JC}}^{\downarrow}(R)
        &=
        \frac{
            4R
            \left|
                g_{L,\mathrm{JC}}^{(1)}
            \right|^2
            \gamma_R
        }{
            \left(
                \wp_R
                +
                \gamma_R
            \right)
            \left[
                \wp_R
                +
                \gamma_R
                +
                (2R-1)\kappa_c
            \right]
        }.
    \end{aligned}
    \label{eq:supp_local_JC_down}
\end{equation}

For the AJC+relaxation realization of the movable local connector, the auxiliary relaxation rate in the generic AJC description is identified as $\gamma_{\mathrm a}\rightarrow\gamma_R$.
We denote the corresponding effective bosonic population-transfer rates of the local connector by $\mathcal R_{\mathrm{L,AJC}}^{\uparrow}(R)$ and $\mathcal R_{\mathrm{L,AJC}}^{\downarrow}(R)$, for the transitions $\ket{R-1}\rightarrow\ket{R}$ and $\ket{R}\rightarrow\ket{R-1}$, respectively.
Under the mapping $m=R-1$ and $q=1$ in Eq.~\eqref{eq:supp_local_mapping}, the forward rate is identified with $\mathcal R_{\mathrm{AJC}}(R-1,1)$, while the reduced bosonic dynamics in Eq.~\eqref{eq:supp_AJC_reduced_population_dynamics} give $\mathcal R_{\mathrm{L,AJC}}^{\downarrow}(R)\simeq0$.
Substituting the local mapping into the effective bosonic population-transfer rate in Eq.~\eqref{eq:supp_AJC_effective_rate} gives
\begin{equation}
    \begin{aligned}
        \mathcal R_{\mathrm{L,AJC}}^{\uparrow}(R)
        &=
        \mathcal R_{\mathrm{AJC}}(R-1,1)
        \\
        &=
        \frac{
            W_{\mathrm{AJC}}^{(R-1,1)}
            \left[
                \gamma_R
                +
                R\kappa_c
            \right]
        }{
            W_{\mathrm{AJC}}^{(R-1,1)}
            +
            \gamma_R
            +
            R\kappa_c
        },
        \\
        \mathcal R_{\mathrm{L,AJC}}^{\downarrow}(R)
        &\simeq
        0.
    \end{aligned}
    \label{eq:supp_local_AJC_rates}
\end{equation}
The corresponding population-exchange rate follows from Eq.~\eqref{eq:supp_AJC_effective_exchange_rate}.
Under the local mapping, it becomes
\begin{equation}
    W_{\mathrm{AJC}}^{(R-1,1)}
    =
    \frac{
        2R
        \left|
            g_{L,\mathrm{AJC}}^{(1)}
        \right|^2
        \Gamma_{\mathrm{AJC}}^{(R-1,1)}
    }{
        \left[
            \Gamma_{\mathrm{AJC}}^{(R-1,1)}
        \right]^2
        +
        \left[
            \delta_{\mathrm{AJC}}^{(R-1,1)}
        \right]^2
    }.
    \label{eq:supp_local_AJC_exchange_rate}
\end{equation}
Here, $\delta_{\mathrm{AJC}}^{(R-1,1)}$ and $\Gamma_{\mathrm{AJC}}^{(R-1,1)}$ are, respectively, the detuning and total coherence-decay rate of the selected local AJC transition $\ket{R-1;g_R}\leftrightarrow\ket{R;e_R}$.
At resonance, the selected local AJC transition satisfies $\delta_{\mathrm{AJC}}^{(R-1,1)}=0$.
For the AJC+relaxation realization of the movable local connector, the corresponding total coherence-decay rate follows from Eq.~\eqref{eq:supp_AJC_total_coherence_decay} as
\begin{equation}
    \Gamma_{\mathrm{AJC}}^{(R-1,1)}
    =
    \frac{
        \gamma_R
        +
        (2R-1)\kappa_c
    }{2}.
    \label{eq:supp_local_AJC_coherence_decay}
\end{equation}
The resonant population-exchange rate is therefore
\begin{equation}
    W_{\mathrm{AJC}}^{(R-1,1)}
    =
    \frac{
        4R
        \left|
            g_{L,\mathrm{AJC}}^{(1)}
        \right|^2
    }{
        \gamma_R
        +
        (2R-1)\kappa_c
    }.
    \label{eq:supp_local_AJC_exchange_rate_resonance}
\end{equation}
Substituting Eq.~\eqref{eq:supp_local_AJC_exchange_rate_resonance} into Eq.~\eqref{eq:supp_local_AJC_rates} gives the explicit upward effective bosonic population-transfer rate
\begin{equation}
    \begin{aligned}
        \mathcal R_{\mathrm{L,AJC}}^{\uparrow}(R)
        &=
        \frac{
            4R
            \left|
                g_{L,\mathrm{AJC}}^{(1)}
            \right|^2
            \left[
                \gamma_R
                +
                R\kappa_c
            \right]
        }{
            4R
            \left|
                g_{L,\mathrm{AJC}}^{(1)}
            \right|^2
            +
            \left[
                \gamma_R
                +
                (2R-1)\kappa_c
            \right]
            \left[
                \gamma_R
                +
                R\kappa_c
            \right]
        }.
    \end{aligned}
    \label{eq:supp_local_AJC_effective_rate}
\end{equation}

For the global and local AJC+relaxation realizations considered above, the identification of the corresponding effective bosonic population-transfer rates is based on the reduced bosonic dynamics in Eq.~\eqref{eq:supp_AJC_reduced_population_dynamics}, which should be understood as a reduced-rate representation of the net bosonic population current obtained after eliminating the auxiliary degree of freedom.
Accordingly, the vanishing reverse rates $\mathcal R_{\mathrm{G,AJC}}^{\downarrow}\simeq0$ and $\mathcal R_{\mathrm{L,AJC}}^{\downarrow}(R)\simeq0$ indicate the absence of an independent reverse term in this reduced bosonic description, rather than the complete absence of microscopic coherent back-transfer within the selected AJC transition.

Importantly, the effective global and local rates derived above are valid either for the time-dependent dynamics under the corresponding adiabatic-elimination conditions or for the stationary state.
For the stationary solution of the reduced spectrally selected model, the corresponding adiabatic-elimination conditions are not required, because the eliminated dynamical variables directly satisfy their stationary equations.

To formulate the subsequent bosonic population dynamics in a unified form across the different microscopic implementations, we introduce unified notation for the effective connector rates.
For a given realization of the global connector, we define $\mathcal R_{\mathrm G}^{d}$ with $d\in\{\uparrow,\downarrow\}$ as
\begin{equation}
    \mathcal R_{\mathrm G}^{d}
    \equiv
    \begin{cases}
        \mathcal R_{\mathrm{G,JC}}^{d},
        & \text{pumped $N_s$-JC realization},
        \\
        \mathcal R_{\mathrm{G,drv}}^{d},
        & \text{$N_s$-quantum-drive realization},
        \\
        \mathcal R_{\mathrm{G,AJC}}^{d},
        & \text{$N_s$-AJC+relaxation realization}.
    \end{cases}
    \label{eq:supp_global_effective_rate_definition}
\end{equation}
Similarly, for a given realization of the movable local connector, we define
$\mathcal R_{\mathrm L}^{d}(R)$ with
$d\in\{\uparrow,\downarrow\}$ as
\begin{equation}
    \mathcal R_{\mathrm L}^{d}(R)
    \equiv
    \begin{cases}
        \mathcal R_{\mathrm{L,JC}}^{d}(R),
        & \text{pumped-JC realization},
        \\
        \mathcal R_{\mathrm{L,AJC}}^{d}(R),
        & \text{AJC+relaxation realization}.
    \end{cases}
    \label{eq:supp_local_effective_rate_definition}
\end{equation}
Here, the superscripts $\uparrow$ and $\downarrow$ denote population transfer along and opposite to the directions $\ket{0}\rightarrow\ket{N_s}$ for the global connector and $\ket{R-1}\rightarrow\ket{R}$ for the movable local connector, respectively.
The three global realizations and two local realizations admit six microscopic combinations. The selected transitions and allowed target positions are specified in the main text (Fig.~1d,e).
Within the reduced bosonic population description, for any allowed microscopic combination, the microscopic realization enters through the corresponding values of $\mathcal R_{\mathrm G}^{\uparrow}$, $\mathcal R_{\mathrm G}^{\downarrow}$, $\mathcal R_{\mathrm L}^{\uparrow}(R)$, and $\mathcal R_{\mathrm L}^{\downarrow}(R)$, so that the population equations need not be rederived for each combination.

\section{Effective connector dynamics with dispersive spectral selection}
\label{sec:supp_dispersive_connectors}

Supplementary Sec.~\ref{sec:supp_effective_connectors} derived the effective connector dynamics under a Kerr-resolved spectral-selection mechanism, in which the intrinsic anharmonicity of the bosonic mode spectrally isolates the desired Fock-state transition.
Here we show that the same Fock-space-current architecture can also be realized for an intrinsically harmonic bosonic mode, $K_c=0$, with the required Fock-state selectivity instead generated by dispersive coupling to the auxiliary TLSs participating in the global and local connectors.
The resulting dispersively dressed joint boson--TLS level structure gives rise to Fock-state-dependent transition frequencies, thereby allowing the desired transition to be spectrally isolated.
Such a dispersive realization is particularly relevant to bosonic platforms in which the intrinsic Kerr nonlinearity is weak.
For example, the intrinsic Kerr nonlinearity of magnon modes can be extremely small, making direct Kerr-based spectral resolution of neighboring Fock-state transitions challenging~\cite{Zhang2019Kerr,Weng2026}.

The dispersive construction considered here applies to the TLS-assisted microscopic mechanisms based on an incoherently pumped $q$-quantum JC coupling and a $q$-quantum AJC coupling with fast auxiliary-TLS relaxation.
The direct coherent $N_s$-quantum-drive realization of the global connector considered in the main text is not included here, since its Fock-state selectivity in the present architecture relies on the intrinsic Kerr anharmonicity of the bosonic mode.

In the dispersive construction, the transition frequencies associated with one auxiliary TLS generally depend on the state of the other TLS through the corresponding dispersive shifts.
We therefore distinguish the TLS directly involved in a given selected transition as the active TLS and the other TLS as the spectator TLS, and retain the spectator-state dependence explicitly in the derivation below.
We then derive the corresponding dispersive resonance conditions and effective transition rates and show how they yield the same effective connector structure introduced in Supplementary Sec.~\ref{sec:supp_effective_connectors}.

\subsection{Dispersive spectral selection of a generic Fock-state transition}
\label{subsec:supp_disp_generic_selection}

We first establish the spectral-selection mechanism generated by dispersive coupling for a generic $q$-quantum Fock-state transition $\ket{m}\leftrightarrow\ket{m+q}$, where $m$ denotes the Fock number of the lower state and $q\geq1$ is the number of bosonic excitations exchanged in the selected transition.

More specifically, the TLS whose state is changed by the selected $q$-quantum JC or AJC interaction is referred to as the active TLS, whereas the second TLS, whose state remains unchanged during that transition, is referred to as the spectator TLS.
We denote their ground and excited states by $\ket{g_{\mathrm{act}}}$, $\ket{e_{\mathrm{act}}}$, and $\ket{g_{\mathrm{sp}}}$, $\ket{e_{\mathrm{sp}}}$, respectively.
The corresponding raising and lowering operators are
\begin{equation}
    \hat{\sigma}_{\mathrm{act}}^{+}
    =
    \ket{e_{\mathrm{act}}}
    \bra{g_{\mathrm{act}}},
    \qquad
    \hat{\sigma}_{\mathrm{act}}^{-}
    =
    \ket{g_{\mathrm{act}}}
    \bra{e_{\mathrm{act}}},
\end{equation}
for the active TLS, and
\begin{equation}
    \hat{\sigma}_{\mathrm{sp}}^{+}
    =
    \ket{e_{\mathrm{sp}}}
    \bra{g_{\mathrm{sp}}},
    \qquad
    \hat{\sigma}_{\mathrm{sp}}^{-}
    =
    \ket{g_{\mathrm{sp}}}
    \bra{e_{\mathrm{sp}}},
\end{equation}
for the spectator TLS.
The Pauli-$z$ operators of the active and spectator TLSs are defined,
respectively, as
\begin{equation}
    \hat{\sigma}_{\mathrm{act}}^{z}
    =
    \ket{e_{\mathrm{act}}}
    \bra{e_{\mathrm{act}}}
    -
    \ket{g_{\mathrm{act}}}
    \bra{g_{\mathrm{act}}},
\end{equation}
and
\begin{equation}
    \hat{\sigma}_{\mathrm{sp}}^{z}
    =
    \ket{e_{\mathrm{sp}}}
    \bra{e_{\mathrm{sp}}}
    -
    \ket{g_{\mathrm{sp}}}
    \bra{g_{\mathrm{sp}}}.
\end{equation}
The eigenvalues of $\hat{\sigma}_{\mathrm{act}}^{z}$ and $\hat{\sigma}_{\mathrm{sp}}^{z}$ are denoted by $z_{\mathrm{act}}$ and $z_{\mathrm{sp}}$, respectively, with
\begin{equation}
    z_{\mathrm{act}},z_{\mathrm{sp}}
    \in
    \{-1,+1\}.
\end{equation}
Here $z_{\mathrm{act}}=+1$ ($-1$) corresponds to $\ket{e_{\mathrm{act}}}$ ($\ket{g_{\mathrm{act}}}$), while $z_{\mathrm{sp}}=+1$ ($-1$) corresponds to $\ket{e_{\mathrm{sp}}}$ ($\ket{g_{\mathrm{sp}}}$).

Let $\omega_{\mathrm{act}}$ and $\omega_{\mathrm{sp}}$ denote the bare transition frequencies of the active and spectator TLSs, respectively.
For the $q$-quantum transition considered below, the active TLS is described in a frame rotating at $q\omega_r$, with detuning
\begin{equation}
    \Delta_{\mathrm{act}}
    =
    \omega_{\mathrm{act}}
    -
    q\omega_r.
\end{equation}
We denote by $\omega_{\mathrm{sp}}^{\mathrm{rf}}$ the rotation frequency of the frame used for the spectator TLS; its explicit form is fixed by the corresponding global or local realization specified below.
The spectator-TLS detuning is then defined as
\begin{equation}
    \Delta_{\mathrm{sp}}
    =
    \omega_{\mathrm{sp}}
    -
    \omega_{\mathrm{sp}}^{\mathrm{rf}}.
\end{equation}
The dispersive coupling strengths of the active and spectator TLSs to the bosonic mode are denoted by $\chi_{\mathrm{act}}$ and $\chi_{\mathrm{sp}}$, respectively.
The rotating-frame Hamiltonian governing the dispersively dressed level structure is therefore
\begin{align}
    \hat{H}_{0}^{\mathrm{disp}}
    ={}&
    \Delta_c\hat{n}_c
    +
    \Delta_{\mathrm{act}}
    \hat{\sigma}_{\mathrm{act}}^{+}
    \hat{\sigma}_{\mathrm{act}}^{-}
    +
    \Delta_{\mathrm{sp}}
    \hat{\sigma}_{\mathrm{sp}}^{+}
    \hat{\sigma}_{\mathrm{sp}}^{-}
    \nonumber\\
    &+
    \chi_{\mathrm{act}}
    \hat{n}_c
    \hat{\sigma}_{\mathrm{act}}^{z}
    +
    \chi_{\mathrm{sp}}
    \hat{n}_c
    \hat{\sigma}_{\mathrm{sp}}^{z}.
    \label{eq:supp_disp_generic_H0}
\end{align}

The eigenstates of $\hat{H}_{0}^{\mathrm{disp}}$ can be written as
\begin{equation}
    \ket{n;z_{\mathrm{act}},z_{\mathrm{sp}}}
    =
    \ket{n}_c
    \otimes
    \ket{z_{\mathrm{act}}}_{\mathrm{act}}
    \otimes
    \ket{z_{\mathrm{sp}}}_{\mathrm{sp}},
\end{equation}
where $\ket{n}_c$ denotes the $n$-th Fock state of the bosonic mode, and $\ket{z_{\mathrm{act}}}_{\mathrm{act}}$ and $\ket{z_{\mathrm{sp}}}_{\mathrm{sp}}$ denote the eigenstates of $\hat{\sigma}_{\mathrm{act}}^{z}$ and $\hat{\sigma}_{\mathrm{sp}}^{z}$ corresponding to the eigenvalues $z_{\mathrm{act}}$ and $z_{\mathrm{sp}}$, respectively.
The corresponding rotating-frame eigenenergy is
\begin{align}
    \mathcal E_n
    (z_{\mathrm{act}},z_{\mathrm{sp}})
    ={}&
    n\Delta_c
    +
    \frac{1+z_{\mathrm{act}}}{2}
    \Delta_{\mathrm{act}}
    +
    \frac{1+z_{\mathrm{sp}}}{2}
    \Delta_{\mathrm{sp}}
    \nonumber\\
    &+
    nz_{\mathrm{act}}\chi_{\mathrm{act}}
    +
    nz_{\mathrm{sp}}\chi_{\mathrm{sp}}.
    \label{eq:supp_disp_generic_energy}
\end{align}

We now use Eq.~\eqref{eq:supp_disp_generic_energy} to determine how the dispersive interactions distinguish different $q$-quantum Fock-state transitions.

\subsubsection{JC-assisted spectral selection}
\label{subsubsec:supp_disp_generic_JC_selection}

We first consider a generic $q$-quantum JC interaction,
\begin{equation}
    \hat{H}_{\mathrm{JC}}^{(q)}
    =
    g_{\mathrm{JC}}^{(q)}
    \left[
        (\hat{a}_c^{\dagger})^{q}
        \hat{\sigma}_{\mathrm{act}}^{-}
        +
        \hat{a}_c^{q}
        \hat{\sigma}_{\mathrm{act}}^{+}
    \right],
    \label{eq:supp_disp_generic_JC_H}
\end{equation}
where $g_{\mathrm{JC}}^{(q)}$ denotes the corresponding $q$-quantum JC coupling strength.
For a fixed spectator-TLS eigenvalue $z_{\mathrm{sp}}$, the interaction in Eq.~\eqref{eq:supp_disp_generic_JC_H} couples $\ket{m;+1,z_{\mathrm{sp}}}\leftrightarrow\ket{m+q;-1,z_{\mathrm{sp}}}$.
The spectator TLS remains unchanged during this transition.
We define the corresponding JC transition mismatch as the energy of $\ket{m+q;-1,z_{\mathrm{sp}}}$ relative to that of $\ket{m;+1,z_{\mathrm{sp}}}$,
\begin{align}
    \nu_{\mathrm{JC}}^{(m,q)}(z_{\mathrm{sp}})
    \equiv{}&
    \mathcal E_{m+q}(-1,z_{\mathrm{sp}})
    -
    \mathcal E_m(+1,z_{\mathrm{sp}}).
    \label{eq:supp_disp_generic_JC_mismatch_def}
\end{align}
Substituting Eq.~\eqref{eq:supp_disp_generic_energy} into Eq.~\eqref{eq:supp_disp_generic_JC_mismatch_def} gives
\begin{equation}
    \nu_{\mathrm{JC}}^{(m,q)}(z_{\mathrm{sp}})
    =
    q\Delta_c
    +
    qz_{\mathrm{sp}}\chi_{\mathrm{sp}}
    -
    (2m+q)\chi_{\mathrm{act}}
    -
    \Delta_{\mathrm{act}}.
    \label{eq:supp_disp_generic_JC_mismatch}
\end{equation}

We now select $\ket{m_0}\leftrightarrow\ket{m_0+q}$ as the resonant bosonic transition, where $m_0$ denotes the selected lower Fock-state index.  
Because the transition frequency also depends on the spectator TLS, we introduce a reference spectator eigenvalue $\bar z_{\mathrm{sp}}\in\{-1,+1\}$ and impose
\begin{equation}
    \nu_{\mathrm{JC}}^{(m_0,q)}
    (\bar z_{\mathrm{sp}})
    =
    0.
    \label{eq:supp_disp_generic_JC_resonance_condition}
\end{equation}
Using Eqs.~\eqref{eq:supp_disp_generic_JC_mismatch} and \eqref{eq:supp_disp_generic_JC_resonance_condition}, the active-TLS detuning that satisfies this resonance condition, denoted by $\Delta_{\mathrm{act}}^{\mathrm{JC}}$, is
\begin{equation}
    \Delta_{\mathrm{act}}^{\mathrm{JC}}
    =
    q\Delta_c
    +
    q\bar z_{\mathrm{sp}}\chi_{\mathrm{sp}}
    -
    (2m_0+q)\chi_{\mathrm{act}}.
    \label{eq:supp_disp_generic_JC_resonance}
\end{equation}
Substituting the resonant value $\Delta_{\mathrm{act}}=\Delta_{\mathrm{act}}^{\mathrm{JC}}$ from Eq.~\eqref{eq:supp_disp_generic_JC_resonance} into Eq.~\eqref{eq:supp_disp_generic_JC_mismatch} gives
\begin{align}
    \nu_{\mathrm{JC}}^{(m,q)}(z_{\mathrm{sp}})
    ={}&
    -2(m-m_0)\chi_{\mathrm{act}}
    \nonumber\\
    &+
    q
    (z_{\mathrm{sp}}-\bar z_{\mathrm{sp}})
    \chi_{\mathrm{sp}}.
    \label{eq:supp_disp_generic_JC_compensated}
\end{align}
When the spectator TLS is in the reference state associated with $\bar z_{\mathrm{sp}}$, so that $z_{\mathrm{sp}}=\bar z_{\mathrm{sp}}$, Eq.~\eqref{eq:supp_disp_generic_JC_compensated} reduces to
\begin{equation}
    \nu_{\mathrm{JC}}^{(m,q)}
    (\bar z_{\mathrm{sp}})
    =
    -2(m-m_0)\chi_{\mathrm{act}}.
    \label{eq:supp_disp_generic_JC_reference}
\end{equation}
Thus, although the bare bosonic spectrum is harmonic, the JC-assisted dispersive coupling renders transitions with different lower Fock-state indices spectrally distinguishable.
In particular, when the spectator TLS is in the reference state, the selected transition and its neighboring $q$-quantum transition are separated by
\begin{equation}
    \left|
    \nu_{\mathrm{JC}}^{(m_0+1,q)}(\bar z_{\mathrm{sp}})
    -
    \nu_{\mathrm{JC}}^{(m_0,q)}(\bar z_{\mathrm{sp}})
    \right|
    =
    2|\chi_{\mathrm{act}}|.
    \label{eq:supp_disp_generic_JC_spacing}
\end{equation}

\subsubsection{AJC-assisted spectral selection}
\label{subsubsec:supp_disp_generic_AJC_selection}

We next consider a generic $q$-quantum AJC interaction,
\begin{equation}
    \hat{H}_{\mathrm{AJC}}^{(q)}
    =
    g_{\mathrm{AJC}}^{(q)}
    \left[
        (\hat{a}_c^{\dagger})^{q}
        \hat{\sigma}_{\mathrm{act}}^{+}
        +
        \hat{a}_c^{q}
        \hat{\sigma}_{\mathrm{act}}^{-}
    \right],
    \label{eq:supp_disp_generic_AJC_H}
\end{equation}
where $g_{\mathrm{AJC}}^{(q)}$ denotes the corresponding $q$-quantum AJC coupling strength.
For a fixed spectator-TLS eigenvalue $z_{\mathrm{sp}}$, the interaction in Eq.~\eqref{eq:supp_disp_generic_AJC_H} couples $\ket{m;-1,z_{\mathrm{sp}}}\leftrightarrow\ket{m+q;+1,z_{\mathrm{sp}}}$.
We define the corresponding AJC transition mismatch as the energy of $\ket{m+q;+1,z_{\mathrm{sp}}}$ relative to that of $\ket{m;-1,z_{\mathrm{sp}}}$,
\begin{align}
    \nu_{\mathrm{AJC}}^{(m,q)}(z_{\mathrm{sp}})
    \equiv{}&
    \mathcal E_{m+q}(+1,z_{\mathrm{sp}})
    -
    \mathcal E_m(-1,z_{\mathrm{sp}}).
    \label{eq:supp_disp_generic_AJC_mismatch_def}
\end{align}
Substituting Eq.~\eqref{eq:supp_disp_generic_energy} into Eq.~\eqref{eq:supp_disp_generic_AJC_mismatch_def} gives
\begin{equation}
    \nu_{\mathrm{AJC}}^{(m,q)}(z_{\mathrm{sp}})
    =
    q\Delta_c
    +
    qz_{\mathrm{sp}}\chi_{\mathrm{sp}}
    +
    (2m+q)\chi_{\mathrm{act}}
    +
    \Delta_{\mathrm{act}}.
    \label{eq:supp_disp_generic_AJC_mismatch}
\end{equation}

For the same selected resonant transition $\ket{m_0}\leftrightarrow\ket{m_0+q}$ and reference spectator eigenvalue $\bar z_{\mathrm{sp}}$ introduced above, we impose
\begin{equation}
    \nu_{\mathrm{AJC}}^{(m_0,q)}
    (\bar z_{\mathrm{sp}})
    =
    0.
    \label{eq:supp_disp_generic_AJC_resonance_condition}
\end{equation}
Using Eqs.~\eqref{eq:supp_disp_generic_AJC_mismatch} and \eqref{eq:supp_disp_generic_AJC_resonance_condition}, the corresponding resonant active-TLS detuning, denoted by $\Delta_{\mathrm{act}}^{\mathrm{AJC}}$, is
\begin{equation}
    \Delta_{\mathrm{act}}^{\mathrm{AJC}}
    =
    -
    q\Delta_c
    -
    q\bar z_{\mathrm{sp}}\chi_{\mathrm{sp}}
    -
    (2m_0+q)\chi_{\mathrm{act}}.
    \label{eq:supp_disp_generic_AJC_resonance}
\end{equation}
Setting $\Delta_{\mathrm{act}}=\Delta_{\mathrm{act}}^{\mathrm{AJC}}$ in Eq.~\eqref{eq:supp_disp_generic_AJC_mismatch} yields
\begin{align}
    \nu_{\mathrm{AJC}}^{(m,q)}(z_{\mathrm{sp}})
    ={}&
    2(m-m_0)\chi_{\mathrm{act}}
    \nonumber\\
    &+
    q
    (z_{\mathrm{sp}}-\bar z_{\mathrm{sp}})
    \chi_{\mathrm{sp}}.
    \label{eq:supp_disp_generic_AJC_compensated}
\end{align}
When the spectator TLS is in the reference state, so that $z_{\mathrm{sp}}=\bar z_{\mathrm{sp}}$, Eq.~\eqref{eq:supp_disp_generic_AJC_compensated} reduces to
\begin{equation}
    \nu_{\mathrm{AJC}}^{(m,q)}
    (\bar z_{\mathrm{sp}})
    =
    2(m-m_0)\chi_{\mathrm{act}}.
    \label{eq:supp_disp_generic_AJC_reference}
\end{equation}
Hence, as in the JC case, neighboring $q$-quantum transitions are
separated by
\begin{equation}
    \left|
    \nu_{\mathrm{AJC}}^{(m_0+1,q)}(\bar z_{\mathrm{sp}})
    -
    \nu_{\mathrm{AJC}}^{(m_0,q)}(\bar z_{\mathrm{sp}})
    \right|
    =
    2|\chi_{\mathrm{act}}|.
    \label{eq:supp_disp_generic_AJC_spacing}
\end{equation}

\subsubsection{Comparison with Kerr spectral selection}
\label{subsubsec:supp_disp_generic_comparison}

The generic dispersive result can now be compared directly with the Kerr-induced selection derived in Supplementary Sec.~\ref{sec:supp_effective_connectors}. 
As shown in Eq.~\eqref{eq:supp_selective_detuning}, for a selected $q$-quantum transition $\ket{m_0}\leftrightarrow\ket{m_0+q}$, the Kerr nonlinearity produces the relative spectral detuning
\begin{equation}
    \delta_q(m;m_0)
    =
    2qK_c(m-m_0)
\end{equation}
for the $q$-quantum transition with lower Fock-state index $m$.
Consequently, neighboring $q$-quantum transitions are spectrally separated by $2q|K_c|$.
As shown in Eqs.~\eqref{eq:supp_disp_generic_JC_reference} and \eqref{eq:supp_disp_generic_AJC_reference}, when the spectator TLS is in the reference state, the JC- and AJC-assisted dispersive couplings produce the transition mismatches
\begin{equation}
    \nu_{\mathrm{JC}}^{(m,q)}(\bar z_{\mathrm{sp}})
    =
    -2(m-m_0)\chi_{\mathrm{act}},
\end{equation}
and
\begin{equation}
    \nu_{\mathrm{AJC}}^{(m,q)}(\bar z_{\mathrm{sp}})
    =
    2(m-m_0)\chi_{\mathrm{act}},
\end{equation}
respectively.
In both cases, neighboring $q$-quantum transitions are therefore spectrally separated by $2|\chi_{\mathrm{act}}|$.
Thus, unlike the Kerr-induced spectral separation $2q|K_c|$, which scales linearly with the transition order $q$, the dispersive separation $2|\chi_{\mathrm{act}}|$ generated by the active TLS is independent of $q$.

A further distinction is that the dispersive transition mismatch also depends on the state of the spectator TLS.
As seen in Eqs.~\eqref{eq:supp_disp_generic_JC_compensated} and \eqref{eq:supp_disp_generic_AJC_compensated}, this dependence enters through the term $q(z_{\mathrm{sp}}-\bar z_{\mathrm{sp}})\chi_{\mathrm{sp}}$, which has no counterpart in the Kerr-only realization.
When the spectator TLS is in the eigenstate opposite to the reference state $z_{\mathrm{sp}}=-\bar z_{\mathrm{sp}}$, the selected transition acquires an additional mismatch of magnitude $2q|\chi_{\mathrm{sp}}|$.

The generic results above specialize to the global connector according to
\begin{equation}
    m_0=0,
    \qquad
    q=N_s,
    \qquad
    \mathrm{act}\mapsto \mathrm s,
    \qquad
    \mathrm{sp}\mapsto \mathrm R,
    \label{eq:supp_disp_generic_to_global}
\end{equation}
where $\mathrm{act}\mapsto \mathrm s$ and $\mathrm{sp}\mapsto \mathrm R$ indicate that the TLS labeled by $\mathrm s$ serves as the active TLS, while the TLS labeled by $R$ serves as the spectator TLS.
For the movable local connector, the corresponding specialization is
\begin{equation}
    m_0=R-1,
    \qquad
    q=1,
    \qquad
    \mathrm{act}\mapsto \mathrm R,
    \qquad
    \mathrm{sp}\mapsto \mathrm s.
    \label{eq:supp_disp_generic_to_local}
\end{equation}
Here $\mathrm{act}\mapsto \mathrm R$ and $\mathrm{sp}\mapsto \mathrm s$ indicate that the TLS labeled by $R$ serves as the active TLS, while the TLS labeled by $\mathrm s$ serves as the spectator TLS.
These two cases are considered explicitly below.

\subsection{Effective dynamics of a pumped JC realization}
\label{subsec:supp_disp_effective_JC}

Throughout this subsection, the active-TLS detuning is fixed at the value $\Delta_{\mathrm{act}}=\Delta_{\mathrm{act}}^{\mathrm{JC}}$ defined in Eq.~\eqref{eq:supp_disp_generic_JC_resonance}, which makes the selected transition $\ket{m_0}\leftrightarrow\ket{m_0+q}$ resonant in the reference spectator-TLS subspace $z_{\mathrm{sp}}=\bar z_{\mathrm{sp}}$.
The lower Fock-state index $m$ is nevertheless kept generic in the following reduction.
For a fixed spectator-TLS eigenvalue $z_{\mathrm{sp}}$, the $q$-quantum JC interaction couples the joint states
\begin{equation}
    \ket{A_{z_{\mathrm{sp}}}}
    =
    \ket{m;+1,z_{\mathrm{sp}}},
    \quad
    \ket{B_{z_{\mathrm{sp}}}}
    =
    \ket{m+q;-1,z_{\mathrm{sp}}}.
    \label{eq:supp_disp_JC_states}
\end{equation}
The corresponding coherent coupling is
\begin{equation}
        \Omega_{\mathrm{JC}}^{(m,q)}
        \equiv
        \bra{B_{z_{\mathrm{sp}}}}
        \hat{H}_{\mathrm{JC}}^{(q)}
        \ket{A_{z_{\mathrm{sp}}}}
        =
        g_{\mathrm{JC}}^{(q)}
        \sqrt{
            \frac{(m+q)!}{m!}
        }.
    \label{eq:supp_disp_JC_matrix}
\end{equation}
Because the spectator TLS is unchanged by $\hat{H}_{\mathrm{JC}}^{(q)}$, $\Omega_{\mathrm{JC}}^{(m,q)}$ is independent of $z_{\mathrm{sp}}$.

We denote the incoherent pumping and relaxation rates of the active TLS by $\wp_{\mathrm{act}}$ and $\gamma_{\mathrm{act}}$, respectively. 
For a fixed spectator-TLS eigenvalue $z_{\mathrm{sp}}$, we consider the dynamics within the corresponding spectator-TLS subspace and exclude processes that change the spectator-TLS state. 
We refer to this description, in which the dynamics are retained separately for each spectator-TLS eigenvalue $z_{\mathrm{sp}}$, as spectator resolved.
Let $\hat{\rho}_{\mathrm{joint}}$ denote the joint density operator of the bosonic mode and the two TLSs. 
Including single-quantum bosonic loss, the dynamics are governed by
\begin{equation}
    \begin{aligned}
        \frac{d\hat{\rho}_{\mathrm{joint}}}{dt}
        =&
        -i
        \left[
            \hat{H}_{0}^{\mathrm{disp}}
            +
            \hat{H}_{\mathrm{JC}}^{(q)},
            \hat{\rho}_{\mathrm{joint}}
        \right]
        +
        \kappa_c
        \mathcal D[\hat{a}_c]
        \hat{\rho}_{\mathrm{joint}}
        \\
        &+
        \wp_{\mathrm{act}}
        \mathcal D[\hat{\sigma}_{\mathrm{act}}^{+}]
        \hat{\rho}_{\mathrm{joint}}
        +
        \gamma_{\mathrm{act}}
        \mathcal D[\hat{\sigma}_{\mathrm{act}}^{-}]
        \hat{\rho}_{\mathrm{joint}}.
    \end{aligned}
    \label{eq:supp_disp_JC_joint_master_equation}
\end{equation}
We introduce the populations of these two joint states,
\begin{equation}
    \begin{aligned}
        P_A(z_{\mathrm{sp}})
        &\equiv
        \bra{A_{z_{\mathrm{sp}}}}
        \hat{\rho}_{\mathrm{joint}}
        \ket{A_{z_{\mathrm{sp}}}},
        \\
        P_B(z_{\mathrm{sp}})
        &\equiv
        \bra{B_{z_{\mathrm{sp}}}}
        \hat{\rho}_{\mathrm{joint}}
        \ket{B_{z_{\mathrm{sp}}}},
    \end{aligned}
    \label{eq:supp_disp_JC_populations}
\end{equation}
and the coherence between them
\begin{equation}
    C_{\mathrm{JC}}^{(m,q)}(z_{\mathrm{sp}})
    \equiv
    \bra{A_{z_{\mathrm{sp}}}}
    \hat{\rho}_{\mathrm{joint}}
    \ket{B_{z_{\mathrm{sp}}}}.
    \label{eq:supp_disp_JC_coherence}
\end{equation}

To identify the contribution of each term in Eq.~\eqref{eq:supp_disp_JC_joint_master_equation} to the evolution of the coherence $C_{\mathrm{JC}}^{(m,q)}(z_{\mathrm{sp}})$, we project the master equation onto the matrix element $\bra{A_{z_{\mathrm{sp}}}}\hat{\rho}_{\mathrm{joint}}\ket{B_{z_{\mathrm{sp}}}}$.
The resulting equation of motion can be decomposed as
\begin{equation}
    \begin{aligned}
        \dot C_{\mathrm{JC}}^{(m,q)}(z_{\mathrm{sp}})
        =&
        \left.
            \dot C_{\mathrm{JC}}^{(m,q)}(z_{\mathrm{sp}})
        \right|_{\mathrm{free}}
        +
        \left.
            \dot C_{\mathrm{JC}}^{(m,q)}(z_{\mathrm{sp}})
        \right|_{\mathrm{JC}}
        \\
        &+
        \left.
            \dot C_{\mathrm{JC}}^{(m,q)}(z_{\mathrm{sp}})
        \right|_{\mathrm{pump}}
        +
        \left.
            \dot C_{\mathrm{JC}}^{(m,q)}(z_{\mathrm{sp}})
        \right|_{\mathrm{relax}}
        \\
        &+
        \left.
            \dot C_{\mathrm{JC}}^{(m,q)}(z_{\mathrm{sp}})
        \right|_{\mathrm{loss}}.
    \end{aligned}
    \label{eq:supp_disp_JC_coherence_decomposition}
\end{equation}
The contribution from the dispersively dressed free Hamiltonian is
\begin{equation}
    \begin{aligned}
        \left.
            \dot C_{\mathrm{JC}}^{(m,q)}(z_{\mathrm{sp}})
        \right|_{\mathrm{free}}
        &=
        -i
        \bra{A_{z_{\mathrm{sp}}}}
        \left[
            \hat{H}_{0}^{\mathrm{disp}},
            \hat{\rho}_{\mathrm{joint}}
        \right]
        \ket{B_{z_{\mathrm{sp}}}}
        \\
        &=
        i
        \nu_{\mathrm{JC}}^{(m,q)}(z_{\mathrm{sp}})
        C_{\mathrm{JC}}^{(m,q)}(z_{\mathrm{sp}}).
    \end{aligned}
    \label{eq:supp_disp_JC_coherence_free}
\end{equation}
Here, $\nu_{\mathrm{JC}}^{(m,q)}(z_{\mathrm{sp}})$ is the transition mismatch derived in Eq.~\eqref{eq:supp_disp_generic_JC_compensated}. 
The contribution from the $q$-quantum JC interaction is
\begin{equation}
    \begin{aligned}
        \left.
            \dot C_{\mathrm{JC}}^{(m,q)}(z_{\mathrm{sp}})
        \right|_{\mathrm{JC}}
        &=
        -i
        \bra{A_{z_{\mathrm{sp}}}}
        \left[
            \hat{H}_{\mathrm{JC}}^{(q)},
            \hat{\rho}_{\mathrm{joint}}
        \right]
        \ket{B_{z_{\mathrm{sp}}}}
        \\
        &=
        -i
        \Omega_{\mathrm{JC}}^{(m,q)}
        \left[
            P_B(z_{\mathrm{sp}})
            -
            P_A(z_{\mathrm{sp}})
        \right].
    \end{aligned}
    \label{eq:supp_disp_JC_coherence_interaction}
\end{equation}
Here, $\Omega_{\mathrm{JC}}^{(m,q)}$ is the coherent coupling defined in Eq.~\eqref{eq:supp_disp_JC_matrix}.
The incoherent pumping channel of the active TLS contributes
\begin{equation}
    \begin{aligned}
        \left.
            \dot C_{\mathrm{JC}}^{(m,q)}(z_{\mathrm{sp}})
        \right|_{\mathrm{pump}}
        &=
        \wp_{\mathrm{act}}
        \bra{A_{z_{\mathrm{sp}}}}
        \mathcal D[\hat{\sigma}_{\mathrm{act}}^{+}]
        \hat{\rho}_{\mathrm{joint}}
        \ket{B_{z_{\mathrm{sp}}}}
        \\
        &=
        -
        \frac{\wp_{\mathrm{act}}}{2}
        C_{\mathrm{JC}}^{(m,q)}(z_{\mathrm{sp}}).
    \end{aligned}
    \label{eq:supp_disp_JC_coherence_pump}
\end{equation}
The relaxation channel of the active TLS gives
\begin{equation}
    \begin{aligned}
        \left.
            \dot C_{\mathrm{JC}}^{(m,q)}(z_{\mathrm{sp}})
        \right|_{\mathrm{relax}}
        &=
        \gamma_{\mathrm{act}}
        \bra{A_{z_{\mathrm{sp}}}}
        \mathcal D[\hat{\sigma}_{\mathrm{act}}^{-}]
        \hat{\rho}_{\mathrm{joint}}
        \ket{B_{z_{\mathrm{sp}}}}
        \\
        &=
        -
        \frac{\gamma_{\mathrm{act}}}{2}
        C_{\mathrm{JC}}^{(m,q)}(z_{\mathrm{sp}}).
    \end{aligned}
    \label{eq:supp_disp_JC_coherence_relax}
\end{equation}
The single-quantum bosonic-loss contribution is
\begin{equation}
    \begin{aligned}
        \left.
            \dot C_{\mathrm{JC}}^{(m,q)}(z_{\mathrm{sp}})
        \right|_{\mathrm{loss}}
        =&
        \kappa_c
        \bra{A_{z_{\mathrm{sp}}}}
        \mathcal D[\hat{a}_c]
        \hat{\rho}_{\mathrm{joint}}
        \ket{B_{z_{\mathrm{sp}}}}
        \\
        =&
        \kappa_c
        \sqrt{(m+1)(m+q+1)}
        C_{\mathrm{JC}}^{(m+1,q)}(z_{\mathrm{sp}})
        \\
        &-
        \frac{(2m+q)\kappa_c}{2}
        C_{\mathrm{JC}}^{(m,q)}(z_{\mathrm{sp}}),
    \end{aligned}
    \label{eq:supp_disp_JC_coherence_loss}
\end{equation}
where
\begin{equation}
    \begin{aligned}
        C_{\mathrm{JC}}^{(m+1,q)}(z_{\mathrm{sp}})
        \equiv{}&
        \bra{m+1;+1,z_{\mathrm{sp}}}
        \hat{\rho}_{\mathrm{joint}}
        \ket{m+q+1;-1,z_{\mathrm{sp}}}
    \end{aligned}
    \label{eq:supp_disp_JC_neighboring_coherence}
\end{equation}
denotes the neighboring coherence associated with the transition $\ket{m+1;+1,z_{\mathrm{sp}}}\leftrightarrow\ket{m+q+1;-1,z_{\mathrm{sp}}}$.

Combining Eqs.~\eqref{eq:supp_disp_JC_coherence_free}--\eqref{eq:supp_disp_JC_coherence_loss}, the equation of motion for the coherence $C_{\mathrm{JC}}^{(m,q)}(z_{\mathrm{sp}})$ becomes
\begin{equation}
    \begin{aligned}
        \dot C_{\mathrm{JC}}^{(m,q)}(z_{\mathrm{sp}})
        =&
        -
        \left[
            \Gamma_{\mathrm{JC}}^{(m,q)}
            -
            i\nu_{\mathrm{JC}}^{(m,q)}(z_{\mathrm{sp}})
        \right]
        C_{\mathrm{JC}}^{(m,q)}(z_{\mathrm{sp}})
        \\
        &-
        i
        \Omega_{\mathrm{JC}}^{(m,q)}
        \left[
            P_B(z_{\mathrm{sp}})
            -
            P_A(z_{\mathrm{sp}})
        \right]
        \\
        &+
        \kappa_c
        \sqrt{(m+1)(m+q+1)}
        C_{\mathrm{JC}}^{(m+1,q)}(z_{\mathrm{sp}}),
    \end{aligned}
    \label{eq:supp_disp_JC_coherence_full}
\end{equation}
where the total decay rate of this coherence is
\begin{equation}
    \Gamma_{\mathrm{JC}}^{(m,q)}
    \equiv
    \frac{
        \wp_{\mathrm{act}}
        +
        \gamma_{\mathrm{act}}
        +
        (2m+q)\kappa_c
    }{2}.
    \label{eq:supp_disp_JC_coherence_decay}
\end{equation}
The dispersive interactions modify the transition mismatch but do not introduce an additional irreversible decay channel at this level.
Thus, Eq.~\eqref{eq:supp_disp_JC_coherence_full} retains the form of Eq.~\eqref{eq:supp_JC_coherence_full}, with the Kerr-induced detuning replaced by $\nu_{\mathrm{JC}}^{(m,q)}(z_{\mathrm{sp}})$.

If pure dephasing of the active TLS and the bosonic mode is taken into account, we introduce the corresponding dephasing rates $\gamma_{\mathrm{act},\phi}$ and $\kappa_{c,\phi}$, respectively. 
The additional dissipative terms in the joint master equation are
\begin{equation}
    \gamma_{\mathrm{act},\phi}
    \mathcal D[\hat{\sigma}_{\mathrm{act}}^{z}]
    \hat{\rho}_{\mathrm{joint}}
    +
    \kappa_{c,\phi}
    \mathcal D[\hat{a}_c^{\dagger}\hat{a}_c]
    \hat{\rho}_{\mathrm{joint}}.
    \label{eq:supp_disp_JC_dephasing_channels}
\end{equation}
The pure-dephasing channel of the active TLS contributes
\begin{equation}
    \begin{aligned}
        \left.
            \dot C_{\mathrm{JC}}^{(m,q)}(z_{\mathrm{sp}})
        \right|_{\mathrm{deph,act}}
        &=
        \gamma_{\mathrm{act},\phi}
        \bra{A_{z_{\mathrm{sp}}}}
        \mathcal D[\hat{\sigma}_{\mathrm{act}}^{z}]
        \hat{\rho}_{\mathrm{joint}}
        \ket{B_{z_{\mathrm{sp}}}}
        \\
        &=
        -2\gamma_{\mathrm{act},\phi}
        C_{\mathrm{JC}}^{(m,q)}(z_{\mathrm{sp}}).
    \end{aligned}
    \label{eq:supp_disp_JC_dephasing_act}
\end{equation}
The pure-dephasing channel of the bosonic mode contributes
\begin{equation}
    \begin{aligned}
        \left.
            \dot C_{\mathrm{JC}}^{(m,q)}(z_{\mathrm{sp}})
        \right|_{\mathrm{deph,c}}
        &=
        \kappa_{c,\phi}
        \bra{A_{z_{\mathrm{sp}}}}
        \mathcal D[
            \hat{a}_c^{\dagger}\hat{a}_c
        ]
        \hat{\rho}_{\mathrm{joint}}
        \ket{B_{z_{\mathrm{sp}}}}
        \\
        &=
        -
        \frac{q^2\kappa_{c,\phi}}{2}
        C_{\mathrm{JC}}^{(m,q)}(z_{\mathrm{sp}}).
    \end{aligned}
    \label{eq:supp_disp_JC_dephasing_c}
\end{equation}
When both pure-dephasing channels are included, their effects can be incorporated into the total coherence-decay rate by replacing Eq.~\eqref{eq:supp_disp_JC_coherence_decay} with
\begin{equation}
        \Gamma_{\mathrm{JC}}^{(m,q)}
        =
        \frac{
            \wp_{\mathrm{act}}
            +
            \gamma_{\mathrm{act}}
            +
            (2m+q)\kappa_c
        }{2}
        +
        2\gamma_{\mathrm{act},\phi}
        +
        \frac{q^2\kappa_{c,\phi}}{2}.
    \label{eq:supp_disp_JC_coherence_decay_dephasing}
\end{equation}
Equation~\eqref{eq:supp_disp_JC_coherence_full} therefore retains the same form, with $\Gamma_{\mathrm{JC}}^{(m,q)}$ understood as the total coherence-decay rate given in
Eq.~\eqref{eq:supp_disp_JC_coherence_decay_dephasing}.

The last term in Eq.~\eqref{eq:supp_disp_JC_coherence_full} describes the bosonic-loss-induced feeding from the neighboring coherence $C_{\mathrm{JC}}^{(m+1,q)}(z_{\mathrm{sp}})$.
Neighboring transitions are spectrally separated by $2|\chi_{\mathrm{act}}|$, as established above. 
For a general spectator-TLS subspace, however, the suppression of the neighboring coherence is controlled by its actual mismatch $\nu_{\mathrm{JC}}^{(m+1,q)}(z_{\mathrm{sp}})$.
Moreover, since its contribution to Eq.~\eqref{eq:supp_disp_JC_coherence_full} is weighted by $\kappa_c\sqrt{(m+1)(m+q+1)}$, the corresponding feeding remains a small correction when this coefficient is sufficiently small compared with the coherence-decay rate of the retained transition.
A convenient set of sufficient conditions for neglecting this coupling is
\begin{equation}
    \begin{aligned}
        &\left|
            \nu_{\mathrm{JC}}^{(m+1,q)}(z_{\mathrm{sp}})
        \right|
        \gg
        \max
        \left\{
            \left|
                \Omega_{\mathrm{JC}}^{(m+1,q)}
            \right|,
            \Gamma_{\mathrm{JC}}^{(m+1,q)}
        \right\},
        \\
       &\kappa_c
        \sqrt{(m+1)(m+q+1)}
        \ll
        \Gamma_{\mathrm{JC}}^{(m,q)}.
    \end{aligned}
    \label{eq:supp_disp_JC_neighboring_condition}
\end{equation}
For the resonant reference channel $m=m_0$ and $z_{\mathrm{sp}}=\bar z_{\mathrm{sp}}$, the first inequality reduces to
\begin{equation}
    2|\chi_{\mathrm{act}}|
    \gg
    \max
    \left\{
        \left|
            \Omega_{\mathrm{JC}}^{(m_0+1,q)}
        \right|,
        \Gamma_{\mathrm{JC}}^{(m_0+1,q)}
    \right\}.
\end{equation}
When the conditions in Eq.~\eqref{eq:supp_disp_JC_neighboring_condition} are satisfied, Eq.~\eqref{eq:supp_disp_JC_coherence_full} reduces to
\begin{equation}
    \begin{aligned}
        \dot C_{\mathrm{JC}}^{(m,q)}(z_{\mathrm{sp}})
        =&
        -
        \left[
            \Gamma_{\mathrm{JC}}^{(m,q)}
            -
            i\nu_{\mathrm{JC}}^{(m,q)}(z_{\mathrm{sp}})
        \right]
        C_{\mathrm{JC}}^{(m,q)}(z_{\mathrm{sp}})
        \\
        &-
        i
        \Omega_{\mathrm{JC}}^{(m,q)}
        \left[
            P_B(z_{\mathrm{sp}})
            -
            P_A(z_{\mathrm{sp}})
        \right].
    \end{aligned}
    \label{eq:supp_disp_JC_coherence_closed}
\end{equation}

For the time-dependent dynamics, $C_{\mathrm{JC}}^{(m,q)}(z_{\mathrm{sp}})$ can be adiabatically eliminated when its evolution is fast compared with the population dynamics. 
A convenient sufficient condition for this time-scale separation is
\begin{equation}
    \sqrt{
        \left[
            \Gamma_{\mathrm{JC}}^{(m,q)}
        \right]^2
        +
        \left[
            \nu_{\mathrm{JC}}^{(m,q)}(z_{\mathrm{sp}})
        \right]^2
    }
    \gg
    \left|
        \Omega_{\mathrm{JC}}^{(m,q)}
    \right|.
    \label{eq:supp_disp_JC_adiabatic_condition}
\end{equation}
Near resonance, this condition reduces to
\begin{equation}
    \Gamma_{\mathrm{JC}}^{(m,q)}
    \gg
    \left|
        \Omega_{\mathrm{JC}}^{(m,q)}
    \right|.
    \label{eq:supp_disp_JC_adiabatic_condition_resonant}
\end{equation}
Under this condition, setting $\dot C_{\mathrm{JC}}^{(m,q)}(z_{\mathrm{sp}})\simeq 0$ in Eq.~\eqref{eq:supp_disp_JC_coherence_closed} gives
\begin{equation}
        C_{\mathrm{JC}}^{(m,q)}(z_{\mathrm{sp}})
        \simeq
        -
        \frac{
            i\Omega_{\mathrm{JC}}^{(m,q)}
        }{
            \Gamma_{\mathrm{JC}}^{(m,q)}
            -
            i\nu_{\mathrm{JC}}^{(m,q)}(z_{\mathrm{sp}})
        }
        \left[
            P_B(z_{\mathrm{sp}})
            -
            P_A(z_{\mathrm{sp}})
        \right].
    \label{eq:supp_disp_JC_coherence_eliminated}
\end{equation}

For the stationary solution of the reduced model described by Eq.~\eqref{eq:supp_disp_JC_coherence_closed}, no adiabatic condition is required. 
Setting $\dot C_{\mathrm{JC}}^{(m,q)}(z_{\mathrm{sp}})=0$ gives the exact stationary relation
\begin{equation}
        C_{\mathrm{JC}}^{(m,q)}(z_{\mathrm{sp}})
        =
        -
        \frac{
            i\Omega_{\mathrm{JC}}^{(m,q)}
        }{
            \Gamma_{\mathrm{JC}}^{(m,q)}
            -
            i\nu_{\mathrm{JC}}^{(m,q)}(z_{\mathrm{sp}})
        }
        \left[
            P_B(z_{\mathrm{sp}})
            -
            P_A(z_{\mathrm{sp}})
        \right].
    \label{eq:supp_disp_JC_coherence_stationary}
\end{equation}
Thus, the time-dependent and stationary treatments yield the same algebraic form for the eliminated coherence, but the former relies on the adiabatic approximation, whereas the latter is exact within the reduced model. 

We next determine the coherent population transfer between these two joint states $\ket{A_{z_{\mathrm{sp}}}}$ and $\ket{B_{z_{\mathrm{sp}}}}$ by projecting the Hamiltonian part of Eq.~\eqref{eq:supp_disp_JC_joint_master_equation} onto the corresponding diagonal matrix elements.
Since both states are eigenstates of $\hat{H}_{0}^{\mathrm{disp}}$, the dispersive free evolution does not change their populations. 
The Hamiltonian contributions are therefore
\begin{equation}
    \begin{aligned}
        \left.
            \dot P_A(z_{\mathrm{sp}})
        \right|_{\mathrm{free+JC}}
        &=
        -i
        \bra{A_{z_{\mathrm{sp}}}}
        \left[
            \hat{H}_{0}^{\mathrm{disp}}
            +
            \hat{H}_{\mathrm{JC}}^{(q)},
            \hat{\rho}_{\mathrm{joint}}
        \right]
        \ket{A_{z_{\mathrm{sp}}}}
        \\
        &=
        -i
        \bra{A_{z_{\mathrm{sp}}}}
        \left[
            \hat{H}_{\mathrm{JC}}^{(q)},
            \hat{\rho}_{\mathrm{joint}}
        \right]
        \ket{A_{z_{\mathrm{sp}}}}
        \\
        &=
        -2
        \Omega_{\mathrm{JC}}^{(m,q)}
        \operatorname{Im}
        C_{\mathrm{JC}}^{(m,q)}(z_{\mathrm{sp}}),
        \\
        \left.
            \dot P_B(z_{\mathrm{sp}})
        \right|_{\mathrm{free+JC}}
        &=
        -i
        \bra{B_{z_{\mathrm{sp}}}}
        \left[
            \hat{H}_{0}^{\mathrm{disp}}
            +
            \hat{H}_{\mathrm{JC}}^{(q)},
            \hat{\rho}_{\mathrm{joint}}
        \right]
        \ket{B_{z_{\mathrm{sp}}}}
        \\
        &=
        -i
        \bra{B_{z_{\mathrm{sp}}}}
        \left[
            \hat{H}_{\mathrm{JC}}^{(q)},
            \hat{\rho}_{\mathrm{joint}}
        \right]
        \ket{B_{z_{\mathrm{sp}}}}
        \\
        &=
        2
        \Omega_{\mathrm{JC}}^{(m,q)}
        \operatorname{Im}
        C_{\mathrm{JC}}^{(m,q)}(z_{\mathrm{sp}}).
    \end{aligned}
    \label{eq:supp_disp_JC_population_H}
\end{equation}
Here, the label $\mathrm{free+JC}$ denotes the contribution generated by the full Hamiltonian part of the joint master equation. 
In obtaining the final equalities, we have used the coherent coupling $\Omega_{\mathrm{JC}}^{(m,q)}$ defined in Eq.~\eqref{eq:supp_disp_JC_matrix}, together with $\bra{B_{z_{\mathrm{sp}}}}\hat{\rho}_{\mathrm{joint}}\ket{A_{z_{\mathrm{sp}}}}=[C_{\mathrm{JC}}^{(m,q)}(z_{\mathrm{sp}})]^* $.

The Hamiltonian-induced population transfer between $\ket{A_{z_{\mathrm{sp}}}}$ and $\ket{B_{z_{\mathrm{sp}}}}$ can then be written
in the rate form
\begin{equation}
    \begin{aligned}
        \left.
            \dot P_A(z_{\mathrm{sp}})
        \right|_{\mathrm{free+JC}}
        &=
        W_{\mathrm{JC}}^{(m,q)}(z_{\mathrm{sp}})
        \left[
            P_B(z_{\mathrm{sp}})
            -
            P_A(z_{\mathrm{sp}})
        \right],
        \\
        \left.
            \dot P_B(z_{\mathrm{sp}})
        \right|_{\mathrm{free+JC}}
        &=
        W_{\mathrm{JC}}^{(m,q)}(z_{\mathrm{sp}})
        \left[
            P_A(z_{\mathrm{sp}})
            -
            P_B(z_{\mathrm{sp}})
        \right],
    \end{aligned}
    \label{eq:supp_disp_JC_population_exchange}
\end{equation}
where $W_{\mathrm{JC}}^{(m,q)}(z_{\mathrm{sp}})$ denotes the population-exchange rate for a fixed spectator-TLS eigenvalue $z_{\mathrm{sp}}$.
For the time-dependent dynamics, $W_{\mathrm{JC}}^{(m,q)}(z_{\mathrm{sp}})$ follows from Eq.~\eqref{eq:supp_disp_JC_coherence_eliminated}, whereas for the stationary solution it follows from the exact relation Eq.~\eqref{eq:supp_disp_JC_coherence_stationary}. 
In both cases, the resulting exchange rate has the same form,
\begin{equation}
    W_{\mathrm{JC}}^{(m,q)}(z_{\mathrm{sp}})
    =
    \frac{
        2
        \left|
            \Omega_{\mathrm{JC}}^{(m,q)}
        \right|^2
        \Gamma_{\mathrm{JC}}^{(m,q)}
    }{
        \left[
            \Gamma_{\mathrm{JC}}^{(m,q)}
        \right]^2
        +
        \left[
            \nu_{\mathrm{JC}}^{(m,q)}(z_{\mathrm{sp}})
        \right]^2
    }.
    \label{eq:supp_disp_JC_exchange_rate}
\end{equation}
The two terms in Eq.~\eqref{eq:supp_disp_JC_population_exchange} are equal in magnitude and opposite in sign, so the Hamiltonian evolution transfers population between the two joint states without changing their total population.

To reduce the joint-state population exchange to an effective dynamics of the bosonic mode and the spectator TLS, we eliminate the active TLS.
Its isolated pump--relaxation dynamics are identical in form to those of the auxiliary TLS considered in Supplementary Sec.~\ref{sec:supp_effective_connectors}.
Therefore, its stationary populations follow directly from Eq.~\eqref{eq:supp_auxiliary_stationary_populations}, with the replacements $\mathrm a\rightarrow{\mathrm{act}}$, $\wp_{\mathrm a}\rightarrow\wp_{\mathrm{act}}$, and $\gamma_{\mathrm a}\rightarrow\gamma_{\mathrm{act}}$.
We define the reduced density operator
\begin{equation}
    \hat{\rho}_{\mathrm{c,sp}}
    \equiv
    {\operatorname{Tr}}_{\mathrm{act}}
    \hat{\rho}_{\mathrm{joint}},
\end{equation}
and the spectator-resolved bosonic Fock-state populations
\begin{equation}
    P_n(z_{\mathrm{sp}})
    \equiv
    \bra{n;z_{\mathrm{sp}}}
    \hat{\rho}_{\mathrm{c,sp}}
    \ket{n;z_{\mathrm{sp}}}.
    \label{eq:supp_disp_JC_conditional_population}
\end{equation}
For a given bosonic Fock state and spectator-TLS subspace, let $p_{e,\mathrm{act}|n,z_{\mathrm{sp}}}$ and $p_{g,\mathrm{act}|n,z_{\mathrm{sp}}}$ denote the conditional probabilities for the active TLS to be in its excited and ground states, respectively.
The joint-state populations can then be written exactly as
\begin{equation}
    \begin{aligned}
        P_A(z_{\mathrm{sp}})
        &=
        P_m(z_{\mathrm{sp}})
        p_{e,\mathrm{act}|m,z_{\mathrm{sp}}},
        \\
        P_B(z_{\mathrm{sp}})
        &=
        P_{m+q}(z_{\mathrm{sp}})
        p_{g,\mathrm{act}|m+q,z_{\mathrm{sp}}}.
    \end{aligned}
    \label{eq:supp_disp_JC_conditional_decomposition}
\end{equation}
We assume that the active-TLS pump--relaxation dynamics are sufficiently fast compared with the reduced bosonic population dynamics and that the population correlations induced by the $q$-quantum JC interaction remain weak.
The conditional active-TLS populations are then approximately independent of the bosonic Fock state and the spectator-TLS subspace,
\begin{equation}
    \begin{aligned}
        p_{e,\mathrm{act}|n,z_{\mathrm{sp}}}
        &\simeq
        p_{e,\mathrm{act}}^{\mathrm{ss}},
        \\
        p_{g,\mathrm{act}|n,z_{\mathrm{sp}}}
        &\simeq
        p_{g,\mathrm{act}}^{\mathrm{ss}}.
    \end{aligned}
    \label{eq:supp_disp_JC_conditional_TLS_populations}
\end{equation}
Therefore, the relevant joint-state populations can then be approximated as
\begin{equation}
    \begin{aligned}
        P_A(z_{\mathrm{sp}})
        &\simeq
        P_m(z_{\mathrm{sp}})
        p_{e,\mathrm{act}}^{\mathrm{ss}},
        \\
        P_B(z_{\mathrm{sp}})
        &\simeq
        P_{m+q}(z_{\mathrm{sp}})
        p_{g,\mathrm{act}}^{\mathrm{ss}},
    \end{aligned}
    \label{eq:supp_disp_JC_factorized_populations}
\end{equation}
with
\begin{equation}
    p_{e,\mathrm{act}}^{\mathrm{ss}}
    =
    \frac{\wp_{\mathrm{act}}}
    {\wp_{\mathrm{act}}+\gamma_{\mathrm{act}}},
    \qquad
    p_{g,\mathrm{act}}^{\mathrm{ss}}
    =
    \frac{\gamma_{\mathrm{act}}}
    {\wp_{\mathrm{act}}+\gamma_{\mathrm{act}}},
    \label{eq:supp_disp_JC_active_stationary_populations}
\end{equation}
as obtained directly from Eq.~\eqref{eq:supp_auxiliary_stationary_populations}.

Each population-transfer event from $\ket{A_{z_{\mathrm{sp}}}}=\ket{m;+1,z_{\mathrm{sp}}}$ to $\ket{B_{z_{\mathrm{sp}}}}=\ket{m+q;-1,z_{\mathrm{sp}}}$ changes the bosonic state from $\ket{m}$ to $\ket{m+q}$, while the reverse event produces the opposite transition. 
Substituting Eq.~\eqref{eq:supp_disp_JC_factorized_populations} into Eq.~\eqref{eq:supp_disp_JC_population_exchange}, the Hamiltonian-induced contributions to the reduced populations can be written as
\begin{equation}
    \begin{aligned}
        \left.
            \dot P_m(z_{\mathrm{sp}})
        \right|_{\mathrm{free+JC}}
        =&
        -
        \mathcal R_{\mathrm{JC}}^{\uparrow}(m,q;z_{\mathrm{sp}})
        P_m(z_{\mathrm{sp}})
        \\
        &+
        \mathcal R_{\mathrm{JC}}^{\downarrow}(m,q;z_{\mathrm{sp}})
        P_{m+q}(z_{\mathrm{sp}}),
        \\
        \left.
            \dot P_{m+q}(z_{\mathrm{sp}})
        \right|_{\mathrm{free+JC}}
        =&
        \mathcal R_{\mathrm{JC}}^{\uparrow}(m,q;z_{\mathrm{sp}})
        P_m(z_{\mathrm{sp}})
        \\
        &-
        \mathcal R_{\mathrm{JC}}^{\downarrow}(m,q;z_{\mathrm{sp}})
        P_{m+q}(z_{\mathrm{sp}}).
    \end{aligned}
    \label{eq:supp_disp_JC_conditional_population_dynamics}
\end{equation}
Here, $\mathcal R_{\mathrm{JC}}^{\uparrow}(m,q;z_{\mathrm{sp}})$ and $\mathcal R_{\mathrm{JC}}^{\downarrow}(m,q;z_{\mathrm{sp}})$ denote the effective population-transfer rates for the bosonic transitions $\ket{m}\rightarrow\ket{m+q}$ and $\ket{m+q}\rightarrow\ket{m}$, respectively, for a fixed spectator-TLS eigenvalue $z_{\mathrm{sp}}$. 
They are given by
\begin{equation}
    \begin{aligned}
        \mathcal R_{\mathrm{JC}}^{\uparrow}(m,q;z_{\mathrm{sp}})
        &=
        W_{\mathrm{JC}}^{(m,q)}(z_{\mathrm{sp}})
        p_{e,\mathrm{act}}^{\mathrm{ss}}
        \\
        &=
        W_{\mathrm{JC}}^{(m,q)}(z_{\mathrm{sp}})
        \frac{
            \wp_{\mathrm{act}}
        }{
            \wp_{\mathrm{act}}
            +
            \gamma_{\mathrm{act}}
        },
        \\
        \mathcal R_{\mathrm{JC}}^{\downarrow}(m,q;z_{\mathrm{sp}})
        &=
        W_{\mathrm{JC}}^{(m,q)}(z_{\mathrm{sp}})
        p_{g,\mathrm{act}}^{\mathrm{ss}}
        \\
        &=
        W_{\mathrm{JC}}^{(m,q)}(z_{\mathrm{sp}})
        \frac{
            \gamma_{\mathrm{act}}
        }{
            \wp_{\mathrm{act}}
            +
            \gamma_{\mathrm{act}}
        }.
    \end{aligned}
    \label{eq:supp_disp_JC_conditional_rates}
\end{equation}
Thus, the spectator-TLS state affects the magnitudes of both population-transfer rates through $W_{\mathrm{JC}}^{(m,q)}(z_{\mathrm{sp}})$, while their ratio is
\begin{equation}
    \frac{
        \mathcal R_{\mathrm{JC}}^{\uparrow}(m,q;z_{\mathrm{sp}})
    }{
        \mathcal R_{\mathrm{JC}}^{\downarrow}(m,q;z_{\mathrm{sp}})
    }
    =
    \frac{
        p_{e,\mathrm{act}}^{\mathrm{ss}}
    }{
        p_{g,\mathrm{act}}^{\mathrm{ss}}
    }
    =
    \frac{
        \wp_{\mathrm{act}}
    }{
        \gamma_{\mathrm{act}}
    }.
    \label{eq:supp_disp_JC_directionality}
\end{equation}
Hence, as in Eq.~\eqref{eq:supp_JC_directionality}, the pumping-to-relaxation ratio of the active TLS determines the directional bias of the effective bosonic population transfer, whereas the dispersive spectator dependence enters only through the common exchange rate $W_{\mathrm{JC}}^{(m,q)}(z_{\mathrm{sp}})$.

For the resonant reference channel $m=m_0$ and $z_{\mathrm{sp}}=\bar z_{\mathrm{sp}}$, and in the absence of additional pure dephasing, the dependence of the upward transfer rate on the active-TLS pumping rate has the same form as in the Kerr-resolved pumped-JC realization.
The corresponding optimal-pumping and strong-pumping behaviors therefore follow directly from Eqs.~\eqref{eq:supp_JC_upward_rate_pump_dependence}--\eqref{eq:supp_JC_upward_rate_strong_pump} and are not repeated here.

\subsection{Effective dynamics of an AJC+relaxation realization}
\label{subsec:supp_disp_effective_AJC}

We next consider the effective dynamics generated by a $q$-quantum AJC interaction followed by relaxation of the active TLS.
For the active AJC TLS we assume $\wp_{\mathrm{act}}=0$ and $\gamma_{\mathrm{act}}>0$; this does not constrain the pumping of a separate spectator TLS.
Throughout this subsection, the active-TLS detuning is fixed at the value $\Delta_{\mathrm{act}}=\Delta_{\mathrm{act}}^{\mathrm{AJC}}$ defined in Eq.~\eqref{eq:supp_disp_generic_AJC_resonance}, which makes the selected bosonic transition $\ket{m_0}\leftrightarrow\ket{m_0+q}$ resonant in the reference spectator-TLS subspace $z_{\mathrm{sp}}=\bar z_{\mathrm{sp}}$.
The lower Fock-state index $m$ is nevertheless kept generic in the following reduction.

We retain the spectator-resolved description introduced in the preceding subsection. 
For a fixed spectator-TLS eigenvalue $z_{\mathrm{sp}}$, the $q$-quantum AJC interaction couples the joint states
\begin{equation}
    \ket{D_{z_{\mathrm{sp}}}}
    =
    \ket{m;-1,z_{\mathrm{sp}}},
    \qquad
    \ket{E_{z_{\mathrm{sp}}}}
    =
    \ket{m+q;+1,z_{\mathrm{sp}}}.
    \label{eq:supp_disp_AJC_states}
\end{equation}
Relaxation of the active TLS subsequently connects $\ket{E_{z_{\mathrm{sp}}}}$ to
\begin{equation}
    \ket{B_{z_{\mathrm{sp}}}}
    =
    \ket{m+q;-1,z_{\mathrm{sp}}},
    \label{eq:supp_disp_AJC_relaxed_state}
\end{equation}
without changing the bosonic Fock number.
The resulting sequence
\begin{equation}
    \ket{D_{z_{\mathrm{sp}}}}
    \leftrightarrow
    \ket{E_{z_{\mathrm{sp}}}}
    \rightarrow
    \ket{B_{z_{\mathrm{sp}}}}
    \label{eq:supp_disp_AJC_sequence}
\end{equation}
therefore provides a pathway from the bosonic Fock state $\ket{m}$ toward $\ket{m+q}$.
The corresponding coherent coupling is
\begin{equation}
        \Omega_{\mathrm{AJC}}^{(m,q)}
        \equiv
        \bra{E_{z_{\mathrm{sp}}}}
        \hat{H}_{\mathrm{AJC}}^{(q)}
        \ket{D_{z_{\mathrm{sp}}}}
       =
        g_{\mathrm{AJC}}^{(q)}
        \sqrt{
            \frac{(m+q)!}{m!}
        }.
    \label{eq:supp_disp_AJC_coupling}
\end{equation}
Because the spectator TLS is unchanged by $\hat{H}_{\mathrm{AJC}}^{(q)}$, $\Omega_{\mathrm{AJC}}^{(m,q)}$ is independent of $z_{\mathrm{sp}}$.

We denote the relaxation rate of the active TLS by $\gamma_{\mathrm{act}}$.
For each fixed $z_{\mathrm{sp}}$, we work within the corresponding spectator-TLS subspace and exclude processes that change the spectator-TLS state.
Including intrinsic single-quantum bosonic loss, the joint dynamics are governed by
\begin{equation}
    \begin{aligned}
        \frac{d\hat{\rho}_{\mathrm{joint}}}{dt}
        =&
        -i
        \left[
            \hat{H}_{0}^{\mathrm{disp}}
            +
            \hat{H}_{\mathrm{AJC}}^{(q)},
            \hat{\rho}_{\mathrm{joint}}
        \right]
        +
        \kappa_c
        \mathcal D[\hat{a}_c]
        \hat{\rho}_{\mathrm{joint}}
        \\
        &+
        \gamma_{\mathrm{act}}
        \mathcal D[\hat{\sigma}_{\mathrm{act}}^{-}]
        \hat{\rho}_{\mathrm{joint}}.
    \end{aligned}
    \label{eq:supp_disp_AJC_joint_master_equation}
\end{equation}
We introduce the populations of the two joint states,
\begin{equation}
    \begin{aligned}
        P_D(z_{\mathrm{sp}})
        &\equiv
        \bra{D_{z_{\mathrm{sp}}}}
        \hat{\rho}_{\mathrm{joint}}
        \ket{D_{z_{\mathrm{sp}}}},
        \\
        P_E(z_{\mathrm{sp}})
        &\equiv
        \bra{E_{z_{\mathrm{sp}}}}
        \hat{\rho}_{\mathrm{joint}}
        \ket{E_{z_{\mathrm{sp}}}},
    \end{aligned}
    \label{eq:supp_disp_AJC_DE_populations}
\end{equation}
and the coherence between them,
\begin{equation}
    C_{\mathrm{AJC}}^{(m,q)}(z_{\mathrm{sp}})
    \equiv
    \bra{D_{z_{\mathrm{sp}}}}
    \hat{\rho}_{\mathrm{joint}}
    \ket{E_{z_{\mathrm{sp}}}}.
    \label{eq:supp_disp_AJC_coherence}
\end{equation}
Projecting Eq.~\eqref{eq:supp_disp_AJC_joint_master_equation} onto this coherence gives the decomposition
\begin{equation}
    \begin{aligned}
        \dot C_{\mathrm{AJC}}^{(m,q)}(z_{\mathrm{sp}})
        =&
        \left.
            \dot C_{\mathrm{AJC}}^{(m,q)}(z_{\mathrm{sp}})
        \right|_{\mathrm{free}}
        +
        \left.
            \dot C_{\mathrm{AJC}}^{(m,q)}(z_{\mathrm{sp}})
        \right|_{\mathrm{AJC}}
        \\
        &+
        \left.
            \dot C_{\mathrm{AJC}}^{(m,q)}(z_{\mathrm{sp}})
        \right|_{\mathrm{relax}}
        +
        \left.
            \dot C_{\mathrm{AJC}}^{(m,q)}(z_{\mathrm{sp}})
        \right|_{\mathrm{loss}}.
    \end{aligned}
    \label{eq:supp_disp_AJC_coherence_decomposition}
\end{equation}
The contribution from the dispersively dressed free Hamiltonian is
\begin{equation}
    \begin{aligned}
        \left.
            \dot C_{\mathrm{AJC}}^{(m,q)}(z_{\mathrm{sp}})
        \right|_{\mathrm{free}}
        &=
        -i
        \bra{D_{z_{\mathrm{sp}}}}
        \left[
            \hat{H}_{0}^{\mathrm{disp}},
            \hat{\rho}_{\mathrm{joint}}
        \right]
        \ket{E_{z_{\mathrm{sp}}}}
        \\
        &=
        i
        \nu_{\mathrm{AJC}}^{(m,q)}(z_{\mathrm{sp}})
        C_{\mathrm{AJC}}^{(m,q)}(z_{\mathrm{sp}}).
    \end{aligned}
    \label{eq:supp_disp_AJC_coherence_free}
\end{equation}
Here, $\nu_{\mathrm{AJC}}^{(m,q)}(z_{\mathrm{sp}})$ is the transition
mismatch derived in Eq.~\eqref{eq:supp_disp_generic_AJC_compensated}. 
The contribution from the $q$-quantum AJC interaction is
\begin{equation}
    \begin{aligned}
        \left.
            \dot C_{\mathrm{AJC}}^{(m,q)}(z_{\mathrm{sp}})
        \right|_{\mathrm{AJC}}
        &=
        -i
        \bra{D_{z_{\mathrm{sp}}}}
        \left[
            \hat{H}_{\mathrm{AJC}}^{(q)},
            \hat{\rho}_{\mathrm{joint}}
        \right]
        \ket{E_{z_{\mathrm{sp}}}}
        \\
        &=
        -i
        \Omega_{\mathrm{AJC}}^{(m,q)}
        \left[
            P_E(z_{\mathrm{sp}})
            -
            P_D(z_{\mathrm{sp}})
        \right].
    \end{aligned}
    \label{eq:supp_disp_AJC_coherence_interaction}
\end{equation}
Here, $\Omega_{\mathrm{AJC}}^{(m,q)}$ is the coherent coupling defined in Eq.~\eqref{eq:supp_disp_AJC_coupling}.
Relaxation of the active TLS contributes
\begin{equation}
    \begin{aligned}
        \left.
            \dot C_{\mathrm{AJC}}^{(m,q)}(z_{\mathrm{sp}})
        \right|_{\mathrm{relax}}
        &=
        \gamma_{\mathrm{act}}
        \bra{D_{z_{\mathrm{sp}}}}
        \mathcal D[\hat{\sigma}_{\mathrm{act}}^{-}]
        \hat{\rho}_{\mathrm{joint}}
        \ket{E_{z_{\mathrm{sp}}}}
        \\
        &=
        -
        \frac{\gamma_{\mathrm{act}}}{2}
        C_{\mathrm{AJC}}^{(m,q)}(z_{\mathrm{sp}}).
    \end{aligned}
    \label{eq:supp_disp_AJC_coherence_relax}
\end{equation}
The single-quantum bosonic-loss contribution is
\begin{equation}
    \begin{aligned}
        \left.
            \dot C_{\mathrm{AJC}}^{(m,q)}(z_{\mathrm{sp}})
        \right|_{\mathrm{loss}}
        =&
        \kappa_c
        \bra{D_{z_{\mathrm{sp}}}}
        \mathcal D[\hat{a}_c]
        \hat{\rho}_{\mathrm{joint}}
        \ket{E_{z_{\mathrm{sp}}}}
        \\
        =&
        \kappa_c
        \sqrt{(m+1)(m+q+1)}
        C_{\mathrm{AJC}}^{(m+1,q)}(z_{\mathrm{sp}})
        \\
        &-
        \frac{(2m+q)\kappa_c}{2}
        C_{\mathrm{AJC}}^{(m,q)}(z_{\mathrm{sp}}),
    \end{aligned}
    \label{eq:supp_disp_AJC_coherence_loss}
\end{equation}
where
\begin{equation}
        C_{\mathrm{AJC}}^{(m+1,q)}(z_{\mathrm{sp}})
        \equiv
        \bra{m+1;-1,z_{\mathrm{sp}}}
        \hat{\rho}_{\mathrm{joint}}
        \ket{m+q+1;+1,z_{\mathrm{sp}}}
    \label{eq:supp_disp_AJC_neighboring_coherence}
\end{equation}
denotes the neighboring coherence associated with the transition $\ket{m+1;-1,z_{\mathrm{sp}}}\leftrightarrow\ket{m+q+1;+1,z_{\mathrm{sp}}}$.

Combining Eqs.~\eqref{eq:supp_disp_AJC_coherence_free}--\eqref{eq:supp_disp_AJC_coherence_loss}, the equation of motion for the AJC coherence becomes
\begin{equation}
    \begin{aligned}
        \dot C_{\mathrm{AJC}}^{(m,q)}(z_{\mathrm{sp}})
        =&
        -
        \left[
            \Gamma_{\mathrm{AJC}}^{(m,q)}
            -
            i\nu_{\mathrm{AJC}}^{(m,q)}(z_{\mathrm{sp}})
        \right]
        C_{\mathrm{AJC}}^{(m,q)}(z_{\mathrm{sp}})
        \\
        &-
        i
        \Omega_{\mathrm{AJC}}^{(m,q)}
        \left[
            P_E(z_{\mathrm{sp}})
            -
            P_D(z_{\mathrm{sp}})
        \right]
        \\
        &+
        \kappa_c
        \sqrt{(m+1)(m+q+1)}
        C_{\mathrm{AJC}}^{(m+1,q)}(z_{\mathrm{sp}}),
    \end{aligned}
    \label{eq:supp_disp_AJC_coherence_full}
\end{equation}
where the total coherence-decay rate is
\begin{equation}
    \Gamma_{\mathrm{AJC}}^{(m,q)}
    \equiv
    \frac{
        \gamma_{\mathrm{act}}
        +
        (2m+q)\kappa_c
    }{2}.
    \label{eq:supp_disp_AJC_coherence_decay}
\end{equation}
The dispersive interactions modify the transition mismatch but do not introduce an additional irreversible decay channel at this level.
Thus, Eq.~\eqref{eq:supp_disp_AJC_coherence_full} retains the form of Eq.~\eqref{eq:supp_AJC_coherence_full}, with the Kerr-induced detuning replaced by $\nu_{\mathrm{AJC}}^{(m,q)}(z_{\mathrm{sp}})$.

If pure dephasing of the active TLS and the bosonic mode is included with rates $\gamma_{\mathrm{act},\phi}$ and $\kappa_{c,\phi}$, respectively, the additional dissipative terms are
\begin{equation}
    \gamma_{\mathrm{act},\phi}
    \mathcal D[\hat{\sigma}_{\mathrm{act}}^{z}]
    \hat{\rho}_{\mathrm{joint}}
    +
    \kappa_{c,\phi}
    \mathcal D[\hat{a}_c^{\dagger}\hat{a}_c]
    \hat{\rho}_{\mathrm{joint}}.
    \label{eq:supp_disp_AJC_dephasing_channels}
\end{equation}
Their contributions to the AJC coherence are
\begin{equation}
    \begin{aligned}
        \left.
            \dot C_{\mathrm{AJC}}^{(m,q)}(z_{\mathrm{sp}})
        \right|_{\mathrm{deph,act}}
        &=
        -2\gamma_{\mathrm{act},\phi}
        C_{\mathrm{AJC}}^{(m,q)}(z_{\mathrm{sp}}),
        \\
        \left.
            \dot C_{\mathrm{AJC}}^{(m,q)}(z_{\mathrm{sp}})
        \right|_{\mathrm{deph,c}}
        &=
        -
        \frac{q^2\kappa_{c,\phi}}{2}
        C_{\mathrm{AJC}}^{(m,q)}(z_{\mathrm{sp}}).
    \end{aligned}
    \label{eq:supp_disp_AJC_dephasing_contributions}
\end{equation}
Accordingly, the total coherence-decay rate becomes
\begin{equation}
    \begin{aligned}
        \Gamma_{\mathrm{AJC}}^{(m,q)}
        =&
        \frac{
            \gamma_{\mathrm{act}}
            +
            (2m+q)\kappa_c
        }{2}
        \\
        &+
        2\gamma_{\mathrm{act},\phi}
        +
        \frac{q^2\kappa_{c,\phi}}{2}.
    \end{aligned}
    \label{eq:supp_disp_AJC_coherence_decay_dephasing}
\end{equation}
Equation~\eqref{eq:supp_disp_AJC_coherence_full} therefore retains the same form, with $\Gamma_{\mathrm{AJC}}^{(m,q)}$ understood as the total coherence-decay rate in Eq.~\eqref{eq:supp_disp_AJC_coherence_decay_dephasing}.

The last term in Eq.~\eqref{eq:supp_disp_AJC_coherence_full} describes the bosonic-loss-induced feeding from the neighboring coherence $C_{\mathrm{AJC}}^{(m+1,q)}(z_{\mathrm{sp}})$.
Neighboring AJC transitions are spectrally separated by $2|\chi_{\mathrm{act}}|$.
For a general spectator-TLS subspace, however, the suppression of the neighboring coherence is controlled by its actual mismatch $\nu_{\mathrm{AJC}}^{(m+1,q)}(z_{\mathrm{sp}})$.
Moreover, its feeding into $C_{\mathrm{AJC}}^{(m,q)}(z_{\mathrm{sp}})$ remains a small correction when the corresponding loss coefficient is sufficiently small compared with the coherence-decay rate.
A convenient set of sufficient conditions is
\begin{equation}
    \begin{aligned}
        &\left|
            \nu_{\mathrm{AJC}}^{(m+1,q)}(z_{\mathrm{sp}})
        \right|
        \gg
        \max
        \left\{
            \left|
                \Omega_{\mathrm{AJC}}^{(m+1,q)}
            \right|,
            \Gamma_{\mathrm{AJC}}^{(m+1,q)}
        \right\},
        \\
        &\kappa_c
        \sqrt{(m+1)(m+q+1)}
        \ll
        \Gamma_{\mathrm{AJC}}^{(m,q)}.
    \end{aligned}
    \label{eq:supp_disp_AJC_neighboring_condition}
\end{equation}
For the resonant reference channel $m=m_0$ and $z_{\mathrm{sp}}=\bar z_{\mathrm{sp}}$, the first condition reduces to
\begin{equation}
    2|\chi_{\mathrm{act}}|
    \gg
    \max
    \left\{
        \left|
            \Omega_{\mathrm{AJC}}^{(m_0+1,q)}
        \right|,
        \Gamma_{\mathrm{AJC}}^{(m_0+1,q)}
    \right\}.
    \label{eq:supp_disp_AJC_reference_selectivity_condition}
\end{equation}
When the conditions in Eq.~\eqref{eq:supp_disp_AJC_neighboring_condition} are satisfied, Eq.~\eqref{eq:supp_disp_AJC_coherence_full} reduces to
\begin{equation}
    \begin{aligned}
        \dot C_{\mathrm{AJC}}^{(m,q)}(z_{\mathrm{sp}})
        =&
        -
        \left[
            \Gamma_{\mathrm{AJC}}^{(m,q)}
            -
            i\nu_{\mathrm{AJC}}^{(m,q)}(z_{\mathrm{sp}})
        \right]
        C_{\mathrm{AJC}}^{(m,q)}(z_{\mathrm{sp}})
        \\
        &-
        i
        \Omega_{\mathrm{AJC}}^{(m,q)}
        \left[
            P_E(z_{\mathrm{sp}})
            -
            P_D(z_{\mathrm{sp}})
        \right].
    \end{aligned}
    \label{eq:supp_disp_AJC_coherence_closed}
\end{equation}

For the time-dependent dynamics, $C_{\mathrm{AJC}}^{(m,q)}(z_{\mathrm{sp}})$ can be adiabatically eliminated when its evolution is fast compared with the population dynamics. 
A convenient sufficient condition for this time-scale separation is
\begin{equation}
    \sqrt{
        \left[
            \Gamma_{\mathrm{AJC}}^{(m,q)}
        \right]^2
        +
        \left[
            \nu_{\mathrm{AJC}}^{(m,q)}(z_{\mathrm{sp}})
        \right]^2
    }
    \gg
    \left|
        \Omega_{\mathrm{AJC}}^{(m,q)}
    \right|.
    \label{eq:supp_disp_AJC_coherence_adiabatic_condition}
\end{equation}
Near resonance, this condition reduces to
\begin{equation}
    \Gamma_{\mathrm{AJC}}^{(m,q)}
    \gg
    \left|
        \Omega_{\mathrm{AJC}}^{(m,q)}
    \right|.
    \label{eq:supp_disp_AJC_coherence_adiabatic_condition_resonant}
\end{equation}
Under this condition, setting $\dot C_{\mathrm{AJC}}^{(m,q)}(z_{\mathrm{sp}})\simeq0$ in Eq.~\eqref{eq:supp_disp_AJC_coherence_closed} gives
\begin{equation}
        C_{\mathrm{AJC}}^{(m,q)}(z_{\mathrm{sp}})
        \simeq
        -
        \frac{
            i\Omega_{\mathrm{AJC}}^{(m,q)}
        }{
            \Gamma_{\mathrm{AJC}}^{(m,q)}
            -
            i\nu_{\mathrm{AJC}}^{(m,q)}(z_{\mathrm{sp}})
        }
        \left[
            P_E(z_{\mathrm{sp}})
            -
            P_D(z_{\mathrm{sp}})
        \right].
    \label{eq:supp_disp_AJC_coherence_eliminated}
\end{equation}

For the stationary solution of the reduced model described by Eq.~\eqref{eq:supp_disp_AJC_coherence_closed}, no adiabatic condition is required. 
Setting $\dot C_{\mathrm{AJC}}^{(m,q)}(z_{\mathrm{sp}})=0$ gives the exact stationary relation
\begin{equation}
        C_{\mathrm{AJC}}^{(m,q)}(z_{\mathrm{sp}})
        =
        -
        \frac{
            i\Omega_{\mathrm{AJC}}^{(m,q)}
        }{
            \Gamma_{\mathrm{AJC}}^{(m,q)}
            -
            i\nu_{\mathrm{AJC}}^{(m,q)}(z_{\mathrm{sp}})
        }
        \left[
            P_E(z_{\mathrm{sp}})
            -
            P_D(z_{\mathrm{sp}})
        \right].
    \label{eq:supp_disp_AJC_coherence_stationary}
\end{equation}

We next determine the coherent population transfer between $\ket{D_{z_{\mathrm{sp}}}}$ and $\ket{E_{z_{\mathrm{sp}}}}$ by projecting the Hamiltonian part of Eq.~\eqref{eq:supp_disp_AJC_joint_master_equation} onto the corresponding diagonal matrix elements.
Since both states are eigenstates of $\hat{H}_{0}^{\mathrm{disp}}$, the dispersive free evolution does not change their populations. 
The Hamiltonian contributions are therefore
\begin{equation}
    \begin{aligned}
        \left.
            \dot P_D(z_{\mathrm{sp}})
        \right|_{\mathrm{free+AJC}}
        &=
        -i
        \bra{D_{z_{\mathrm{sp}}}}
        \left[
            \hat{H}_{0}^{\mathrm{disp}}
            +
            \hat{H}_{\mathrm{AJC}}^{(q)},
            \hat{\rho}_{\mathrm{joint}}
        \right]
        \ket{D_{z_{\mathrm{sp}}}}
        \\
        &=
        -2
        \Omega_{\mathrm{AJC}}^{(m,q)}
        \operatorname{Im}
        C_{\mathrm{AJC}}^{(m,q)}(z_{\mathrm{sp}}),
        \\
        \left.
            \dot P_E(z_{\mathrm{sp}})
        \right|_{\mathrm{free+AJC}}
        &=
        -i
        \bra{E_{z_{\mathrm{sp}}}}
        \left[
            \hat{H}_{0}^{\mathrm{disp}}
            +
            \hat{H}_{\mathrm{AJC}}^{(q)},
            \hat{\rho}_{\mathrm{joint}}
        \right]
        \ket{E_{z_{\mathrm{sp}}}}
        \\
        &=
        2
        \Omega_{\mathrm{AJC}}^{(m,q)}
        \operatorname{Im}
        C_{\mathrm{AJC}}^{(m,q)}(z_{\mathrm{sp}}).
    \end{aligned}
    \label{eq:supp_disp_AJC_population_H}
\end{equation}
Here, the label $\mathrm{free+AJC}$ denotes the contribution generated by the full Hamiltonian part of Eq.~\eqref{eq:supp_disp_AJC_joint_master_equation}, namely $\hat{H}_{0}^{\mathrm{disp}}+\hat{H}_{\mathrm{AJC}}^{(q)}$.
In obtaining the final equalities, we have used the coherent coupling defined in Eq.~\eqref{eq:supp_disp_AJC_coupling}, together with $\bra{E_{z_{\mathrm{sp}}}}\hat{\rho}_{\mathrm{joint}}\ket{D_{z_{\mathrm{sp}}}}=\left[C_{\mathrm{AJC}}^{(m,q)}(z_{\mathrm{sp}})\right]^*$.

The Hamiltonian-induced population transfer can then be written in the rate form
\begin{equation}
    \begin{aligned}
        \left.
            \dot P_D(z_{\mathrm{sp}})
        \right|_{\mathrm{free+AJC}}
        &=
        W_{\mathrm{AJC}}^{(m,q)}(z_{\mathrm{sp}})
        \left[
            P_E(z_{\mathrm{sp}})
            -
            P_D(z_{\mathrm{sp}})
        \right],
        \\
        \left.
            \dot P_E(z_{\mathrm{sp}})
        \right|_{\mathrm{free+AJC}}
        &=
        W_{\mathrm{AJC}}^{(m,q)}(z_{\mathrm{sp}})
        \left[
            P_D(z_{\mathrm{sp}})
            -
            P_E(z_{\mathrm{sp}})
        \right],
    \end{aligned}
    \label{eq:supp_disp_AJC_population_exchange}
\end{equation}
where $W_{\mathrm{AJC}}^{(m,q)}(z_{\mathrm{sp}})$ denotes the population-exchange rate for a fixed spectator-TLS eigenvalue $z_{\mathrm{sp}}$ and is given by
\begin{equation}
    W_{\mathrm{AJC}}^{(m,q)}(z_{\mathrm{sp}})
    =
    \frac{
        2
        \left|
            \Omega_{\mathrm{AJC}}^{(m,q)}
        \right|^2
        \Gamma_{\mathrm{AJC}}^{(m,q)}
    }{
        \left[
            \Gamma_{\mathrm{AJC}}^{(m,q)}
        \right]^2
        +
        \left[
            \nu_{\mathrm{AJC}}^{(m,q)}(z_{\mathrm{sp}})
        \right]^2
    }.
    \label{eq:supp_disp_AJC_exchange_rate}
\end{equation}
For the time-dependent dynamics, Eq.~\eqref{eq:supp_disp_AJC_exchange_rate} follows from Eq.~\eqref{eq:supp_disp_AJC_coherence_eliminated}, whereas for the stationary solution it follows from Eq.~\eqref{eq:supp_disp_AJC_coherence_stationary}.
The Hamiltonian-induced population changes of $P_D(z_{\mathrm{sp}})$ and $P_E(z_{\mathrm{sp}})$ in Eq.~\eqref{eq:supp_disp_AJC_population_exchange} are equal in magnitude and opposite in sign. 
Consequently, $P_D(z_{\mathrm{sp}})+P_E(z_{\mathrm{sp}})$ is conserved by the Hamiltonian evolution, which only redistributes population between the two joint states.

We next include relaxation of the active TLS.
We define
\begin{equation}
    P_B(z_{\mathrm{sp}})
    \equiv
    \bra{B_{z_{\mathrm{sp}}}}
    \hat{\rho}_{\mathrm{joint}}
    \ket{B_{z_{\mathrm{sp}}}}.
    \label{eq:supp_disp_AJC_B_population}
\end{equation}
Within the same spectator-resolved description, we define the reduced density operator of the bosonic mode and the spectator TLS for the present AJC dynamics as
\begin{equation}
    \hat{\rho}_{\mathrm{c,sp}}
    \equiv
    \operatorname{Tr}_{\mathrm{act}}
    \hat{\rho}_{\mathrm{joint}},
    \label{eq:supp_disp_AJC_reduced_density}
\end{equation}
and the corresponding bosonic Fock-state populations as
\begin{equation}
    P_n(z_{\mathrm{sp}})
    \equiv
    \bra{n;z_{\mathrm{sp}}}
    \hat{\rho}_{\mathrm{c,sp}}
    \ket{n;z_{\mathrm{sp}}}.
    \label{eq:supp_disp_AJC_bosonic_population}
\end{equation}
The populations associated with the two relevant bosonic Fock states can be decomposed as
\begin{equation}
    \begin{aligned}
        &P_m(z_{\mathrm{sp}})
        =
        P_D(z_{\mathrm{sp}})
        +
        P_{m,+1}(z_{\mathrm{sp}})
        \simeq
        P_D(z_{\mathrm{sp}}),
        \\
        &P_{m+q}(z_{\mathrm{sp}})
        =
        P_E(z_{\mathrm{sp}})
        +
        P_B(z_{\mathrm{sp}}),
    \end{aligned}
    \label{eq:supp_disp_AJC_bosonic_joint_relation}
\end{equation}
where
\begin{equation}
    P_{n,+1}(z_{\mathrm{sp}})
    \equiv
    \bra{n;+1,z_{\mathrm{sp}}}
    \hat{\rho}_{\mathrm{joint}}
    \ket{n;+1,z_{\mathrm{sp}}}
    \label{eq:supp_disp_AJC_excited_population}
\end{equation}
denotes the population of the joint state $\ket{n;+1,z_{\mathrm{sp}}}$.
Within the reduced selected-transition description, we neglect the population of active-TLS-excited states outside the retained AJC pathway. 
In particular, we assume $P_{m,+1}(z_{\mathrm{sp}})\simeq0$, so that $P_m(z_{\mathrm{sp}})\simeq P_D(z_{\mathrm{sp}})$.
This approximation is consistent with the relaxation-dominated regime considered below, in which population transferred to such off-pathway active-TLS-excited states is rapidly removed by active-TLS relaxation.
The second relation follows directly from tracing over the active TLS.

For the three relevant joint-state populations, we write
\begin{equation}
    \left.
        \dot P_{\mu}(z_{\mathrm{sp}})
    \right|_{\mathrm{free+AJC+relax}}
    =
    \left.
        \dot P_{\mu}(z_{\mathrm{sp}})
    \right|_{\mathrm{free+AJC}}
    +
    \left.
        \dot P_{\mu}(z_{\mathrm{sp}})
    \right|_{\mathrm{relax}},
    \label{eq:supp_disp_AJC_population_decomposition}
\end{equation}
with $\mu=D,E,B$.
The Hamiltonian contributions to $P_D(z_{\mathrm{sp}})$ and $P_E(z_{\mathrm{sp}})$ have already been obtained in Eq.~\eqref{eq:supp_disp_AJC_population_exchange}.
By contrast, $\ket{B_{z_{\mathrm{sp}}}}=\ket{m+q;-1,z_{\mathrm{sp}}}$ does not participate in the retained resonant AJC transition and therefore has no Hamiltonian-induced population change within the selected-transition description,
\begin{equation}
    \left.
        \dot P_B(z_{\mathrm{sp}})
    \right|_{\mathrm{free+AJC}}
    \simeq 0.
    \label{eq:supp_disp_AJC_population_B_H}
\end{equation}
For $\ket{D_{z_{\mathrm{sp}}}}=\ket{m;-1,z_{\mathrm{sp}}}$, the active-TLS relaxation contribution is
\begin{equation}
    \begin{aligned}
        \left.
            \dot P_D(z_{\mathrm{sp}})
        \right|_{\mathrm{relax}}
        &=
        \gamma_{\mathrm{act}}
        \bra{D_{z_{\mathrm{sp}}}}
        \mathcal D[\hat{\sigma}_{\mathrm{act}}^-]
        \hat{\rho}_{\mathrm{joint}}
        \ket{D_{z_{\mathrm{sp}}}}
        \\
        &=
        \gamma_{\mathrm{act}}
        P_{m,+1}(z_{\mathrm{sp}})
        \simeq 0.
    \end{aligned}
    \label{eq:supp_disp_AJC_relax_D}
\end{equation}
For the intermediate state $\ket{E_{z_{\mathrm{sp}}}}=\ket{m+q;+1,z_{\mathrm{sp}}}$, the corresponding relaxation contribution is
\begin{equation}
    \begin{aligned}
        \left.
            \dot P_E(z_{\mathrm{sp}})
        \right|_{\mathrm{relax}}
        &=
        \gamma_{\mathrm{act}}
        \bra{E_{z_{\mathrm{sp}}}}
        \mathcal D[\hat{\sigma}_{\mathrm{act}}^-]
        \hat{\rho}_{\mathrm{joint}}
        \ket{E_{z_{\mathrm{sp}}}}
        \\
        &=
        -\gamma_{\mathrm{act}}
        P_E(z_{\mathrm{sp}}).
    \end{aligned}
    \label{eq:supp_disp_AJC_relax_E}
\end{equation}
For $\ket{B_{z_{\mathrm{sp}}}}=\ket{m+q;-1,z_{\mathrm{sp}}}$, the relaxation contribution is
\begin{equation}
    \begin{aligned}
        \left.
            \dot P_B(z_{\mathrm{sp}})
        \right|_{\mathrm{relax}}
        &=
        \gamma_{\mathrm{act}}
        \bra{B_{z_{\mathrm{sp}}}}
        \mathcal D[\hat{\sigma}_{\mathrm{act}}^-]
        \hat{\rho}_{\mathrm{joint}}
        \ket{B_{z_{\mathrm{sp}}}}
        \\
        &=
        \gamma_{\mathrm{act}}
        P_E(z_{\mathrm{sp}}).
    \end{aligned}
    \label{eq:supp_disp_AJC_relax_B}
\end{equation}
Thus, active-TLS relaxation removes population from $\ket{E_{z_{\mathrm{sp}}}}$ and transfers it to $\ket{B_{z_{\mathrm{sp}}}}$ without changing the bosonic Fock number.

Combining the Hamiltonian contributions in Eqs.~\eqref{eq:supp_disp_AJC_population_exchange} and \eqref{eq:supp_disp_AJC_population_B_H} with the active-TLS relaxation contributions in Eqs.~\eqref{eq:supp_disp_AJC_relax_D}--\eqref{eq:supp_disp_AJC_relax_B} gives
\begin{equation}
    \begin{aligned}
        \left.
            \dot P_D(z_{\mathrm{sp}})
        \right|_{\mathrm{free+AJC+relax}}
        &\simeq
        W_{\mathrm{AJC}}^{(m,q)}(z_{\mathrm{sp}})
        \left[
            P_E(z_{\mathrm{sp}})
            -
            P_D(z_{\mathrm{sp}})
        \right],
        \\
        \left.
            \dot P_E(z_{\mathrm{sp}})
        \right|_{\mathrm{free+AJC+relax}}
        &=
        W_{\mathrm{AJC}}^{(m,q)}(z_{\mathrm{sp}})
        \left[
            P_D(z_{\mathrm{sp}})
            -
            P_E(z_{\mathrm{sp}})
        \right]
        \\
        &\quad
        -
        \gamma_{\mathrm{act}}
        P_E(z_{\mathrm{sp}}),
        \\
        \left.
            \dot P_B(z_{\mathrm{sp}})
        \right|_{\mathrm{free+AJC+relax}}
        &\simeq
        \gamma_{\mathrm{act}}
        P_E(z_{\mathrm{sp}}).
    \end{aligned}
    \label{eq:supp_disp_AJC_joint_population_dynamics}
\end{equation}

Using Eq.~\eqref{eq:supp_disp_AJC_bosonic_joint_relation}, the corresponding population exchange between the two bosonic Fock states is
\begin{equation}
    \begin{aligned}
        &\left.
            \dot P_m(z_{\mathrm{sp}})
        \right|_{\mathrm{free+AJC+relax}}
        \simeq
        -
        W_{\mathrm{AJC}}^{(m,q)}(z_{\mathrm{sp}})
        \left[
            P_D(z_{\mathrm{sp}})
            -
            P_E(z_{\mathrm{sp}})
        \right],
        \\
        &\left.
            \dot P_{m+q}(z_{\mathrm{sp}})
        \right|_{\mathrm{free+AJC+relax}}
        \simeq
        W_{\mathrm{AJC}}^{(m,q)}(z_{\mathrm{sp}})
        \left[
            P_D(z_{\mathrm{sp}})
            -
            P_E(z_{\mathrm{sp}})
        \right].
    \end{aligned}
    \label{eq:supp_disp_AJC_bosonic_population_exchange}
\end{equation}
The loss and gain terms generated by active-TLS relaxation cancel in $P_{m+q}(z_{\mathrm{sp}})=P_E(z_{\mathrm{sp}})+P_B(z_{\mathrm{sp}})$.

The population-exchange equations above still depend on the intermediate-state population $P_E(z_{\mathrm{sp}})$ and therefore do not yet provide a closed description of the bosonic population dynamics. 
To obtain an effective bosonic population-transfer rate expressed solely in terms of the retained bosonic populations, we next eliminate $P_E(z_{\mathrm{sp}})$ by relating it to $P_D(z_{\mathrm{sp}})\simeq P_m(z_{\mathrm{sp}})$.

For this purpose, the intrinsic single-quantum loss of the bosonic mode must also be included in the dynamics of $\ket{E_{z_{\mathrm{sp}}}}=\ket{m+q;+1,z_{\mathrm{sp}}}$.
Projecting the full joint master equation~\eqref{eq:supp_disp_AJC_joint_master_equation} onto $\bra{E_{z_{\mathrm{sp}}}}\hat{\rho}_{\mathrm{joint}}\ket{E_{z_{\mathrm{sp}}}}$ gives
\begin{equation}
    \dot P_E(z_{\mathrm{sp}})
    =
    \left.
        \dot P_E(z_{\mathrm{sp}})
    \right|_{\mathrm{free+AJC+relax}}
    +
    \left.
        \dot P_E(z_{\mathrm{sp}})
    \right|_{\mathrm{loss}}.
    \label{eq:supp_disp_AJC_intermediate_full}
\end{equation}
The intrinsic single-quantum-loss channel contributes
\begin{equation}
    \begin{aligned}
        \left.
            \dot P_E(z_{\mathrm{sp}})
        \right|_{\mathrm{loss}}
        &=
        \kappa_c
        \bra{E_{z_{\mathrm{sp}}}}
        \mathcal D[\hat{a}_c]
        \hat{\rho}_{\mathrm{joint}}
        \ket{E_{z_{\mathrm{sp}}}}
        \\
        &=
        (m+q+1)\kappa_c
        P_{m+q+1,+1}(z_{\mathrm{sp}})
        \\
        &\quad
        -
        (m+q)\kappa_c
        P_E(z_{\mathrm{sp}}).
    \end{aligned}
    \label{eq:supp_disp_AJC_intermediate_loss}
\end{equation}
The first term in Eq.~\eqref{eq:supp_disp_AJC_intermediate_loss} describes population feeding into $\ket{E_{z_{\mathrm{sp}}}}$ from $\ket{m+q+1;+1,z_{\mathrm{sp}}}$, whereas the second term describes loss from $\ket{E_{z_{\mathrm{sp}}}}$ to $\ket{m+q-1;+1,z_{\mathrm{sp}}}$.
If the pure-dephasing channels introduced above are included, they do not directly modify the population equation for $P_E(z_{\mathrm{sp}})$; their effect is already contained in $W_{\mathrm{AJC}}^{(m,q)}(z_{\mathrm{sp}})$ through the total coherence-decay rate.

Substituting the Hamiltonian and active-TLS-relaxation contribution from Eq.~\eqref{eq:supp_disp_AJC_joint_population_dynamics} and the single-quantum-loss contribution from Eq.~\eqref{eq:supp_disp_AJC_intermediate_loss} into Eq.~\eqref{eq:supp_disp_AJC_intermediate_full} gives
\begin{equation}
    \begin{aligned}
        \dot P_E(z_{\mathrm{sp}})
        =&
        W_{\mathrm{AJC}}^{(m,q)}(z_{\mathrm{sp}})
        P_D(z_{\mathrm{sp}})
        \\
        &+
        (m+q+1)\kappa_c
        P_{m+q+1,+1}(z_{\mathrm{sp}})
        \\
        &-
        \left[
            W_{\mathrm{AJC}}^{(m,q)}(z_{\mathrm{sp}})
            +
            \gamma_{\mathrm{act}}
            +
            (m+q)\kappa_c
        \right]
        P_E(z_{\mathrm{sp}}).
    \end{aligned}
    \label{eq:supp_disp_AJC_intermediate_population}
\end{equation}
The feeding term proportional to $P_{m+q+1,+1}(z_{\mathrm{sp}})$ originates from single-quantum bosonic loss from $\ket{m+q+1;+1,z_{\mathrm{sp}}}$.
Since this state lies outside the retained resonant AJC pathway, we assume that its population is sufficiently small that this feeding term is negligible compared with the AJC-induced population transfer into $\ket{E_{z_{\mathrm{sp}}}}$, such that
\begin{equation}
    (m+q+1)\kappa_c
    P_{m+q+1,+1}(z_{\mathrm{sp}})
    \ll
    W_{\mathrm{AJC}}^{(m,q)}(z_{\mathrm{sp}})
    P_D(z_{\mathrm{sp}}).
    \label{eq:supp_disp_AJC_upper_feeding_condition}
\end{equation}
Equation~\eqref{eq:supp_disp_AJC_intermediate_population} then reduces to
\begin{equation}
    \begin{aligned}
        \dot P_E(z_{\mathrm{sp}})
        =&
        W_{\mathrm{AJC}}^{(m,q)}(z_{\mathrm{sp}})
        P_D(z_{\mathrm{sp}})
        \\
        &-
        \left[
            W_{\mathrm{AJC}}^{(m,q)}(z_{\mathrm{sp}})
            +
            \gamma_{\mathrm{act}}
            +
            (m+q)\kappa_c
        \right]
        P_E(z_{\mathrm{sp}}).
    \end{aligned}
    \label{eq:supp_disp_AJC_intermediate_population_reduced}
\end{equation}

For the time-dependent dynamics, $P_E(z_{\mathrm{sp}})$ can be adiabatically eliminated when its relaxation is fast compared with the population dynamics of the retained bosonic states. 
A convenient sufficient condition is
\begin{equation}
    \gamma_{\mathrm{act}}
    +
    (m+q)\kappa_c
    \gg
    W_{\mathrm{AJC}}^{(m,q)}(z_{\mathrm{sp}}).
    \label{eq:supp_disp_AJC_population_adiabatic_condition}
\end{equation}
Under this separation of time scales, setting $\dot P_E(z_{\mathrm{sp}})\simeq0$ in Eq.~\eqref{eq:supp_disp_AJC_intermediate_population_reduced} gives
\begin{equation}
    P_E(z_{\mathrm{sp}})
    \simeq
    \frac{
        W_{\mathrm{AJC}}^{(m,q)}(z_{\mathrm{sp}})
    }{
        W_{\mathrm{AJC}}^{(m,q)}(z_{\mathrm{sp}})
        +
        \gamma_{\mathrm{act}}
        +
        (m+q)\kappa_c
    }
    P_D(z_{\mathrm{sp}}).
    \label{eq:supp_disp_AJC_intermediate_elimination}
\end{equation}
For the stationary solution of the reduced population model, the same relation follows directly from $\dot P_E(z_{\mathrm{sp}})=0$ and therefore does not require the time-scale-separation condition in Eq.~\eqref{eq:supp_disp_AJC_population_adiabatic_condition}.
Under the condition in Eq.~\eqref{eq:supp_disp_AJC_population_adiabatic_condition}, Eq.~\eqref{eq:supp_disp_AJC_intermediate_elimination} further reduces to
\begin{equation}
    \begin{aligned}
        P_E(z_{\mathrm{sp}})
        &\simeq
        \frac{
            W_{\mathrm{AJC}}^{(m,q)}(z_{\mathrm{sp}})
        }{
            \gamma_{\mathrm{act}}
            +
            (m+q)\kappa_c
        }
        P_D(z_{\mathrm{sp}})
        \\
        &\simeq
        \frac{
            W_{\mathrm{AJC}}^{(m,q)}(z_{\mathrm{sp}})
        }{
            \gamma_{\mathrm{act}}
            +
            (m+q)\kappa_c
        }
        P_m(z_{\mathrm{sp}}).
    \end{aligned}
    \label{eq:supp_disp_AJC_intermediate_fast}
\end{equation}

Substituting the eliminated intermediate-state population in Eq.~\eqref{eq:supp_disp_AJC_intermediate_elimination} into the bosonic population-exchange equations in Eq.~\eqref{eq:supp_disp_AJC_bosonic_population_exchange}, and using $P_D(z_{\mathrm{sp}})\simeq P_m(z_{\mathrm{sp}})$, gives
\begin{equation}
    \begin{aligned}
        &\left.
            \dot P_m(z_{\mathrm{sp}})
        \right|_{\mathrm{free+AJC+relax}}
        \simeq
        -
        \mathcal R_{\mathrm{AJC}}^{\uparrow}
        (m,q;z_{\mathrm{sp}})
        P_m(z_{\mathrm{sp}}),
        \\
        &\left.
            \dot P_{m+q}(z_{\mathrm{sp}})
        \right|_{\mathrm{free+AJC+relax}}
        \simeq
        \mathcal R_{\mathrm{AJC}}^{\uparrow}
        (m,q;z_{\mathrm{sp}})
        P_m(z_{\mathrm{sp}}).
    \end{aligned}
    \label{eq:supp_disp_AJC_effective_population_dynamics}
\end{equation}
Here, $\mathcal R_{\mathrm{AJC}}^{\uparrow}(m,q;z_{\mathrm{sp}})$ denotes the effective forward bosonic population-transfer rate from $\ket{m}$ to $\ket{m+q}$ for a fixed spectator-TLS eigenvalue $z_{\mathrm{sp}}$, and is given by
\begin{equation}
    \mathcal R_{\mathrm{AJC}}^{\uparrow}
    (m,q;z_{\mathrm{sp}})
    =
    \frac{
        W_{\mathrm{AJC}}^{(m,q)}(z_{\mathrm{sp}})
        \left[
            \gamma_{\mathrm{act}}
            +
            (m+q)\kappa_c
        \right]
    }{
        W_{\mathrm{AJC}}^{(m,q)}(z_{\mathrm{sp}})
        +
        \gamma_{\mathrm{act}}
        +
        (m+q)\kappa_c
    }.
    \label{eq:supp_disp_AJC_forward_rate}
\end{equation}
No term proportional to $P_{m+q}(z_{\mathrm{sp}})$ that would generate reverse transfer toward $P_m(z_{\mathrm{sp}})$ arises from the reduced AJC+relaxation contribution at this order. 
We denote the corresponding effective reverse bosonic population-transfer rate from $\ket{m+q}$ to $\ket{m}$, for a fixed spectator-TLS eigenvalue $z_{\mathrm{sp}}$, by $\mathcal R_{\mathrm{AJC}}^{\downarrow}(m,q;z_{\mathrm{sp}})$. 
Accordingly, this rate vanishes to leading order,
\begin{equation}
    \mathcal R_{\mathrm{AJC}}^{\downarrow}
    (m,q;z_{\mathrm{sp}})
    \simeq 0.
    \label{eq:supp_disp_AJC_reverse_rate}
\end{equation}
The $q$-quantum AJC interaction itself allows coherent population exchange in both directions between $\ket{D_{z_{\mathrm{sp}}}}$ and $\ket{E_{z_{\mathrm{sp}}}}$.
After eliminating the intermediate state, this bidirectional exchange is absorbed into the effective forward rate $\mathcal R_{\mathrm{AJC}}^{\uparrow}(m,q;z_{\mathrm{sp}})$, which therefore describes the net bosonic population-transfer rate.
The direct intrinsic single-quantum bosonic-loss term remains explicit in the reduced bosonic dynamics, while its depletion of the intermediate state enters $\mathcal R_{\mathrm{AJC}}^{\uparrow}(m,q;z_{\mathrm{sp}})$ through the elimination of $P_E(z_{\mathrm{sp}})$.

It is useful to distinguish the effective forward rate in Eq.~\eqref{eq:supp_disp_AJC_forward_rate} from the rate at which population completes the specific sequential pathway
\begin{equation}
    \ket{D_{z_{\mathrm{sp}}}}
    \rightarrow
    \ket{E_{z_{\mathrm{sp}}}}
    \rightarrow
    \ket{B_{z_{\mathrm{sp}}}}.
    \label{eq:supp_disp_AJC_directional_path}
\end{equation}
The latter is determined by the active-TLS-relaxation flux out of the intermediate state. 
Using Eq.~\eqref{eq:supp_disp_AJC_relax_B} together with Eq.~\eqref{eq:supp_disp_AJC_intermediate_elimination} and $P_D(z_{\mathrm{sp}})\simeq P_m(z_{\mathrm{sp}})$ gives
\begin{equation}
    \left.
        \dot P_B(z_{\mathrm{sp}})
    \right|_{\mathrm{relax}}
    \simeq
    \mathcal R_{\mathrm{AJC}}^{\mathrm{path}}
    (m,q;z_{\mathrm{sp}})
    P_m(z_{\mathrm{sp}}),
    \label{eq:supp_disp_AJC_path_flux}
\end{equation}
where
\begin{equation}
    \mathcal R_{\mathrm{AJC}}^{\mathrm{path}}
    (m,q;z_{\mathrm{sp}})
    =
    \frac{
        \gamma_{\mathrm{act}}
        W_{\mathrm{AJC}}^{(m,q)}(z_{\mathrm{sp}})
    }{
        W_{\mathrm{AJC}}^{(m,q)}(z_{\mathrm{sp}})
        +
        \gamma_{\mathrm{act}}
        +
        (m+q)\kappa_c
    }.
    \label{eq:supp_disp_AJC_path_rate}
\end{equation}
The two rates are therefore related by
\begin{equation}
    \mathcal R_{\mathrm{AJC}}^{\mathrm{path}}
    (m,q;z_{\mathrm{sp}})
    =
    \frac{
        \gamma_{\mathrm{act}}
    }{
        \gamma_{\mathrm{act}}
        +
        (m+q)\kappa_c
    }
    \mathcal R_{\mathrm{AJC}}^{\uparrow}
    (m,q;z_{\mathrm{sp}}).
    \label{eq:supp_disp_AJC_path_rate_relation}
\end{equation}
The prefactor
\begin{equation}
    \frac{
        \gamma_{\mathrm{act}}
    }{
        \gamma_{\mathrm{act}}
        +
        (m+q)\kappa_c
    }
    \label{eq:supp_disp_AJC_branching_fraction}
\end{equation}
is the branching fraction of the irreversible removal from the intermediate state that proceeds through active-TLS relaxation rather than through the competing single-quantum bosonic-loss channel.
The relaxation branch transfers the population to $\ket{B_{z_{\mathrm{sp}}}}$ and thereby completes the desired sequential pathway.

Under the time-scale-separation condition in Eq.~\eqref{eq:supp_disp_AJC_population_adiabatic_condition}, the effective forward bosonic population-transfer rate reduces to
\begin{equation}
    \mathcal R_{\mathrm{AJC}}^{\uparrow}
    (m,q;z_{\mathrm{sp}})
    \simeq
    W_{\mathrm{AJC}}^{(m,q)}(z_{\mathrm{sp}}),
    \label{eq:supp_disp_AJC_forward_rate_fast}
\end{equation}
whereas the rate for completing the desired sequential pathway becomes
\begin{equation}
    \mathcal R_{\mathrm{AJC}}^{\mathrm{path}}
    (m,q;z_{\mathrm{sp}})
    \simeq
    \frac{
        \gamma_{\mathrm{act}}
    }{
        \gamma_{\mathrm{act}}
        +
        (m+q)\kappa_c
    }
    W_{\mathrm{AJC}}^{(m,q)}(z_{\mathrm{sp}}).
    \label{eq:supp_disp_AJC_path_rate_fast}
\end{equation}
For the AJC-assisted process to predominantly follow the desired pathway, active-TLS relaxation must further dominate the competing bosonic loss from the intermediate state,
\begin{equation}
    \gamma_{\mathrm{act}}
    \gg
    (m+q)\kappa_c.
    \label{eq:supp_disp_AJC_relaxation_dominance}
\end{equation}
In this limit, the branching fraction approaches unity and the two rates coincide to leading order,
\begin{equation}
    \mathcal R_{\mathrm{AJC}}^{\mathrm{path}}
    (m,q;z_{\mathrm{sp}})
    \simeq
    \mathcal R_{\mathrm{AJC}}^{\uparrow}
    (m,q;z_{\mathrm{sp}})
    \simeq
    W_{\mathrm{AJC}}^{(m,q)}(z_{\mathrm{sp}}).
    \label{eq:supp_disp_AJC_directional_limit}
\end{equation}

We finally specialize to the resonant reference channel $m=m_0$ and $z_{\mathrm{sp}}=\bar z_{\mathrm{sp}}$.
Since $\nu_{\mathrm{AJC}}^{(m_0,q)}(\bar z_{\mathrm{sp}})=0$, Eq.~\eqref{eq:supp_disp_AJC_exchange_rate} gives
\begin{equation}
    W_{\mathrm{AJC}}^{(m_0,q)}(\bar z_{\mathrm{sp}})
    =
    \frac{
        2
        \left|
            \Omega_{\mathrm{AJC}}^{(m_0,q)}
        \right|^2
    }{
        \Gamma_{\mathrm{AJC}}^{(m_0,q)}
    }.
    \label{eq:supp_disp_AJC_exchange_rate_reference}
\end{equation}
For the opposite spectator-TLS eigenvalue, the same bosonic transition acquires the mismatch
\begin{equation}
    \left|
        \nu_{\mathrm{AJC}}^{(m_0,q)}
        (-\bar z_{\mathrm{sp}})
    \right|
    =
    2q|\chi_{\mathrm{sp}}|.
    \label{eq:supp_disp_AJC_opposite_spectator_mismatch}
\end{equation}
Consequently, the population-exchange rate for the opposite spectator-TLS eigenvalue is suppressed relative to that for the reference eigenvalue according to
\begin{equation}
    \frac{
        W_{\mathrm{AJC}}^{(m_0,q)}(-\bar z_{\mathrm{sp}})
    }{
        W_{\mathrm{AJC}}^{(m_0,q)}(\bar z_{\mathrm{sp}})
    }
    =
    \frac{
        \left[
            \Gamma_{\mathrm{AJC}}^{(m_0,q)}
        \right]^2
    }{
        \left[
            \Gamma_{\mathrm{AJC}}^{(m_0,q)}
        \right]^2
        +
        4q^2\chi_{\mathrm{sp}}^2
    }.
    \label{eq:supp_disp_AJC_spectator_suppression}
\end{equation}
Thus, the spectator-TLS state controls the magnitude of the population exchange through $W_{\mathrm{AJC}}^{(m,q)}(z_{\mathrm{sp}})$ and therefore also controls the effective forward rate $\mathcal R_{\mathrm{AJC}}^{\uparrow}(m,q;z_{\mathrm{sp}})$.
Under the time-scale-separation condition in Eq.~\eqref{eq:supp_disp_AJC_population_adiabatic_condition}, $\mathcal R_{\mathrm{AJC}}^{\uparrow}\simeq W_{\mathrm{AJC}}$, so the spectator dependence of the forward rate directly follows the Lorentzian suppression in Eq.~\eqref{eq:supp_disp_AJC_spectator_suppression}.

For the resonant reference channel, the further relaxation-dominated reduction has the same form as in the Kerr-resolved AJC+relaxation realization.
In particular, increasing the active-TLS relaxation rate suppresses the intermediate-state population and coherent back-transfer, whereas excessively strong relaxation ultimately reduces the transfer rate through coherence broadening.
Under the same limiting assumptions used in Supplementary Sec.~\ref{sec:supp_effective_connectors}, the corresponding analytical results follow directly from Eqs.~\eqref{eq:supp_AJC_relaxation_dominated_coherence}--\eqref{eq:supp_AJC_eff_inverse_relaxation_scaling} and are not repeated here.

\subsection{Mapping onto the global and local connectors}
\label{subsec:supp_disp_mapping}

The dispersive $q$-quantum pumped-JC and AJC+relaxation realizations derived above yield spectator-resolved effective bosonic population-transfer rates for selected transitions of the form $\ket{m}\leftrightarrow\ket{m+q}$.
For the $q$-quantum pumped-JC realization, the reduced population dynamics are given in Eq.~\eqref{eq:supp_disp_JC_conditional_population_dynamics}, with the upward and downward population-transfer rates $\mathcal R_{\mathrm{JC}}^{\uparrow}(m,q;z_{\mathrm{sp}})$ and $\mathcal R_{\mathrm{JC}}^{\downarrow}(m,q;z_{\mathrm{sp}})$ given explicitly in Eq.~\eqref{eq:supp_disp_JC_conditional_rates}.
For the $q$-quantum AJC+relaxation realization, the reduced population dynamics are given in Eq.~\eqref{eq:supp_disp_AJC_effective_population_dynamics}, with the forward and reverse population-transfer rates $\mathcal R_{\mathrm{AJC}}^{\uparrow}(m,q;z_{\mathrm{sp}})$ and $\mathcal R_{\mathrm{AJC}}^{\downarrow}(m,q;z_{\mathrm{sp}})$ obtained from Eqs.~\eqref{eq:supp_disp_AJC_forward_rate} and \eqref{eq:supp_disp_AJC_reverse_rate}.
We now map these spectator-resolved effective population-transfer rates onto the selective global and movable local connector rates.
When both auxiliary TLSs are present, the global mapping identifies
$(\chi_{\mathrm{act}},\chi_{\mathrm{sp}},\Delta_{\mathrm{act}})
=(\chi_s,\chi_R,\Delta_s)$, whereas the local mapping identifies
$(\chi_{\mathrm{act}},\chi_{\mathrm{sp}},\Delta_{\mathrm{act}})
=(\chi_R,\chi_s,\Delta_R)$, in the notation of the main text.
The generic amplitudes $g_{\mathrm{JC}}^{(q)}$ and
$g_{\mathrm{AJC}}^{(q)}$ correspond to the main-text amplitudes
$g_{G,\mathrm{JC}}^{(N_s)}$ and $g_{G,\mathrm{AJC}}^{(N_s)}$
for the global connector, or $g_{L,\mathrm{JC}}^{(1)}$
and $g_{L,\mathrm{AJC}}^{(1)}$ for the local connector.
These identifications do not alter the spectator-subspace conditions stated below.

For the global connector, the selected bosonic transition is $\ket{0}\leftrightarrow\ket{N_s}$.
Accordingly, the lower Fock-state index and the exchanged-quantum number are
\begin{equation}
    m=0,
    \qquad
    q=N_s.
    \label{eq:supp_disp_global_selected_transition}
\end{equation}
Retaining the spectator-TLS eigenvalue explicitly, the global connector rates $\mathcal R_{\mathrm G}^{d}(z_{\mathrm{sp}})$, with $d\in\{\uparrow,\downarrow\}$, are defined from the effective rates derived in Eq.~\eqref{eq:supp_disp_JC_conditional_rates} and Eqs.~\eqref{eq:supp_disp_AJC_forward_rate}--\eqref{eq:supp_disp_AJC_reverse_rate} as
\begin{equation}
    \mathcal R_{\mathrm G}^{d}(z_{\mathrm{sp}})
    \equiv
    \begin{cases}
        \mathcal R_{\mathrm{JC}}^{d}
        (0,N_s;z_{\mathrm{sp}}),
        &
        \text{pumped $N_s$-JC},
        \\[6pt]
        \mathcal R_{\mathrm{AJC}}^{d}
        (0,N_s;z_{\mathrm{sp}}),
        &
        \text{$N_s$-AJC+relaxation}.
    \end{cases}
    \label{eq:supp_disp_global_connector_mapping}
\end{equation}

Similarly, the movable local connector targeting $\ket{R}$ corresponds to the transition $\ket{R-1}\leftrightarrow\ket{R}$, with
\begin{equation}
    m=R-1,
    \qquad
    q=1.
    \label{eq:supp_disp_local_selected_transition}
\end{equation}
Upon retaining the spectator-TLS eigenvalue explicitly and evaluating the generic transition rates at the indices specified in Eq.~\eqref{eq:supp_disp_local_selected_transition}, the corresponding spectator-resolved local-connector rates $\mathcal R_{\mathrm L}^{d}(R;z_{\mathrm{sp}})$, with $d\in\{\uparrow,\downarrow\}$, take the form
\begin{equation}
    \mathcal R_{\mathrm L}^{d}(R;z_{\mathrm{sp}})
    \equiv
    \begin{cases}
        \mathcal R_{\mathrm{JC}}^{d}(R-1,1;z_{\mathrm{sp}}),
        & \text{pumped-JC},
        \\[4pt]
        \mathcal R_{\mathrm{AJC}}^{d}(R-1,1;z_{\mathrm{sp}}),
        & \text{AJC+relaxation}.
    \end{cases}
    \label{eq:supp_disp_local_connector_mapping}
\end{equation}

The rates in Eqs.~\eqref{eq:supp_disp_global_connector_mapping} and \eqref{eq:supp_disp_local_connector_mapping} remain conditional on the spectator-TLS eigenvalue $z_{\mathrm{sp}}$.
To realize the selected $q$-quantum transition $\ket{m_0}\leftrightarrow\ket{m_0+q}$, the active-TLS detuning is chosen such that the transition is resonant in a reference spectator-TLS subspace $z_{\mathrm{sp}}=\bar z_{\mathrm{sp}}$.
The reference eigenvalues $\bar z_{\mathrm{sp}}$ for the global and local connectors are denoted by $\bar z_{\mathrm{sp}}^{\mathrm G}$ and $\bar z_{\mathrm{sp}}^{\mathrm L}$, respectively. 
The former refers to the local spectator TLS and the latter to the global spectator TLS; their values need not coincide.
Evaluating the conditional rates in their respective resonant spectator-TLS subspaces defines the connector rates used in the reduced Fock-space model as
\begin{equation}
    \mathcal R_{\mathrm G}^{d}
    \equiv
    \mathcal R_{\mathrm G}^{d}
    (\bar z_{\mathrm{sp}}^{\mathrm G}),
    \qquad
    \mathcal R_{\mathrm L}^{d}(R)
    \equiv
    \mathcal R_{\mathrm L}^{d}
    (R;\bar z_{\mathrm{sp}}^{\mathrm L}).
    \label{eq:supp_disp_reference_connector_rates}
\end{equation}
Within the selected-transition and effective-rate approximations, the reduction to the spectator-independent connector rates defined in Eq.~\eqref{eq:supp_disp_reference_connector_rates} is exact when the spectator-TLS population is confined to the resonant reference subspace $z_{\mathrm{sp}}=\bar z_{\mathrm{sp}}$.
This reduction remains a good approximation when the population-weighted contribution of the nonresonant subspace $z_{\mathrm{sp}}=-\bar z_{\mathrm{sp}}$ is negligible, and its accuracy deteriorates as this contribution increases.

In the opposite spectator-TLS subspace $z_{\mathrm{sp}}=-\bar z_{\mathrm{sp}}$, the selected $q$-quantum transition acquires a mismatch of magnitude $2q|\chi_{\mathrm{sp}}|$ relative to the resonant reference subspace, as follows from Eqs.~\eqref{eq:supp_disp_generic_JC_compensated} and \eqref{eq:supp_disp_generic_AJC_compensated}.
This mismatch suppresses the population-transfer rate through its Lorentzian dependence, as described by Eq.~\eqref{eq:supp_disp_JC_exchange_rate} for the $q$-quantum JC implementation and by Eq.~\eqref{eq:supp_disp_AJC_spectator_suppression} for the $q$-quantum AJC+relaxation implementation.

To reduce the spectator-resolved dynamics to the boson-only connector description, we consider an operating regime in which the stationary population of the spectator TLS is strongly concentrated in the selected reference state $z_{\mathrm{sp}}=\bar z_{\mathrm{sp}}$. 
Such a population bias can be established in an appropriate operating regime by choosing its incoherent pumping and relaxation rates relative to the other relevant dynamical scales.
The contribution associated with the opposite spectator state $z_{\mathrm{sp}}=-\bar z_{\mathrm{sp}}$ is then suppressed by both its small stationary population and the dispersive detuning of the corresponding transition and can therefore be neglected to leading order.
Under this condition, evaluating the spectator-resolved rates in the resonant reference subspace $z_{\mathrm{sp}}=\bar z_{\mathrm{sp}}$ gives the effective rates $\mathcal R_{\mathrm G}^{d}$ and $\mathcal R_{\mathrm L}^{d}(R)$ entering the global and local currents defined in Eqs.~\eqref{eq:supp_JG} and \eqref{eq:supp_JL}, respectively.
The microscopic realizations therefore modify the values of the effective connector rates while leaving the unified Fock-state population model unchanged.
The resulting bosonic populations then obey Eq.~\eqref{eq:supp_population_equation}.

\section{Unified effective Fock-space population model and steady-state solution}
\label{sec:supp_steady_state}

\begin{table*}[t]

\caption{Unified mapping of the microscopic implementations with Kerr-resolved and dispersive spectral selection onto the effective global and movable local connector rates, with $d\in\{\uparrow,\downarrow\}$ denoting the direction of bosonic population transfer.
For the dispersive implementations, the rates entering the reduced Fock-space model are evaluated in the resonant reference subspace $z_{\mathrm{sp}}=\bar z_{\mathrm{sp}}$.
}
\label{tab:supp_connector_rate_summary}

\centering
\small

\begin{ruledtabular}

\begin{tabular}{
@{}
p{1.15cm}
p{4.9cm}
p{2.85cm}
p{4.75cm}
p{2.25cm}
@{}
}

Connector
&
Microscopic implementation
&
Selected indices and source
&
Effective-rate identification
&
Rate source
\\

\colrule

\multicolumn{5}{c}{
\textit{Intrinsic Kerr spectral selection}
}
\\

\colrule

\multirow{3}{*}{Global}
&
pumped $N_s$-quantum JC
&
\multirow{3}{2.85cm}{
$(m,q)=(0,N_s)$,
Eq.~\eqref{eq:supp_global_mapping}
}
&
$\mathcal R_{\mathrm G}^{d}
\equiv
\mathcal R_{\mathrm{G,JC}}^{d}$
&
Eq.~\eqref{eq:supp_global_JC_general_rates}
\\

&
coherent $N_s$-quantum drive
&
&
$\mathcal R_{\mathrm G}^{\uparrow}
=
\mathcal R_{\mathrm G}^{\downarrow}
\equiv
W_{\mathrm{drv}}^{(0,N_s)}$
&
Eq.~\eqref{eq:supp_global_drive_rates}
\\

&
$N_s$-quantum AJC+relaxation
&
&
$\mathcal R_{\mathrm G}^{\uparrow}
\equiv
\mathcal R_{\mathrm{G,AJC}}^{\uparrow}$,
$\mathcal R_{\mathrm G}^{\downarrow}\simeq0$
&
Eq.~\eqref{eq:supp_global_AJC_rates}
\\

\cline{1-5}

\multirow{2}{*}{Local}
&
pumped single-quantum JC
&
\multirow{2}{2.85cm}{
$(m,q)=(R-1,1)$,
Eq.~\eqref{eq:supp_local_mapping}
}
&
$\mathcal R_{\mathrm L}^{d}(R)
\equiv
\mathcal R_{\mathrm{L,JC}}^{d}(R)$
&
Eq.~\eqref{eq:supp_local_JC_general_rates}
\\

&
single-quantum AJC+relaxation
&
&
$\mathcal R_{\mathrm L}^{\uparrow}(R)
\equiv
\mathcal R_{\mathrm{L,AJC}}^{\uparrow}(R)$,
$\mathcal R_{\mathrm L}^{\downarrow}(R)\simeq0$
&
Eq.~\eqref{eq:supp_local_AJC_rates}
\\

\colrule

\multicolumn{5}{c}{
\textit{Dispersive spectral selection}
}
\\

\colrule

\multirow{2}{*}{Global}
&
pumped $N_s$-quantum JC
&
\multirow{2}{2.85cm}{
$(m,q)=(0,N_s)$,
Eq.~\eqref{eq:supp_disp_global_selected_transition}
}
&
$\mathcal R_{\mathrm G}^{d}
\equiv
\mathcal R_{\mathrm{JC}}^{d}
(0,N_s;\bar z_{\mathrm{sp}}^{\mathrm G})$
&
Eq.~\eqref{eq:supp_disp_JC_conditional_rates}
\\

&
$N_s$-quantum AJC+relaxation
&
&
$\mathcal R_{\mathrm G}^{\uparrow}
\equiv
\mathcal R_{\mathrm{AJC}}^{\uparrow}
(0,N_s;\bar z_{\mathrm{sp}}^{\mathrm G})$,
$\mathcal R_{\mathrm G}^{\downarrow}\simeq0$
&
Eqs.~\eqref{eq:supp_disp_AJC_forward_rate}
and \eqref{eq:supp_disp_AJC_reverse_rate}
\\

\cline{1-5}

\multirow{2}{*}{Local}
&
pumped single-quantum JC
&
\multirow{2}{2.85cm}{
$(m,q)=(R-1,1)$,
Eq.~\eqref{eq:supp_disp_local_selected_transition}
}
&
$\mathcal R_{\mathrm L}^{d}(R)
\equiv
\mathcal R_{\mathrm{JC}}^{d}
(R-1,1;\bar z_{\mathrm{sp}}^{\mathrm L})$
&
Eq.~\eqref{eq:supp_disp_JC_conditional_rates}
\\

&
single-quantum AJC+relaxation
&
&
$\mathcal R_{\mathrm L}^{\uparrow}(R)
\equiv
\mathcal R_{\mathrm{AJC}}^{\uparrow}
(R-1,1;\bar z_{\mathrm{sp}}^{\mathrm L})$,
$\mathcal R_{\mathrm L}^{\downarrow}(R)\simeq0$
&
Eqs.~\eqref{eq:supp_disp_AJC_forward_rate}
and \eqref{eq:supp_disp_AJC_reverse_rate}
\\[-0.50ex]
\end{tabular}

\end{ruledtabular}

\end{table*}

Having reduced the microscopic realizations of the global and local connectors to the unified effective bosonic population-transfer rates derived above, we now combine these engineered processes with the intrinsic linear loss of the bosonic mode.
At this level, the microscopic realization of each connector enters only through the corresponding effective rates $\mathcal R_{\mathrm G}^{\uparrow}$, $\mathcal R_{\mathrm G}^{\downarrow}$, $\mathcal R_{\mathrm L}^{\uparrow}(R)$, and $\mathcal R_{\mathrm L}^{\downarrow}(R)$.
By construction, these effective bosonic population-transfer rates are nonnegative.
The position $R$ of the movable local connector determines where the corresponding local transition appears in Fock space.
Accordingly, the population dynamics can be formulated in a unified rate-matrix form, with the dependence on $R$ entering only through the corresponding matrix elements.

For the reduced bosonic density operator $\hat{\rho}_c$, we denote the Fock-state populations by
\begin{equation}
    P_n
    \equiv
    \bra{n}
        \hat{\rho}_c
    \ket{n},
    \qquad
    \sum_{n=0}^{\infty}P_n
    =
    1.
    \label{eq:supp_Pm}
\end{equation}
The net population current generated by the global connector is defined as
\begin{equation}
    J_{\mathrm G}
    \equiv
    \mathcal R_{\mathrm G}^{\uparrow}P_0
    -
    \mathcal R_{\mathrm G}^{\downarrow}P_{N_s},
    \label{eq:supp_JG}
\end{equation}
where positive $J_{\mathrm G}$ corresponds to net population transfer from $\ket{0}$ to $\ket{N_s}$.
Similarly, the net population current generated by the movable local connector is
\begin{equation}
    J_{\mathrm L}(R)
    \equiv
    \mathcal R_{\mathrm L}^{\uparrow}(R)P_{R-1}
    -
    \mathcal R_{\mathrm L}^{\downarrow}(R)P_R,
    \label{eq:supp_JL}
\end{equation}
where positive $J_{\mathrm L}(R)$ corresponds to net population transfer from $\ket{R-1}$ to $\ket{R}$.

All microscopic implementations enter the unified effective Fock-space population model through the effective connector rates summarized in Supplementary Table~\ref{tab:supp_connector_rate_summary}.

We next include the intrinsic linear loss of the bosonic mode.
Its contribution to the population of the Fock state $\ket n$ follows by projecting the Lindblad term $\kappa_c\mathcal D[\hat{a}_c]\hat{\rho}_c$ onto $\ket n$,
\begin{align}
    \left.
        \dot P_n
    \right|_{\mathrm{loss}}
    &=
    \kappa_c
    \bra n
        \mathcal D[\hat{a}_c]
        \hat{\rho}_c
    \ket n
    \nonumber\\
    &=
    \kappa_c
    \bra n
        \hat{a}_c
        \hat{\rho}_c
        \hat{a}_c^{\dagger}
    \ket n
    -
    \frac{\kappa_c}{2}
    \bra n
        \left\{
            \hat{a}_c^{\dagger}\hat{a}_c,
            \hat{\rho}_c
        \right\}
    \ket n
    \nonumber\\
    &=
    (n+1)\kappa_cP_{n+1}
    -
    n\kappa_cP_n.
    \label{eq:supp_population_loss_contribution}
\end{align}
Here, the first term describes population feeding into $\ket n$ through the intrinsic transition $\ket{n+1}\rightarrow\ket n$, whereas the second term describes population loss from $\ket n$ through $\ket n\rightarrow\ket{n-1}$.
The global connector contributes only to the populations of $\ket0$ and $\ket{N_s}$.
Using the net current defined in Eq.~\eqref{eq:supp_JG}, its contribution can be written as
\begin{equation}
    \left.
        \dot P_n
    \right|_{\mathrm G}
    =
    \left(
        \delta_{n,N_s}
        -
        \delta_{n,0}
    \right)
    J_{\mathrm G},
    \label{eq:supp_population_global_contribution}
\end{equation}
where $\delta_{i,j}$ denotes the Kronecker delta.
Similarly, the movable local connector contributes only to the populations of $\ket{R-1}$ and $\ket R$, so that Eq.~\eqref{eq:supp_JL} gives
\begin{equation}
    \left.
        \dot P_n
    \right|_{\mathrm L}
    =
    \left(
        \delta_{n,R}
        -
        \delta_{n,R-1}
    \right)
    J_{\mathrm L}(R).
    \label{eq:supp_population_local_contribution}
\end{equation}
Here, $\left.\dot P_n\right|_{\mathrm{loss}}$, $\left.\dot P_n\right|_{\mathrm G}$, and $\left.\dot P_n\right|_{\mathrm L}$ denote the contributions to the time derivative of $P_n$ arising from intrinsic linear loss, the effective global connector, and the effective movable local connector, respectively.
Combining Eqs.~\eqref{eq:supp_population_loss_contribution},
\eqref{eq:supp_population_global_contribution}, and
\eqref{eq:supp_population_local_contribution}, the population dynamics take the unified form
\begin{align}
    \dot P_n
    &=
    (n+1)\kappa_cP_{n+1}
    -
    n\kappa_cP_n
    \nonumber\\
    &\quad
    +
    \left(
        \delta_{n,N_s}
        -
        \delta_{n,0}
    \right)
    J_{\mathrm G}
    \nonumber\\
    &\quad
    +
    \left(
        \delta_{n,R}
        -
        \delta_{n,R-1}
    \right)
    J_{\mathrm L}(R).
    \label{eq:supp_population_equation}
\end{align}
Equation~\eqref{eq:supp_population_equation} applies to all allowed positions of the movable local connector.
Changing $R$ changes only the Fock-space location of the local population-transfer terms.

To obtain a finite-dimensional representation of Eq.~\eqref{eq:supp_population_equation}, we truncate the Fock space at the maximum Fock-state index
\begin{equation}
    n_{\max}
    \equiv
    N_{\mathrm{cut}}-1,
    \label{eq:supp_fock_cutoff}
\end{equation}
where $N_{\mathrm{cut}}$ is the bosonic Hilbert-space dimension:
the retained Fock basis is $\ket{0},\ldots,\ket{N_{\mathrm{cut}}-1}$.
The cutoff is chosen such that
\begin{equation}
    n_{\max}
    \geq
    \max
    \left(
        N_s,
        R
    \right),
    \label{eq:supp_fock_cutoff_condition}
\end{equation}
so that all Fock states directly involved in the global and local connectors are contained in the truncated space.
The value of $N_{\mathrm{cut}}$ is a Hilbert-space dimension rather than a physical boundary of the Fock-space model; it can be systematically increased to verify convergence of the stationary solution.

Within this truncated Fock space, we introduce the population vector
\begin{equation}
    \mathbf P
    \equiv
    \left(
        P_0,
        P_1,
        \ldots,
        P_{n_{\max}}
    \right)^{\mathsf T}.
    \label{eq:supp_population_vector}
\end{equation}
Equation~\eqref{eq:supp_population_equation} can then be written as the linear rate equation
\begin{equation}
    \dot{\mathbf P}
    =
    \mathsf M(R)
    \mathbf P,
    \label{eq:supp_rate_matrix_equation}
\end{equation}
where $\mathsf M(R)$ is the $(n_{\max}+1)\times(n_{\max}+1)$ effective Fock-space rate matrix.
For $0\leq m,n\leq n_{\max}$, its matrix elements are
\begin{align}
    \mathsf M_{mn}(R)
    &=
    n\kappa_c
    \left(
        \delta_{m,n-1}
        -
        \delta_{m,n}
    \right)
    \nonumber\\
    &\quad
    +
    \mathcal R_{\mathrm G}^{\uparrow}
    \left(
        \delta_{m,N_s}
        -
        \delta_{m,0}
    \right)
    \delta_{n,0}
    \nonumber\\
    &\quad
    +
    \mathcal R_{\mathrm G}^{\downarrow}
    \left(
        \delta_{m,0}
        -
        \delta_{m,N_s}
    \right)
    \delta_{n,N_s}
    \nonumber\\
    &\quad
    +
    \mathcal R_{\mathrm L}^{\uparrow}(R)
    \left(
        \delta_{m,R}
        -
        \delta_{m,R-1}
    \right)
    \delta_{n,R-1}
    \nonumber\\
    &\quad
    +
    \mathcal R_{\mathrm L}^{\downarrow}(R)
    \left(
        \delta_{m,R-1}
        -
        \delta_{m,R}
    \right)
    \delta_{n,R}.
    \label{eq:supp_rate_matrix_elements}
\end{align}
The first line describes the intrinsic downward loss $\ket{n}\rightarrow\ket{n-1}$ at rate $n\kappa_c$.
The second and third lines describe the upward and downward population transfer generated by the global connector, respectively, while the fourth and fifth lines describe the corresponding population transfer generated by the movable local connector.

The conservation of the total population follows directly from Eq.~\eqref{eq:supp_population_equation}.
Summing the population equations over the truncated Fock space and using $P_{n_{\max}+1}=0$, we obtain
\begin{equation}
    \frac{d}{dt}
    \sum_{n=0}^{n_{\max}}
    P_n
    =
    0.
    \label{eq:supp_total_population_conservation}
\end{equation}
Here, the intrinsic-loss contributions cancel between neighboring Fock states, while the gain and loss contributions associated with the global and local population-transfer currents cancel separately.
Accordingly, the vanishing column sums of the rate matrix can be expressed as
\begin{equation}
    \sum_{m=0}^{n_{\max}}
    \mathsf M_{mn}(R)
    =
    0,
    \qquad
    0\leq n\leq n_{\max}.
    \label{eq:supp_rate_matrix_conservation}
\end{equation}
In vector notation, this condition can equivalently be written as
\begin{equation}
    \mathbf 1^{\mathsf T}
    \mathsf M(R)
    =
    0,
    \label{eq:supp_rate_matrix_left_null}
\end{equation}
where $\mathbf 1=(1,1,\ldots,1)^{\mathsf T}$.

Equation~\eqref{eq:supp_rate_matrix_left_null} shows that $\mathsf M(R)$ has a nontrivial left null vector and is therefore singular.
Since a singular square matrix has a zero eigenvalue, $\mathsf M(R)$ possesses at least one zero eigenvalue.
The stationary population vector is therefore a normalized right eigenvector associated with a zero eigenvalue of $\mathsf M(R)$.
We denote it by $\mathbf P^{\mathrm{ss}}\equiv(P_0^{\mathrm{ss}},P_1^{\mathrm{ss}},\ldots,P_{n_{\max}}^{\mathrm{ss}})^{\mathsf T}$.
It is obtained from the corresponding right null space and satisfies
\begin{equation}
    \mathsf M(R)
    \mathbf P^{\mathrm{ss}}
    =
    0,
    \qquad
    \mathbf 1^{\mathsf T}
    \mathbf P^{\mathrm{ss}}
    =
    1.
    \label{eq:supp_stationary_nullspace}
\end{equation}
The first relation imposes stationarity, while the second fixes the normalization of the population distribution.
Thus, different positions of the movable local connector are incorporated through the $R$ dependence of $\mathsf M(R)$, while the stationary solution is obtained from the same linear-algebraic formulation.

Since the rate matrix $\mathsf M(R)$ is singular, the stationary-state equation $\mathsf M(R)\mathbf P^{\mathrm{ss}}=0$ cannot be solved by direct matrix inversion.
We therefore replace one redundant stationary equation by the normalization condition $\sum_n P_n^{\mathrm{ss}}=1$, thereby obtaining an explicit linear system for the
normalized stationary distribution. 
Specifically, we define the modified matrix $\widetilde{\mathsf M}(R)$ as
\begin{equation}
    \widetilde{\mathsf M}_{mn}(R)
    \equiv
    \begin{cases}
        \mathsf M_{mn}(R),
        &
        0\leq m<n_{\max},
        \\[4pt]
        1,
        &
        m=n_{\max},
    \end{cases}
    \label{eq:supp_normalized_rate_matrix}
\end{equation}
together with the corresponding right-hand-side vector
\begin{equation}
    \mathbf b
    \equiv
    (0,\ldots,0,1)^{\mathsf T},
    \qquad
    b_m=\delta_{m,n_{\max}}.
    \label{eq:supp_stationary_source_vector}
\end{equation}
The stationary conditions in Eq.~\eqref{eq:supp_stationary_nullspace} are then equivalent to the linear system
\begin{equation}
    \widetilde{\mathsf M}(R)
    \mathbf P^{\mathrm{ss}}
    =
    \mathbf b.
    \label{eq:supp_stationary_linear_system}
\end{equation}
When the stationary state is unique, the linear system in Eq.~\eqref{eq:supp_stationary_linear_system} has a unique solution.
Accordingly, $\widetilde{\mathsf M}(R)$ is nonsingular, and the stationary population vector is
\begin{equation}
    \mathbf P^{\mathrm{ss}}
    =
    \left[
        \widetilde{\mathsf M}(R)
    \right]^{-1}
    \mathbf b.
    \label{eq:supp_stationary_solution}
\end{equation}
Accordingly, $P_R^{\mathrm{ss}}$ is the component of $\mathbf P^{\mathrm{ss}}$ associated with the Fock state $\ket R$.
Since the right-hand-side vector defined in Eq.~\eqref{eq:supp_stationary_source_vector} satisfies $b_m=\delta_{m,n_{\max}}$, multiplication by $\mathbf b$ selects the column of $\widetilde{\mathsf M}^{-1}(R)$ labeled by $n_{\max}$.
The stationary population of the target Fock state is therefore
\begin{equation}
    P_R^{\mathrm{ss}}
    =
    \left[
        \widetilde{\mathsf M}^{-1}(R)
    \right]_{R,n_{\max}},
    \label{eq:supp_stationary_target_matrix_element}
\end{equation}
where the rows and columns of the matrix are indexed by $n=0,1,\ldots,n_{\max}$, corresponding to the Fock states $\ket n$.

Thus, for a specified microscopic realization of the global and local connectors, the stationary target-state population is obtained by substituting the corresponding four effective connector rates into Eq.~\eqref{eq:supp_stationary_target_matrix_element}.
For stabilization of the single-excitation Fock state $\ket{1}$, both the configuration with the coherent $N_s$-quantum drive retained and the configuration without this drive can realize the desired stabilization.
In the latter case, the global connector is absent, and the same rate-matrix formulation applies with $R=1$ and $\mathcal R_{\mathrm G}^{\uparrow}=\mathcal R_{\mathrm G}^{\downarrow}=0$.

The convergence of the stationary solution with respect to the Fock-space truncation is verified by systematically increasing $N_{\mathrm{cut}}$.
For a given connector configuration and target state $\ket R$, we denote the corresponding stationary target-state population by $P_R^{\mathrm{ss}}(N_{\mathrm{cut}})$ to make its dependence on the truncation explicit.
Convergence is reached when $P_R^{\mathrm{ss}}(N_{\mathrm{cut}})$ remains unchanged upon a further increase of $N_{\mathrm{cut}}$.
The same convergence check is applied to all steady-state calculations, so that $N_{\mathrm{cut}}$ serves solely as a numerical truncation parameter and does not enter the physical description of the effective Fock-space dynamics.

\section{Closed-form stationary populations}
\label{sec:supp_closed_form_stationary}

In the preceding section, we formulated the population dynamics of the bosonic mode in matrix form, which provides both the full time evolution of the Fock-state populations and their stationary solution. 
For the stationary problem, however, the population-flow equation possesses a convenient structure that allows the stationary populations to be obtained directly in closed form. 
In this section, we derive these explicit stationary populations from Eq.~\eqref{eq:supp_population_equation} without relying on the matrix solution of Eq.~\eqref{eq:supp_stationary_solution}. 
The essential step is to convert the individual stationary population equations into cumulative population-balance relations in Fock space. 

\subsection{Cumulative stationary population balance}
\label{sec:supp_cumulative_stationary_balance}

For each Fock state $\ket{n}$, we denote its stationary population by $P_n^{\mathrm{ss}}$.
We likewise denote the stationary values of the global and local connector currents defined in Eqs.~\eqref{eq:supp_JG} and \eqref{eq:supp_JL} by $J_{\mathrm G}^{\mathrm{ss}}$ and $J_{\mathrm L}^{\mathrm{ss}}(R)$, respectively. 
Throughout this section, the superscript $\mathrm{ss}$ indicates that the corresponding current is evaluated using the stationary populations $P_n^{\mathrm{ss}}$.
For later use, we introduce the discrete step function 
\begin{equation} 
\Theta_{\mathrm d}(j) \equiv 
\begin{cases} 
1, & j\geq0, \\[2pt] 0, & j<0. 
\end{cases} 
\label{eq:supp_discrete_step} 
\end{equation}

For a fixed nonvacuum Fock state $n=1,\ldots,n_{\max}$, we consider the total population contained in the upper part of the truncated Fock space, $\{\ket n,\ket{n+1},\ldots,\ket{n_{\max}}\}$.
At stationarity, the total population of this set is time independent.
Therefore, summing the stationary population equation over the Fock-state index $j=n,\ldots,n_{\max}$ yields
\begin{align}
    0
    &=
    \sum_{j=n}^{n_{\max}}
    \dot P_j^{\mathrm{ss}}
    \nonumber\\
    &=
    \sum_{j=n}^{n_{\max}}
    \left[
        (j+1)\kappa_c
        P_{j+1}^{\mathrm{ss}}
        -
        j\kappa_c
        P_j^{\mathrm{ss}}
    \right]
    \nonumber\\
    &\quad
    +
    J_{\mathrm G}^{\mathrm{ss}}
    \sum_{j=n}^{n_{\max}}
    \left(
        \delta_{j,N_s}
        -
        \delta_{j,0}
    \right)
    \nonumber\\
    &\quad
    +
    J_{\mathrm L}^{\mathrm{ss}}(R)
    \sum_{j=n}^{n_{\max}}
    \left(
        \delta_{j,R}
        -
        \delta_{j,R-1}
    \right).
     \label{eq:supp_stationary_cumulative_sum}
\end{align}

We now evaluate the three contributions in Eq.~\eqref{eq:supp_stationary_cumulative_sum} separately.
For the intrinsic single-quantum loss, reindexing the first sum according to $\ell=j+1$ gives
\begin{align}
    &
    \sum_{j=n}^{n_{\max}}
    \left[
        (j+1)\kappa_c
        P_{j+1}^{\mathrm{ss}}
        -
        j\kappa_c
        P_j^{\mathrm{ss}}
    \right]
    \nonumber\\
    &\qquad
    =
    \kappa_c
    \left[
        \sum_{\ell=n+1}^{n_{\max}+1}
        \ell P_{\ell}^{\mathrm{ss}}
        -
        \sum_{j=n}^{n_{\max}}
        jP_j^{\mathrm{ss}}
    \right]
    \nonumber\\
    &\qquad
    =
    \kappa_c
    \left[
        (n_{\max}+1)
        P_{n_{\max}+1}^{\mathrm{ss}}
        -
        nP_n^{\mathrm{ss}}
    \right].
    \label{eq:supp_stationary_loss_telescoping}
\end{align}
All intermediate loss terms cancel pairwise.
Using the upper boundary condition $P_{n_{\max}+1}^{\mathrm{ss}}=0$, Eq.~\eqref{eq:supp_stationary_loss_telescoping} reduces to
\begin{equation}
    \sum_{j=n}^{n_{\max}}
    \left[
        (j+1)\kappa_c
        P_{j+1}^{\mathrm{ss}}
        -
        j\kappa_c
        P_j^{\mathrm{ss}}
    \right]
    =
    -
    n\kappa_c
    P_n^{\mathrm{ss}}.
    \label{eq:supp_stationary_loss_boundary}
\end{equation}
Thus, after summing over the entire upper Fock subspace, the only remaining intrinsic-loss contribution is the population current leaving this subspace through the transition $\ket n\rightarrow\ket{n-1}$.
The global-connector contribution follows directly from the summation range.
Because the summation starts from $n\geq1$, the index value $j=0$ is excluded from the summation range, and hence
\begin{equation}
    \sum_{j=n}^{n_{\max}}
    \delta_{j,0}
    =
    0.
\end{equation}
On the other hand, the state $\ket{N_s}$ belongs to the summed subspace if and only if $n\leq N_s$.
Since the truncation satisfies $n_{\max}\geq N_s$,
\begin{align}
    \sum_{j=n}^{n_{\max}}
    \left(
        \delta_{j,N_s}
        -
        \delta_{j,0}
    \right)
    &=
    \begin{cases}
        1,
        &
        n\leq N_s,
        \\[2pt]
        0,
        &
        n>N_s,
    \end{cases}
    \nonumber\\
    &=
    \Theta_{\mathrm d}
    \left(
        N_s-n
    \right).
    \label{eq:supp_stationary_global_sum}
\end{align}
Therefore, the global current contributes to the cumulative population balance whenever its upper endpoint $\ket{N_s}$ lies inside the chosen upper Fock-space subspace.
For the local connector, using $n_{\max}\geq R$ gives
\begin{align}
    &
    \sum_{j=n}^{n_{\max}}
    \left(
        \delta_{j,R}
        -
        \delta_{j,R-1}
    \right)
    \nonumber\\
    &\qquad
    =
    \Theta_{\mathrm d}(R-n)
    -
    \Theta_{\mathrm d}(R-1-n)
    \nonumber\\
    &\qquad
    =
    \delta_{n,R}.
    \label{eq:supp_stationary_local_sum}
\end{align}
This result can be understood directly from the position of the two states connected by the local process.
For $n\leq R-1$, both $\ket{R-1}$ and $\ket R$ belong to the summed subspace, so the population removed from one state and added to the other cancel in the total population balance.
For $n>R$, neither state belongs to the subspace, and the local connector again gives no contribution.
Only for $n=R$ does the boundary of the summed subspace lie between $\ket{R-1}$ and $\ket R$; the local connector then transfers population across this boundary and contributes $J_{\mathrm L}^{\mathrm{ss}}(R)$.

Combining Eqs.~\eqref{eq:supp_stationary_loss_boundary}, \eqref{eq:supp_stationary_global_sum}, and \eqref{eq:supp_stationary_local_sum}, we obtain the cumulative stationary population balance
\begin{equation}
    0
    =
    -
    n\kappa_c
    P_n^{\mathrm{ss}}
    +
    \Theta_{\mathrm d}
    \left(
        N_s-n
    \right)
    J_{\mathrm G}^{\mathrm{ss}}
    +
    \delta_{n,R}
    J_{\mathrm L}^{\mathrm{ss}}(R),
    \label{eq:supp_stationary_cumulative_balance}
\end{equation}
where $n=1,\ldots,n_{\max}$.
Equation~\eqref{eq:supp_stationary_cumulative_balance} has a direct current-balance interpretation: the intrinsic loss current crossing the lower boundary of the subspace, $n\kappa_cP_n^{\mathrm{ss}}$, must be supplied by the engineered connector currents that cross the same boundary.
The global connector crosses this boundary for every $n\leq N_s$, whereas the local connector crosses it only for $n=R$.
Solving Eq.~\eqref{eq:supp_stationary_cumulative_balance} for the population gives
\begin{equation}
    P_n^{\mathrm{ss}}
    =
    \frac{
        \Theta_{\mathrm d}
        \left(
            N_s-n
        \right)
        J_{\mathrm G}^{\mathrm{ss}}
        +
        \delta_{n,R}
        J_{\mathrm L}^{\mathrm{ss}}(R)
    }{
        n\kappa_c
    },
    \quad
    n=1,\ldots,n_{\max}.
    \label{eq:supp_stationary_population_current_form}
\end{equation}
Thus, the coupled stationary populations of all nonvacuum Fock states are reduced to two scalar quantities, $J_{\mathrm G}^{\mathrm{ss}}$ and $J_{\mathrm L}^{\mathrm{ss}}(R)$.
According to their definitions in Eqs.~\eqref{eq:supp_JG} and \eqref{eq:supp_JL}, both $J_{\mathrm G}^{\mathrm{ss}}$ and $J_{\mathrm L}^{\mathrm{ss}}(R)$ still depend explicitly on the stationary populations and are therefore not yet determined.
The remaining task is therefore to determine these two currents self-consistently from the connector-current relations and the normalization condition.

\subsection{Closed-form solution for nonzero global stationary current}
\label{sec:supp_nonzero_global_current_solution}

We first consider the configurations in which the global connector is retained, for which $\mathcal R_{\mathrm G}^{\uparrow}>0$ according to the effective-rate definitions given above.
For parameter regimes in which $J_{\mathrm G}^{\mathrm{ss}}\neq0$, it is convenient to characterize the local stationary connector current relative to the global one through the dimensionless ratio
\begin{equation}
    \zeta_R
    \equiv
    \frac{
        J_{\mathrm L}^{\mathrm{ss}}(R)
    }{
        J_{\mathrm G}^{\mathrm{ss}}
    }.
    \label{eq:supp_stationary_current_ratio}
\end{equation}
The exceptional coincident configuration $N_s=R=1$ can admit $J_{\mathrm G}^{\mathrm{ss}}=0$ even when the global connector is retained and is treated separately below.

The determination of $\zeta_R$ naturally separates into the cases $R\geq2$ and $R=1$.
According to the local-current definition in Eq.~\eqref{eq:supp_JL}, $J_{\mathrm L}^{\mathrm{ss}}(R)$ depends on the two populations $P_{R-1}^{\mathrm{ss}}$ and $P_R^{\mathrm{ss}}$.
For $R\geq2$, both states connected by the local process are nonvacuum Fock states, with $R-1\geq1$, so both populations are directly given by Eq.~\eqref{eq:supp_stationary_population_current_form}, which was derived only for the nonvacuum subspace $n\geq1$.
For $R=1$, however, $R-1=0$ lies outside this range, so that $P_{R-1}^{\mathrm{ss}}=P_0^{\mathrm{ss}}$ cannot be obtained from Eq.~\eqref{eq:supp_stationary_population_current_form}. 
The case $R=1$ therefore requires a separate elimination of the vacuum population.
This distinction originates solely from the vacuum boundary of the cumulative stationary solution and does not represent a different physical mechanism.

We first consider
\begin{equation}
    2\leq R\leq N_s+1.
\end{equation}
Because $R-1\leq N_s$ throughout this range, Eq.~\eqref{eq:supp_stationary_population_current_form} gives
\begin{align}
    P_{R-1}^{\mathrm{ss}}
    &=
    \frac{
        J_{\mathrm G}^{\mathrm{ss}}
    }{
        (R-1)\kappa_c
    },
    \\
    P_R^{\mathrm{ss}}
    &=
    \frac{
        \Theta_{\mathrm d}
        \left(
            N_s-R
        \right)
        +
        \zeta_R
    }{
        R\kappa_c
    }
    J_{\mathrm G}^{\mathrm{ss}}.
\end{align}
Substituting these two populations into the local-current definition in Eq.~\eqref{eq:supp_JL} and using $J_{\mathrm L}^{\mathrm{ss}}(R)=\zeta_RJ_{\mathrm G}^{\mathrm{ss}}$ gives
\begin{equation}
        \zeta_R
        =
        \frac{
            R
            \mathcal R_{\mathrm L}^{\uparrow}(R)
            -
            (R-1)
            \Theta_{\mathrm d}
            \left(
                N_s-R
            \right)
            \mathcal R_{\mathrm L}^{\downarrow}(R)
        }{
            (R-1)
            \left[
                R\kappa_c
                +
                \mathcal R_{\mathrm L}^{\downarrow}(R)
            \right]
        }
    ,
    \label{eq:supp_stationary_current_ratio_general}
\end{equation}
with $2\leq R\leq N_s+1$.
The discrete step function in Eq.~\eqref{eq:supp_stationary_current_ratio_general} distinguishes whether the upper state $\ket R$ of the local connector lies within the population cascade sustained by the global connector.
For $R\leq N_s$, $\ket R$ carries a global-current contribution, whereas for $R=N_s+1$ this contribution is absent.
Within $2\leq R\leq N_s+1$, the dependence on the position of the local connector relative to the upper endpoint $\ket{N_s}$ of the global connector is therefore encoded in $\Theta_{\mathrm d}(N_s-R)$.

For $R\geq N_s+2$, both states addressed by the local connector lie above the global loss cascade. Equation~\eqref{eq:supp_stationary_population_current_form} then gives $P_{R-1}^{\mathrm{ss}}=0$ and $R\kappa_c P_R^{\mathrm{ss}}=J_{\mathrm L}^{\mathrm{ss}}(R)$. 
Substituting these relations into the local-current definition in Eq.~\eqref{eq:supp_JL} yields
\begin{equation}
    \left[
        R\kappa_c
        +
        \mathcal R_{\mathrm L}^{\downarrow}(R)
    \right]
    P_R^{\mathrm{ss}}
    =
    0.
\end{equation}
Since $\kappa_c>0$ and $\mathcal R_{\mathrm L}^{\downarrow}(R)\geq0$, we obtain $P_R^{\mathrm{ss}}=J_{\mathrm L}^{\mathrm{ss}}(R)=0$, and hence $\zeta_R=0$ for $J_{\mathrm G}^{\mathrm{ss}}\neq0$. Consequently, the stationary population distribution reduces to the global-only solution for all $R\geq N_s+2$.

We now turn to
\begin{equation}
    R=1.
\end{equation}
In this case, Eq.~\eqref{eq:supp_stationary_population_current_form} still determines the relevant nonvacuum populations.
In particular, one obtains
\begin{equation}
    P_{N_s}^{\mathrm{ss}}
    =
    \frac{
        \left(
            1
            +
            \zeta_1
            \delta_{N_s,1}
        \right)
        J_{\mathrm G}^{\mathrm{ss}}
    }{
        N_s\kappa_c
    },
    \qquad
    P_1^{\mathrm{ss}}
    =
    \frac{
        \left(
            1+\zeta_1
        \right)
        J_{\mathrm G}^{\mathrm{ss}}
    }{
        \kappa_c
    }.
    \label{eq:supp_stationary_R1_populations}
\end{equation}
The vacuum population can then be obtained from the global-current relation.
Substituting the first relation in Eq.~\eqref{eq:supp_stationary_R1_populations} into Eq.~\eqref{eq:supp_JG} gives
\begin{equation}
    P_0^{\mathrm{ss}}
    =
    \tau_0(1)
    J_{\mathrm G}^{\mathrm{ss}},
    \label{eq:supp_stationary_P0_R1}
\end{equation}
where
\begin{equation}
    \tau_0(1)
    \equiv
    \frac{1}{
        \mathcal R_{\mathrm G}^{\uparrow}
    }
    \left[
        1
        +
        \frac{
            \mathcal R_{\mathrm G}^{\downarrow}
        }{
            N_s\kappa_c
        }
        \left(
            1
            +
            \zeta_1
            \delta_{1,N_s}
        \right)
    \right].
    \label{eq:supp_tau0_R1}
\end{equation}
For $N_s>1$, the Kronecker-delta contribution vanishes.
For $N_s=1$, it accounts for the fact that the upper endpoint of the global connector coincides with the upper state of the local connector, so that the local current also contributes to $P_{N_s}^{\mathrm{ss}}=P_1^{\mathrm{ss}}$.

The local-current definition in Eq.~\eqref{eq:supp_JL} can now be evaluated entirely in terms of $J_{\mathrm G}^{\mathrm{ss}}$ and $\zeta_1$.
Substituting Eqs.~\eqref{eq:supp_stationary_R1_populations} and \eqref{eq:supp_stationary_P0_R1} into Eq.~\eqref{eq:supp_JL} for $R=1$, and using $J_{\mathrm L}^{\mathrm{ss}}(1)=\zeta_1J_{\mathrm G}^{\mathrm{ss}}$, gives
\begin{equation}
    \zeta_1
    =
    \frac{
        \dfrac{
            \kappa_c
            \mathcal R_{\mathrm L}^{\uparrow}(1)
        }{
            \mathcal R_{\mathrm G}^{\uparrow}
        }
        \left(
            1
            +
            \dfrac{
                \mathcal R_{\mathrm G}^{\downarrow}
            }{
                N_s\kappa_c
            }
        \right)
        -
        \mathcal R_{\mathrm L}^{\downarrow}(1)
    }{
        \kappa_c
        +
        \mathcal R_{\mathrm L}^{\downarrow}(1)
        -
        \delta_{N_s,1}
        \dfrac{
            \mathcal R_{\mathrm L}^{\uparrow}(1)
            \mathcal R_{\mathrm G}^{\downarrow}
        }{
            N_s
            \mathcal R_{\mathrm G}^{\uparrow}
        }
    }.
    \label{eq:supp_stationary_current_ratio_R1}
\end{equation}

Equation~\eqref{eq:supp_stationary_current_ratio_R1} therefore applies for $R=1$ throughout the range $N_s\geq1$ whenever $J_{\mathrm G}^{\mathrm{ss}}\neq0$.
For $N_s=1$, the global and local connectors address the same pair of Fock states, and the current-ratio parametrization becomes singular at parameter values for which $J_{\mathrm G}^{\mathrm{ss}}=0$.
This exceptional case is more directly described by the two-state stationary solution obtained from the combined transition rates.
Together, Eqs.~\eqref{eq:supp_stationary_current_ratio_general} and \eqref{eq:supp_stationary_current_ratio_R1} determine $\zeta_R$ for the stabilization range $1\leq R\leq N_s+1$, while the result $\zeta_R=0$ extends the stationary solution to $R\geq N_s+2$, provided $J_{\mathrm G}^{\mathrm{ss}}\neq0$.

Having determined the relative current $J_{\mathrm L}^{\mathrm{ss}}(R)/J_{\mathrm G}^{\mathrm{ss}}$ through $\zeta_R$, it remains to determine the overall scale of the
stationary currents.
Evaluating Eq.~\eqref{eq:supp_stationary_population_current_form} at $n=N_s$ gives
\begin{equation}
    P_{N_s}^{\mathrm{ss}}
    =
    \frac{
        1
        +
        \zeta_R
        \delta_{R,N_s}
    }{
        N_s\kappa_c
    }
    J_{\mathrm G}^{\mathrm{ss}}.
    \label{eq:supp_stationary_PNs_general}
\end{equation}
Substitution of Eq.~\eqref{eq:supp_stationary_PNs_general} into the global-current definition in Eq.~\eqref{eq:supp_JG} expresses the vacuum population in the same current scale,
\begin{equation}
    P_0^{\mathrm{ss}}
    =
    \tau_0(R)
    J_{\mathrm G}^{\mathrm{ss}},
    \label{eq:supp_stationary_P0_general}
\end{equation}
where
\begin{equation}
    \tau_0(R)
    \equiv
    \frac{1}{
        \mathcal R_{\mathrm G}^{\uparrow}
    }
    \left[
        1
        +
        \frac{
            \mathcal R_{\mathrm G}^{\downarrow}
        }{
            N_s\kappa_c
        }
        \left(
            1
            +
            \zeta_R
            \delta_{R,N_s}
        \right)
    \right].
    \label{eq:supp_tau0}
\end{equation}
This definition reduces to the previously introduced $\tau_0(1)$ in Eq.~\eqref{eq:supp_tau0_R1}  when $R=1$.
The Kronecker-delta contribution in Eq.~\eqref{eq:supp_tau0} appears only for $R=N_s$.
In that configuration, the upper state of the local connector coincides with the upper endpoint of the global connector, and the local current therefore modifies $P_{N_s}^{\mathrm{ss}}$, which enters the reverse global process directly.

The remaining current scale is fixed by normalization.
We introduce the harmonic number
\begin{equation}
    H_{N_s}
    \equiv
    \sum_{n=1}^{N_s}
    \frac{1}{n}.
    \label{eq:supp_harmonic_number}
\end{equation}
Summing Eq.~\eqref{eq:supp_stationary_population_current_form} over all nonvacuum Fock states gives
\begin{equation}
    \sum_{n=1}^{n_{\max}}
    P_n^{\mathrm{ss}}
    =
    J_{\mathrm G}^{\mathrm{ss}}
    \left[
        \frac{
            H_{N_s}
        }{
            \kappa_c
        }
        +
        \frac{
            \zeta_R
        }{
            R\kappa_c
        }
    \right].
\end{equation}
Combining this result with Eq.~\eqref{eq:supp_stationary_P0_general} and imposing the normalization condition in Eq.~\eqref{eq:supp_Pm} yields
\begin{equation}
    1
    =
    J_{\mathrm G}^{\mathrm{ss}}
    \left[
        \tau_0(R)
        +
        \frac{
            H_{N_s}
        }{
            \kappa_c
        }
        +
        \frac{
            \zeta_R
        }{
            R\kappa_c
        }
    \right].
    \label{eq:supp_stationary_normalization_current}
\end{equation}
It is therefore useful to define
\begin{equation}
    \mathcal T_R
    \equiv
    \tau_0(R)
    +
    \frac{
        H_{N_s}
    }{
        \kappa_c
    }
    +
    \frac{
        \zeta_R
    }{
        R\kappa_c
    },
    \label{eq:supp_stationary_normalization_factor}
\end{equation}
in terms of which the two stationary connector currents take the closed forms
\begin{equation}
    J_{\mathrm G}^{\mathrm{ss}}
    =
    \frac{1}{\mathcal T_R},
    \qquad
    J_{\mathrm L}^{\mathrm{ss}}(R)
    =
    \frac{
        \zeta_R
    }{
        \mathcal T_R
    }.
    \label{eq:supp_stationary_currents_closed_form}
\end{equation}
The normalization factor $\mathcal T_R$ has the dimension of time and admits a direct population-flow interpretation.
Its three terms encode, per unit stationary global current, the vacuum contribution $\tau_0(R)$, the harmonic global-loss-cascade contribution $H_{N_s}/\kappa_c$, and the target-specific local-current correction $\zeta_R/(R\kappa_c)$, respectively.
Accordingly, $J_{\mathrm G}^{\mathrm{ss}}=1/\mathcal T_R$ identifies $\mathcal T_R$ as the total stationary population weight per unit global current.
The appearance of the harmonic number $H_{N_s}$ reflects the $1/n$ population profile generated by the number-dependent intrinsic loss.

Within the parameter regime in which $J_{\mathrm G}^{\mathrm{ss}}\neq0$, substitution of Eq.~\eqref{eq:supp_stationary_currents_closed_form} into Eqs.~\eqref{eq:supp_stationary_population_current_form} and \eqref{eq:supp_stationary_P0_general} gives the stationary Fock-state distribution
\begin{equation}
        P_0^{\mathrm{ss}}
        =
        \frac{
            \tau_0(R)
        }{
            \mathcal T_R
        },
        \qquad
        P_n^{\mathrm{ss}}
        =
        \frac{
            \Theta_{\mathrm d}
            \left(
                N_s-n
            \right)
            +
            \zeta_R\delta_{n,R}
        }{
            n\kappa_c\mathcal T_R
        },
    \label{eq:supp_stationary_population_closed_form}
\end{equation}
with $1\leq n\leq n_{\max}$.
In particular, the stationary population of the target Fock state $\ket R$ is
\begin{equation}
        P_R^{\mathrm{ss}}
        =
        \frac{
            \Theta_{\mathrm d}
            \left(
                N_s-R
            \right)
            +
            \zeta_R
        }{
            R\kappa_c\mathcal T_R
        }.
    \label{eq:supp_closed_form_target_population}
\end{equation}

\subsection{Properties of the nonzero-current stationary solution}
\label{sec:supp_stationary_distribution_structure}

To isolate the population profile generated by the global stationary current, we next consider the special case in which the local connector is removed.
Specifically, we set
\begin{equation}
    \mathcal R_{\mathrm L}^{\uparrow}(R)
    =
    \mathcal R_{\mathrm L}^{\downarrow}(R)
    =
    0,
\end{equation}
which implies $J_{\mathrm L}^{\mathrm{ss}}(R)=0$.
Since no local transition is retained, no particular Fock state is selected as a local target.
Equation~\eqref{eq:supp_stationary_population_current_form} then reduces to
\begin{equation}
    P_n^{\mathrm{ss}}
    =
    \frac{
        J_{\mathrm G}^{\mathrm{ss}}
    }{
        n\kappa_c
    },
    \qquad
    1\leq n\leq N_s,
    \label{eq:supp_global_only_population_profile}
\end{equation}
while $P_n^{\mathrm{ss}}=0$ for $n>N_s$.
The vacuum population follows from the global-current relation in Eq.~\eqref{eq:supp_JG} as
\begin{equation}
    P_0^{\mathrm{ss}}
    =
    \tau_{\mathrm G}
    J_{\mathrm G}^{\mathrm{ss}},
    \label{eq:supp_global_only_P0}
\end{equation}
where
\begin{equation}
    \tau_{\mathrm G}
    \equiv
    \frac{1}{
        \mathcal R_{\mathrm G}^{\uparrow}
    }
    \left(
        1
        +
        \frac{
            \mathcal R_{\mathrm G}^{\downarrow}
        }{
            N_s\kappa_c
        }
    \right).
    \label{eq:supp_global_only_tau}
\end{equation}
Using the harmonic number defined in Eq.~\eqref{eq:supp_harmonic_number}, the normalization condition in Eq.~\eqref{eq:supp_Pm} gives
\begin{equation}
    J_{\mathrm G}^{\mathrm{ss}}
    =
    \frac{1}{
        \mathcal T_{\mathrm G}
    },
    \qquad
    \mathcal T_{\mathrm G}
    \equiv
    \tau_{\mathrm G}
    +
    \frac{
        H_{N_s}
    }{
        \kappa_c
    }.
    \label{eq:supp_global_only_current}
\end{equation}
The complete global-only stationary distribution is therefore
\begin{equation}
    P_0^{\mathrm{ss}}
    =
    \frac{
        \tau_{\mathrm G}
    }{
        \mathcal T_{\mathrm G}
    },
    \qquad
    P_n^{\mathrm{ss}}
    =
    \frac{1}{
        n\kappa_c\mathcal T_{\mathrm G}
    },
    \quad
    1\leq n\leq N_s.
    \label{eq:supp_global_only_stationary_distribution}
\end{equation}
Equation~\eqref{eq:supp_global_only_stationary_distribution} shows that, in the absence of the local connector, the interplay between the global connector and intrinsic loss generates a harmonic $1/n$ population profile over the nonvacuum Fock states.
Indeed, the stationary populations of any two nonvacuum Fock states with $1\leq n,m\leq N_s$ satisfy
\begin{equation}
    \frac{
        P_n^{\mathrm{ss}}
    }{
        P_m^{\mathrm{ss}}
    }
    =
    \frac{m}{n}.
    \label{eq:supp_global_only_population_ratio}
\end{equation}
This population ratio is independent of the global connector rates and of the overall intrinsic-loss scale $\kappa_c$.
Equivalently, after normalization within the nonvacuum subspace,
\begin{equation}
    \frac{
        P_n^{\mathrm{ss}}
    }{
        \displaystyle
        \sum_{m=1}^{N_s}
        P_m^{\mathrm{ss}}
    }
    =
    \frac{1}{
        nH_{N_s}
    },
    \qquad
    1\leq n\leq N_s.
    \label{eq:supp_global_only_normalized_harmonic_profile}
\end{equation}
Thus, the global connector rates together with intrinsic loss determine the overall stationary population current and the total weight of the nonvacuum subspace, whereas the relative population distribution within that subspace is fixed solely by the number dependence of the intrinsic loss.
The same stationary global current is carried by every downward transition $\ket{n}\rightarrow\ket{n-1}$ for $1\leq n\leq N_s$, so that
\begin{equation}
    n\kappa_cP_n^{\mathrm{ss}}
    =
    J_{\mathrm G}^{\mathrm{ss}}
\end{equation}
enforces the harmonic $1/n$ profile.
For $N_s=1$, the nonvacuum subspace contains only the single state $\ket1$, so Eqs.~\eqref{eq:supp_global_only_population_ratio} and \eqref{eq:supp_global_only_normalized_harmonic_profile} reduce to a trivial one-state relation.
A nontrivial harmonic hierarchy among different nonvacuum Fock states emerges for $N_s\geq2$.

Having established the global-only distribution, we now return to the general stationary solution with the local connector present.
Equations~\eqref{eq:supp_stationary_population_closed_form} and \eqref{eq:supp_closed_form_target_population} show how the local connector modifies the harmonic background sustained by the interplay between the global stationary current and intrinsic loss.
For every non-target state within the global-loss cascade, $1\leq n\leq N_s$ and $n\neq R$, the stationary population remains
\begin{equation}
    P_n^{\mathrm{ss}}
    =
    \frac{1}{
        n\kappa_c\mathcal T_R
    },
    \qquad
    n\kappa_cP_n^{\mathrm{ss}}
    =
    J_{\mathrm G}^{\mathrm{ss}}.
    \label{eq:supp_stationary_background_profile}
\end{equation}
Thus, the local connector preserves the harmonic $1/n$ dependence of the non-target populations within the global-loss cascade, while changing their common population scale through the stationary global current $J_{\mathrm G}^{\mathrm{ss}}=1/\mathcal T_R$.
Its target-specific effect enters through the additional local-current contribution proportional to $\zeta_R$ in Eq.~\eqref{eq:supp_closed_form_target_population}.

For $R\leq N_s$, the target lies inside the global-loss cascade and Eq.~\eqref{eq:supp_closed_form_target_population} reduces to
\begin{equation}
    P_R^{\mathrm{ss}}
    =
    \frac{
        1+\zeta_R
    }{
        R\kappa_c\mathcal T_R
    }.
    \label{eq:supp_target_inside_cascade}
\end{equation}
The unity in the numerator represents the contribution associated with the harmonic background sustained by the global current, whereas the term $\zeta_R$ represents the additional contribution associated with the local stationary current.
Accordingly, a positive $\zeta_R$ enhances the target population relative to the harmonic background at the same Fock number.
For $R=N_s+1$, by contrast, the target lies outside the global-loss cascade and therefore has no global-current background contribution,
\begin{equation}
    P_{N_s+1}^{\mathrm{ss}}
    =
    \frac{
        \zeta_{N_s+1}
    }{
        (N_s+1)
        \kappa_c
        \mathcal T_{N_s+1}
    }.
    \label{eq:supp_target_above_cascade}
\end{equation}
In this case, the target population is supported entirely by the local stationary current.
The two cases therefore differ in whether the target state already participates in the harmonic global-loss cascade before the local-current contribution is included.

The closed-form solution also provides a direct criterion for whether the target state is dominant over the nonvacuum harmonic background sustained by the global stationary current and intrinsic loss.
For $R=1$ and $N_s\geq2$, the largest non-target population in the $1/n$ background is $P_2^{\mathrm{ss}}$, and
\begin{equation}
    \frac{
        P_1^{\mathrm{ss}}
    }{
        P_2^{\mathrm{ss}}
    }
    =
    2
    \left(
        1+\zeta_1
    \right).
    \label{eq:supp_target_to_background_ratio_R1}
\end{equation}
Thus, $\ket1$ exceeds every other nonvacuum state in the global-loss cascade when
\begin{equation}
    \zeta_1
    >
    -\frac{1}{2}.
    \label{eq:supp_target_dominance_condition_R1}
\end{equation}
Notably, this condition does not require a positive local stationary current.
Even for $\zeta_1=0$, the intrinsic $1/n$ loss-cascade profile already gives $P_1^{\mathrm{ss}}=2P_2^{\mathrm{ss}}$.
For $R=1$ and $N_s=1$, no non-target nonvacuum state belongs to the global-loss cascade, and therefore no separate nonvacuum-dominance condition is required.
For $2\leq R\leq N_s$, the largest non-target population in the $1/n$ background is $P_1^{\mathrm{ss}}$, and
\begin{equation}
    \frac{
        P_R^{\mathrm{ss}}
    }{
        P_1^{\mathrm{ss}}
    }
    =
    \frac{
        1+\zeta_R
    }{
        R
    }.
    \label{eq:supp_target_to_background_ratio}
\end{equation}
The target state therefore exceeds every other nonvacuum state in the global-loss cascade when
\begin{equation}
    \zeta_R>R-1,
    \qquad
    2\leq R\leq N_s.
    \label{eq:supp_target_dominance_condition_inside}
\end{equation}
For the target immediately above the global connector, $R=N_s+1$, the corresponding ratio is
\begin{equation}
    \frac{
        P_{N_s+1}^{\mathrm{ss}}
    }{
        P_1^{\mathrm{ss}}
    }
    =
    \frac{
        \zeta_{N_s+1}
    }{
        N_s+1
    },
    \label{eq:supp_target_to_background_ratio_above}
\end{equation}
so that domination over the nonvacuum global-loss background requires
\begin{equation}
    \zeta_{N_s+1}
    >
    N_s+1.
    \label{eq:supp_target_dominance_condition_above}
\end{equation}
The $R=1$ case is therefore qualitatively distinct from the higher-Fock-state targets: the intrinsic $1/n$ loss-cascade profile already favors $\ket{1}$ over all other nonvacuum states, and a positive local stationary current can further enhance this existing population advantage.
For $R\geq2$, by contrast, the target must overcome the larger background populations at lower Fock numbers.
Because the intrinsic decay rate of $\ket R$ is $R\kappa_c$, this requires a progressively larger local-to-global stationary-current ratio as the target Fock number increases.

Dominance over the nonvacuum loss ladder alone does not guarantee that the target is the most populated Fock state, because a finite stationary weight may remain in the vacuum.
Using Eq.~\eqref{eq:supp_stationary_P0_general}, the target-to-vacuum population ratio is
\begin{equation}
    \frac{
        P_R^{\mathrm{ss}}
    }{
        P_0^{\mathrm{ss}}
    }
    =
    \frac{
        \Theta_{\mathrm d}
        \left(
            N_s-R
        \right)
        +
        \zeta_R
    }{
        R\kappa_c\tau_0(R)
    }.
    \label{eq:supp_target_to_vacuum_ratio}
\end{equation}
Hence, the target population exceeds the vacuum population when
\begin{equation}
    \Theta_{\mathrm d}
    \left(
        N_s-R
    \right)
    +
    \zeta_R
    >
    R\kappa_c\tau_0(R).
    \label{eq:supp_target_over_vacuum_condition}
\end{equation}
The quantities $\zeta_R$ and $\tau_0(R)$ therefore characterize two distinct requirements for Fock-state stabilization: the former determines the local enhancement of the target relative to the loss cascade, whereas the latter determines the residual weight accumulated in the vacuum.

The closed form further reveals a special limiting behavior when the target coincides with the upper endpoint of the global connector.
For $R\neq N_s$, $\tau_0(R)$ is independent of $\zeta_R$, and, within the effective rate description, increasingly strong local capture gives
\begin{equation}
    \lim_{\zeta_R\rightarrow\infty}
    P_R^{\mathrm{ss}}
    =
    1,
    \qquad
    R\neq N_s.
    \label{eq:supp_strong_local_capture_generic}
\end{equation}
For $R=N_s\geq2$, however, the target state also participates directly in the reverse global process.
In this case, the $\zeta_R$-dependent term in $\tau_0(R)$ survives in the same limit and gives
\begin{equation}
    \lim_{\zeta_{N_s}\rightarrow\infty}
    P_{N_s}^{\mathrm{ss}}
    =
    \frac{
        \mathcal R_{\mathrm G}^{\uparrow}
    }{
        \mathcal R_{\mathrm G}^{\uparrow}
        +
        \mathcal R_{\mathrm G}^{\downarrow}
    }.
    \label{eq:supp_strong_local_capture_overlap}
\end{equation}
Thus, for $R=N_s\geq2$, the reverse global process imposes an additional ceiling on the achievable target population even in the strong-local-capture limit.
This ceiling disappears for a fully directional global connector with $\mathcal R_{\mathrm G}^{\downarrow}=0$.
The coincident case $N_s=R=1$ is qualitatively different because the global and local connectors act on the same Fock-state transition.
In this case, the current-ratio parametrization in terms of $\zeta_1$ can become singular when $J_{\mathrm G}^{\mathrm{ss}}=0$, and the corresponding two-state stationary solution is treated separately below.

The closed-form distribution also makes the role of the Fock-space truncation transparent.
For $n>\max(N_s,R)$, the global-current contribution vanishes because $\Theta_{\mathrm d}(N_s-n)=0$, and the local-current contribution vanishes because $\delta_{n,R}=0$.
Equation~\eqref{eq:supp_stationary_population_closed_form} therefore gives
\begin{equation}
    P_n^{\mathrm{ss}}
    =
    0,
    \qquad
    n>
    \max
    \left(
        N_s,
        R
    \right).
    \label{eq:supp_stationary_population_upper_zero}
\end{equation}
Consequently, once the truncation condition in Eq.~\eqref{eq:supp_fock_cutoff_condition} is satisfied, increasing $N_{\mathrm{cut}}$ only appends Fock states whose stationary populations vanish identically.
The closed-form stationary solution is therefore independent of any further increase of the numerical Fock-space cutoff.

\subsection{Classification of zero-global-current stationary solutions}
\label{sec:supp_zero_global_current_solutions}

The current-ratio parametrization introduced in Eq.~\eqref{eq:supp_stationary_current_ratio} and used to obtain the closed-form representation above is restricted to the branch with $J_{\mathrm G}^{\mathrm{ss}}\neq0$.
To complete the stationary analysis, we now consider the complementary branch
\begin{equation}
    J_{\mathrm G}^{\mathrm{ss}}
    =
    0.
    \label{eq:supp_zero_global_current}
\end{equation}
Setting $J_{\mathrm G}^{\mathrm{ss}}=0$ in the cumulative stationary balance, Eq.~\eqref{eq:supp_stationary_cumulative_balance}, gives
\begin{equation}
    P_n^{\mathrm{ss}}
    =
    \frac{
        \delta_{n,R}
        J_{\mathrm L}^{\mathrm{ss}}(R)
    }{
        n\kappa_c
    },
    \qquad
    n=1,\ldots,n_{\max}.
    \label{eq:supp_zero_global_current_population}
\end{equation}
Therefore, all non-target nonvacuum populations vanish, while the target-state population satisfies
\begin{equation}
    J_{\mathrm L}^{\mathrm{ss}}(R)
    =
    R\kappa_c
    P_R^{\mathrm{ss}}.
    \label{eq:supp_zero_global_current_target_balance}
\end{equation}

We first consider $R\geq2$.
Equation~\eqref{eq:supp_zero_global_current_population} then gives $P_{R-1}^{\mathrm{ss}}=0$, so that the local-current definition in Eq.~\eqref{eq:supp_JL} reduces to
\begin{equation}
    J_{\mathrm L}^{\mathrm{ss}}(R)
    =
    -
    \mathcal R_{\mathrm L}^{\downarrow}(R)
    P_R^{\mathrm{ss}}.
    \label{eq:supp_zero_global_current_local_Rge2}
\end{equation}
Combining Eqs.~\eqref{eq:supp_zero_global_current_target_balance} and \eqref{eq:supp_zero_global_current_local_Rge2} yields
\begin{equation}
    \left[
        R\kappa_c
        +
        \mathcal R_{\mathrm L}^{\downarrow}(R)
    \right]
    P_R^{\mathrm{ss}}
    =
    0.
    \label{eq:supp_zero_global_current_Rge2_condition}
\end{equation}
For $\kappa_c>0$ and nonnegative connector rates, Eq.~\eqref{eq:supp_zero_global_current_Rge2_condition} requires $P_R^{\mathrm{ss}}=0$.
Equation~\eqref{eq:supp_zero_global_current_population} then gives $P_n^{\mathrm{ss}}=0$ for every $n\geq1$, and normalization requires $P_0^{\mathrm{ss}}=1$.
The global-current definition in Eq.~\eqref{eq:supp_JG} consequently gives
\begin{equation}
    J_{\mathrm G}^{\mathrm{ss}}
    =
    \mathcal R_{\mathrm G}^{\uparrow}.
\end{equation}
Hence, the condition $J_{\mathrm G}^{\mathrm{ss}}=0$ requires $\mathcal R_{\mathrm G}^{\uparrow}=0$ for $R\geq2$, and the corresponding stationary state is just the vacuum.
In particular, no nontrivial stationary state with $J_{\mathrm G}^{\mathrm{ss}}=0$ exists for $R\geq2$ when the global connector is retained with $\mathcal R_{\mathrm G}^{\uparrow}>0$.

We next consider $R=1$.
Equation~\eqref{eq:supp_zero_global_current_population} gives
\begin{equation}
    P_n^{\mathrm{ss}}
    =
    0,
    \qquad
    n\geq2,
\end{equation}
and Eq.~\eqref{eq:supp_zero_global_current_target_balance} becomes
\begin{equation}
    J_{\mathrm L}^{\mathrm{ss}}(1)
    =
    \kappa_c
    P_1^{\mathrm{ss}}.
    \label{eq:supp_zero_global_current_R1_balance}
\end{equation}
For $N_s\geq2$, one has $P_{N_s}^{\mathrm{ss}}=0$, and the global-current definition in Eq.~\eqref{eq:supp_JG} therefore reduces to
\begin{equation}
    J_{\mathrm G}^{\mathrm{ss}}
    =
    \mathcal R_{\mathrm G}^{\uparrow}
    P_0^{\mathrm{ss}}.
    \label{eq:supp_zero_global_current_R1_Nsge2}
\end{equation}
If the global connector is retained with $\mathcal R_{\mathrm G}^{\uparrow}>0$, the condition $J_{\mathrm G}^{\mathrm{ss}}=0$ would require $P_0^{\mathrm{ss}}=0$ and hence $P_1^{\mathrm{ss}}=1$.
The local-current definition would then give
\begin{equation}
    J_{\mathrm L}^{\mathrm{ss}}(1)
    =
    -
    \mathcal R_{\mathrm L}^{\downarrow}(1),
\end{equation}
which is incompatible with Eq.~\eqref{eq:supp_zero_global_current_R1_balance} for $\kappa_c>0$.
Thus, for $R=1$ and $N_s\geq2$, the zero-global-current condition requires $\mathcal R_{\mathrm G}^{\uparrow}=0$.
Since all populations with $n\geq2$ vanish in this subspace, the reverse global process proportional to $\mathcal R_{\mathrm G}^{\downarrow}$ is inactive because $P_{N_s}^{\mathrm{ss}}=0$.
The stationary solution is therefore identical to that of the strictly local-only configuration, which we represent below by setting both global connector rates to zero.

The only configuration in which $J_{\mathrm G}^{\mathrm{ss}}$ can vanish while the global connector is retained with $\mathcal R_{\mathrm G}^{\uparrow}>0$ is therefore the coincident case $N_s=R=1$.
Here, the global and local connectors act on the same transition $\ket0\leftrightarrow\ket1$, allowing their stationary currents to redistribute while their sum continues to balance the intrinsic loss.
We treat the local-only configuration and this coincident configuration explicitly below.

\subsection{Local-only stabilization of the single-excitation state}
\label{sec:supp_local_only_R1}

We first consider the local-only configuration for $R=1$, obtained by removing the global connector,
\begin{equation}
    \mathcal R_{\mathrm G}^{\uparrow}
    =
    \mathcal R_{\mathrm G}^{\downarrow}
    =
    0.
    \label{eq:supp_R1_local_only_global_rates}
\end{equation}
The global-current definition in Eq.~\eqref{eq:supp_JG} then gives $J_{\mathrm G}^{\mathrm{ss}}=0$ identically.
Equation~\eqref{eq:supp_zero_global_current_population} gives
\begin{equation}
    P_n^{\mathrm{ss}}
    =
    0,
    \qquad
    2\leq n\leq n_{\max},
    \label{eq:supp_R1_local_only_upper_populations}
\end{equation}
so that the stationary problem reduces to the two states $\ket0$ and $\ket1$.
Moreover, Eq.~\eqref{eq:supp_zero_global_current_R1_balance} gives
\begin{equation}
    J_{\mathrm L}^{\mathrm{ss}}(1)
    =
    \kappa_c
    P_1^{\mathrm{ss}}.
\end{equation}
Using the local-current definition in Eq.~\eqref{eq:supp_JL} then yields
\begin{equation}
    \mathcal R_{\mathrm L}^{\uparrow}(1)
    P_0^{\mathrm{ss}}
    =
    \left[
        \kappa_c
        +
        \mathcal R_{\mathrm L}^{\downarrow}(1)
    \right]
    P_1^{\mathrm{ss}}.
    \label{eq:supp_R1_local_only_balance}
\end{equation}
Together with
\begin{equation}
    P_0^{\mathrm{ss}}
    +
    P_1^{\mathrm{ss}}
    =
    1,
\end{equation}
this gives
\begin{equation}
        P_0^{\mathrm{ss}}
        =
        \frac{
            \kappa_c
            +
            \mathcal R_{\mathrm L}^{\downarrow}(1)
        }{
            \mathcal R_{\mathrm L}^{\uparrow}(1)
            +
            \mathcal R_{\mathrm L}^{\downarrow}(1)
            +
            \kappa_c
        },
        \quad
        P_1^{\mathrm{ss}}
        =
        \frac{
            \mathcal R_{\mathrm L}^{\uparrow}(1)
        }{
            \mathcal R_{\mathrm L}^{\uparrow}(1)
            +
            \mathcal R_{\mathrm L}^{\downarrow}(1)
            +
            \kappa_c
        }.
    \label{eq:supp_R1_local_only_stationary}
\end{equation}

Equation~\eqref{eq:supp_R1_local_only_stationary} makes the stabilization condition for the local-only configuration explicit.
The target population can equivalently be written as
\begin{equation}
    P_1^{\mathrm{ss}}
    =
    \frac{1}{
        1
        +
        \dfrac{
            \kappa_c
            +
            \mathcal R_{\mathrm L}^{\downarrow}(1)
        }{
            \mathcal R_{\mathrm L}^{\uparrow}(1)
        }
    }.
    \label{eq:supp_R1_local_only_stationary_interpretation}
\end{equation}
Thus, the desired upward local-transfer rate competes directly with two depletion channels of the target state: intrinsic single-quantum loss at rate $\kappa_c$ and the reverse local process at rate $\mathcal R_{\mathrm L}^{\downarrow}(1)$.
A high stationary population of the target state therefore requires
\begin{equation}
    \mathcal R_{\mathrm L}^{\uparrow}(1)
    \gg
    \kappa_c
    +
    \mathcal R_{\mathrm L}^{\downarrow}(1),
    \label{eq:supp_R1_local_only_near_unit_population_condition}
\end{equation}
under which $P_1^{\mathrm{ss}}\rightarrow1$.
This result also shows that the reverse local process directly competes with the desired upward transfer and therefore reduces the stationary population of the target state.

\subsection{Coincident global and local connectors for $N_s=R=1$}
\label{sec:supp_coincident_Ns1_R1}

We next consider the coincident configuration $N_s=R=1$ with both the global and local connectors retained.
In this case, the two connectors act on the same Fock-state transition $\ket0\leftrightarrow\ket1$, and no process populates Fock states above $\ket1$.
The stationary dynamics therefore reduce exactly to the two-state subspace spanned by $\ket0$ and $\ket1$.
Setting $n=1$, $N_s=1$, and $R=1$ in the cumulative stationary population balance, Eq.~\eqref{eq:supp_stationary_cumulative_balance}, gives
\begin{equation}
    0
    =
    J_{\mathrm G}^{\mathrm{ss}}
    +
    J_{\mathrm L}^{\mathrm{ss}}(1)
    -
    \kappa_cP_1^{\mathrm{ss}}.
    \label{eq:supp_Ns1_R1_current_balance}
\end{equation}
Using the definitions of the global and local connector currents in Eqs.~\eqref{eq:supp_JG} and \eqref{eq:supp_JL}, Eq.~\eqref{eq:supp_Ns1_R1_current_balance} becomes
\begin{equation}
    0
    =
    \left[
        \mathcal R_{\mathrm G}^{\uparrow}
        +
        \mathcal R_{\mathrm L}^{\uparrow}(1)
    \right]
    P_0^{\mathrm{ss}}
    -
    \left[
        \kappa_c
        +
        \mathcal R_{\mathrm G}^{\downarrow}
        +
        \mathcal R_{\mathrm L}^{\downarrow}(1)
    \right]
    P_1^{\mathrm{ss}}.
    \label{eq:supp_Ns1_R1_stationary_balance}
\end{equation}
Together with $P_0^{\mathrm{ss}}+P_1^{\mathrm{ss}}=1$, this gives
\begin{align}
    P_0^{\mathrm{ss}}
    &=
    \frac{
        \kappa_c
        +
        \mathcal R_{\mathrm G}^{\downarrow}
        +
        \mathcal R_{\mathrm L}^{\downarrow}(1)
    }{
        \mathcal R_{\mathrm G}^{\uparrow}
        +
        \mathcal R_{\mathrm L}^{\uparrow}(1)
        +
        \kappa_c
        +
        \mathcal R_{\mathrm G}^{\downarrow}
        +
        \mathcal R_{\mathrm L}^{\downarrow}(1)
    },
    \nonumber\\
    P_1^{\mathrm{ss}}
    &=
    \frac{
        \mathcal R_{\mathrm G}^{\uparrow}
        +
        \mathcal R_{\mathrm L}^{\uparrow}(1)
    }{
        \mathcal R_{\mathrm G}^{\uparrow}
        +
        \mathcal R_{\mathrm L}^{\uparrow}(1)
        +
        \kappa_c
        +
        \mathcal R_{\mathrm G}^{\downarrow}
        +
        \mathcal R_{\mathrm L}^{\downarrow}(1)
    }.
    \label{eq:supp_Ns1_R1_stationary_populations}
\end{align}

Because the global and local connectors act on the same Fock-state transition, Eq.~\eqref{eq:supp_Ns1_R1_current_balance} constrains only the sum of their stationary currents rather than the two currents separately.
Consequently, the global stationary current is not constrained to remain positive in this coincident configuration and can vanish or reverse while the total engineered current continues to balance the intrinsic loss.
Using Eq.~\eqref{eq:supp_Ns1_R1_stationary_populations}, the stationary global current is
\begin{equation}
    J_{\mathrm G}^{\mathrm{ss}}
    =
    \frac{
        \mathcal R_{\mathrm G}^{\uparrow}
        \left[
            \kappa_c
            +
            \mathcal R_{\mathrm L}^{\downarrow}(1)
        \right]
        -
        \mathcal R_{\mathrm G}^{\downarrow}
        \mathcal R_{\mathrm L}^{\uparrow}(1)
    }{
        \mathcal R_{\mathrm G}^{\uparrow}
        +
        \mathcal R_{\mathrm L}^{\uparrow}(1)
        +
        \kappa_c
        +
        \mathcal R_{\mathrm G}^{\downarrow}
        +
        \mathcal R_{\mathrm L}^{\downarrow}(1)
    }.
    \label{eq:supp_Ns1_R1_global_current}
\end{equation}
Because the denominator in Eq.~\eqref{eq:supp_Ns1_R1_global_current} is positive, the direction of the stationary global current is determined entirely by its numerator.
The global current is positive when
\begin{equation}
    \mathcal R_{\mathrm G}^{\uparrow}
    \left[
        \kappa_c
        +
        \mathcal R_{\mathrm L}^{\downarrow}(1)
    \right]
    >
    \mathcal R_{\mathrm G}^{\downarrow}
    \mathcal R_{\mathrm L}^{\uparrow}(1),
\end{equation}
and it is negative when the opposite inequality holds.
The boundary between these two current directions is therefore determined by
\begin{equation}
    \mathcal R_{\mathrm G}^{\uparrow}
    \left[
        \kappa_c
        +
        \mathcal R_{\mathrm L}^{\downarrow}(1)
    \right]
    =
    \mathcal R_{\mathrm G}^{\downarrow}
    \mathcal R_{\mathrm L}^{\uparrow}(1),
    \label{eq:supp_Ns1_R1_zero_global_current}
\end{equation}
at which
\begin{equation}
    J_{\mathrm G}^{\mathrm{ss}}
    =
    0.
\end{equation}
Thus, the zero-global-current point separates stationary regimes with opposite net population flow through the global connector.

At the zero-global-current point, Eq.~\eqref{eq:supp_JG} gives
\begin{equation}
    \mathcal R_{\mathrm G}^{\uparrow}
    P_0^{\mathrm{ss}}
    =
    \mathcal R_{\mathrm G}^{\downarrow}
    P_1^{\mathrm{ss}}.
\end{equation}
Together with $P_0^{\mathrm{ss}}+P_1^{\mathrm{ss}}=1$, this yields
\begin{equation}
    P_0^{\mathrm{ss}}
    =
    \frac{
        \mathcal R_{\mathrm G}^{\downarrow}
    }{
        \mathcal R_{\mathrm G}^{\uparrow}
        +
        \mathcal R_{\mathrm G}^{\downarrow}
    },
    \qquad
    P_1^{\mathrm{ss}}
    =
    \frac{
        \mathcal R_{\mathrm G}^{\uparrow}
    }{
        \mathcal R_{\mathrm G}^{\uparrow}
        +
        \mathcal R_{\mathrm G}^{\downarrow}
    }.
    \label{eq:supp_Ns1_R1_zero_global_current_populations}
\end{equation}
Moreover, Eq.~\eqref{eq:supp_Ns1_R1_current_balance} reduces to
\begin{equation}
    J_{\mathrm L}^{\mathrm{ss}}(1)
    =
    \kappa_c
    P_1^{\mathrm{ss}},
\end{equation}
showing that, although the global connector carries no net stationary population current at this point, the local stationary current continues to supply the intrinsic loss from the target state.

Under the condition in Eq.~\eqref{eq:supp_Ns1_R1_zero_global_current}, the ratio $\zeta_1=J_{\mathrm L}^{\mathrm{ss}}(1)/J_{\mathrm G}^{\mathrm{ss}}$ becomes singular even though the stationary populations remain finite.
The current-ratio parametrization introduced in Eq.~\eqref{eq:supp_stationary_current_ratio} is therefore not suitable for describing the full parameter range of the coincident configuration $N_s=R=1$.

Within the effective rate description, the direct two-state solution also makes the strong-local-transfer limit transparent.
For fixed $\mathcal R_{\mathrm G}^{\uparrow}$, $\mathcal R_{\mathrm G}^{\downarrow}$, $\mathcal R_{\mathrm L}^{\downarrow}(1)$, and $\kappa_c$, it gives
\begin{equation}
    \lim_{
        \mathcal R_{\mathrm L}^{\uparrow}(1)
        \rightarrow\infty
    }
    P_1^{\mathrm{ss}}
    =
    1.
    \label{eq:supp_Ns1_R1_strong_local_limit}
\end{equation}
Thus, unlike the case $R=N_s\geq2$, where the reverse global process imposes the ceiling in Eq.~\eqref{eq:supp_strong_local_capture_overlap}, no analogous ceiling occurs for $N_s=R=1$.
Here, increasing the upward local-transfer rate directly increases the total upward transfer rate on the same transition and can drive the stationary population of $\ket1$ arbitrarily close to unity.
The local-only configuration discussed above is recovered immediately by setting $\mathcal R_{\mathrm G}^{\uparrow}=\mathcal R_{\mathrm G}^{\downarrow}=0$ in Eq.~\eqref{eq:supp_Ns1_R1_stationary_populations}.
Although the coincident configuration $N_s=R=1$ is included in the general effective population model, it is distinct from the separate-connector examples emphasized in main-text Fig.~1a,e; the current-balance equations nevertheless cover this overlap.

\section{Parameter dependence of Kerr-resolved Fock-state stabilization}
\label{sec:supp_figS1_parameter_dependence}

\begin{figure}[t]
\centering
\includegraphics[width=1.0\linewidth]{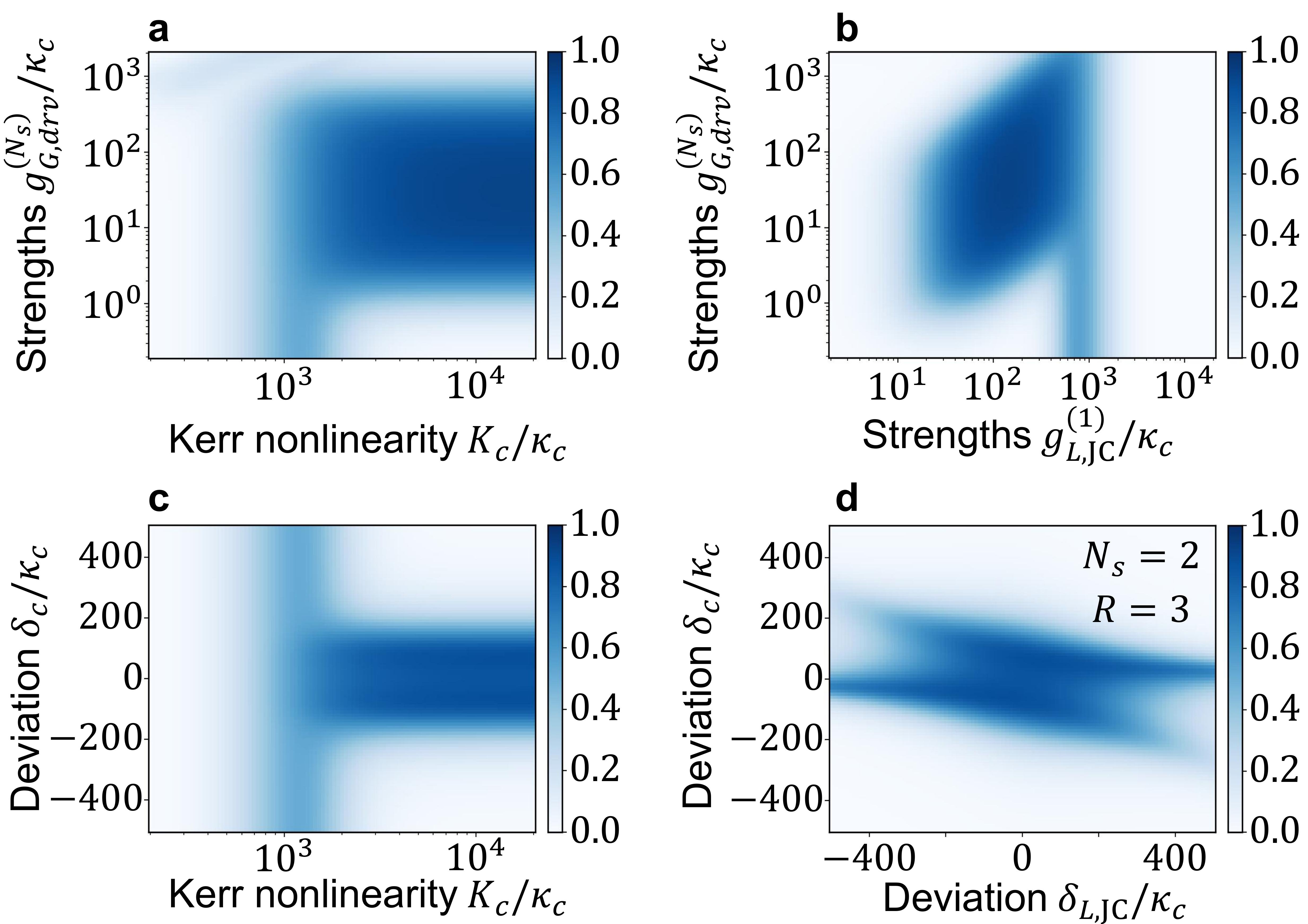}
\caption{\textbf{Parameter dependence of Kerr-resolved Fock-state stabilization.}
Full microscopic Lindblad steady-state target populations $P_3^{\mathrm{ss}}$ for $N_s=2$ and $R=3$, using a coherent two-quantum global drive and a pumped JC local connector.
\textbf{a,} Dependence on the Kerr nonlinearity $K_c$ and global drive strength $g_{G,\mathrm{drv}}^{(N_s)}$.
\textbf{b,} Dependence on the local JC coupling $g_{L,\mathrm{JC}}^{(1)}$ and global drive strength $g_{G,\mathrm{drv}}^{(N_s)}$.
\textbf{c,} Dependence on $K_c$ and the residual detuning parameter $\delta_c$.
\textbf{d,} Dependence on the residual detuning parameters $\delta_{L,\mathrm{JC}}$ and $\delta_c$.
In \textbf{a,b,} $\delta_c=\delta_{L,\mathrm{JC}}=0$; in \textbf{c}, $\delta_{L,\mathrm{JC}}=0$.
Unless varied, $K_c/\kappa_c=10^4$, $g_{G,\mathrm{drv}}^{(N_s)}/\kappa_c=5$, $g_{L,\mathrm{JC}}^{(1)}/\kappa_c=100$, and $(\wp_R,\gamma_R)/\kappa_c=(300,1)$.
}
\label{figS1}
\end{figure}

Supplementary Fig.~\ref{figS1} considers Kerr-resolved spectral selection with $N_s=2$ and the target Fock state $R=3$.
The global connector is implemented by a coherent two-quantum drive, which targets $\ket{0}\leftrightarrow\ket{2}$ without a global auxiliary TLS.
The movable local connector is implemented by a single-quantum JC interaction assisted by incoherent pumping of the local TLS, addressing $\ket{2}\leftrightarrow\ket{3}$.
For this configuration, substituting the corresponding global and local connector realizations into Eq.~\eqref{eq:supp_general_master} yields the microscopic master equation
\begin{align}
\frac{d\hat\rho}{dt}
=&-i[\hat H_{\mathrm{S1}},\hat\rho]
+\kappa_c\mathcal D[\hat a_c]\hat\rho+\wp_R\mathcal D[\hat\sigma_R^+]\hat\rho
\nonumber\\&
+\gamma_R\mathcal D[\hat\sigma_R^-]\hat\rho,
\label{eq:supp_figS1_master}
\end{align}
where the Hamiltonian is given by
\begin{align}
\hat H_{\mathrm{S1}}
=&\Delta_c\hat n_c+K_c\hat n_c^2
+\Delta_R\hat\sigma_R^+\hat\sigma_R^-
\nonumber\\
&+g_{G,\mathrm{drv}}^{(2)}
\bigl[(\hat a_c^\dagger)^2+\hat a_c^2\bigr]
\nonumber\\
&+g_{L,\mathrm{JC}}^{(1)}
\bigl(\hat a_c^\dagger\hat\sigma_R^-
+\hat a_c\hat\sigma_R^+\bigr).
\label{eq:supp_figS1_hamiltonian}
\end{align}

The detuning conditions used in the parameter scans are determined by the energy differences of the selected global and local transitions.
For the resonance analysis, we define the bare-state energies of the Fock--TLS states $\ket{n;g_R}$ and $\ket{n;e_R}$ by the diagonal matrix elements of $\hat H_{\mathrm{S1}}$, giving $E_{n;g_R}=n\Delta_c+K_c n^2$ and $E_{n;e_R}=n\Delta_c+K_c n^2+\Delta_R$, respectively.
The global two-quantum drive couples $\ket{0;g_R}$ and $\ket{2;g_R}$ without changing the local TLS state, whereas the local JC interaction couples $\ket{2;e_R}$ and $\ket{3;g_R}$.

We introduce $\delta_c$ as the deviation of $\Delta_c$ from the resonance condition for the global $\ket{0}\leftrightarrow\ket{2}$ transition driven by the coherent two-quantum drive, and $\delta_{L,\mathrm{JC}}$ as the signed energy mismatch of the selected local JC transition $\ket{2;e_R}\leftrightarrow\ket{3;g_R}$.
These quantities are defined by
\begin{align}
2\delta_c
&=E_{2;g_R}-E_{0;g_R}
=2\Delta_c+4K_c,\\
\delta_{L,\mathrm{JC}}
&=E_{3;g_R}-E_{2;e_R}
=\Delta_c+5K_c-\Delta_R.
\end{align}
Solving these relations gives the detunings used at each scan point,
\begin{align}
\Delta_c&=-2K_c+\delta_c,\\
\Delta_R&=3K_c+\delta_c-\delta_{L,\mathrm{JC}}.
\label{eq:supp_figS1_detunings}
\end{align}
Thus, $\delta_c=0$ corresponds to the resonant global two-quantum drive, while $\delta_{L,\mathrm{JC}}=0$ corresponds to the resonant local JC transition.
When $K_c$ or $\delta_c$ is varied with $\delta_{L,\mathrm{JC}}=0$, the local-TLS detuning $\Delta_R$ is adjusted accordingly to maintain the selected local resonance.

The parameter scans in Supplementary Fig.~\ref{figS1}a--d examine how the stationary target population $P_3^{\mathrm{ss}}$ depends on the Kerr nonlinearity, the global and local coupling strengths, and the residual detuning parameters $\delta_c$ and $\delta_{L,\mathrm{JC}}$.

Supplementary Fig.~\ref{figS1}a shows that increasing $K_c$ opens a broad region of enhanced $P_3^{\mathrm{ss}}$ over a finite range of global-drive strengths.
This reflects the role of Kerr nonlinearity in separating the selected transitions from other Fock-state transitions.
Weak global driving provides insufficient population replenishment to sustain a high target-state population, whereas excessively strong driving can also reduce the target-state population as off-resonant excitation becomes more appreciable.
Thus, high target populations require an appropriate global-drive strength and sufficiently resolved Fock-state transitions.

In Supplementary Fig.~\ref{figS1}b, $P_3^{\mathrm{ss}}$ exhibits a nonmonotonic dependence on the global and local coupling strengths.
Weak local JC coupling results in inefficient population transfer to the target state, whereas excessively strong local coupling can also reduce $P_3^{\mathrm{ss}}$ at a given global-drive strength.
The resulting region of high $P_3^{\mathrm{ss}}$ indicates that effective stabilization requires an appropriate combination of the two coupling strengths rather than simply maximizing either one.

The detuning scans further characterize the operating region.
In Supplementary Fig.~\ref{figS1}c, the local JC transition remains on resonance, so varying $\delta_c$ probes the response to the global two-quantum energy mismatch $2\delta_c$.
For sufficiently resolved Fock-state transitions, a finite high-population window persists around $\delta_c=0$, whereas increasing $|\delta_c|$ weakens population replenishment through the global connector.
When both $\delta_c$ and $\delta_{L,\mathrm{JC}}$ are varied [Supplementary Fig.~\ref{figS1}d], the region of high $P_3^{\mathrm{ss}}$ forms a tilted band near the point where both selected transitions are resonant, indicating correlated sensitivity to the two detunings.
The finite width of this high-population band indicates a degree of tolerance to deviations from exact resonance, with the achievable target population depending on the combination of $\delta_c$ and $\delta_{L,\mathrm{JC}}$.

The Kerr-resolved configuration combining a coherent two-photon drive and an incoherently pumped single-quantum JC interaction has previously been investigated for single-photon Fock-state stabilization ($R=1$), with an additional two-photon-loss channel~\cite{Liu2026WignerNegativity}.
The present analysis extends the investigation of this Kerr-resolved stabilization scheme from the previously studied single-photon case ($R=1$) to the higher target Fock state $\ket{3}$ ($R=3$).

\section{Parameter dependence of dispersively resolved Fock-state stabilization}
\label{sec:supp_figS2_parameter_dependence}

\begin{figure}[t]
\centering
\includegraphics[width=1.0\linewidth]{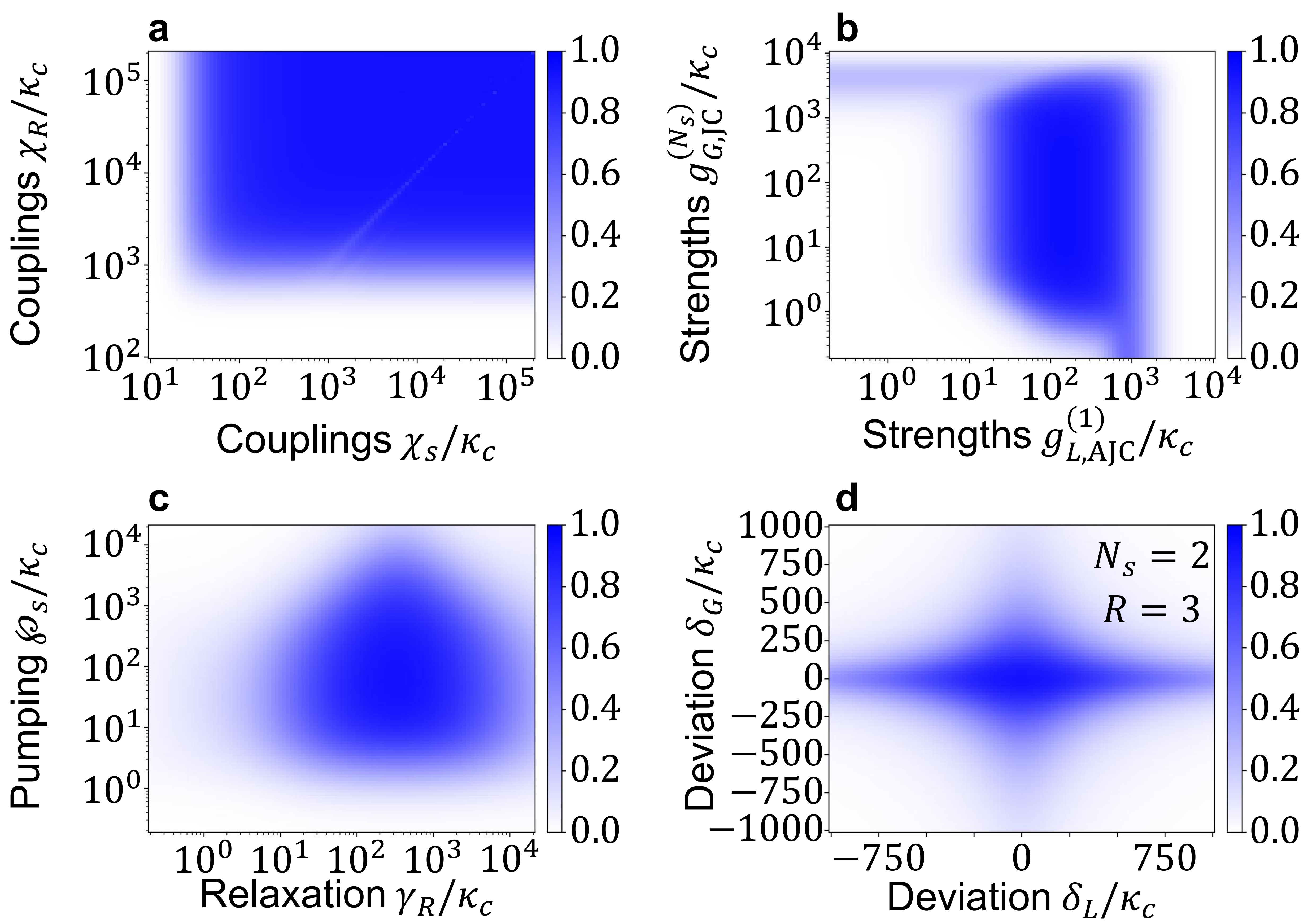}
\caption{\textbf{Parameter dependence of dispersively resolved Fock-state stabilization.}
Full microscopic Lindblad steady-state target populations $P_3^{\mathrm{ss}}$ for $N_s=2$ and $R=3$, using a pumped two-quantum JC global connector and a relaxation-assisted AJC local connector, with no Kerr nonlinearity ($K_c=0$).
\textbf{a,} Dependence on the global and local dispersive couplings $\chi_s$ and $\chi_R$.
\textbf{b,} Dependence on the local AJC coupling $g_{L,\mathrm{AJC}}^{(1)}$ and global JC coupling $g_{G,\mathrm{JC}}^{(N_s)}$.
\textbf{c,} Dependence on the local-TLS relaxation rate $\gamma_R$ and global-TLS pumping rate $\wp_s$.
\textbf{d,} Dependence on the selected-transition energy mismatches $\delta_L$ and $\delta_G$.
In \textbf{a--c}, $\delta_G=\delta_L=0$, and $\Delta_c=0$ in all panels.
Unless scanned, $\chi_s/\kappa_c=3\times10^4$, $\chi_R/\kappa_c=1.2\times10^4$, $g_{G,\mathrm{JC}}^{(N_s)}/\kappa_c=5$, $g_{L,\mathrm{AJC}}^{(1)}/\kappa_c=100$, $\wp_s/\kappa_c=100$, $\gamma_s/\kappa_c=1$, and $\gamma_R/\kappa_c=300$, with $\wp_R=0$.
}
\label{figS2}
\end{figure}

Supplementary Fig.~\ref{figS2} examines dispersively resolved Fock-state stabilization for $N_s=2$ and $R=3$, with no intrinsic Kerr nonlinearity ($K_c=0$).
The global connector is implemented by a pumped two-quantum JC interaction addressing $\ket{0}\leftrightarrow\ket{2}$, whereas the movable local connector employs a single-quantum AJC interaction assisted by relaxation of the local TLS, addressing $\ket{2}\leftrightarrow\ket{3}$.
For this AJC+relaxation realization, the local-TLS pumping rate is set to $\wp_R=0$.
Substituting these realizations into Eq.~\eqref{eq:supp_general_master} gives the microscopic master equation
\begin{align}
\frac{d\hat\rho}{dt}
=&-i[\hat H_{\mathrm{S2}},\hat\rho]
+\kappa_c\mathcal D[\hat a_c]\hat\rho
+\wp_s\mathcal D[\hat\sigma_s^+]\hat\rho
\nonumber\\
&+\gamma_s\mathcal D[\hat\sigma_s^-]\hat\rho
+\gamma_R\mathcal D[\hat\sigma_R^-]\hat\rho,
\label{eq:supp_figS2_master}
\end{align}
where
\begin{align}
\hat H_{\mathrm{S2}}
=&\Delta_c\hat n_c
+\chi_s\hat n_c\hat\sigma_s^z
+\chi_R\hat n_c\hat\sigma_R^z
%\nonumber\\&
+\Delta_s\hat\sigma_s^+\hat\sigma_s^-
+\Delta_R\hat\sigma_R^+\hat\sigma_R^-
\nonumber\\
&+g_{G,\mathrm{JC}}^{(2)}
\bigl[(\hat a_c^\dagger)^2\hat\sigma_s^-
+\hat a_c^2\hat\sigma_s^+\bigr]
\nonumber\\&
+g_{L,\mathrm{AJC}}^{(1)}
\bigl(\hat a_c^\dagger\hat\sigma_R^+
+\hat a_c\hat\sigma_R^-\bigr).
\label{eq:supp_figS2_hamiltonian}
\end{align}

The detunings are set by the selected transitions in the specified spectator-TLS subspaces.
For this two-TLS configuration, we write $\ket{n;q_s,q_R}\equiv\ket{n}\otimes\ket{q_s}\otimes\ket{q_R}$, where $q_s\in\{g_s,e_s\}$ and $q_R\in\{g_R,e_R\}$.
The global JC interaction couples $\ket{0;e_s,g_R}\leftrightarrow\ket{2;g_s,g_R}$ with the local TLS in $\ket{g_R}$, while the local AJC interaction couples $\ket{2;e_s,g_R}\leftrightarrow\ket{3;e_s,e_R}$ with the global TLS in $\ket{e_s}$.
The bare-state energies are defined as $E_{n;q_s,q_R}=\bra{n;q_s,q_R}\hat H_{\mathrm{S2}}\ket{n;q_s,q_R}$.
We define $\delta_G$ and $\delta_L$ as the signed energy mismatches of the selected global JC and local AJC transitions, respectively
\begin{align}
\delta_G
&=E_{2;g_s,g_R}-E_{0;e_s,g_R}
\nonumber\\&
=2\Delta_c-2(\chi_s+\chi_R)-\Delta_s,\\
\delta_L
&=E_{3;e_s,e_R}-E_{2;e_s,g_R}
\nonumber\\
&=\Delta_c+\chi_s+5\chi_R+\Delta_R.
\end{align}
In Supplementary Sec.~\ref{sec:supp_figS1_parameter_dependence}, $\delta_c$ denotes the cavity-detuning deviation, so that the total two-quantum transition mismatch is $2\delta_c$, whereas $\delta_G$ here denotes the full global-transition energy mismatch.
With $\Delta_c=0$, the detunings used at each scan point are
\begin{align}
\Delta_s&=-2(\chi_s+\chi_R)-\delta_G,\\
\Delta_R&=-\chi_s-5\chi_R+\delta_L.
\label{eq:supp_figS2_detunings}
\end{align}
Thus, $\delta_G=\delta_L=0$ places both selected transitions on resonance in their respective spectator-TLS subspaces.
In Supplementary Fig.~\ref{figS2}a--c, the selected-transition energy mismatches $\delta_G$ and $\delta_L$ are both kept at zero, with $\Delta_s$ and $\Delta_R$ adjusted as needed when the dispersive couplings are varied.

The parameter scans in Supplementary Fig.~\ref{figS2}a--d examine how $P_3^{\mathrm{ss}}$ depends on the dispersive couplings, global and local interaction strengths, auxiliary-TLS dissipative rates, and the selected-transition energy mismatches $\delta_G$ and $\delta_L$.

Supplementary Fig.~\ref{figS2}a shows that high target-state populations occur over a broad region when both dispersive couplings provide sufficient spectral resolution and unwanted transitions remain off resonance.
Weakening either $\chi_s$ or $\chi_R$ reduces the target-state population, consistent with diminished spectral selectivity.
The relative values of $\chi_s$ and $\chi_R$ also matter, as indicated by a faint diagonal region of reduced target-state population near $\chi_s=\chi_R$ in Supplementary Fig.~\ref{figS2}a.
This feature is associated with an additional unwanted global two-quantum JC transition, $\ket{2;e_s,e_R}\leftrightarrow\ket{4;g_s,e_R}$, which becomes resonant in the excited local-TLS subspace when $\chi_s=\chi_R$, thereby weakening spectral selectivity.
The high-population region demonstrates that selective stabilization can be achieved through dispersive spectral selection even when $K_c=0$.

In Supplementary Fig.~\ref{figS2}b, $P_3^{\mathrm{ss}}$ exhibits a nonmonotonic dependence on the global JC and local AJC coupling strengths.
Weak global JC or local AJC coupling is insufficient to maintain a high target-state population, whereas excessively strong coupling can also reduce $P_3^{\mathrm{ss}}$.
As in the Kerr-resolved case, the high-population region therefore requires suitable global and local coupling strengths rather than simply maximizing either one.

Supplementary Fig.~\ref{figS2}c examines the global-TLS pumping rate $\wp_s$ and local-TLS relaxation rate $\gamma_R$.
Global-TLS pumping supports population replenishment through the two-quantum JC connector, whereas local-TLS relaxation resets the auxiliary TLS following the AJC-mediated transition from $\ket{2}$ to $\ket{3}$, thereby favoring net population transfer toward the target state.
High $P_3^{\mathrm{ss}}$ occurs over finite ranges of both rates; very weak or excessively strong pumping or relaxation reduces the target-state population.
The pumping and relaxation rates used in the main-text simulations are chosen to favor directional population transport in the corresponding JC- and AJC-assisted implementations.

Finally, Supplementary Fig.~\ref{figS2}d examines the dependence of $P_3^{\mathrm{ss}}$ on the selected-transition energy mismatches $\delta_G$ and $\delta_L$.
The high-population region is centered near $\delta_G=\delta_L=0$ and is more extended along the $\delta_L$ direction than along the $\delta_G$ direction. 
This indicates that, for the parameters considered, the stabilization is more tolerant to local-transition detuning than to global-transition detuning.

\section{Symmetric global population transport under balanced pumping and relaxation}
\label{sec:supp_figS3_balanced_transport}

\begin{figure}[t]
\centering
\includegraphics[width=1.0\linewidth]{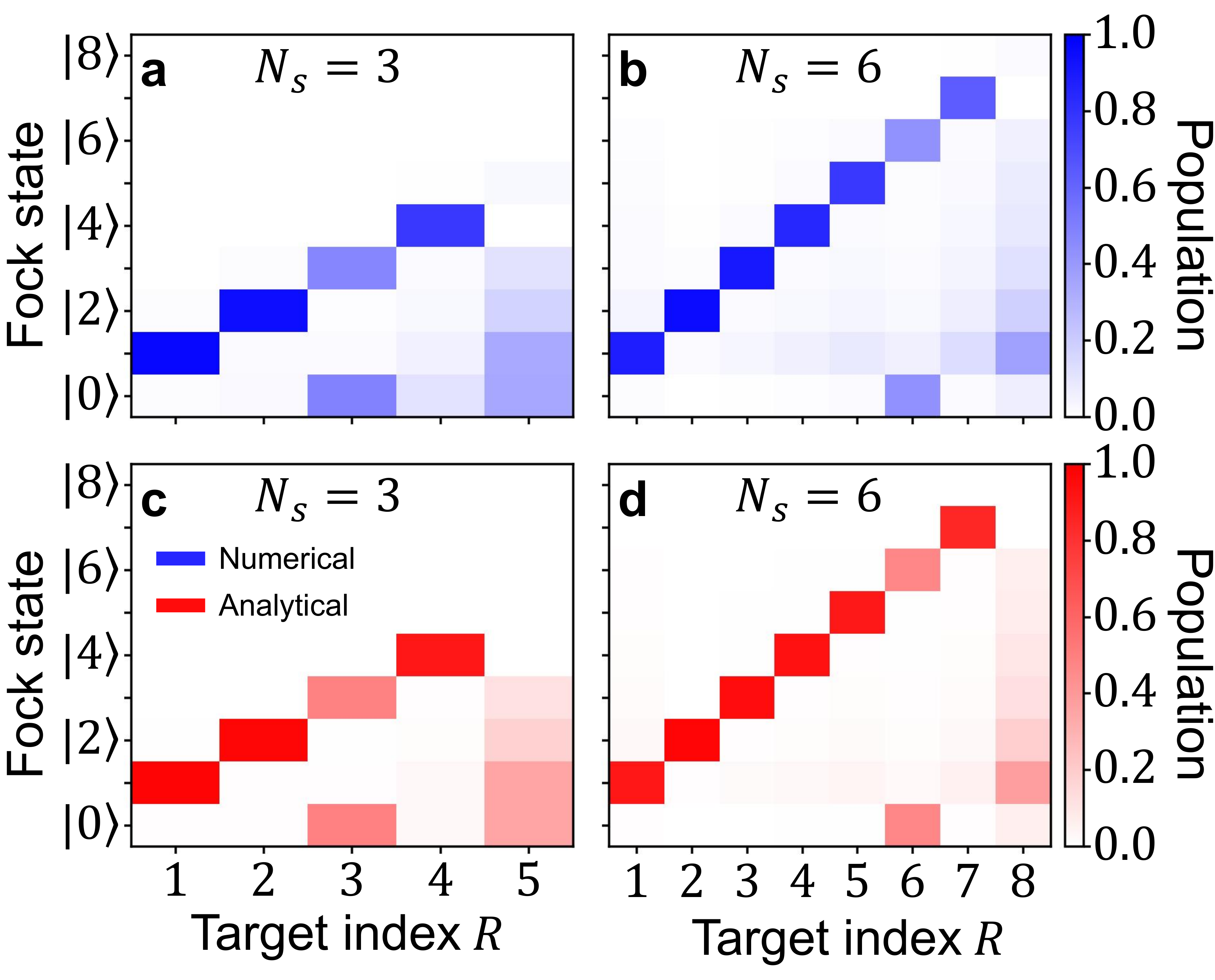}
\caption{\textbf{Symmetric global population transport with balanced pumping and relaxation.}
Steady-state Fock-state populations $P_n^{\mathrm{ss}}$ for a Kerr-resolved system with a pumped $N_s$-quantum JC global connector under balanced TLS pumping and relaxation ($\wp_s=\gamma_s$), combined with a pumped JC local connector.
\textbf{a,b,} Full microscopic Lindblad results for $N_s=3$ and $N_s=6$, respectively.
\textbf{c,d,} Corresponding analytical populations obtained from the closed-form stationary solution [Eq.~\eqref{eq:supp_stationary_population_closed_form}].
Parameters are $K_c/\kappa_c=10^4$, $g_{G,\mathrm{JC}}^{(N_s)}/\kappa_c=5$, $g_{L,\mathrm{JC}}^{(1)}/\kappa_c=100$, $(\wp_s,\gamma_s)/\kappa_c=(100,100)$, and $(\wp_R,\gamma_R)/\kappa_c=(300,1)$.
The detunings are set to $\Delta_c/\kappa_c=-10^4N_s$, $\Delta_s=0$, and $\Delta_R/\kappa_c=10^4(2R-1-N_s)$, ensuring resonance of the selected global and local transitions.
}
\label{figS3}
\end{figure}

Supplementary Fig.~\ref{figS3} considers Kerr-resolved Fock-state stabilization with a pumped $N_s$-quantum JC global connector and a pumped single-quantum JC local connector, taking equal pumping and relaxation rates for the global auxiliary TLS, $\wp_s=\gamma_s$.
With $\wp_s=\gamma_s$, the pumped $N_s$-quantum JC global connector has equal effective upward and downward population-transfer rates, as does a coherent $N_s$-quantum drive.
This explains why the stationary Fock-state population distributions in Supplementary Fig.~\ref{figS3} closely resemble those obtained with coherent global driving and pumped local JC coupling in the upper row of Fig.~3c of the main text.

According to Eqs.~\eqref{eq:supp_JC_general_rates} and \eqref{eq:supp_auxiliary_stationary_populations}, the effective upward and downward population-transfer rates of the pumped global JC connector are
\begin{align}
\mathcal R_{\mathrm{G,JC}}^{\uparrow}
&=
W_{\mathrm{JC}}^{(0,N_s)}
\frac{\wp_s}{\wp_s+\gamma_s},
\\
\mathcal R_{\mathrm{G,JC}}^{\downarrow}
&=
W_{\mathrm{JC}}^{(0,N_s)}
\frac{\gamma_s}{\wp_s+\gamma_s}.
\label{eq:supp_figS3_JC_rates}
\end{align}
When $\wp_s=\gamma_s$, the stationary ground- and excited-state populations of the isolated global TLS become equal, yielding
\begin{equation}
\mathcal R_{\mathrm{G,JC}}^{\uparrow}
=
\mathcal R_{\mathrm{G,JC}}^{\downarrow}
=
\frac{1}{2}W_{\mathrm{JC}}^{(0,N_s)}.
\label{eq:supp_figS3_balanced_rates}
\end{equation}
Thus, balanced global-TLS pumping and relaxation eliminate the directional bias of the effective global connector.
This is analogous to the coherent $N_s$-quantum drive, for which Eq.~\eqref{eq:supp_coherent_symmetric_rates} gives
\begin{equation}
\mathcal R_{\mathrm{G,drv}}^{\uparrow}
=\mathcal R_{\mathrm{G,drv}}^{\downarrow}
=W_{\mathrm{drv}}^{(0,N_s)}.
\end{equation}
Both realizations therefore generate symmetric effective population transfer between $\ket{0}$ and $\ket{N_s}$.
Consequently, both realizations enter the effective population equation through the same symmetric global population-transfer structure, although their effective exchange rates may differ in magnitude.

Supplementary Figs.~\ref{figS3}a,b show the full microscopic Lindblad steady-state populations for $N_s=3$ and $N_s=6$, respectively, while Supplementary Figs.~\ref{figS3}c,d present the corresponding analytical populations obtained from the closed-form stationary solution in Eq.~\eqref{eq:supp_stationary_population_closed_form}.
The agreement between the numerical and analytical populations confirms that the effective-current description captures the stationary population distributions for the parameters considered.
For $R=N_s+2$, the local connector lies beyond the global loss cascade, and the analytical effective stationary population distribution reduces to the global-only solution.
These results demonstrate that similar stationary Fock-state population distributions can arise from distinct microscopic global-connector realizations through their common effective population-transfer structure.

\section{Fock-selective population depletion by reversed local transport}
\label{sec:supp_figS4_depletion}

\begin{figure}[t]
\centering
\includegraphics[width=1.0\linewidth]{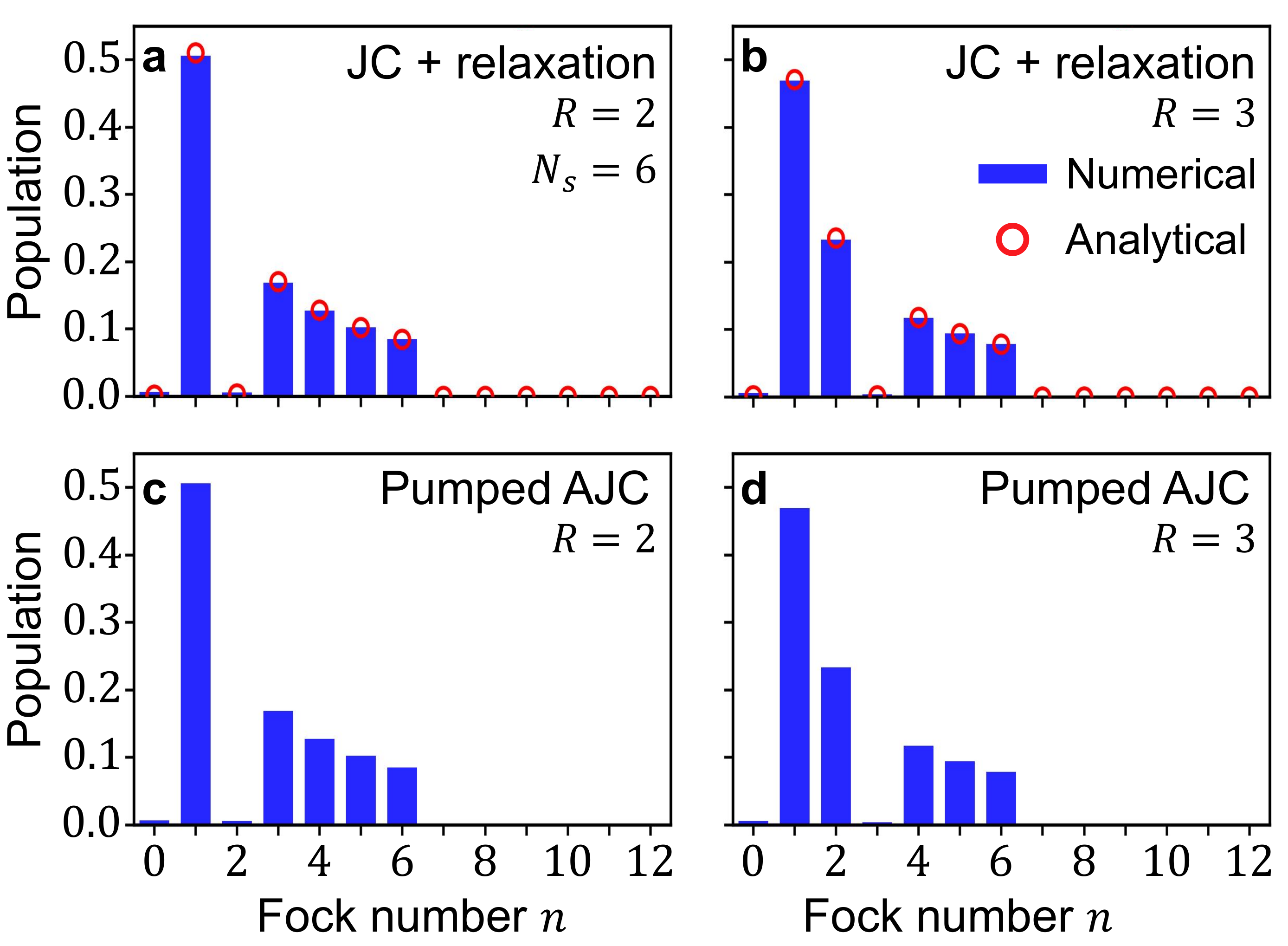}
\caption{\textbf{Fock-selective population depletion by reversed local transport.}
Steady-state Fock-state populations $P_n^{\mathrm{ss}}$ for a Kerr-resolved system with a pumped $N_s$-quantum JC global connector ($N_s=6$).
\textbf{a,b,} Local JC with relaxation-dominated TLS dissipation at $R=2$ and $R=3$, respectively. 
Blue bars show full microscopic Lindblad results, while red open circles represent analytical populations obtained from Eq.~\eqref{eq:supp_stationary_population_closed_form}.
\textbf{c,d,} Corresponding local AJC with pumping-dominated TLS dissipation at $R=2$ and $R=3$, respectively; only microscopic Lindblad results are displayed.
The parameters are $K_c/\kappa_c=10^4$, $g_{G,\mathrm{JC}}^{(6)}/\kappa_c=5$, $g_{L,\mathrm{JC}}^{(1)}/\kappa_c=g_{L,\mathrm{AJC}}^{(1)}/\kappa_c=100$, and $(\wp_s,\gamma_s)/\kappa_c=(100,1)$.
For local JC, $(\wp_R^{\mathrm{JC}},\gamma_R^{\mathrm{JC}})/\kappa_c=(1,300)$; for local AJC, $(\wp_R^{\mathrm{AJC}},\gamma_R^{\mathrm{AJC}})/\kappa_c=(300,1)$.
The detunings are set to $\Delta_c/\kappa_c=-6\times10^4$, $\Delta_s=0$, $\Delta_R^{\mathrm{JC}}/\kappa_c=10^4(2R-7)$, and $\Delta_R^{\mathrm{AJC}}=-\Delta_R^{\mathrm{JC}}$, ensuring that the selected global and local transitions are on resonance.
}
\label{figS4}
\end{figure}

Supplementary Fig.~\ref{figS4} examines Fock-selective population depletion in the Kerr-resolved model.
The global connector remains a pumped $N_s$-quantum JC interaction with $N_s=6$, while the local connector addresses $\ket{R-1}\leftrightarrow\ket R$ for $R=2$ and $R=3$.
In contrast to local capture, the local connector is configured to favor population transfer from $\ket R$ to $\ket{R-1}$, thereby reducing the stationary population $P_R^{\mathrm{ss}}$.

In Supplementary Figs.~\ref{figS4}a,b, the local connector is a single-quantum JC interaction with relaxation-dominated TLS dissipation.
The JC interaction couples $\ket{R;g_R}\leftrightarrow\ket{R-1;e_R}$, while subsequent TLS relaxation transfers population from $\ket{R-1;e_R}$ to $\ket{R-1;g_R}$, favoring net downward bosonic transport.
According to Eq.~\eqref{eq:supp_local_JC_directionality}, the ratio of the upward and downward effective local rates is $\mathcal R_{\mathrm{L,JC}}^{\uparrow}(R)/\mathcal R_{\mathrm{L,JC}}^{\downarrow}(R)=\wp_R/\gamma_R$.
For the relaxation-dominated parameters considered here, Eq.~\eqref{eq:supp_stationary_current_ratio_general} gives $\zeta_R<0$ at both $R=2$ and $R=3$, corresponding to a negative stationary local current $J_{\mathrm L}^{\mathrm{ss}}(R)<0$.
The numerical results consequently exhibit pronounced population depletion at $n=R$, consistent with the analytical predictions obtained from the closed-form stationary solution in Eq.~\eqref{eq:supp_stationary_population_closed_form}.

Supplementary Figs.~\ref{figS4}c,d demonstrate that the same downward population transport can also be realized using a single-quantum AJC interaction with pumping-dominated TLS dissipation.
In this case, the AJC interaction couples $\ket{R;e_R}\leftrightarrow\ket{R-1;g_R}$, while incoherent pumping transfers the local TLS from $\ket{R-1;g_R}$ to $\ket{R-1;e_R}$.
Thus, TLS pumping resets the auxiliary system after the AJC-assisted downward transition, allowing population to be repeatedly transferred from $\ket R$ to $\ket{R-1}$.
This reverses the effective population-transfer direction of the AJC+relaxation realization considered earlier, where the active TLS has no incoherent pumping.

The analytical correspondence between the two local realizations can be established by interchanging the ground and excited states of the local TLS.
Specifically, we introduce the transformation operator $\hat U_R=\hat\sigma_R^++\hat\sigma_R^-$, which acts on the local TLS states as $\hat U_R\ket{g_R}=\ket{e_R}$ and $\hat U_R\ket{e_R}=\ket{g_R}$.
The corresponding transformations of the TLS operators are
\begin{align}
\hat U_R\hat\sigma_R^-\hat U_R^\dagger
&=\hat\sigma_R^+,
\\
\hat U_R\hat\sigma_R^+\hat U_R^\dagger
&=\hat\sigma_R^-,
\\
\hat U_R(\hat\sigma_R^+\hat\sigma_R^-)\hat U_R^\dagger
&=\hat I_R-\hat\sigma_R^+\hat\sigma_R^-,
\end{align}
where $\hat I_R$ is the identity operator of the local TLS.

The first two relations transform the local JC interaction $\hat a_c^\dagger\hat\sigma_R^-+\hat a_c\hat\sigma_R^+$ into the AJC interaction $\hat a_c^\dagger\hat\sigma_R^++\hat a_c\hat\sigma_R^-$.
Correspondingly, the JC transition $\ket{R;g_R}\leftrightarrow\ket{R-1;e_R}$ followed by TLS relaxation $\ket{R-1;e_R}\to\ket{R-1;g_R}$ is mapped onto the AJC transition $\ket{R;e_R}\leftrightarrow\ket{R-1;g_R}$ followed by TLS pumping $\ket{R-1;g_R}\to\ket{R-1;e_R}$.
The pumping and relaxation dissipators are therefore interchanged, requiring $\wp_R^{\mathrm{AJC}}=\gamma_R^{\mathrm{JC}}$ and $\gamma_R^{\mathrm{AJC}}=\wp_R^{\mathrm{JC}}$, where $\wp_R^{\mathrm{JC}}$ and $\gamma_R^{\mathrm{JC}}$ denote the local-TLS incoherent pumping and relaxation rates in the JC realization, respectively, while $\wp_R^{\mathrm{AJC}}$ and $\gamma_R^{\mathrm{AJC}}$ denote the corresponding rates in the AJC realization.
Denoting the local-TLS detunings in the JC and AJC realizations by $\Delta_R^{\mathrm{JC}}$ and $\Delta_R^{\mathrm{AJC}}$, respectively, the third relation transforms the local-TLS Hamiltonian term $\Delta_R^{\mathrm{JC}}\hat\sigma_R^+\hat\sigma_R^-$ into $-\Delta_R^{\mathrm{JC}}\hat\sigma_R^+\hat\sigma_R^-$ up to a constant energy shift, giving the following relation $\Delta_R^{\mathrm{AJC}}=-\Delta_R^{\mathrm{JC}}$.

For equal local coupling strengths and the parameter relations established above, the correspondence can be demonstrated directly at the level of the full microscopic master equation.
Let $\hat\rho_{\mathrm{JC}}(t)$ denote the density operator for the local JC realization and define the transformed density operator
\begin{equation}
\hat\rho_{\mathrm{AJC}}(t)
=\hat U_R\hat\rho_{\mathrm{JC}}(t)\hat U_R^\dagger,
\label{eq:supp_figS4_density_map}
\end{equation}
where $\hat U_R$ acts only on the local TLS, with identity operators on the bosonic mode and the global TLS understood.
The transformation maps the local JC interaction to the AJC interaction, while the transformed local-TLS Hamiltonian differs from its AJC counterpart only by a constant energy shift, which does not affect the commutator.
Moreover, since $\hat U_R$ is unitary ($\hat U_R\hat U_R^\dagger=\hat I_R$), the Lindblad dissipator satisfies
\begin{equation}
\mathcal D[\hat U_R\hat O\hat U_R^\dagger]
\bigl(\hat U_R\hat\rho\hat U_R^\dagger\bigr)
=
\hat U_R\mathcal D[\hat O]\hat\rho\hat U_R^\dagger.
\label{eq:supp_figS4_dissipator_map}
\end{equation}
For $\hat O=\hat\sigma_R^\pm$, this identity exchanges the pumping and relaxation terms according to the parameter relations specified above.
The local Liouvillian therefore satisfies
\begin{align}
\hat U_R\bigl[\mathcal L_L^{\mathrm{JC}}(R)
\hat\rho\bigr]\hat U_R^\dagger
&=\mathcal L_L^{\mathrm{AJC}}(R)
\bigl(\hat U_R\hat\rho\hat U_R^\dagger\bigr),
\label{eq:supp_figS4_local_map}
\end{align}
where $\mathcal L_L^{\mathrm{JC}}(R)$ and $\mathcal L_L^{\mathrm{AJC}}(R)$ denote the local Liouvillians for the JC and AJC realizations, respectively.
Since $\hat U_R$ acts only on the local TLS, it leaves the Kerr Hamiltonian, global connector, and intrinsic bosonic loss unchanged, so their corresponding Liouvillian terms retain the same form under this transformation.
Consequently, the local-JC master equation and its transformed counterpart obey
\begin{align}
\frac{d\hat\rho_{\mathrm{JC}}}{dt}
&=\left[\mathcal L_c+\mathcal L_G
+\mathcal L_L^{\mathrm{JC}}(R)\right]
\hat\rho_{\mathrm{JC}},\\
\frac{d\hat\rho_{\mathrm{AJC}}}{dt}
&=\hat U_R
\frac{d\hat\rho_{\mathrm{JC}}}{dt}
\hat U_R^\dagger
\nonumber\\
&=\left[\mathcal L_c+\mathcal L_G
+\mathcal L_L^{\mathrm{AJC}}(R)\right]
\hat\rho_{\mathrm{AJC}}.
\label{eq:supp_figS4_master_map}
\end{align}
Thus, the transformed density operator in Eq.~\eqref{eq:supp_figS4_density_map} is a solution of the pumped AJC master equation whenever $\hat\rho_{\mathrm{JC}}(t)$ solves the JC master equation, provided that their initial density operators are related by the same transformation.
Furthermore, since $\hat U_R$ acts only on the local TLS, the partial trace over the auxiliary TLSs is invariant under this unitary transformation. 
Using Eq.~\eqref{eq:supp_figS4_density_map}, we obtain
\begin{align}
\operatorname{Tr}_{\mathrm{TLS}}
\hat\rho_{\mathrm{AJC}}(t)
&=
\operatorname{Tr}_{\mathrm{TLS}}
\left[
\hat U_R\hat\rho_{\mathrm{JC}}(t)\hat U_R^\dagger
\right]
\nonumber\\
&=
\operatorname{Tr}_{\mathrm{TLS}}
\left[
\hat\rho_{\mathrm{JC}}(t)\hat U_R^\dagger\hat U_R
\right]
\nonumber\\
&=
\operatorname{Tr}_{\mathrm{TLS}}
\hat\rho_{\mathrm{JC}}(t).
\end{align}
We define the steady-state Fock-state population of the bosonic mode for each local-connector realization as
\begin{equation}
P_{n,\lambda}^{\mathrm{ss}}
=
\langle n|\operatorname{Tr}_{\mathrm{TLS}}
\hat\rho_{\lambda}^{\mathrm{ss}}|n\rangle,
\end{equation}
where $\lambda\in\{\mathrm{JC},\mathrm{AJC}\}$ labels the local-connector realization, and $\hat\rho_{\lambda}^{\mathrm{ss}}$ denotes the corresponding steady-state density operator of the full system.
The equality of the reduced density matrices then implies
\begin{equation}
P_{n,\mathrm{AJC}}^{\mathrm{ss}}
=
P_{n,\mathrm{JC}}^{\mathrm{ss}}.
\label{eq:supp_figS4_population_map}
\end{equation}
Therefore, the local JC and pumped AJC realizations produce identical stationary bosonic populations under the specified parameter transformation, explaining the agreement between Supplementary Figs.~\ref{figS4}a,c and Supplementary Figs.~\ref{figS4}b,d.

The unitary correspondence established above also relates the effective local transition rates of the two realizations. 
Using the JC results in Eqs.~\eqref{eq:supp_local_JC_up} and \eqref{eq:supp_local_JC_down}, the effective upward and downward rates can be written explicitly as
\begin{align}
\mathcal R_{\mathrm{L,JC}}^{\uparrow}(R)
&=
\frac{4R|g_{L,\mathrm{JC}}^{(1)}|^2}
{\wp_R^{\mathrm{JC}}+\gamma_R^{\mathrm{JC}}+(2R-1)\kappa_c}
\frac{\wp_R^{\mathrm{JC}}}
{\wp_R^{\mathrm{JC}}+\gamma_R^{\mathrm{JC}}},
\\
\mathcal R_{\mathrm{L,JC}}^{\downarrow}(R)
&=
\frac{4R|g_{L,\mathrm{JC}}^{(1)}|^2}
{\wp_R^{\mathrm{JC}}+\gamma_R^{\mathrm{JC}}+(2R-1)\kappa_c}
\frac{\gamma_R^{\mathrm{JC}}}
{\wp_R^{\mathrm{JC}}+\gamma_R^{\mathrm{JC}}}.
\end{align}
Substituting the parameter relations $\wp_R^{\mathrm{JC}}=\gamma_R^{\mathrm{AJC}}$ and $\gamma_R^{\mathrm{JC}}=\wp_R^{\mathrm{AJC}}$, together with equal local coupling strengths $|g_{L,\mathrm{JC}}^{(1)}|=|g_{L,\mathrm{AJC}}^{(1)}|$, we obtain the effective rates for the pumped AJC realization at the corresponding local resonance,
\begin{align}
\mathcal R_{\mathrm{L,AJC}}^{\uparrow}(R)
&=
W_{\mathrm{L,AJC}}^{\mathrm{pump}}(R)
\frac{\gamma_R^{\mathrm{AJC}}}
{\wp_R^{\mathrm{AJC}}+\gamma_R^{\mathrm{AJC}}},
\\
\mathcal R_{\mathrm{L,AJC}}^{\downarrow}(R)
&=
W_{\mathrm{L,AJC}}^{\mathrm{pump}}(R)
\frac{\wp_R^{\mathrm{AJC}}}
{\wp_R^{\mathrm{AJC}}+\gamma_R^{\mathrm{AJC}}},
\label{eq:supp_figS4_pumped_AJC_rates}
\end{align}
where the effective exchange-rate factor is
\begin{equation}
W_{\mathrm{L,AJC}}^{\mathrm{pump}}(R)
=
\frac{
4R|g_{L,\mathrm{AJC}}^{(1)}|^2
}{
\wp_R^{\mathrm{AJC}}+\gamma_R^{\mathrm{AJC}}
+(2R-1)\kappa_c
}.
\label{eq:supp_figS4_pumped_AJC_exchange}
\end{equation}
The superscript $\mathrm{pump}$ distinguishes this exchange-rate factor from that of the zero-pumping AJC+relaxation realization derived earlier.
In the pumping-dominated regime $\wp_R^{\mathrm{AJC}}\gg\gamma_R^{\mathrm{AJC}}$, the effective rates satisfy $\mathcal R_{\mathrm{L,AJC}}^{\downarrow}(R)\gg\mathcal R_{\mathrm{L,AJC}}^{\uparrow}(R)$, favoring downward population transfer from $\ket R$ to $\ket{R-1}$ and thereby enabling Fock-selective population depletion.

For the parameter pairs used in Supplementary Fig.~\ref{figS4}, the local JC and pumped AJC realizations satisfy the parameter mapping established above, with interchanged TLS pumping and relaxation rates, equal local coupling strengths, and opposite local-TLS detunings.
Consequently, their effective upward and downward transition rates coincide, yielding the same closed-form stationary Fock-state populations within the effective population model.
The exact unitary correspondence further guarantees identical reduced bosonic steady states in the full microscopic Lindblad description, consistent with the numerical results in Supplementary Figs.~\ref{figS4}a,c and \ref{figS4}b,d.
The analytical populations are therefore shown only in the JC panels, since the corresponding AJC results are identical.

\bibliographystyle{apsrev4-2-bookfix-fixed}
\bibliography{WF-sub}